\documentclass[fleqn,12pt]{book}
\pdfoutput=1
\usepackage{setspace}
\usepackage{tocbibind}
\usepackage{times}
\usepackage[T1]{fontenc}
\usepackage[font=footnotesize,labelfont=bf]{caption}
\usepackage{amssymb}
\usepackage{amsmath}
\usepackage{graphicx}
\usepackage{microtype}
\usepackage{booktabs}
\usepackage[english]{babel}
\usepackage[round, authoryear]{natbib} %previously also sort, numbers
\usepackage
[a4paper,pdftex,twoside,hmargin=21mm,vmargin=29.7mm,bindingoffset=7mm]
{geometry}
\usepackage[grey]{quotchap}
\usepackage{fancyhdr}
\usepackage[colorlinks =true,hypertexnames=false,
plainpages=false,
breaklinks = true,linktocpage,
linkcolor=black,urlcolor=black,citecolor=black,anchorcolor=black
,hyperfigures,pdfpagelabels
]{hyperref} 

\bibpunct{(}{)}{;}{a}{}{,}

\fancypagestyle{plain}{
  \fancyhf{}
}

\fancypagestyle{tocstyle}{
  \fancyhf{}
  \fancyhead[LE,RO]{\small{\thepage}}

}

\fancypagestyle{mainstyle}{
  \fancyhf{}
  
   \fancyhead[LE,RO]{\small{\thepage}}
  \fancyhead[RE]{\textit{\small{\nouppercase\leftmark}}}
  \fancyhead[LO]{\textit{\small{\thesection~\nouppercase\rightmark}}}
}

\makeatletter
\def\blfootnote{\xdef\@thefnmark{}\@footnotetext}
\makeatother

\newcommand{\mnras}{MNRAS}

\newcommand{\shorteq}[1]{
\begin{equation}
#1
\end{equation}}

\newcommand{\mean}[1]{{\langle #1 \rangle}}

\newcommand{\pdif}[2]{{\frac{\partial #1}{\partial #2}}}
\newcommand{\tdif}[2]{{\frac{\mathrm{d} #1}{\mathrm{d} #2}}}

\newcommand{\qdif}[2]{\partial #1/\partial #2}
\newcommand{\udif}[2]{\mathrm{d}#1/\mathrm{d}#2}

\newcommand{\stars}{\textsc{stars}}
\newcommand{\rose}{\textsc{rose}}
\newcommand{\uvp}{\texorpdfstring{$U$--$V$ plane}{U-V plane}}

\newcommand{\chem}[2]{{}^{#2}\text{#1}}

\newcommand{\mixl}{\ell_\text{m}}

\newcommand{\mchead}[1]{\multicolumn{2}{c}{#1}}
\newcommand{\sci}[2]{#1\times10^{#2}}
\newcommand{\st}[1]{_\text{#1}}

\newcommand{\Htu}{\text{H}_2}
\newcommand{\rb}{r\st{B}}
\newcommand{\rs}{r\st{S}}
\newcommand{\Lbh}{L\st{BH}}
\newcommand{\Mbh}{M\st{BH}}
\newcommand{\dMbh}{\dot{M}\st{BH}}

\newcommand{\bracfrac}[2]{\left(\frac{#1}{#2}\right)}
\newcommand{\sq}{\mkern-6mu }

\newcommand{\pack}[1]{\texttt{#1}}

\newcommand{\K}{\rm\thinspace K}
\newcommand{\Gpc}{\rm\thinspace Gpc}

\newcommand{\km}{\rm\thinspace km}

\newcommand{\cm}{\rm\thinspace cm}
\newcommand{\AU}{\rm\thinspace AU}

\newcommand{\cmsq}{\hbox{$\cm^2\,$}}

\newcommand{\pcmcu}{\hbox{$\cm^{-3}\,$}}

\newcommand{\Rsun}{\hbox{$\rm\thinspace \text{R}_{\odot}$}}
\newcommand{\yr}{\rm\thinspace yr}
\newcommand{\Myr}{\rm\thinspace Myr}

\newcommand{\s}{\rm\thinspace s}

\newcommand{\g}{\rm\thinspace g}
\newcommand{\gpcm}{\hbox{$\g\cm^{-3}\,$}}

\newcommand{\Msun}{\hbox{$\rm\thinspace \text{M}_{\odot}$}}

\newcommand{\Msunpyr}{\hbox{$\Msun\yr^{-1}\,$}}
\newcommand{\erg}{\rm\thinspace erg}

\newcommand{\ergps}{\hbox{$\erg\s^{-1}\,$}}

\newcommand{\dyn}{\rm\thinspace dyn}
\newcommand{\dynpcmsq}{\hbox{$\dyn\cm^{-2}\,$}}
\newcommand{\Lsun}{\hbox{$\rm\thinspace L_{\odot}$}}

\newcommand{\pg}{\hbox{$\g^{-1}\,$}}

\newcommand{\ps}{\hbox{$\s^{-1}\,$}}

\newcommand{\Title}{Quasi-stars and the Sch\"onberg--Chan\-dra\-sek\-har limit}
\newcommand{\Author}{Warrick Heinz Ball}
\newcommand{\Date}{May 29, 2012}

\hypersetup{
pdfauthor = {\Author{}},
pdftitle = {\Title{}},
}

\title{\Title{}}
\author{\Author{}}

\begin{document}

\pagestyle{empty}

%\setlength{\parskip}{\parskip plus 1pt plus 1pt plus 1pt minus 1pt minus 1pt minus 1pt}

%\flushbottom
%\baselineskip=1\baselineskip plus 1pt plus 1pt minus 1pt minus 1pt\relax
%\showthe\baselineskip
%\maketitle

\frontmatter
\pagenumbering{roman}
\begin{titlepage}
\null\vfill
\begin{flushright}
{ \Huge {{Quasi-stars and the\\Sch\"onberg--Chandrasekhar limit}} \par}
\null\vfill
\end{flushright}
% %{\large \vspace*{25mm} 
% %{{\includegraphics[bb = 0 0 292 336, width=30mm]{cam}} \par} 
\begin{minipage}{0.3\textwidth}
\begin{center}
{{\includegraphics[width=30mm]{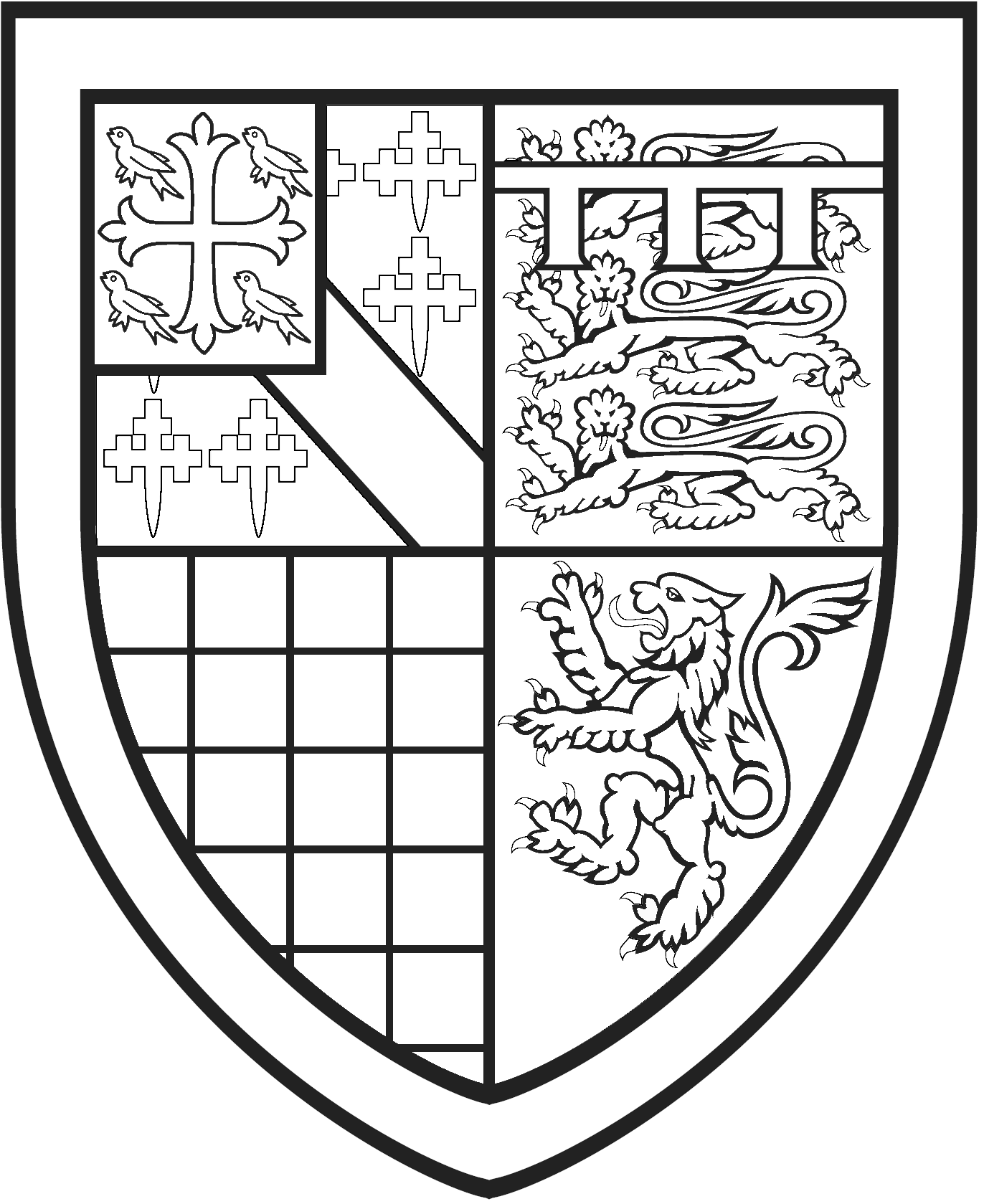}\vspace*{1ex}}
{\includegraphics[width=30mm]{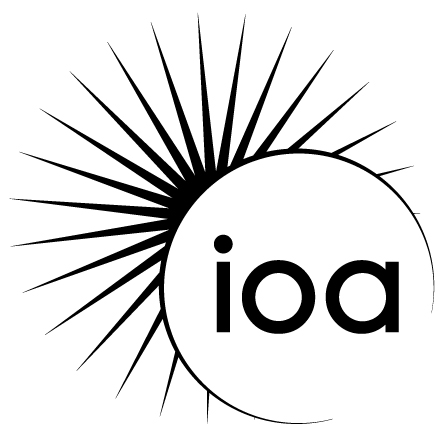}\vspace*{2ex}}
{\includegraphics[width=30mm]{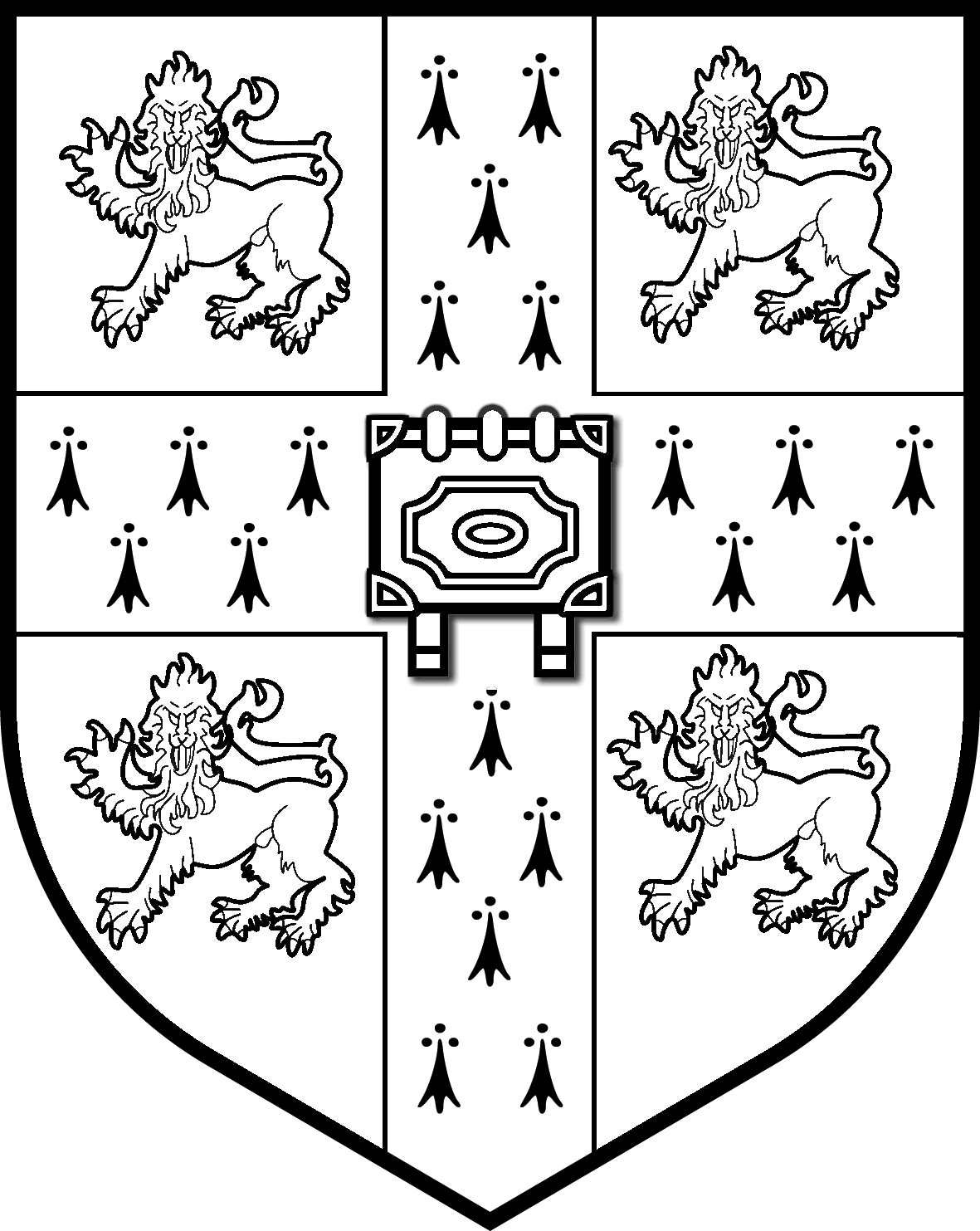}} \par}
\end{center}
\end{minipage}
% %\vspace*{20mm}}
\begin{minipage}{0.7\textwidth}
\null\vfill
\begin{flushright}
{{\large \Author{}} \par}{\normalsize \vspace*{1ex}
{{St Edmund's College} \par}\vspace*{1ex}
{{and} \par}\vspace*{1ex}
{{Institute of Astronomy} \par}\vspace*{1ex}
% {{University of Cambridge} \par}\vspace*{25mm}}
{{University of Cambridge}}}
\end{flushright}
\end{minipage}
\null\vfill
\begin{flushright}
{{{This dissertation is submitted for the degree of}} \par}\vspace*{1ex}
{\it {Doctor of Philosophy} \par}\vspace*{2ex}
% % {\Date{}}
% % {{This dissertation is submitted for the degree of}
% % {\it Doctor of Philosophy} \par}\vspace*{2ex}
% }
\end{flushright}

\null\vfill

\end{titlepage}

\chapter*{Declaration}
\begin{quote}

  I hereby declare that my dissertation entitled \emph{\Title{}} is
  not substantially the same as any that I have submitted for a degree
  or diploma or other qualification at any other University.

  I further state that no part of my dissertation has already been or
  is being concurrently submitted for any such degree, diploma or
  other qualification.

  This dissertation is the result of my own work and includes nothing
  which is the outcome of work done in collaboration except as
  specified in the text. Those parts of this dissertation that have
  been published or accepted for publication are as follows.

  \begin{itemize}
  \item Parts of Chapters 1 and 2 and most of Chapter 3 have been
    published as 

    Ball, W.~H., Tout, C.~A., {\.Z}ytkow, A.~N., \& Eldridge, J.~J.
    \\ 2011, \mnras, 414, 2751

  \item Material from Chapters 5 and 6 and Appendix A has been
    published as

    Ball, W.~H., Tout, C.~A., {\.Z}ytkow, A.~N.,
    \\ 2012, \mnras, 421, 2713
  \end{itemize}

  This dissertation contains fewer than 60\,000 words.

%  \vskip2cm
  \vspace{\fill}

  \Author{}

  Cambridge, \Date

  \vspace*{\fill}

\end{quote}
 % Declaration
\chapter*{Abstract}
\addcontentsline{toc}{chapter}{Abstract}
\begin{quotation}
The mechanism by which the supermassive black holes that power bright
quasars at high redshift form remains unknown.  One possibility is
that, if fragmentation is prevented, the monolithic collapse of a
massive protogalactic disc proceeds via a cascade of triaxial
instabilities and leads to the formation of a \emph{quasi-star}: a
growing black hole, initially of typical stellar-mass, embedded in a
hydrostatic giant-like envelope.  Quasi-stars are the main object of
study in this dissertation.  Their envelopes satisfy the equations of
stellar structure so the Cambridge \stars{} code is modified to model
them.  Analysis of the models leads to an extension of the classical
Sch\"onberg--Chandrasekhar limit and an exploration of the
implications of this extension for the evolution of main-sequence
stars into giants.

In Chapter \ref{cint}, I introduce the problem posed by the
supermassive black holes that power high-redshift quasars.  I discuss
potential solutions and describe the conditions under which a
quasi-star might form.  In Chapter \ref{cstars}, I outline the
Cambridge \stars{} code and the modifications that are made to model
quasi-star envelopes.

In Chapter \ref{cqs1}, I present models of quasi-stars where the base
of the envelope is located at the Bondi radius of the black hole.  The
black holes in these models are subject to a robust upper fractional
mass limit of about one tenth.  In addition, the final black hole mass
is sensitive to the choice of the inner boundary radius of the
envelope.  In Chapter \ref{cqs2}, I construct alternative models of
quasi-stars by drawing from work on convection- and
advection-dominated accretion flows around black holes.  To improve
the accuracy of my models, I incorporate corrections owing to special
and general relativity into a variant of the \stars{} code that
includes rotation.  The evolution of these quasi-stars is
qualitatively different from those described in Chapter \ref{cqs1}.
Most notably, the core black holes are no longer subject to a
fractional mass limit and ultimately accrete all of the material in
their envelopes.

In Chapter \ref{cscl}, I demonstrate that the fractional mass limit
found in Chapter \ref{cqs1}, for the black holes in quasi-stars, is in
essence the same as the Sch\"onberg--Chandrasekhar limit.  The
analysis demonstrates how other similar limits are related and that
limits exist under a wider range of circumstances than previously
thought.  A test is provided that determines whether a composite
polytrope is at a fractional mass limit.  In Chapter \ref{crg}, I
apply this test to realistic stellar models and find evidence that the
existence of fractional mass limits is connected to the evolution of
stars into the red giants.
\end{quotation}
 % Abstract
\chapter*{}
\vspace*{\fill}
\begin{center}
%\emph{For the giants upon whose shoulders I have stood.}
\emph{For dad, mom and Rudi.}
\end{center}
\vspace{\fill}
\mbox{}
\vspace*{\fill}
 % Dedication
\chapter*{Acknowledgements}
\addcontentsline{toc}{chapter}{Acknowledgements}
\begin{quotation}

  First, I owe deep thanks to my supervisor, Chris Tout.  His
  scientific insights are always invaluable.  If I ever left our
  discussions without all the answers, I at least left with the right
  questions.  I am also tremendously grateful to Anna \.Zytkow for her
  keen interest and extensive input.  Her insistence on precise
  language has not only helped me keep my writing clear but my
  thinking too.  I have also appreciated the continuing support of my
  MSc supervisor, Kinwah Wu, during my PhD and I owe him substantial
  thanks for his prompt responses to my various needs.

  I would also like to thank the members of the Cambridge stellar
  evolution group who have come, been and gone during my time at the
  Institute.  Particular thanks go to John Eldridge for being a
  sounding board during Chris' sabbatical early in my studies, for
  helping me get to grips with the \stars{} code and for his support
  in my applications to academic positions.  I'd also like to thank my
  comrade-in-arms, Adrian Potter, not least for lending me his code,
  \rose{}, but, owing to our common supervisor, being a willing ear
  for discussions regarding italicized axis labels and their units,
  hanging participles, and the Oxford comma.  Thanks also to Richard
  Stancliffe for taking the time to respond to my confused emails,
  which were usually about a problem that existed between the seat and
  the keyboard.

  On a personal note, I am tremendously thankful for my friendships,
  forged in Cambridge, in London or at home, that have endured my
  infrequent communication.  I am also grateful for my fellow graduate
  students at the Institute who have come and gone during my degree.
  If anything, they have endured my \emph{frequent} communication.
  Finally, to my family, thank you for providing inspiration,
  motivation and all manner of unconditional support to this marathon
  of mine.

% There are two people I will single out.  Raoul, you've always been
% an ass-\\
% \noindent iduous
%  and I look forward to exchanging jokes, insults and ideas
% as we continue to level up as scientistes.  To my kitchen dance
% partner, external memory bank and so much more, Em, thank you for
% tolerating my eccentricities and absent-mindedness, and the times that
% I seem to lose the distinction between carefree and careless.
%The good news is that our rollercoaster ride is only starting!
% and I'm glad I haven't mananged to put you off yet.

% Secondly, Raoul, you've always been such an... asset.  I consider
% myself lucky to have had an alter ego with whom to share our
% development as scientistes for so long.  I look forward to comparing
% FORTRAN woes and thoughts on the wonders of science for many years to
% come.  I apologise for not making more trips to The Other Place.  A
% few hours on the X5 will seem like nothing compared with crossing the
% Atlantic!

% Em, for not getting annoyed more often than when I mowed daffodils
% Raoulie, for, liek, everything. Schwa, bru.

% FAMILY

  \vspace{\fill}

  \noindent\Author{}

  \noindent Cambridge, \Date

  \vspace*{\fill}

\end{quotation}

 % Acknowledgements
\clearpage
\thispagestyle{empty}%\clearpage

\blfootnote{ \noindent\textbf{Technical credits.} The work presented
  in this dissertation was performed using a large amount of free and
  open-source software.  Code and written work was edited with
  \textsc{GNU emacs} and code compiled by the \textsc{fortran}
  compiler in the \textsc{GNU} Compiler Collection.  Day-to-day
  plotting was done using \pack{gnuplot}.  All the figures in this
  dissertation other than Figs 1.1, 1.2, 3.1 and 3.4 were made using
  \textsc{veusz} by Jeremy Sanders.

  \noindent\qquad\textbf{Production notes.} This dissertation was
  compiled directly to PDF from \LaTeX{} source written in the
  \pack{book} class with the \pack{times} font family and the
  \pack{microtype} package.  Margins are specified with the
  \pack{geometry} package and chapter headings are set using
  \pack{quotchap}.  
  % The title page was drawn from the dissertation
  % template available from the Department of Engineering.  
  Extra table layout features and caption formatting controls were
  included with the \pack{booktabs} and \pack{caption} packages,
  respectively.}
 % Technical thanks

\pagestyle{tocstyle}
\tableofcontents
\listoffigures
\listoftables
\clearpage\thispagestyle{tocstyle}

\pagestyle{mainstyle}
\mainmatter
\pagenumbering{arabic}
\begin{savequote}[80mm]
  `Begin at the beginning,' the King said gravely, `and go on till you
  come to the end: then stop.'  
  \qauthor{from \emph{Alice's Adventures in Wonderland}, \\
  Lewis Carroll, 1865}
\end{savequote}

%\chapter{Introduction}
\chapter[Supermassive black holes in the early Universe]{Supermassive
  black holes \\in the early Universe}
\label{cint}

Over the last decade, high-redshift surveys have detected bright
quasars at redshifts $z\gtrsim 6$ \citep{fan+06,jiang+08,willott+10}.
Such observations imply that black holes (BHs) of more than
$10^9\Msun$ were present less than $10^9\yr$ after the Big Bang.  A
simple open question remains: how did these objects become so massive
so quickly? Despite a large and growing body of investigation into
the problem, no clear solution has yet been found 
\citep[see][for a review]{volonteri10}. 

\citet*[hereinafter BVR06]{begelman06} proposed that the direct
collapse of baryonic gas in a massive dark matter (DM) halo can lead to an
isolated structure comprising an initially stellar-mass BH embedded in
a hydrostatic envelope.  Such structures were dubbed
\emph{quasi-stars}.  At the centre of a quasi-star, a BH can grow
faster than its own Eddington-limited rate.  Quasi-stars can thus leave
massive BH remnants that subsequently grow into the supermassive black
holes (SMBHs) that power high-redshift quasars.

The structure of the gas around the BH is expected to obey the same
equations as the envelopes of supergiant stars.  In both cases,
hydrostatic material surrounds a dense core.  Giant envelopes are
supported by radiation from nuclear reactions in or around the core
whereas quasi-star envelopes are supported by radiation from accretion
on to the BH.  By choosing suitable interior conditions to describe the
interaction of the BH and the envelope, it is possible to model a
quasi-star with software packages designed to calculate stellar
structure and evolution.  Such an undertaking was the initial aim of
the work described in this dissertation and the results ultimately
shed new light on the structure of giant stars.

% The purpose of this chapter is to provide the background for the work
% undertaken in this thesis. 
The purpose of this chapter is to provide the background for my work on
quasi-stars.  In Section \ref{ssmbh}, I provide a synopsis of progress
in explaining the existence of high-redshift BHs and I explain
the place of quasi-stars within our present understanding of the early
Universe.  In Section \ref{s1stlum}, I describe the structure and
evolution of the first luminous objects in the Universe and the
remnants they are expected to leave.
% In Section \ref{sqsg}, I explain the connection between quasi-stars
% and giants. 
Finally, in Section \ref{soutline}, I outline the layout of the rest
of this dissertation.

\section{Supermassive black hole formation}
\label{ssmbh}

It is broadly accepted that the intense radiation from quasars is
produced by material falling on to massive BHs
\citep{salpeter64}.  To understand how and where such objects came to
be, I first outline the current understanding of cosmic structure
formation and then describe how baryonic material evolves after it
decouples from the DM.

\subsection{The early Universe}
\label{sslcdm}

The currently accepted cosmological model is $\Lambda$CDM
\citep[see][for a review]{peebles03}.
%, often referred to as the \emph{concordance} model.
It is characterised by a non-zero cosmological constant $\Lambda$ and
matter dominance of material with low average kinetic energy (or cold
material) that interacts only via gravity, known as dark matter (DM).
The model fits and is well-constrained by observations of anisotropies
of the cosmic microwave background \citep{larson+11}, high-redshift
supernovae \citep{kowalski08} and baryon acoustic oscillations
\citep{percival+10}.  Quantitatively, $\Lambda$CDM describes an
expanding universe that has no spatial curvature, a scale-invariant
fluctuation spectrum and a matter-energy content of, to within about 1
part in 20 of each parameter, 72.5 per cent dark energy, 22.9 per cent
DM and 4.6 per cent baryonic matter \citep{komatsu+11}.

The $\Lambda$CDM cosmology implies that large-scale structure
formation is hierarchical.  DM in small density perturbations first
condenses and then merges to form larger halos.  During each merger, the
identities of the merging halos is lost, leading to a self-similar
distribution of halos over all masses \citep{press74}.  This
theoretical scenario is supported by large-scale simulations of
structure formation \citep[e.g.][]{springel+05, kim+09}.  More recent
results indicate that such simulations are well-converged and probably
do describe how DM structure formed to the limit of current theory
\citep{boylan-kolchin09}.

The baryonic matter, which I refer to just as gas, is shock-heated
during these mergers until it reaches a temperature above which it can
cool by radiating.  Once the cooling timescale becomes shorter than the
dynamical timescale, the gas contracts towards the centre of the halo
\citep{rees77}, while the dissipationless DM does not.  As larger
halos form, either by merging or condensing, the temperature of the
gas rises until cooling is sufficient to meet the collapse
criterion.  Depending on the mass of the halo and the nature of the
cooling, the material may fragment into smaller objects and form a
protogalaxy \citep{white78}.  The size of the first astrophysical
objects thus depends directly on how the gas is able to cool.  Whether
the gas fragments or undergoes direct monolithic collapse remains an
open question and there are two broad classes of structures for the
first luminous objects and, therefore, the progenitors of SMBHs.

Typically, a cloud of gas cools by emitting radiation via atomic and
fine-structure transitions in metals.  In the early Universe, however,
there were no metals: primordial nucleosynthesis is expected to
produce only hydrogen and helium in meaningful quantities and trace
amounts of other light elements such as deuterium and lithium
\citep{coc+04}.  Atomic hydrogen cooling is only effective down to
$T\approx10^4\,{\rm K}$.  Below this temperature, only molecular
hydrogen is able to cool the gas further and, even then, only to
$T\approx200\,{\rm K}$.  The effectiveness of cooling by $\Htu$ is
complicated by its formation via $\text{H}^-$, which requires an
abundance of free electrons to form.  Both $\Htu$ and $\text{H}^-$ are
easily dissociated by an UV background, which could be created by the
light from the first generation of stars in the same or a nearby halo.
Models of halo collapse indicate that the first objects to form
were %isolated,
metal-free stars in the centres of halos with virial temperatures
$T\st{vir}\approx10^3\,{\rm K}$ and masses $M\approx10^6\Msun$
\citep*{abel02}.  They could provide sufficient ionizing radiation to
suppress or prevent $\Htu$ formation on cosmological length scales.
If collapse in nearby minihalos is foiled \citep*{machacek01}, they
might merge into larger halos that collapse later owing to atomic line
cooling.

\citet{tegmark+97} explored the problem of $\Htu$ with semi-analytic
methods and concluded that, as long as the gas temperature is below
about $10^4\,{\rm K}$, the absence of free electrons suppresses $\Htu$
formation.  Conversely, at higher temperatures, the gas is able to
form enough $\Htu$ to collapse and cool quickly.  In addition, a
sufficient UV background strongly suppresses $\Htu$ formation.  If the
gas can cool efficiently via $\Htu$ \citep{bromm04}, the first
generation of stars would have $M\approx100\Msun$.  If, instead, the
formation of $\Htu$ is suppressed, the gas is unable to cool as
rapidly and probably forms a pressure-supported object with
$M\gtrsim10^4\Msun$ \citep[e.g.][]{regan09a}.

\citet*{schleicher10} showed that collisional dissociation suppresses
$\Htu$ formation for particle densities over about $10^5\pcmcu$.  At
lower densities, $\Htu$ could be photodissociated by an ionizing UV
background.  \citet*{shang10} estimated that the necessary specific
intensity exceeds the expected average in the relevant epoch but
\citet{dijkstra+08} suggested that the inhomogeneous distribution of
ionizing sources provides a sufficiently large ionizing field in a
fraction of halos.  \citet{spaans06} instead argued that the
self-trapping of Ly$\alpha$ radiation during the collapse keeps the
temperature above $10^4\K$ and prevents $\Htu$ from forming at all.
Finally, \citet{begelman09} proposed that the bars-within-bars
mechanism of angular momentum transport sustains enough supersonic
turbulence in the collapsing gas to prevent fragmentation, even if the
gas cools.

The structure of the collapsing gas hinges on whether or not $\Htu$
forms and radiates efficiently and it is unclear which is the case or
occurs more frequently.  The two cases lead to distinct evolutionary
sequences for subsequent baryonic structure formation.  I explain
these in the next two subsections. \\

\subsection{Accretion and mergers of seed black holes}

If the first generation of luminous objects were metal-free stars with
masses in the range $100\Msun\lesssim M\lesssim1000\Msun$, stellar
models predict that objects with masses over about $260\Msun$
underwent pair-instability supernovae and left BHs with about half the
mass of their progenitors \citep*{fryer01}.  Stars with masses in the
range $140\Msun\lesssim M\lesssim260\Msun$ also become pair-unstable
but are expected to be completely disrupted.  Stars smaller than
$140\Msun$ burn oxygen stably and ultimately develop iron cores that
collapse.  The largest objects form at the centres of their host DM
halos where these seed BHs accrete infalling material and merge with
other seeds as the DM halos continue to combine into larger and larger
structures.

The simplest explanation for SMBHs at high redshift would be that a
seed BH accretes gas from its surroundings.  If the radiation is
spherically symmetric, the maximum rate at which the BH can accrete is
reached when the amount of radiation released by the material as it
falls on to the BH matches the gravitational attraction of the BH.
The total luminosity in this state is the \emph{Eddington limit} or
\emph{Eddington luminosity}.  The accretion rate that reproduces the
Eddington luminosity is the Eddington-limited accretion rate.  If a
seed BH accretes at its Eddington-limited rate with a radiative
efficiency $\epsilon=\Lbh/\dMbh c^2=0.1$, a $10\Msun$ seed takes about
$\sci{7}{8}\yr$ to reach a mass of $10^9\Msun$ \citep{haiman01}.
Though this appears to solve the problem, it requires that the BH is
initially sufficiently large, surrounded by a sufficient supply of gas
and able to accrete with constant efficiency at the Eddington limit.
\citet{milosavljevic09a} argue that accretion on to the BHs from the
surrounding gas is self-limiting and
% proceeds at an average accretion efficiency greater
they cannot accrete faster than 60 per cent of the Eddington-limited
rate.  For accretion from a uniform high-density cloud,
\citet{milosavljevic09b} claim that the accretion rate drops as low as
32 per cent of the maximum.  \citet{johnson07} suggest that the
supernova accompanying the formation of a seed BH rarefies the
surrounding gas and delays accretion for up to $10^8\yr$.  More
recently, \citet{clark+11} found that the protostellar accretion disc
around a primordial star is unstable to further fragmentation that
leads to tight multiple systems of lower-mass stars instead of single
isolated massive stars.  \citet{hosokawa+11} found that the
protostellar disc evaporates because of irradiation once the protostar
reaches a few tens of solar masses.  These results show that seed BHs
may have been too small and accreted too slowly to reach the observed
masses by $z=6$, so other solutions have been proposed.

After the first compact objects form, DM halos continue to merge
hierarchically and %we expect that
the baryonic protogalaxies at their centres follow suit
\citep{rees78}.  If the galaxies have core BHs, they too will merge.
This process can be modelled by following sequences of BH mergers
through a hierarchical tree
\citep*[e.g.][]{volonteri03,bromley04,tanaka09}.  In the simplest such
models, seed BHs of a certain mass form in halos of a specified size
(or range of sizes) and the halos are given some probability of
merging.  When they merge, so do the BHs.  This simple picture already
requires a number of free parameters and there are many other physical
processes that can be included and parametrized.  \citet*{sijacki09}
took advantage of progress in general relativistic modelling of BH
mergers, general relativistic magnetohydrodynamic models of BH
accretion and high-resolution simulations of cosmological structure
formation to construct intricate merger trees including these effects.

In general, according to the authors referred to above, the
formation of SMBHs from seeds of a few hundred $\Msun$ is sensitive to
a number of uncertain conditions.  Many models require the BH to
accrete mostly at or near the Eddington-limited rate but it is not
clear that this is possible, for the reasons described above.  Those
models that do successfully form high-mass, high-redshift quasars tend
to overpredict the number of smaller BHs in the nearby Universe
\citep[e.g.][]{bromley04,tanaka09} or require that the first stars
formed at redshifts $z>30$.

Several important factors remain difficult to model. % at all.
First, a BH's spin substantially influences its radiative efficiency.
A non-rotating BH has a radiative efficiency of just 5.7 per cent but
this rises to 42 per cent for a maximally-rotating black hole, making
accretion less efficient.  Consistent accretion from a disc tends to
increase a BH's spin, whether from a standard thin disc or a thick,
radiation-dominated flow \citep*{gammie04}.  Conversely, random,
episodic accretion reduces the BH's spin \citep{wang09}.  Secondly,
feedback %effects 
of the BH's radiation on the protogalaxy is still excluded.  Active
accretion can lead to substantial gas outflows \citep{silk98},
which %could 
starve the BH of material over cosmic times and %further
retard its ability to grow.

There is currently no clear solution that resolves these issues.  One
possibility is an increase in the masses of the seed BHs
\citep{haiman01}.  \citet{volonteri05b} suggest that an
intermediate-mass seed formed in a large halo could undergo a short
period of super-Eddington accretion, effectively increasing the seed
mass for subsequent mergers.  \citet*{spolyar08} suggested that DM
self-annihilation inside primordial stars can significantly heat them
and delay their arrival on the main sequence.  This allows protostars
to accrete more material before they begin to produce ionizing
radiation.  The stars thus achieve larger masses and leave larger seed
BHs \citep{freese+08}.

It is also possible to form larger seed BHs after a first generation
of stars has polluted the interstellar medium with metals.  The second
generation of stars would resemble modern metal-poor populations.
During hierarchical mergers, gas builds up in the cores of halos,
fragments and forms dense stellar clusters \citep*{clark08}.  Such
dense environments can lead to frequent stellar collisions and thence
either directly to a massive BH or to a massive star that leaves a
massive BH as its remnant \citep{devecchi09}.  \citet{mayer+10} also
found that collisions between massive protogalaxies naturally produce
massive nuclear gas discs that rapidly funnel material to sub-parsec
scales and create ideal conditions for collapse into a massive BH.

\begin{figure}[p]\begin{center}
\includegraphics{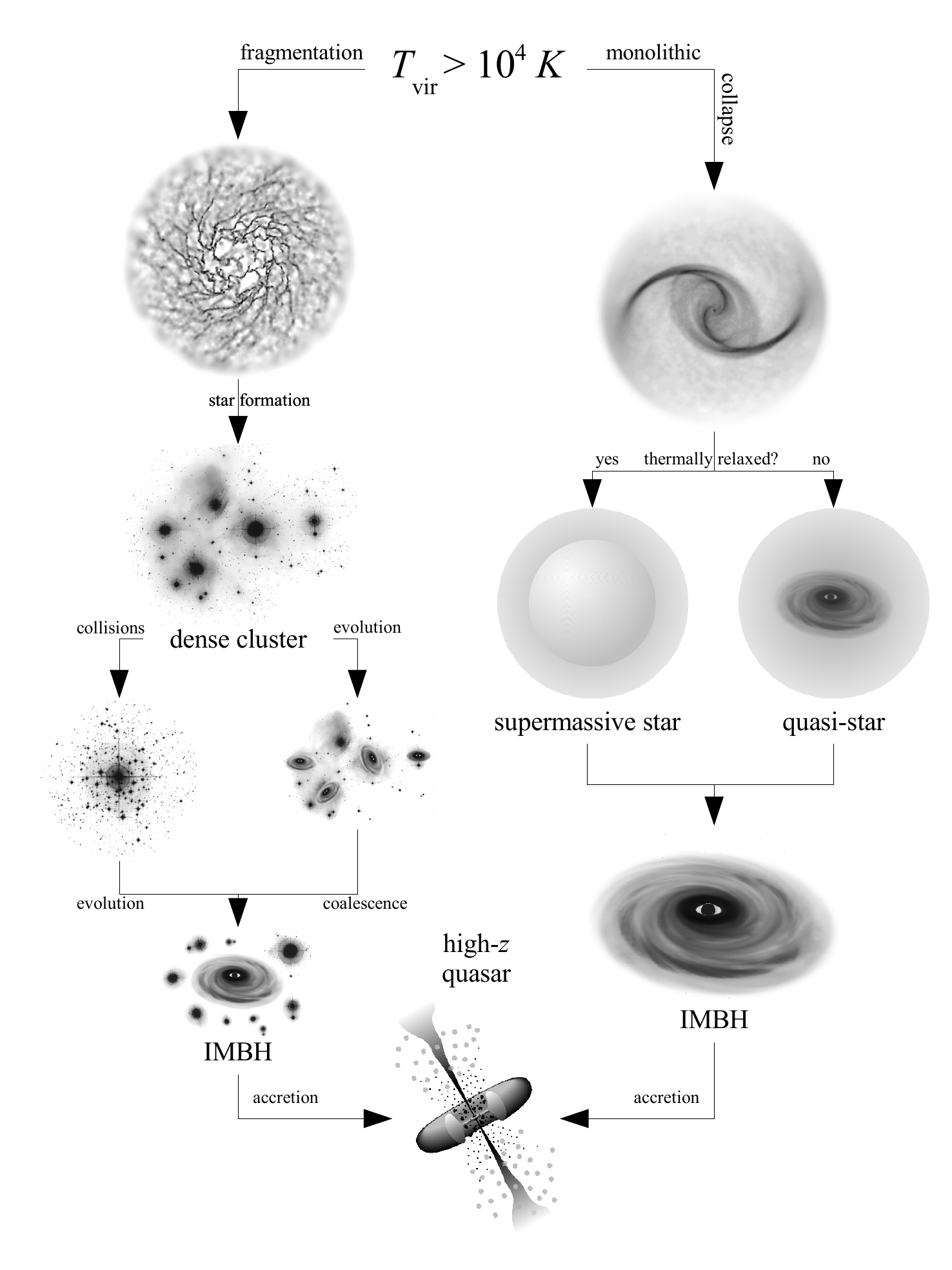}
\caption[Pathways to massive black hole formation.]{A diagram showing
  possible paths to the formation of massive BHs in halos with
  $T\st{vir}>10^4\K$.  If the gas fragments, it can form a dense star
  cluster or high-order multiple system.  The stars can collide and
  form a massive star that leaves an intermediate mass BH remnant or
  the stars can evolve independently after which their remnants can
  coalesce.  If the gas does not fragment, its fate is determined by
  whether the infall of gas is too rapid to establish thermal
  equilibrium in the central hydrostatic core.  If thermal equilibrium
  can be achieved, a supermassive star forms and it collapses directly
  into an intermediate mass BH.  If the infall is greater than about
  $0.1\Msunpyr$, the central star evolves independently and leaves a
  BH embedded in a growing hydrostatic envelope: a quasi-star.}
\label{fbhpaths}\end{center}\end{figure}

In short, the problem of creating massive BHs quickly could be
resolved by any mechanism that allows massive seeds to form.
\citet{sijacki09} found they could reproduce a BH population that fits
observed properties at high- and low-redshifts by using rare seeds
with masses up to about $10^6\Msun$.   Fig.~\ref{fbhpaths} summarizes
several possibilities that lead to intermediate-mass BHs that are
massive enough to grow into SMBHs by redshift $z\approx6$.  In the
next subsection, I discuss how such massive seeds can form through the
monolithic gravitational collapse of massive pregalactic clouds.

\subsection{Direct collapse}
\label{ssdc}

\citet{loeb94} outline a number of obstructions to the direct
formation of a SMBH at the centre of a DM halo.  Among them is
fragmentation (see Section \ref{sslcdm}), which is principally decided
by the efficient formation of $\Htu$ molecules and their ability to
cool.  A further obstruction is the transport of angular momentum.  If
angular momentum is not transported efficiently, a large
self-gravitating disc forms.  %This can be unstable to fragmentation.
If unstable to fragmentation, such an object is more likely to form a
cluster of smaller objects than a single supermassive one.
High-resolution simulations \citep{regan09a,wise07} of material in
massive halos ($M\st{tot}\approx10^7\Msun$) indicate that such discs
are gravitationally stable.  If these discs do not fragment, does
angular momentum still preclude direct collapse?

Numerical simulations have long indicated the existence of triaxial or
\emph{bar} instabilities in self-gravitating discs \citep{ostriker73}.
\citet*{shlosman89} proposed these as a mechanism for effective angular
transport to feed gas to active galactic nuclei.  BVR06 invoke the same
mechanism in pregalactic halos.  They show that, for a variety of
angular momentum distributions, a reasonable number of halos are
susceptible to runaway collapse through a %sequence
series of bar instabilities.  Indeed, \citet*{wise08a} observe such a
cascade of bar instabilities in their simulations of collapsing halos.

% \citet{begelman09} further propose that this mechanism of angular
% momentum loss stabilizes the disc against fragmentation.  The
% \emph{bars-within-bars} mechanism depends on whether the ratio of the
% bulk kinetic energy of material to its gravitational potential exceeds
% a critical factor.  In a conservative system, angular momentum loss
% outpaces potential energy loss, so the material stabilizes.  If the
% material can radiate bulk kinetic energy, the instability is
% sustained.  It is known that the inflowing material becomes supersonic
% \citep{wise07}.  This gives the disk a finite thickness.  This finite
% thickness changes the stability criteria for fragmentation such that a
% turbulent disc always becomes bar-unstable before it fragments because
% it dissipates its turbulent energy into rotational energy.  The
% cascading bar instabilities are only quenched when a massive,
% isolated, pressure-supported object is formed at the centre of the
% disc.

What is the nature of the object that forms? There are two
possibilities.  I shall explore their evolution in more detail in the
next section and provide a simple outline of their structure here.
BVR06 argue that a small protostellar core forms and continues to
accrete material at a rate on the order of $0.1\Msunpyr$.  Because the
gas accumulates so quickly, the envelope of the star does not reach
thermal equilibrium during its lifetime \citep{begelman10}.  After
hydrogen burning is complete, runaway neutrino losses cause the core
to collapse to a stellar mass BH.  The structure is a then stellar
mass BH embedded in and accreting from a giant-like gaseous envelope.
This structure is named a quasi-star.

Alternatively, if thermal equilibrium is established throughout the
protostar prior to collapse, the object resembles a metal-free star,
albeit with a mass exceeding $10^4\Msun$.  If it is sufficiently
massive, an instability from post-Newtonian terms in the equation of
hydrostatic equilibrium leads to gravitational collapse before normal
evolution proceeds to completion \citep{chandra64}.  This leads to a
BH with perhaps 90 per cent of the original stellar mass
\citep{shapiro04}.  In either case, after the stellar (or
quasi-stellar) evolution ends we expect the remnant to be a massive BH
\citep{regan09b} that can then act as a seed for the merging and
accretion processes described in the previous section.

\section{The first luminous objects}
\label{s1stlum}

The gradual collapse of baryonic material in the early Universe can
lead to several different types of objects.  Gas could fragment into
metal-free stars.  If the gas does not break up during the collapse, a
large isolated object could form.  The massive body of gas could
undergo some form of stellar evolution or some fraction of it could
collapse directly into a BH.  In this section, I describe these
various objects and their evolutions.

\subsection{Population III stars}

Metal-free stars, called \emph{Population III} (Pop III) stars, have a
number of properties that distinguish them from Population I and II
stars.  Owing to their importance in the early Universe, an extensive
body of research regarding their structure and evolution now exists.  I
shall discuss here the aspects of metal-free stellar evolution that
distinguish Pop III stars.

% Until recently, consensus settled on theoretical grounds that Pop III
% stars have a top-heavy initial mass function (IMF) \citep{bromm99}.
Based on simple theoretical arguments, Pop III stars have a top-heavy
initial mass function \citep*[IMF,][]{bromm99}.  That is, Pop III stars
tend to have higher initial masses than metal-polluted stars.  In a
cloud of gas the smallest mass unstable to collapse, the \emph{Jeans
  mass}, scales with the temperature of the gas.  The strongest
cooling agent in the early Universe is molecular hydrogen, which only
cools effectively to $T\approx200K$.  Present-day molecular clouds can
cool further because of abundant molecules that include carbon,
nitrogen and oxygen.  This suggests that Pop III stars are on average
more massive than their Pop I or II cousins.  Until recently,
consensus suggested Pop III stars would have masses on the order of
$100\Msun$ so the literature is focused on this case \citep{bromm99}.
Recent work \citep[e.g.][]{clark+11,hosokawa+11,stacy12} suggests that
Pop III stars may have instead had masses on the order of $10\Msun$
because the protostellar disc fragments or evaporates before all of
the material has accreted onto the central protostar.

Massive stars usually burn hydrogen into helium via the catalytic
carbon-nitrogen-oxygen (CNO) cycle.  In the absence of any initial
carbon, nitrogen or oxygen only the proton-proton chain (pp chain) is
available to slow the gradual contraction of a protostar.  The pp chain
scales weakly with temperature and, for such massive stars, fails to
halt the contraction of the star up to core temperatures on the order
of $10^8\,{\rm K}$.  At these temperatures, the $3\alpha$ process
produces a trace amount of carbon, which is quickly converted into an
equilibrium abundance of carbon, nitrogen and oxygen.  The principal
source of energy then shifts from the pp chain to the CNO cycle
%\citep{marigo01,siess02}.  
\citetext{\citealp{marigo+01}; \citealp*{siess02}}.  
In lower-mass stars, sufficient CNO abundances are only reached during
the main sequence but, in stars with $M\gtrsim20\Msun$, the dominance
of the CNO cycle is established earlier.  For stars more massive than
$100\Msun$, the equilibrium mass abundance of carbon, oxygen and
nitrogen is between about $10^{-10}$ and $10^{-9}$ \citep*{bond84}.

These massive stars live short lives.  They are radiation-dominated,
so $L\approx L\st{Edd}$, where $L\st{Edd}$ is the Eddington
luminosity.  Because the main-sequence lifetime $\tau\st{MS}$ scales
as $\tau\st{MS}\propto M/L$ and $L\propto M$, it follows that
$\tau\st{MS}$ is roughly constant.  The lifetime of a $100\Msun$ star
is about $3.14\Myr$ \citep*{marigo03} and decreases only slightly as
the mass increases.  The cores are convective on the main sequence and
remain so during helium burning.  This commences almost immediately
after hydrogen is exhausted so there is no first dredge-up.  After
burning helium, the core burns carbon and then contracts on dynamical
timescales towards a temperature and density at which oxygen burns.

Thereafter, stars with $M\gtrsim140\Msun$ become unstable to an
electron-positron pair-production instability and they are classified
as \emph{very massive objects} \citep{bond84}.  Their core
temperatures are hot enough for photons to spontaneously form
electron-positron pairs.  This pair formation reduces the radiation
pressure so the star contracts and the core temperature increases.
The increase in temperature leads to more pair-production and the
contraction runs away.

The pair-unstable collapse commences after core helium depletion.
During the collapse, oxygen ignites explosively in the core.  For stars
in the range $140\Msun\lesssim M\lesssim260\Msun$, oxygen ignition
releases enough energy to completely disrupt the star in a
pair-instability supernova \citep[PISN,][]{fryer01}.  These events
present an important opportunity to pollute the early Universe with
the metals produced in the first stars because the entire star's worth
of material is expelled \citep{heger+03}.

For stars more massive than $260\Msun$, the collapsing material is so
strongly bound that even the fusion of the entire core into silicon
and iron cannot halt its collapse.  Most of the mass in the core is
burnt to iron-group elements and the core ultimately collapses into a
BH.  For a $300\Msun$ star, the initial BH mass is about $20\Msun$ but
it quickly accretes the remaining $120\Msun$ or so of core material
\citep{fryer01}.  For stars with masses over $260\Msun$, about half of
the star's mass contributes to the near-immediate mass of the BH
\citep{ohkubo+06}.

\begin{figure}[tb]\begin{center}
\includegraphics[angle=270,width=0.8\textwidth]{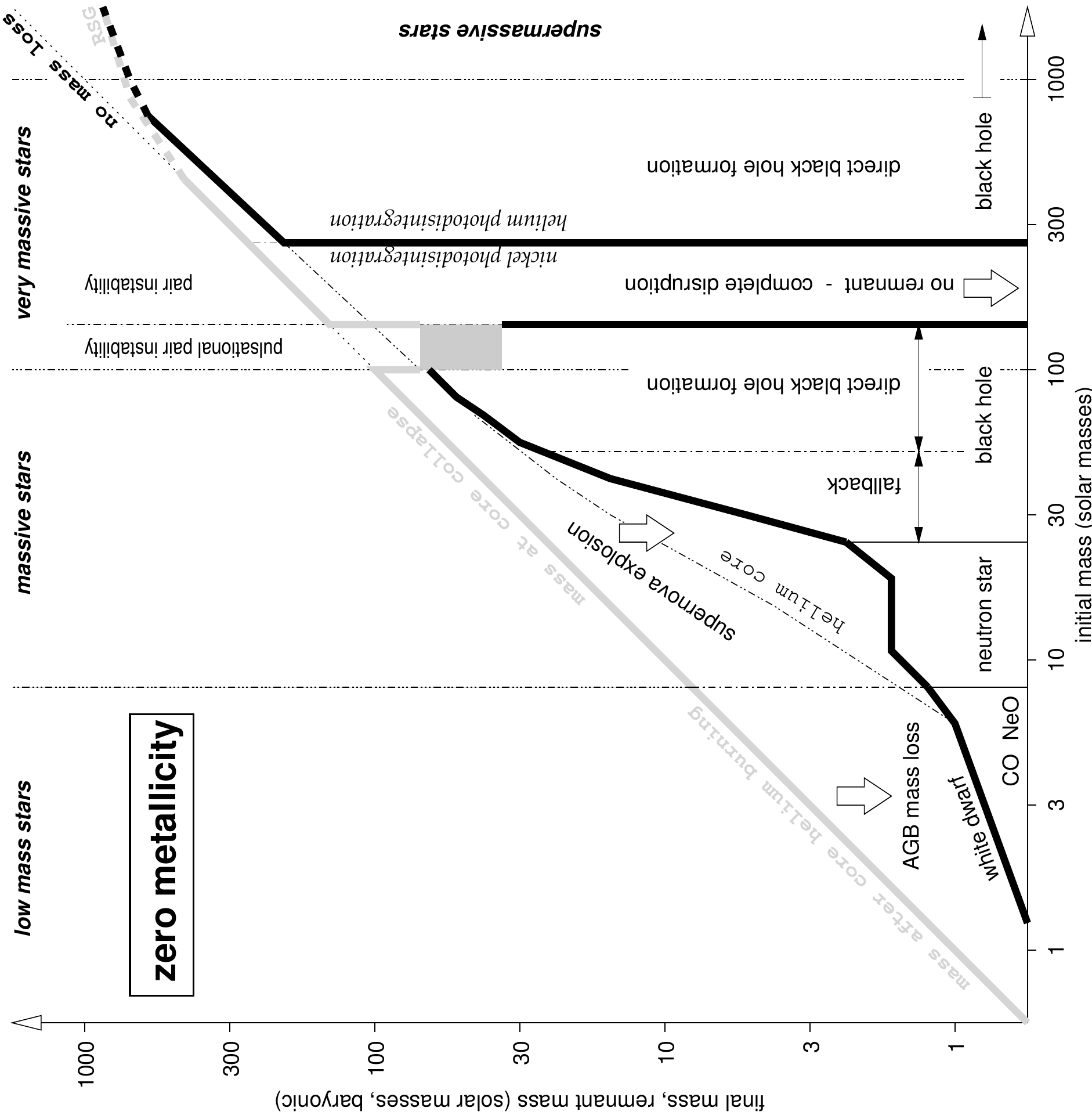}
\caption[Final mass versus initial mass for non-rotating metal-free
(Pop III) objects.]{Plot of the final mass of a Pop III object as a
  function of its initial mass, up to $1000\Msun$.  For stars with
  initial masses smaller than $140\Msun$, the outcomes are broadly
  similar to those for Pop I or II stars.  Between $140$ and
  $260\Msun$, pair-instability leads to total disruption of the star
  and no remnant.  At higher masses, a BH forms directly after
  explosive carbon ignition.  \citep[Figure after][Fig.~2.]{heger02}}
\label{finifin}\end{center}\end{figure}

Fig.~\ref{finifin} summarizes the fates of Pop III stars with masses
up to $1000\Msun$.  The final mass of a stellar remnant is plotted as
a function of the initial mass of its progenitor according to the
models of \citet{heger02}.  The plot includes masses below $140\Msun$,
which I have not discussed.

There are two notable omissions to the evolution described above.
These are mass loss and rotation.  Mass loss from hot stars is usually
driven by atomic transition lines but these are absent in the
metal-free atmospheres of Pop III stars and radiation-driven winds are
very weak \citep{kudritzki02}.  \citet*{baraffe01} and \citet{sonoi12}
additionally found that mass loss via fusion-driven pulsational
instabilities is typically only a few per cent.  Consequently,
mass-loss from the surface of Pop III stars is usually ignored
\citep{marigo+01}.

Angular momentum is imparted to primordial gas during hierarchical
mergers and is preserved in its subsequent collapse.  \citet*{stacy11}
estimated the rotation speeds of primordial stars by following the
velocities and angular momenta of gas in simulations of Pop III star
formation.  They found that there is enough angular momentum for the
stars to be born at near break-up speeds.  If the stars rotate
sufficiently rapidly to shed mass from their surfaces, angular
momentum is quickly lost and further rotational mass loss is limited.
However, rotation can lead to additional mixing processes that bring
metals to the surface and allow line-driven winds to develop.  Extra
mixing also extends core-burning lifetimes and modifies
nucleosynthetic yields.  \citet{chatzopoulos12} calculated
pre-supernova stellar models for a range of masses and rotation rates
without surface mass loss and found that the minimum mass for a PISN
decreases dramatically.  For an initial surface rotation rate of
$0.3\Omega\st{cr}$, where $\Omega\st{cr}$ is the critical Keplerian
rate, the minimum mass is about $85\Msun$.

For massive PISN progenitors ($M\gtrsim260\Msun$), significant
rotation delays accretion by the core BH.  Though the delay is not
itself significant, it leads to the formation of a disc, which
probably launches a jet \citep{fryer01}.  Such a jet could experience
explosive nucleosynthesis before polluting the interstellar (or even
intergalactic) medium with metals.  If the star is more massive than
about $500\Msun$, the jet cannot escape the stellar atmosphere.  The
metals would then pollute the remaining H-rich envelope but the jet
itself would go unseen \citep{ohkubo+06}.

If $\Htu$ cools efficiently, we expect Pop III stars to form at the
centres of small DM halos \citep{wise08a}.  If $\Htu$ cooling is
suppressed, larger objects, that collapse directly into SMBHs, can
form \citep{regan09a}.  Next, I outline two possible structures that
can form in this way.

\subsection{Supermassive stars}

%The existence of stars with masses exceeding $10^5\Msun$ was suggested
%by \citet{hoyle63} as an explanation of 
\citet{hoyle63} suggested that stars with masses exceeding $10^5\Msun$
could explain the tremendous luminosity of quasars.  Though consensus
ultimately settled on quasars being accreting BHs
\citep{salpeter64,lynden-bell69}, these objects were modelled
extensively and much of the work cited here is from that era.  Only
recently have supermassive stars resurfaced, now as the progenitors of
high-redshift SMBHs.

The properties of very massive stars, as described in the previous
subsection, hold for increasing mass until a new instability is
invoked.  In Newtonian gravity, stars with adiabatic index $\gamma$
below the critical value $\gamma\st{crit}=4/3$ are known to be
dynamically unstable.  Purely radiation-dominated stars have precisely
this critical value but pure radiation is not possible because some
gas is always present and raises the adiabatic index.  Real
radiation-dominated stars are thus stable, even if only marginally
so.  However, if we include the post-Newtonian terms of general
relativity (GR) in the equation of hydrostatic equilibrium, the
critical adiabatic index is increased to just over $4/3$
\citep{chandra64} and radiation-dominated stars are susceptible to
the GR instability.  Stars that are expected to be GR unstable are
classified as \emph{supermassive stars} \citep{appenzeller71}.

Because radiation pressure depends on temperature and a star's core
temperature increases over its life, stars of different masses become
GR unstable during different phases of evolution.  In the non-rotating
case, \citet{fricke73} estimated that a zero-metallicity star becomes
unstable during its main-sequence life if $M\gtrsim10^5\Msun$ and
during its He-burning phase if $M\gtrsim\sci{3.4}{4}\Msun$.  These
bounds increase substantially if the star is rotating
\citep{fricke74}.  Once the instability sets in, collapse is certain:
H-ignition in a metal-free star is insufficient to stop it.  Fully
general-relativistic calculations \citep{shibata02} confirm that a BH
forms with a mass of roughly 90 per cent of the star's original mass.

A fundamental feature in the formation of a supermassive star is that
hydrostatic equilibrium is established throughout the protostar.  If
this is not the case, the models above are not valid.  Dynamical,
non-homologous collapse can lead instead to a pressure-supported
protostellar core accreting rapidly from the surrounding material and
I describe this next.

\subsection{Quasi-stars}

One of the greater obstacles to the collapse of a single, supermassive
baryonic object in the early Universe (see Section \ref{ssdc}) is
angular momentum loss \citep{loeb94}.  The collisions and mergers of
DM halos give them angular momentum, which is imparted to the baryonic
material they contain.  BVR06 outline a scenario where the angular
momentum is transported via the bars-within-bars mechanism described
by \citet{shlosman89}.  Whenever the ratio of rotational kinetic
energy of the gas to the gravitational potential exceeds a critical
factor, the disc is liable to form a bar.  The bar transports angular
momentum outwards and material inwards without significant entropy
transport.  Once the infalling material stabilizes, it cools rapidly
and another bar forms.  The result is a cascade of instabilities and
the infall of material on dynamical timescales.

Eventually, the most central material becomes pressure-supported.  The
instability is then quenched because rotational support is no longer
dominant.  The pressure-supported object accretes at a rate on the
order of $0.1\Msunpyr$ and becomes radiation-dominated.  BVR06 claim
that infalling material creates a positive entropy gradient.  This
stabilises the pressure-supported region against convection.  It also
means that the surface layers compress the underlying material and
that there is a core region of a few $\Msun$ beneath the
radiation-dominated envelope where the pressure remains gas-dominated.
Both claims are supported by models of rapidly-accreting massive
protostars presented by \citet*{hosokawa12}, who followed the stars'
evolution up to core hydrogen ignition.

\citet{begelman10} considered the evolution of the growing
thermally-relaxed core.  As in Pop III stars, hydrogen burning begins
through pp chains but the %core continues to contract 
core's contraction continues until a trace
amount of carbon is created through the $3\alpha$ process.  Hydrogen
subsequently burns through the CNO cycle.  The hydrogen-burning phase
lasts a few million years during which the core %continues to grow 
grows but does not incorporate the total mass of the object.  After
the core exhausts its central hydrogen supply, it contracts and heats
up to many $10^8\K$.  At these temperatures, the core collapses owing
to neutrino losses.

The outcome of this evolution is an initially stellar-mass BH embedded
in a giant-like envelope of material.  The BH quickly begins accreting
from the surrounding envelope.  The radiation released in the flow
settles near the Eddington-limited rate of the whole object and drives
convection throughout the envelope, even if it was convectively stable
before core-collapse.  BVR06 called this configuration a quasi-star
% because the envelope has star-like properties but the central energy :LBfix
because the envelope is star-like but the central energy source is
accretion on to the BH rather than nuclear reactions.

BRA08 explored the structure of quasi-stars in some detail.  The
accretion on to the BH is taken to be an optically thick
variant of spherical Bondi accretion \citep{flammang82} reduced by
radiative feedback and the limitations of convective energy transport
from the base of the envelope.  The radiation-dominated envelope is
convective \citep{loeb94} and well-approximated by an $n=3$
polytrope.  At the base of the envelope, $10^5\lesssim
T\st{c}/\text{K}\lesssim10^6$ and
$\rho\st{c}\lesssim10^{-6}\text{ cm}^{-3}$, so nuclear reactions
can safely be ignored.

Above the convective region is a radiative atmosphere.  Initially, the
temperature of the radiative zone is of order $10^4\,{\rm K}$ for a
$10^5\Msun$ quasi-star.  As the BH and its luminosity grow, the
atmosphere expands and the photospheric temperature falls.  BRA08
claim that once $T\st{ph}\approx4000\K$, the entire atmosphere
experiences super-Eddington luminosity.  It expands
more %It = The atmosphere : LBfix
quickly, leading to faster cooling and a runaway process that
disperses the photosphere.  The convection zone becomes unbounded and
is in effect released from the BH.  The life of the quasi-star ends
and leaves a BH of at least a few thousand solar masses.  The models
in this dissertation do not support this scenario.  The BH is either
stopped by reaching a fractional mass limit similar to the
Sch\"onberg--Chandrasekhar limit or it continues to accrete the whole
gaseous envelope.

Angular momentum may play a role in the structure of quasi-stars.  Its
effect is probably minor for the envelope structure but might be
significant for the BH's immediate surroundings.  The angular momentum
of the accreted material leads to the formation of a disc, which
decreases the accretion efficiency.  The emission from the disc
affects the innermost layers of the envelope and ongoing accretion
affects the spin of the BH.

\section{Outline of this dissertation}
\label{soutline}

%The initial aim of my research was to model quasi-star envelopes with
%the Cambridge \stars{} code.  
The remainder of this dissertation describes my work in six chapters.
In Chapter \ref{cstars}, I provide the technical details of the
Cambridge \stars{} code, which was used to calculate models in
Chapters \ref{cqs1}, \ref{cqs2} and \ref{crg}.  In Chapter \ref{cqs1},
I describe models of quasi-stars constructed %by following 
after the example of BRA08 and find two main results.  First, the
models are highly sensitive to the location of inner boundary radius.
Secondly, the models are all subject to a robust limit on the final
mass of the black hole as a fraction of the total mass of the
quasi-star.  In Chapter \ref{cqs2}, I confront the first problem by
constructing a new set of boundary conditions and find that the
evolution of the models is very different.  For all parameter choices,
the BHs ultimately accrete the whole envelope.

In Chapter \ref{cscl}, I explain the fractional mass limits found in
Chapter \ref{cqs1}.  In short, the mass limit is analogous to the
Sch\"onberg--Chandrasekhar limit: the maximum fractional mass that an
isothermal stellar core can achieve when embedded in a polytropic
envelope with index $n=3$.  I extend my work to incorporate other
polytropic limits discussed in the literature.  The description of
these limits leads to the construction of a test that determines
whether a polytropic model is at a fractional mass limit.  In Chapter
\ref{crg}, the test is applied to realistic stellar models.  It
appears that exceeding a fractional mass limit always occurs when a
star evolves into a giant and I introduce and discuss the
long-standing problem of what causes this behaviour.  In Chapter
\ref{ccon}, I summarize my research and propose directions for future
work on SMBH formation and the red giant problem.
 % Introduction
\begin{savequote}[80mm]
%  In the good old days physicists repeated each other's experiments,
%  just to be sure.  Today they stick to FORTRAN, so that they can share
%  each other's programs, bugs included.
  \textsc{fortran}--the "infantile disorder"--, by now nearly 20 years old, is
  hopelessly inadequate for whatever computer application you have in
  mind today: it is now too clumsy, too risky, and too expensive to
  use.
  \qauthor{Edsger W.  Dijkstra, 1975}
\end{savequote}

\chapter{The Cambridge \textsc{stars} code}
\label{cstars}

A triumph of 20th century astrophysics is the development of a
successful stellar model, which describes stars as spherical, static,
self-gravitating fluids in local thermodynamic equilibrium.  This
leads to a system of non-linear differential equations that must be
solved numerically.  With the advent of modern computing after the
second world war, an increasing number of solutions were calculated
and most observable properties of stars and clusters were explained.
Nowadays, these calculations are easily performed using consumer-level
hardware and the theory on which they are based forms a standard body
of knowledge.  The reader should not infer, however, that stellar
structure and evolution is wholly understood.  Though the standard
model is successful, it is also incomplete.  There are a number of
processes, including rotation and surface mass-loss, that are poorly
understood and the subject of ongoing research.

In this chapter, I describe the standard equations of stellar
structure and evolution and the Cambridge \stars{} code, which solves
them.  The \stars{} code was originally written by
\citet{eggleton71,eggleton72,eggleton73}.  It has subsequently been
updated and modified by \citet{han+94}, \citet{pols+95},
\citet{eldridge04}, \citet{stancliffe+05}, \citet{stancliffe08} and
\citet{stancliffe09z} but its distinguishing features remain
unchanged.  The structure, composition and distribution of solution
points are calculated simultaneously using a relaxation method.

I describe here the technical details of the standard version of the
code, which was used to produce the stellar models in Chapter
\ref{crg}.  I also indicate modifications that I made to improve
accuracy and convergence when modelling quasi-stars.  For quantities
that are variables in the parameter input file (usually \texttt{data})
I have given their corresponding names in the \stars{} code in a
typewriter font. e.g. (\texttt{EG}).

% The unmodified code, used to produce the stellar models in Chapter
% \ref{crg}, is described here.  
For the quasi-star models in Chapters \ref{cqs1} and \ref{cqs2}, the
inner boundary conditions were replaced with those of the relevant
models described in each chapter.  For the work described in Chapter
\ref{cqs2}, I extended a variant of the code that includes rotation
\citep*[\rose{},][]{potter12} by adding corrections to the structure
equations from special and general relativity devised by
\citet{thorne77}.  These extensions are described in Section
\ref{sadph}.

\section{Equations of stellar evolution}

The description of a star as a spherical, static, self-gravitating
fluid in local thermodynamic equilibrium leads to a system of
differential equations that allows the calculation of bulk properties
like temperature, density and luminosity as a function of radius in
the star.  These %time-independent equations 
are the \emph{structure} equations.  They depend on the microscopic
properties of the material, which must be provided either through
approximate functions or as interpolations of tabulated data.  These
are the \emph{matter} equations.  Finally, the microscopic properties
depend on the chemical composition of the stellar material.  The
gradual change in a star's structure is caused by changes captured by
the \emph{composition} equations.  These describe how the distribution
of elements changes over time owing to redistribution through
convection and transformation through fusion.  In this section, I
briefly review each set of equations.  For complete derivations, the
reader should consult any standard textbook on stellar structure. e.g.
\citet{kw90}.

\subsection{Structure}

The macroscopic structure of a star is described by four differential
equations.  The spherical approximation requires one independent
variable, which must be monotonic.  The usual choices are the local
radial co-ordinate $r$ or the local mass co-ordinate $m$ and the
differential equations are given below for both.  The first three
equations describe the local conservation of mass,
\shorteq{\pdif{m}{r}=4\pi\rho r^2\text{\quad
    or\quad}\pdif{r}{m}=\frac{1}{4\pi\rho r^2}\text{,}}
%momentum,
hydrostatic equilibrium,
\shorteq{\pdif{p}{r}=-\frac{Gm\rho}{r^2}\text{\quad
    or\quad}\pdif{p}{m}=-\frac{Gm}{4\pi r^4}\text{,}}
and energy generation,
\shorteq{\pdif{L}{r}=4\pi\rho r^2\epsilon\text{\quad
    or\quad}\pdif{L}{m}=\epsilon\text{,}} 
where $G$ is the gravitational constant, $\rho$ the density at $r$,
$p$ the pressure, $T$ the temperature, $L$ the luminosity and
$\epsilon$ the total energy generation rate per unit mass.  The total
energy generation is a combination of contributions from nuclear
reactions $\epsilon\st{nuc}$, neutrino losses $\epsilon_\nu$ and
heating or cooling via contraction or expansion, referred to here as
the thermal energy generation rate $\epsilon\st{th}=T\,\qdif{s}{t}$,
where $s$ is the local specific entropy and $t$ represents time.  The
thermal energy generation $\epsilon\st{th}$ is zero when the star is
in thermal equilibrium.

The fourth structure equation describes how energy is transported through the
star.  In general, we write
\shorteq{\pdif{T}{r}=\nabla\frac{T}{p}\pdif{p}{r}\text{\quad
    or\quad}\pdif{T}{m}=\nabla\frac{T}{p}\pdif{p}{m}\text{,}} where
$\nabla=\qdif{\log T}{\log p}$ depends on whether energy is
transported by radiation or convection.  If the temperature gradient
is due to radiation alone, %then
\shorteq{\nabla=\nabla\st{rad}=\frac{3}{16\pi a
    cG}\frac{p}{T^4}\frac{\kappa L}{m}\text{,}}
where $a$ is the radiation constant, $c$ the speed of light and
$\kappa$ the opacity.

If the radiative temperature gradient $\nabla\st{rad}$ is greater than
the adiabatic temperature gradient, $\nabla\st{ad}=\qdif{\log T}{\log
  p}$ at constant entropy $S$,% then
a parcel of material that is
displaced upward in the star becomes hotter and sparser than the
material around it.  In this unstable situation, the parcel floats
upwards until it dissolves and releases its heat into its
surroundings.  Similarly, a parcel displaced downward is unstable to
sink and cool material beneath it.  The net result is a combination of
upward and downward flows that transport heat outwards.  We call this
process \emph{convection}.  Regions in the star where
$\nabla\st{rad}>\nabla\st{ad}$ are convectively unstable and energy is
transported at least in part through the convective motion of
material.

To calculate the convective temperature gradient, we use
\emph{mixing-length theory} \citep{bohm58}.  Suppose that a parcel of
material rises adiabatically through a radial distance $\mixl$, called
the \emph{mixing length}, before dispersing.  The difference in
internal energy between a parcel of gas and its surroundings is
$\Delta u=c_p\Delta T$, where $u$ is the specific internal energy and
$c_p$ the specific heat capacity at constant pressure.  After rising
one mixing length, the difference between the temperature of the
parcel and its surroundings is
\begin{align}
\Delta T&=\left[\left(\pdif{T}{r}\right)_{\sq S}-\pdif{T}{r}\right]\mixl \\
&=\left[-T\left(\pdif{\log T}{\log p}\right)_{\sq S}\pdif{\log p}{r}+
T\pdif{\log T}{\log p}\pdif{\log p}{r}\right]\mixl \\
&=\left(\nabla-\nabla\st{ad}\right)T\frac{\mixl}{H_p}
\end{align}
where we have defined the pressure scale height $H_p=(\qdif{\log
  p}{r})^{-1}$.  The luminosity from convective heat transport is the
excess heat multiplied by the amount of material that carries it, so
we write 
\shorteq{L\st{con}=4\pi r^2\rho v\st{c}c_p\Delta T\text{,}}
where $v\st{c}$ is the convective velocity.  Buoyancy accelerates the
parcel at a rate 
$a = -g(\Delta\rho/\rho) \approx g(\Delta T/T)$, 
where
$g=Gm/r^2$ is the local acceleration due to gravity.  The blob
accelerates over one mixing length $\mixl$ so its average velocity
over the journey is $v\st{conv}=\sqrt{a\mixl/2}$.  Incorporating this
into the convective luminosity gives \shorteq{L\st{con}=4\pi r^2\rho
  c_pT\left(\frac{\mixl}{H_p}\right)
  \sqrt{\frac{gH_p}{2}}\left(\nabla-\nabla\st{ad}\right)^{3/2}\text{.}}
The total luminosity is given by combining the radiative and
convective components.  To complete the description, the mixing length
must be specified.  In the \stars{} code, the mixing length is
$\mixl=\alpha\st{MLT} H_p$, where $\alpha\st{MLT}$ (\texttt{ALPHA})
takes a default value $2$ based on approximate calibration to a solar
model.

Mixing-length theory is a crude but effective model of convection.  It
only considers local conditions and ignores the known asymmetry
between the expansion of upward flows and the corresponding
contraction of downward flows.  The mixing length can even be larger
than the radial extent of the convective region.  In addition, the
mixing length is a free parameter.  The scale factor $\alpha\st{MLT}$
is chosen so that models agree with observations of the Sun but it is
not clear that the same value of $\alpha\st{MLT}$ should apply to all
stars or all phases of evolution.  Despite these flaws, mixing-length
theory works well.  Near the photosphere, convection is very
inefficient, so the temperature gradient takes its radiative value.
In deep convective regions, such as the convective cores of massive
main-sequence stars, convection is very efficient and the temperature
gradient is nearly adiabatic.  The more difficult intermediate case
exists when convection occurs in the outer envelope as in red giants.
% Fortunately, this is the case in the Sun, where $\alpha\st{MLT}$ is
% calibrated.
% Why did you comment this out?

\subsection{Matter}

To solve the differential equations, we require three equations that
describe the microscopic properties of the stellar material as a
function of bulk properties and the composition.  They are the opacity
law $\kappa$, the nuclear and neutrino energy generation rates
$\epsilon\st{nuc}$ and $\epsilon_\nu$, and an equation of state that
relates the pressure, density and temperature.  All three \emph{matter
  equations} are functions of density, temperature and the elemental
abundances $X_i$.  Simple expressions for these properties do not
generally exist so they are drawn from data produced by either
detailed calculations or experiments.

The opacity $\kappa$ is computed by interpolating in a table of
values.  The most recent tables were compiled by
\citet{eldridge04}.  For each total metal abundance, or
\emph{metallicity}, $Z$, data are tabulated in a 5-dimensional grid of
temperature $T$, density parameter $R=(\rho/\gpcm)/(T/10^6\K)^3$,
hydrogen abundance $X\st{H}$, carbon abundance $X\st{C}$ and oxygen
abundance $X\st{O}$.  The data cover $\log_{10} (T/\K)$ from
$3$ to $10$ in steps of $0.05$, $\log_{10}R$ from $-7$ to $1$ in steps
of $0.5$, $X\st{H}$ at $0$, $0.03$, $0.1$, $0.35$ and $0.7$, and
$X\st{C}$ and $X\st{O}$ at $0$, $0.01$, $0.03$, $0.1$, $0.2$, $0.4$,
$0.6$ and $1$.
The tables are populated in the temperature range $\log_{10} (T/\K)$
from $3.75$ to $8.70$ using data from the OPAL collaboration
\citep{iglesias96}.  At lower temperatures, data are taken from
\citet{alexander94}.  The low temperature tables do not include
enhanced carbon and oxygen abundances so opacity changes due to
changes in $X\st{C}$ and $X\st{O}$ are not calculated when $\log_{10}
(T/\K)<4$.  The high-temperature regime $\log_{10}T/\K>8.70$ is filled
according to \citet{buchler76}.  \citet{stancliffe08}
% added 
replaced the
molecular opacities for $\text{H}_2$, $\text{H}_2\text{O}$,
$\text{OH}$, $\text{CO}$, $\text{CN}$ and $\text{C}_2$ with the
procedure described by \citet{marigo02}.

If the density parameter or temperature takes a value that is not
covered by the opacity table, the subroutine returns the nearest value
of the opacity that is in the table.  This occurred at the innermost
points of the models described in Chapter \ref{cqs2}.  There,
%the temperature is very high and the density very low.  In this case,
%the opacity is dominated by Compton scattering.  
the temperature is very high, the density very low and 
the opacity dominated by Compton scattering.  
%Compton scattering dominates the opacity.
The extrapolation %in effect 
returns a constant opacity although, in reality, the opacity is a %slowly 
decreasing function of temperature.  The affected region
represents less than $0.01\Msun$ of the envelope and has little effect
on the results.

Nuclear reaction rates are taken from the extensive tables of data
compiled by \citet{caughlan88}.  Cooling rates owing to neutrino
losses are drawn from those of \citet{itoh83}, \citet{munkata+87} and
\citet{itoh+89,itoh+92}.  The reaction rates provide both the energy
generated and the rate at which elements are transformed through
nuclear fusion.

The quasi-star models in Chapter \ref{cqs1} include a substantial
fraction of gas mass that is between the quasi-star envelope and the
black hole.  Using a temperature profile $T\propto r^{-1}$
\citep*{narayan00}, I estimated the composition changes owing to pp
chains assuming complete mixing down to $10^{-4}$ times the inner
radius.  I found no significant change to the hydrogen and helium
abundances and conclude that the associated energy generation is also
negligible.  Although the temperatures in these regions are well over
$10^8\K$, the densities are typically only a few $\gpcm$ and decline
rapidly.  This is two orders of magnitude smaller than at the centre
of the Sun and the objects are much shorter-lived.

The same models also neglect heat loss via neutrino emission.  I
estimated total neutrino loss rate using the analytic estimates of
\citet{itoh+96} and integrated them over the interior region in the
fiducial run and found that the neutrino losses are at most 6 per cent
of the total luminosity if the flow extends to the innermost stable
circular orbit.  Such losses would in effect decrease the radiative
efficiency but the structure is principally determined by the
convective efficiency so the envelopes remain stable against
catastrophic neutrino losses.

Finally, the static structure problem is completed by specifying an
equation of state (EoS).  The \stars{} code uses the EoS developed by
\citet{eggleton+73}.  The variable used in the program code is a
parameter $f$ which is related to the electron degeneracy parameter
$\psi$ by
\shorteq{\psi=\log\left(\frac{\sqrt{1+f}-1}{\sqrt{1+f}+1}\right)+2\sqrt{1+f}\text{.}}
In degenerate material, the pressure is calculated by approximating
the Fermi--Dirac integrals as explicit functions of the
parameter $f$.  
%In addition, hydrogen and helium ionisation can be
%expressed explicitly in terms of $f$.
The EoS package provides the pressure $p$ and density $\rho$, along
with derived parameters, in terms of $f$ and temperature $T$.  The
ionisation states of hydrogen and helium are calculated, as is the
molecular hydrogen fraction.  \citet{pols+95} described additional
non-ideal corrections owing to Coulomb interactions and pressure
ionization.  It is presumed that all metals are completely ionized.
In cool metal-free material, the electron fraction %becomes sufficiently close to 
approaches 
zero %that 
and $f$ tends to zero too.  The code
refers to $\log f$ as an intrinsic variable, which tends to negative
infinity.  To avoid associated numerical errors, an insignificant
minimum electron density, $n_{e,0}=10^{-6}\pcmcu$, was added.  I
compared evolutionary tracks for a $1\Msun$ star with and without this
addition and found no discernible difference in the results.

Quasi-stars are strongly radiation-dominated.  The EoS is very close
to a simple combination of ideal gas and radiation,
\shorteq{p=\frac{1}{3}aT^4+\frac{\rho kT}{\mu(X_i)m\st{p}}\text{,}}
where $m_p$ is the mass of a proton and $\mu$ the mean molecular
weight.  The ionisation state of the gas is important to determine
$\mu$ and the detailed EoS in the \stars{} code provides a significant
improvement over previous models of quasi-stars.

\subsection{Composition}

The structural changes that occur during a star's life are driven by
changes in its chemical composition.  The \stars{} code solves for
seven chemical species $i\in\{\chem{H}{1}$, $\chem{He}{3}$,
$\chem{He}{4}$, $\chem{C}{12}$, $\chem{N}{14}$, $\chem{O}{16}$,
$\chem{Ne}{20}$\} in the structure equations.  
% Fixed fractions of material are presumed to be $\chem{Si}{28}$ and $\chem{Fe}{56}$ and
% all other material in the form of $\chem{Mg}{24}$.  
The abundances of $\chem{Si}{28}$ and $\chem{Fe}{56}$ are presumed
to be constant.  The abundance of $\chem{Mg}{24}$ is set so that
all the abundances add up to 1.
Additional equilibrium isotopic abundances can be calculated
explicitly after a solution for the current timestep has been found
but I did not use this functionality.

%\shorteq{\pdif{}{m}\left(\Sigma\pdif{X_i}{m}\right)=\pdif{X_i}{t}+R_{d,i}X_i-R_{c,i}\text{,}}
Each chemical species can be created or destroyed in nuclear
reactions or mixed by convection.  Mixing is treated as a diffusion
process which, combined with the creation and destruction through
fusion, leads to an equation
\shorteq{\pdif{X_i}{t}=R_{\mathrm{c},i}-R_{\mathrm{d},i}X_i+\pdif{}{m}\left(\Sigma\pdif{X_i}{m}\right)\text{}}
for each chemical species.  Here, $X_i$ is the fractional mass
abundance of element $i$, $R_{\mathrm{c},i}$ and $R_{\mathrm{d},i}$ are the rates at
which the element is created and destroyed and $\Sigma$ is a diffusion
coefficient related to the linear diffusion coefficient
$D\st{MLT}=v\st{c}\mixl/3$ by $\Sigma=(4\pi r^2\rho)^2D\st{MLT}$.  In
the code, $\Sigma$ is approximated by
\shorteq{\Sigma=K\st{MLT}\left(\nabla\st{rad}-\nabla\st{ad}\right)^2
  \frac{M_*^2}{t\st{nuc}}\text{,}} 
where $K\st{MLT}=10^8$ (\texttt{RCD}) is a parameter, $M_*$ is the
total mass and $t\st{nuc}$ is the lifetime of the star in its current
evolutionary phase
% current nuclear-burning timescale of the star 
\citep{eggleton72}.

In preliminary calculations, I found that quasi-stars do not undergo
significant composition evolution.  I therefore removed the composition
equations from the calculations in Chapters \ref{cqs1} and \ref{cqs2},
which reduced the code runtime by a factor of about four.

\section{Boundary conditions}

To solve the system of differential equations for a star, we need a
set of boundary conditions.  The four structure equations are
first-order in the dependent variable so we require four boundary
conditions.  In addition, we require two equations that specify the
inner and outer value of the independent variable.  For each
second-order composition equation, two conditions are specified by
requiring no diffusion across the innermost and outermost points.  We
thus require six conditions.  Three are applied at the surface of the
star and three at the centre.

\subsection{The surface}

The surface of the star is defined where the mass co-ordinate is equal
to the total mass $M_*$, which is given as a parameter in the
calculation.  If the total mass of the star is changing because of
accretion or mass loss, the total mass is simply changed over time
consistently with the model that is being calculated.  There is no
provision for additional pressure owing to the velocity of material
leaving from or arriving at the surface.  The total luminosity $L_*$
is specified by assuming that the star radiates into a vacuum as a
black body, which gives \shorteq{L_*=4\pi R_*^2\sigma
  T\st{eff}^4\text{,}} where $R_*$ is the radius of the surface,
$\sigma$ the Stefan--Boltzmann constant and $T\st{eff}$ the
effective temperature.  The gas pressure at the surface $P\st{g}$ is given
by
\shorteq{P\st{g}=\frac{2g}{3\kappa}\left(1-\frac{L_*}{L\st{Edd}}\right)\text{,}}
where $L\st{Edd}=4\pi GcM_*/\kappa$ is the Eddington luminosity.  At
the Eddington luminosity, radiation pressure alone would balance
gravity.  If the luminosity were greater, the force of radiation would
accelerate material away from the star.

\subsection{The centre}

At the centre of the star, the formal boundary conditions are
$L=r=m=0$ but these lead to divergences in the differential
equations.  The central point of the model is calculated as an average
of the formal boundary conditions and the values at the first point
from the centre.  For a specified innermost mass $m\st{c}$, the central radius
$r\st{c}$ is defined by
\shorteq{r\st{c}=\left(\frac{3m\st{c}}{4\pi\rho\st{c}}\right)^{\sq\frac{1}{3}}}
and the central luminosity $L\st{c}$ by
\shorteq{L\st{c}=\epsilon m\st{c}\text{.}}

Quasi-stars are modelled with the \stars{} code by choosing new
boundary conditions for the innermost point.  For example, the inner
mass is at least the central black hole mass.  Because the new inner
boundaries 
%are given 
take
finite values, I omit the averaging process %that is 
used to avoid singularities at the centre.  The innermost meshpoint is
computed directly on the specified boundary.  Different boundary
conditions were used for the models described in Chapters \ref{cqs1}
and \ref{cqs2} so detailed discussions of the boundary conditions in
each set of models are deferred to Sections \ref{sbc1} and \ref{sbc2}.

\section{Implementation}

The structure, matter and composition equations define the
mathematical problem of calculating the structure and evolution of a
star.  To calculate solutions numerically, the variables are
determined at a finite number of meshpoints.  The quasi-star models
described in Chapters \ref{cqs1} and \ref{cqs2} all have 399
meshpoints; the stellar models of Chapter \ref{crg} have 199.  The
structure equations are transformed into difference equations that are
corrected iteratively until the sum of the moduli of the corrections
is smaller than a user-specified tolerance parameter (\texttt{EPS}).
The tolerance parameter is $10^{-6}$ throughout this dissertation.
Once a satisfactory solution is reached, the model is evolved forward
by one timestep and a new solution is computed.  The time evolution is
fully implicit so that the changes in the time derivatives are
computed self-consistently with the %stellar 
structure.

\subsection{Mesh-spacing}

\begin{figure}[tb]\begin{center}
\includegraphics{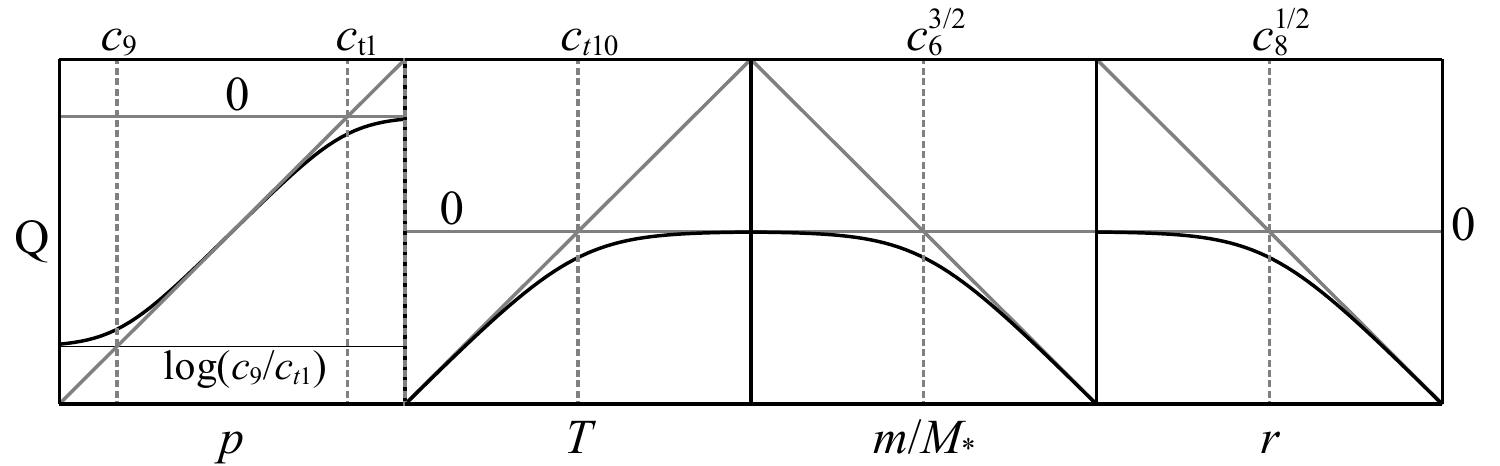}
\caption[Qualitative plots of four terms of the mesh-spacing function
  (equation \ref{emesh}).]{From 
  %top-left to bottom-right, 
    left to right, qualitative plots of the second, fourth, fifth and
    sixth terms of the mesh-spacing function (equation \ref{emesh}).
    Points are preferentially placed where the gradient of $Q$ is
    larger.  All horizontal axes are logarithmic in the given
    variable.  The parameters $c_i$ control the behaviour of each
    component and their values are given in Table \ref{tmesh}.  The
    pressure term concentrates meshpoints between $c_9$ and $c_{t1}$
    and the temperature term distributes them near the photosphere.
    The mass and radius terms both remove points from the centre,
    where less resolution is required.}
\label{fmesh}\end{center}\end{figure}

The independent variable in the structure equations can be any
monotonic variable in the star.  The \stars{} code uses the meshpoint
number $k$ distributed such that a function $Q$ is equally spaced on
the mesh.  In other words, we introduce an equation $\qdif{Q}{k}=C$,
where $C$ is an eigenvalue.  The function $Q$, called the
\emph{mesh-spacing function}, can depend on any of the variables in
the model provided that it remains monotonic in $k$ and has analytic
derivatives with respect to the variables.  By choosing a function
that varies rapidly in regions where greater numerical accuracy is
required, we focus the meshpoints automatically on these regions of
interest.

The present version of the \stars{} code uses a mesh-spacing function
\begin{align}
\nonumber Q&=c_4\log p+c_5\log\left(\frac{p+c_9}{p+c_{t1}}\right)
+c_2\log\left(\frac{p+c_{10}}{p+c_{t1}}\right)
+c_7\log\left(\frac{T}{T+c_{t10}}\right) \\
&\qquad-\log\left(\frac{1}{c_6}\left(\frac{m}{M_*}\right)^{2/3}+1\right)
-c_3\log\left(\frac{r^2}{c_8}+1\right)
\label{emesh}\end{align}
The default parameters $c_i$ are specified in table
\ref{tmesh}\footnote{The labels of the parameters $c_i$ have been
  chosen to match the variables used in the code.} and the qualitative
forms of the second, fourth, fifth and sixth terms are shown in
Fig.~\ref{fmesh}.  Points are concentrated towards where the gradient
of $Q$ is larger.  Each term is chosen so that points are either
concentrated towards regions that require additional resolution or
distributed away from regions that do not.  The default values for the
parameters gave satisfactory results and were used in all models in
this dissertation.

The main contributor the the mesh-spacing is the fifth term, which
distributes points like $(m/M_*)^\frac{2}{3}$ but reduces resolution
at the centre.  The first term spreads points evenly in the pressure
co-ordinate.  The second and third terms allow points to be
concentrated within
% a specific
the pressure ranges defined by $c_9$ and $c_{t1}$ and between $c_{10}$
and $c_{t1}$.  
% In my models, 
When the parameters have their default values,
$c_{t1}$ is so large that the denominator
under the logarithm can be ignored as a constant and
% and points are again shifted away from the centre.  
the third term is ignored (i.e.~$c_2=0$).
% in all my models.  
The fourth term concentrates points towards the upper
atmosphere, where $T\lesssim20\,000\K$ and extra resolution is
required to properly resolve the hydrogen ionization zones.  The sixth
term weakly moves points toward regions with $r\gtrsim10^9\cm$.

\begin{table}\begin{center}
\caption{Default parameters in the mesh-spacing function (equation \ref{emesh}).}
\begin{tabular}{cccccc}
\toprule[1pt]
$c_1$ & $c_2$ & $c_3$ & $c_4$ & $c_5$ & $c_6$ \\
\texttt{C(1)} & \texttt{C(2)} & \texttt{C(3)} & 
\texttt{C(4)} & \texttt{C(5)} & \texttt{C(6)} \\
$9.99$& $0.00$ & $0.05$ & $0.50$ & $0.15$ & $0.02$ \\
\midrule
$c_7$ & $c_8/\cmsq$ & $c_9/\dynpcmsq$ & $c_{10}/\dynpcmsq$ & $c_{t1}/\dynpcmsq$ & $c_{t10}/\K$\\
\texttt{C(7)} & \texttt{C(8)} & \texttt{C(9)} & 
\texttt{C(10)} & \texttt{CT1} & \texttt{CT10} \\
$0.45$ & $\sci{1.0}{18}$ & $\sci{1.0}{15}$ & $\sci{3.0}{19}$ & $10^{10c_1}$ & $\sci{2}{4}$ \\
\bottomrule[1pt]
\end{tabular}
% \begin{tabular}{ccccccc}
% \hline
% \hline
% $c_1$ & $c_2$ & $c_3$ & $c_4$ & $c_5$ & $c_6$ & $c_7$ \\
% \texttt{C(1)} & \texttt{C(2)} & \texttt{C(3)} & \texttt{C(4)} & 
% \texttt{C(5)} & \texttt{C(6)} & \texttt{C(7)} \\
% $9.99$& $0.00$ & $0.05$ & $0.50$ & $0.15$ & $0.02$ & $0.45$ \\
% \end{tabular}
% \begin{tabular}{ccccc}
% \hline
% $c_8/\cmsq$ & $c_9/\dynpcmsq$ & $c_{10}/\dynpcmsq$ & $c_{t1}/\dynpcmsq$ & $c_{t10}/\K$\\
% \texttt{C(8)} & \texttt{C(9)} & 
% \texttt{C(10)} & \texttt{CT1} & \texttt{CT10} \\
% $\sci{1.0}{-4}$ & $\sci{1.0}{15}$ & $\sci{3.0}{19}$ & $10^{10c_1}$ & $\sci{2}{4}$ \\
% \hline
% \end{tabular}
\label{tmesh}
\end{center}\end{table}

%\subsection{Chemical Diffusion}

\subsection{Difference equations}

Each differential equation containing a spatial derivative is recast
as a difference between values at two neighbouring points, denoted by
subscript $k$ and $k+1$.  Averages between two neighbouring points are
denoted by subscript $k+\frac{1}{2}$.  The distribution of points in
the mesh is solved through
\shorteq{m_{k+1}-m_k=\left(\pdif{m}{k}\right)_{\sq
    k+\frac{1}{2}}\text{.}}
% The structure equations for mass, momentum and energy conservation become
The structure equations for mass conservation, hydrostatic equilibrium
and energy generation become
% \shorteq{r_{k+1}^3-r_k^3=\left(\frac{3}{4\pi\rho}
%     \pdif{m}{k}\right)_{k+\frac{1}{2}}\text{,}}
% \shorteq{\log p_{k+1}-\log p_k=-\left(\frac{Gm}{4\pi r^4p}
%     \pdif{m}{k}\right)_{k+\frac{1}{2}}\text{}} 
\begin{gather}
r_{k+1}^3-r_k^3=\left(\frac{3}{4\pi\rho}\pdif{m}{k}\right)_{\sq k+\frac{1}{2}}\text{,} \\
\log p_{k+1}-\log p_k=-\left(\frac{Gm}{4\pi r^4p}\pdif{m}{k}\right)_{\sq k+\frac{1}{2}}\text{}
\end{gather}
and
\shorteq{L_{k+1}-L_k=\left(\epsilon
    \pdif{m}{k}\right)_{\sq k+\frac{1}{2}}\text{.}}
The energy transport equation is
\shorteq{\log T_{k+1}-\log T_k=-\left(\nabla\frac{Gm}{4\pi r^4p}
    \pdif{m}{k}\right)_{\sq k+\frac{1}{2}}\text{.}}
The composition equations become
\begin{align} \nonumber
\Sigma_{k+\frac{1}{2}}&\left(X_{i,k+1}-X_{i,k}\right)
  -\Sigma_{k-\frac{1}{2}}\left(X_{i,k}-X_{i,k-1}\right)=
\left(\frac{X_{i,k}-X_{i,k}^0}{\Delta t}+R_{i,k}\right)\left(\pdif{m}{k}\right)_{\sq k}\\
&\qquad {}-\left(X_{i,k+1}-X_{i,k}\right)\operatorname{Ramp}\left(\pdif{m}{t}\right)_{\sq k}
+\left(X_{i,k}-X_{i,k-1}\right)\operatorname{Ramp}\left(-\pdif{m}{t}\right)_{\sq k}\text{,}
\end{align}
where I have used the ramp function,
$\operatorname{Ramp}(x)=(x+|x|)/2$, and $X_{i,k}^0$ is the abundance
of element $i$ at the previous timestep.  The boundary conditions for
the composition equations are in effect $\Sigma_{k\pm\frac{1}{2}}=0$
as appropriate.  Note that the simultaneous evaluation of the
structure, composition and non-Lagrangian mesh is a defining feature
of the Cambridge \stars{} code.

\subsection{Solution and timestep}

The system of difference equations and their boundary conditions is
solved by a relaxation method \citep{henyey+59}.  Given an initial
approximate solution, the variables are varied by a small amount
(\texttt{DH0}) to determine derivatives of the equations with respect
to the variables.  The matrix of these derivatives is inverted to find
a better solution.  This process is iterated through a Newton--Rhapson
root-finding algorithm until the sum of corrections to the solution is
smaller than some parameter (\texttt{EPS}).  The solution is then
recorded and the next model in time is computed by the same process.

The timestep is calculated by comparing the sum of fractional changes
in each variable at each point with an input parameter (\texttt{DDD}).
The next timestep is chosen so that the two numbers are equal but the
timestep can only be rescaled within a user-specified range
(\texttt{DT1}, \texttt{DT2}).  If the relaxation method fails to
converge on a sufficiently accurate model, the previous model is
abandoned, the code restores the anteprevious model and the timestep
is reduced by a factor $0.2$.  I appended code so that if a value of
\emph{NaN} (not a number) is encountered during the convergence tests,
the run is immediately aborted.

The \stars{} code is versatile and adaptable.  Its mesh-spacing
function allows it to rapidly model many phases of stellar evolution
without interruption and the concise source code (less than 3000 lines
of \textsc{fortran}) makes modification straightforward.  In the next two
chapters, I modify the central boundary conditions to model
quasi-stars with accurate microphysics.
 % STARS 
\clearpage\thispagestyle{tocstyle}
\begin{savequote}[80mm]
  We know that for a long time everything we do will be nothing more
  than the jumping off point for those who have the advantage of
  already being aware of our ultimate results. 
  \qauthor{Norbert Wiener, 1956}
\end{savequote}

\chapter{Bondi-type quasi-stars}
\label{cqs1}

In this chapter, I present models of quasi-star envelopes where the
inner black hole (BH) accretes from a spherically symmetric flow.  The
accretion flow is based on the canonical model of \citet{bondi52} and
these quasi-star models are referred to as \emph{Bondi-type}
quasi-stars in later chapters.  \citet*[][hereinafter
BRA08]{begelman08} studied these structures using analytic estimates
and basic numerical results.  I model the envelope using the Cambridge
\stars{} code, which accurately computes the opacity and ionization
state of the envelope.  In Section \ref{sbc1}, I describe the boundary
conditions used in this chapter.  The results of a fiducial run are
presented and discussed in Section \ref{sfid1} and further runs with
varied parameters are described in Section \ref{smore1}.  In Section
\ref{sbra}, the \stars{} models are compared with the results of
BRA08.

The models in this chapter lead to two main results that are explored
further in subsequent chapters.  First, for a given inner radius, the
models stop evolving once the BH reaches just over one-tenth of the
total mass of the quasi-star regardless of the total mass of the
system or the radiative efficiency of the accretion flow.  Polytropic
models of quasi-star envelopes exhibit the same robust fractional
limit on the BH mass and the mechanism that causes the maximum is thus
a feature of mass conservation and hydrostatic equilibrium.  In
Chapter \ref{cscl}, I show that the BH mass limit found below is
related to the Sch\"onberg--Chandrasekhar limit and, in Chapter
\ref{crg}, I consider some consequences for the evolution of giant
stars.

Secondly, the models are strongly sensitive to the inner boundary
radius.  Roughly speaking, the final mass of the BH is inversely
proportional to the inner radius.  While the boundary conditions used
here are reasonable, the models they produce are unreliable even
though the qualitative evolution may be realistic.  In the next
chapter, I address this by constructing models with a different set of
boundary conditions and find that the evolution is qualitatively very
different from what is described in this one.

\section{Boundary conditions}
\label{sbc1}

Stellar evolution codes normally solve for the interior boundary
conditions $r,\,m,\,L=0$, where $r$ is the radial co-ordinate, $m$ is
the mass within a radius $r$ and $L$ is the luminosity through the
sphere of radius $r$.  To model quasi-stars, these boundary conditions
are replaced with a prescription for the BH's interaction with the
surrounding gas as described below.
% The conditions are derived presuming that the pressure in the
% envelope is dominated by radiation.
\citet{loeb94} showed that a radiation-dominated fluid in hydrostatic
equilibrium and not generating energy must become convective so the
quasi-star envelope should be approximated by a gas with polytropic
index $n=3$.  This presumption is used below but in the subsequent
calculations the adiabatic index ($\gamma=\partial\log
p/\partial\log\rho$ at constant entropy, where $p$ is the pressure and
$\rho$ the density) is determined self-consistently by the
equation-of-state module in the code.  

\begin{figure}\begin{center}
\includegraphics[angle=90,width=13.8cm]{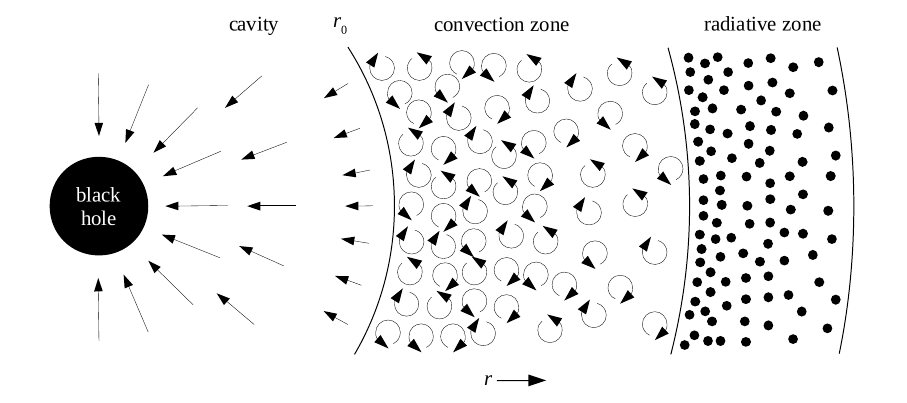}
\caption[Model of the radial structure of Bondi-type quasi-stars.]
{Diagram of the basic model adopted for the radial structure of
  Bondi-type quasi-stars.  The radial co-ordinate increases to the
  right but the structure is not drawn to scale.  Inside the inner
  radius $r_0$, the gravitational potential of the BH overwhelms the
  thermal energy of the gas and material falls on to the central BH.
  Outside $r_0$, there is a convective envelope surrounded by a
  radiative atmosphere.  In models with small BH masses, additional
  intermediate radiative zones can be present.}
\label{fqsd3}
\end{center}\end{figure}

Fig.~\ref{fqsd3} shows the basic components of the model adopted in
this chapter.  The inner radius $r_0$ is a fixed multiple of the Bondi
radius, inside which material falls on to the central BH.  A
substantial amount of mass can be found inside this cavity and its
mass is included in the inner mass boundary condition.  The \stars{}
code models the hydrostatic envelope outside the inner radius.  Near
$r_0$, the envelope is convective but it is necessarily radiative at
the surface because of the surface boundary conditions.  Although not
included in the diagram, additional intermediate radiative zones can
exist (see Fig.~\ref{ffidevol}).

\subsection{Radius}

The radius of the inner boundary of the envelope should be the point
at which some presumption of the code breaks down.  Following BRA08, I
choose the \emph{Bondi radius} $\rb$, at which the thermal energy of
the fluid particles equals their gravitational potential energy with
respect to the BH.  By definition,
\shorteq{\frac{1}{2}mc\st{s}^2=\frac{Gm\Mbh}{\rb}\label{bondidef1}\text{,}}
so 
\shorteq{\rb=\frac{2G\Mbh}{c\st{s}^2}\label{bondidef2}\text{,}}
where $m$ is the mass of a test particle, $c\st{s}=\sqrt{\gamma
  p/\rho}$ the adiabatic sound speed, $\Mbh$ the mass of the BH and
$G$ Newton's gravitational constant.

BRA08 used the inner boundary condition
\shorteq{r_0=\frac{G\Mbh\rho_0}{2p_0}=\frac{1}{4}\gamma\frac{2G\Mbh}{c\st{s}^2}
  =\frac{1}{4}\gamma\rb\text{.}}  
We therefore implement the radial boundary condition
\shorteq{r_0=\frac{1}{b}\rb} 
where $b$ is a parameter that varies the inner radius.  Larger values
of $b$ correspond to smaller inner radii and therefore a stronger
gravitational binding energy there.  Our main results use $b=1$
whereas the boundary condition used by BRA08 corresponds to $b=3$.
% for an $n=3$ polytrope.

\subsection{Mass}
\label{ssmbc1}

For the mass boundary condition, consider the mass of the gas inside
the cavity defined by the Bondi radius (equation \ref{bondidef2}).
By definition,
\shorteq{M\st{cav}=\int^{\rb}_{\rs}4\pi r^2\rho(r)\mathrm{d}r\label{mcav}\text{,}} 
where $\rs$ is the Schwarzschild radius.  Using a general relativistic
form of the equation introduces terms of order $\rs/r_0$
\citep{thorne+77}, which I ignore because all our models have
$\rs\ll r_0$.

To determine the mass of gas inside the cavity, we must assume a
density profile of the material there because the code does not model
this region.  The envelope is supported by radiation pressure and is
expected to radiate near the Eddington limit for the entire
quasi-star.  This is much greater than the same limit for the BH
alone.  The excess flux drives bulk convective motions.  The radial
density profile of the accretion flow then depends on whether angular
momentum is transported outward or inward.  In the former case, the
radial density profile is proportional to $r^{-\frac{3}{2}}$ whereas,
in the latter case, it is proportional to $r^{-\frac{1}{2}}$
\citep*{narayan00,quataert00}.  We presume that the viscosity owing to
small scale magnetic fields is sufficiently large to transport angular
momentum outwards even if convection transports it inwards and thus
take $\rho(r)\propto r^{-\frac{3}{2}}$.  In Section \ref{sscav}, I
construct a model presuming that 
%with
$\rho(r)\propto r^{-\frac{1}{2}}$ and
find that this change to the density profile inside the cavity has
little effect.

Given the density $\rho(r_0)=\rho_0$ at the inner boundary, the 
density profile must be
\shorteq{\rho(r)=\rho_0\bracfrac{r}{r_0}^{\sq-\frac{3}{2}}\sq\text{.}}
Evaluating equation (\ref{mcav}), presuming $\rs\ll r_0$, we obtain
\shorteq{M\st{cav}=\frac{8\pi}{3}\rho_0r_0^3\label{mcavint}\text{.}}

The cavity mass $M\st{cav}$ can be estimated as follows.  
In a radiation-dominated $n=3$ polytrope, 
the pressure and density are related by
\shorteq{p=\bracfrac{k}{\mu m\st{H}}^{\sq\frac{4}{3}}
\bracfrac{3(1-\beta)}{a\beta^4}^{\sq\frac{1}{3}}\rho^\frac{4}{3}
= K\rho^\frac{4}{3}}
\citep{eddington18}, where 
$k$ is Boltzmann's constant, 
$\mu$ the mean molecular weight of the gas, 
$m\st{H}$ the mass of a hydrogen atom and 
$\beta=p\st{g}/p$ the ratio of gas pressure to total pressure.  
Taking the adiabatic sound speed to be $c\st{s}=\sqrt{4p/3\rho}$,
evaluating the Bondi radius using equation (\ref{bondidef2}) and
substituting into equation (\ref{mcavint}), we obtain
\shorteq{M\st{cav}=\frac{8\pi}{3}\bracfrac{3G\Mbh}{2K}^{\sq3}\text{.}}
\citet{fowler64} gives $\beta=4.3(M_*/\Msun)^{-\frac{1}{2}}/\mu$,
where $M_*$ is the total mass of the object.  For a totally ionized
mixture of 70 per cent hydrogen and 30 per cent helium by mass,
$\mu=0.615$, so for a quasi-star of total mass $10^4\Msun$, as in the
fiducial result, $M\st{cav}=\Mbh$ when $\Mbh\approx390\Msun$.  The
cavity mass $M\st{cav}$ must be included in the mass boundary
condition so I use
%The mass boundary condition is therefore
\shorteq{M_0=\Mbh+M\st{cav}\label{mbc}\text{,}}
where $M\st{cav}$ is given by equation (\ref{mcavint}).  

\subsection{Luminosity}
\label{sslbc1}

The luminosity is determined by the mass accretion rate
through the relationship
\shorteq{\Lbh=\epsilon\dot{M}c^2\label{lum1}\text{,}} where $c$ is the
speed of light, $\dot{M}$ the rate of mass flow across the base of
the envelope and $\epsilon$ the radiative efficiency, the fraction
of accreted rest mass that is released as energy.  This fraction is
lost from the system as radiation so the total mass of the quasi-star
decreases over time.  The rate of accretion on to the BH is
$\dMbh\equiv(1-\epsilon)\dot{M}$.  i.e. the amount of infalling matter
less the radiated energy.  The luminosity condition is related to the
BH accretion by
\shorteq{\Lbh=\frac{\epsilon}{1-\epsilon}\dMbh c^2=\epsilon'\dMbh c^2\text{.}}
It is thus implicitly assumed that material travels from the base of the 
envelope to the event horizon within one timestep.  The material actually 
falls inward on a dynamical timescale so this condition is already implied 
by the presumption of hydrostatic equilibrium.

To specify the %mass 
accretion rate, % at the inner boundary 
we begin
with the adiabatic Bondi accretion rate \citep{bondi52},
\shorteq{\dot{M}\st{Bon}=4\pi\lambda\st{c}\frac{(G\Mbh)^2}{c\st{s}^3}\rho_0
  =\pi\lambda\st{c}r_0^2\rho c\st{s}\label{dmbhbondi}\text{,}} 
where $\lambda\st{c}$ is a factor that depends on the adiabatic index
$\gamma$ as described by equation (18) of \citet{bondi52}.  
%For the case of 
When
$\gamma=4/3$, $\lambda\st{c}=1/\sqrt{2}$.  Almost all of this
flux is carried away from the BH by convection.  The convective flux
is on the order of $\rho v\st{c}\mixl \udif{u}{r}$, where
$v\st{c}$ is the convective velocity,
$\mixl$ the mixing length and
$u$ the specific internal energy of the gas \citep{owocki03}.
The convective velocity is at most equal to the adiabatic sound speed
$c\st{s}$ because if material travelled faster it would presumably
rapidly dissipate its energy in shocks.  The mixing length is on the
order of the pressure scale height $p(\udif{p}{r})^{-1}$ and the specific
internal energy gradient is roughly $(\udif{p}{r})/\rho$.  The maximum
convective flux is therefore on the order of $pc\st{s}$ so the maximum
luminosity is
% \begin{align}
% L\st{con,max}&=4\pi r_0^2 pc\st{s} =\frac{4}{\gamma}\pi r_0^2c\st{s}^3\rho \\
% &=\frac{4}{\gamma\lambda\st{c}}\dot{M}\st{Bon}c\st{s}^2\text{.}
% \end{align}
\shorteq{L\st{con,max}=4\pi r_0^2 pc\st{s} =\frac{4}{\gamma}\pi r_0^2c\st{s}^3\rho
  =\frac{4}{\gamma\lambda\st{c}}\dot{M}\st{Bon}c\st{s}^2\text{.}}
In order to limit the luminosity to the convective maximum, the
accretion rate is reduced by a factor
$4c\st{s}^2/\gamma\lambda\st{c}\epsilon' c^2$.  I therefore assume
that the actual convective flux is some fraction of the maximum
computed above and implement the mass accretion rate
\shorteq{\dMbh=16\pi\frac{\eta}{\epsilon'\gamma}\frac{(G\Mbh)^2}{c\st{s}c^2}\rho\text{,}
\label{dmbhbc}}
where $\eta$ is the convective efficiency.  
In the fiducial run, $\eta=\epsilon=0.1$.

\section{Fiducial model}
\label{sfid1}

I begin the exposition of the results by selecting a run that
demonstrates the qualitative features of a model quasi-star's
structure and evolution.  Thereafter, I vary some of the parameters to
explore how the behaviour is affected by such changes.
The results presented in this section describe a model quasi-star with
initial total mass (BH, cavity gas and envelope) $M_*=10^4\Msun$,
initial BH mass $0.0005M_*=5\Msun$ and a uniform composition of 0.7
hydrogen and 0.3 helium by mass.  The envelope is allowed to relax to
thermal equilibrium before the BH begins accreting.

\subsection{Structure}

% \begin{table}
% %\label{tprof1}
% \begin{tabular}{@{}crcr@{}lccr@{}lr@{}lr@{}lr@{}l@{}}
% \hline
% \hline
% $t$        &$\Mbh$ &$\dMbh$           &\mchead{$M\st{cav}$}
% & $L_*$         &$\rho_0$&\mchead{$T_0$}&\mchead{$T\st{eff}$}& 
% \mchead{$\rb$}  & \mchead{$R_*$} \\
% /$10^6\yr$&/\Msun&/$10^{-4}\Msunpyr$&\mchead{/\Msun}&/$10^8\Lsun$
% &/\gpcm &\mchead{/$10^5\K$}&\mchead{/$10^3\K$}& 
% \mchead{/$100\Rsun$}&\mchead{/$10^4\Rsun$} \\
% \hline
% 0.00  &5      &2.14   &    0&.00 &3.48&$\sci{8.41}{-5}$ &40&.3&14&.1&0&.0171&0&.312\\
% 0.51  &100    &1.79   &    3&.80 &2.92&$\sci{5.48}{-8}$ &3&.55&5&.22&3&.66&2&.09 \\
% 1.03  &200    &2.08   &   25&.3  &3.40&$\sci{1.30}{-8}$ &2&.23&4&.77&11&.1&2&.70 \\
% 2.23  &500    &2.96   &  242&    &4.84&$\sci{2.41}{-9}$ &1&.33&4&.55&41&.3&3&.54 \\
% 3.70  &1000   &3.70   & 1380&    &6.06&$\sci{6.42}{-10}$&0&.88&4&.49&115& &4&.08 \\
% 4.22  &1194   &3.56   & 3311&    &5.83&$\sci{3.74}{-10}$&0&.71&4&.51&184& &3&.96 \\
% \hline
% \end{tabular}
% \caption
% {Properties of the fiducial model for increasing values of $\Mbh$.
% The first and last entries correspond to the initial and final models in the run, 
% respectively.  Density profiles are plotted in Fig.~\ref{fprof1}.}
% \end{table}

\begin{table}\begin{center}
  \caption[Properties of the fiducial model as $\Mbh$ increases.]
  {Properties of the fiducial model as $\Mbh$ increases.  The first
    and last entries correspond to the initial and final models in
    the run, respectively.  Density profiles are plotted in
    Fig.~\ref{fprof1}.}
\begin{tabular}{@{}crcr@{}lr@{}lr@{}lr@{}lr@{}l@{}}
\toprule[1pt]
$t$ &$\Mbh$ &$\dMbh$ &\mchead{$M\st{cav}$}&\mchead{$T_0$}&\mchead{$r_0$} \\
/$10^6\yr$&/\Msun&/$10^{-4}\Msunpyr$&\mchead{/\Msun}
&\mchead{/$10^5\K$}&\mchead{/$100\Rsun$}& \\
\midrule
0.00  &5      &2.14&       0&.00 &40&.3 &  0&.0171\\
0.51  &100    &1.79&       3&.80 & 3&.55&  3&.66  \\
1.03  &200    &2.08&      25&.3  & 2&.23& 11&.1   \\
2.23  &500    &2.96&     242&    & 1&.33& 41&.3   \\
3.70  &1000   &3.70&    1380&    & 0&.88&115&     \\
4.22  &1194   &3.56&    3311&    & 0&.71&184&     \\
\midrule
 &  & $L_*$  &\mchead{$\rho_0$}&\mchead{$T\st{eff}$}& \mchead{$R_*$} \\
 &  & /$10^8\Lsun$ &\mchead{/\gpcm} &\mchead{/$10^3\K$}& \mchead{/$10^4\Rsun$} \\
\midrule
0.00  &5      &3.48&\mchead{$\sci{8.41}{-5}$} &14&.1&0&.312 \\
0.51  &100    &2.92&\mchead{$\sci{5.48}{-8}$} &5&.22&2&.09  \\
1.03  &200    &3.40&\mchead{$\sci{1.30}{-8}$} &4&.77&2&.70  \\
2.23  &500    &4.84&\mchead{$\sci{2.41}{-9}$} &4&.55&3&.54  \\
3.70  &1000   &6.06&\mchead{$\sci{6.42}{-10}$}&4&.49&4&.08  \\
4.22  &1194   &5.83&\mchead{$\sci{3.74}{-10}$}&4&.51&3&.96  \\
\bottomrule[1pt]
\end{tabular}
\label{tprof1}
\end{center}\end{table}

\begin{figure}\begin{center}
\includegraphics{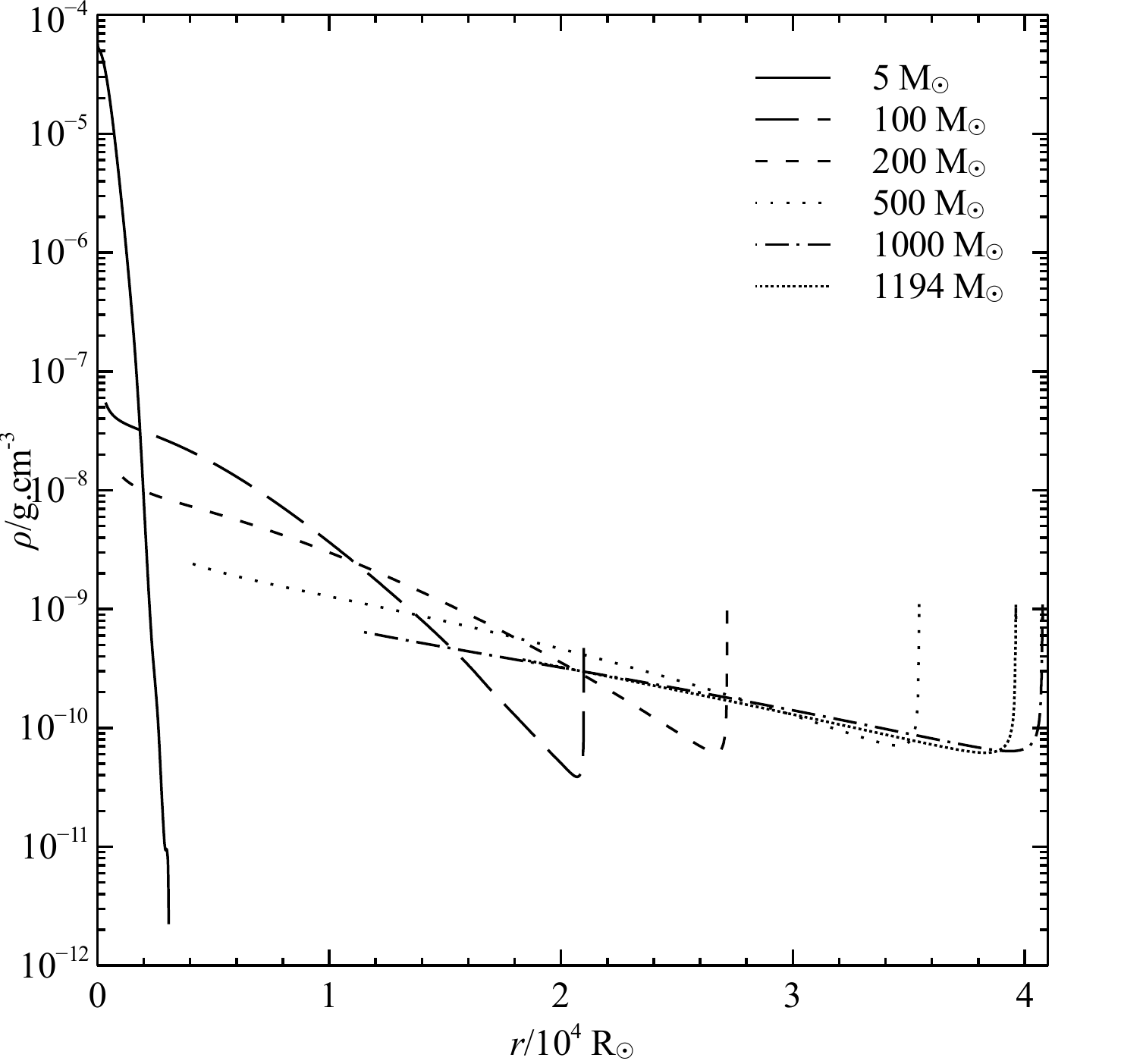}
\caption[Density profiles of the fiducial quasi-star.]
  {Plot of density against radius for models in the fiducial run
  with $\Mbh/\Msun=5,100,200,500,1000$ and $1194$ (see Table
  \ref{tprof1}).  At the base of the envelope the density profile
  steepens because a steeper pressure gradient is required to balance
  the BH gravity.  In the outer layers the density is inverted, as
  discussed by BRA08.  In the initial model, the inner radius of
  $1.66\Rsun$ is too small to be seen.}
\label{fprof1}
\end{center}\end{figure}

In the model, the luminosity is approximately equal to the Eddington
luminosity at the boundary of the innermost convective layer.  The
accretion rate varies between about $\sci{1.8}{-4}$ and
$\sci{3.7}{-4}\Msunpyr$ as the convective boundary moves.  The details
of the variation are described in Section \ref{ssevol1}.  A corollary
of the self-limiting behaviour is that the only major effect of
changing the material composition is to change its opacity and
therefore the Eddington limit.  The accretion rate changes but the
structure is almost entirely unaffected.  In the convective regions,
the envelope has an adiabatic index of about $1.34$ (corresponding to
a polytropic index $n\approx2.90$) confirming that the envelope is
approximated by an $n=3$ polytrope.  The boundaries of the convective
regions depend on the ionization state of the gas but, for most of the
evolution, all but the outermost few $10\Msun$ are convective.

Fig.~\ref{fprof1} shows a sequence of density profiles of the envelope
when $\Mbh/\Msun=5$, $100$, $200$, $500$, $1000$ and $1194$.  Further
parameters are listed in Table \ref{fprof1}.  These profiles
demonstrate two features of the envelope structure.  First, the centre
is condensed as can be seen from the rise in the density at the
innermost radii.  This is clearer for smaller BH masses.  It is caused
by the lack of pressure support at the inner boundary.  To maintain
hydrostatic equilibrium, the equations require
\shorteq{\left.\tdif{p}{r}\right|_{r_0}=-\frac{GM_0\rho}{r_0^2}} 
at the inner boundary so the pressure gradient steepens and the
density gradient follows.  \citet{huntley75} called such structures
\emph{loaded polytropes}.

Secondly, the density in the outer layers is usually inverted as in
all but the first density profile in Fig.~\ref{fprof1}.  The density
inversion appears once the photospheric temperature $T\st{eff}$ drops
below about $8000\K$.  Thereafter, the surface opacity increases owing
to hydrogen recombination, the Eddington luminosity falls and the
quasi-star's luminosity apparently exceeds the limit.
% It is known that hydrostatic models can sustain this super-Eddington luminosity
% through a density inversion \citep{langer97}.  
Density inversions are possible when the temperature gradient becomes
strongly superadiabatic \citep{langer97}, as is the case in regions
where convection occurs but is inefficient.
As an additional check, we calculated the volume-weighted average of
$(3\gamma-4)p$ and found it to be positive, indicating dynamical (but
not pulsational) stability \citep[][p. 1057]{cox68}.
%\citep[][\S27.3b]{cox68}.

\subsection{Evolution}
\label{ssevol1}

\begin{figure}\begin{center}
\includegraphics{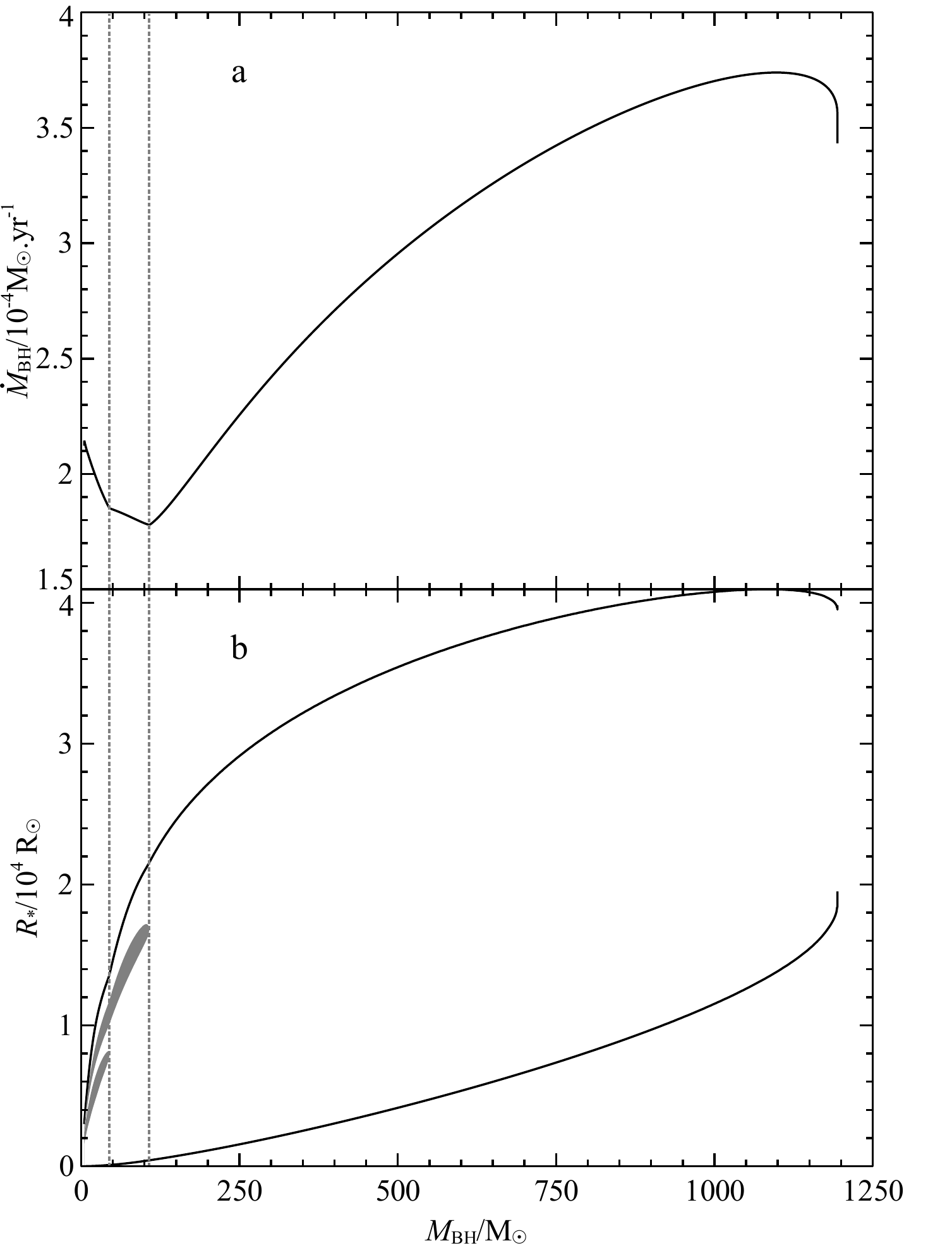}
% \caption[BH accretion rate (a) and radial extent of convection zones
% (b) in the fiducial quasi-star.]  {Plots of (a) the BH accretion rate
%   $\dMbh$ and (b) the radial extent of convective regions against BH
%   mass $\Mbh$ for the fiducial run.  The shaded regions in (b) are the
%   radiative (convectively stable) parts of the envelope.  The solid
%   lines show the extent of the hydrostatic envelope.  The qualitative
%   changes in the locations of convective boundaries causes the
%   discontinuities in the gradient of the accretion rate.  The
%   outermost layer of the envelope is radiative but is too narrow to be
%   seen here.}
\caption[BH accretion rate (a) and structure of the envelope (b) in
the fiducial quasi-star.]{Plots of (a) the BH accretion rate $\dMbh$
  and (b) the radial structure of the envelope against BH mass $\Mbh$
  for the fiducial run.  The lower and upper solid lines in (b) are
  the inner and outer radii of the envelope.  The shaded areas
  correspond to radiatively stable regions, whose disappearance leads
  to the discontinuities in the gradient of the accretion rate, as
  shown by the dashed vertical lines.  The surface layers of the
  envelope are also radiative but too narrow to be seen here.  The
  rest of the envelope is convective.}
\label{ffidevol}
\end{center}\end{figure}

The sequence of density profiles in Fig.~\ref{fprof1} shows the
interior density decreasing over time.  If we consider equation
(\ref{dmbhbc}), ignoring constants and using $p\propto\rho^\gamma$,
then
\shorteq{\dMbh\propto\Mbh^2\rho^\frac{3-\gamma}{2}\text{.}} The
accretion rate $\dMbh$ 
% is kept approximately constant by the Eddington limit and $\Mbh$ is
% always increasing.  Thus,
never increases faster than $\Mbh^2$ so
for any reasonable adiabatic index ($\gamma<3$), $\rho$ decreases at
the inner boundary.  Initially, the density decreases rapidly and the
envelope expands owing to the hydrogen opacity peak at the
surface.  The expansion or contraction of the surface is at most about
$0.1\Rsun\yr^{-1}$, which is five orders of magnitude smaller than the
free-fall velocity.  The models are thus still in hydrostatic
equilibrium.

Fig.~\ref{ffidevol}a shows the accretion rate on to the BH as a
function of its mass.  Fig.~\ref{ffidevol}b shows the locations of
convective boundaries as a function of BH mass and demonstrates how
the rapid changes of the BH accretion rate while $\Mbh<120\Msun$
coincide with the disappearance of radiative regions owing to the
decreasing density throughout the envelope.

Before the end of the evolution the accretion rate achieves a local
maximum.  At the same time, the photospheric temperature reaches a
local minimum and the envelope radius a maximum (see
Fig.~\ref{ffidevol}b).  We do not have a simple explanation for this but
believe it is related to the increasing mass and radius of the inner
cavity.  Whereas a giant expands owing to contraction of the core, the
inner part of the quasi-star is in effect expanding so the envelope
is evolving like a giant in reverse.  Initially, the expansion of the
inner radius is greater than the relative contraction of the surface
but the situation reverses when the surface radius reaches its maximum.

\subsection{Inner mass limit}
\label{sslimit}

In the fiducial run, evolution beyond the final BH mass of $1194\Msun$
is impossible.  The code reduces the timestep below the dynamical
timescale indicating that we cannot construct further models that
satisfy the structure equations.  The non-existence of further
solutions can be demonstrated by constructing polytropic models of
quasi-star envelopes as follows.  A complete derivation of the
Lane--Emden equation and an analysis of its homology-invariant form is
provided in Appendix \ref{auvp} but the relevant details are
reproduced below.

Consider the equations of hydrostatic equilibrium and mass
conservation truncated at some radius $r_0$ and loaded with some mass
$M_0$ interior to that point.  The equations are
\shorteq{\tdif{p}{r}=-\frac{Gm\rho}{r^2}}
and
\shorteq{\tdif{m}{r}=4\pi r^2\rho}
with central boundary condition $\left.m\right|_{r_0}=M_0$,
where $r_0$ and $M_0$ are fixed.  We scale the pressure and density
using the usual polytropic assumptions
\shorteq{p=K\rho^{1+\frac{1}{n}}}
and
\shorteq{\rho=\rho_0\theta^n\text{.}}
We define the dimensionless radius by
\shorteq{r=\alpha\xi\text{,}}
where
\shorteq{\alpha^2=\frac{(n+1)K}{4\pi G}\rho_0^{\frac{1}{n}-1}\text{.}}
We scale the mass interior to a sphere of radius $r$ by defining 
\shorteq{\phi(\xi)=\frac{m}{4\pi\rho_0\alpha^3}\text{.}}
%\citep{huntley75}.  
The non-dimensional form of the equations is then
\shorteq{\tdif{\theta}{\xi}=-\frac{1}{\xi^2}\phi\label{dtheta}}
and
\shorteq{\tdif{\phi}{\xi}=\xi^2\theta^n\label{dphi}}
with boundary conditions $\theta(\xi_0)=1$ and
$\phi(\xi_0)=\phi_0$ (where, by definition, $\xi_0=r_0/\alpha$).\footnote
{If one takes $\xi_0=\phi_0=0$, differentiates equation 
(\ref{dtheta}) and substitutes for $\udif{\phi}{\xi}$ using equation 
(\ref{dphi}), one arrives at the usual Lane--Emden equation.}

The dimensionless BH mass $\phi\st{BH}$ is expressed in terms of
the inner radius by rescaling equation (\ref{bondidef2}) as follows.
\begin{align}
\xi_0=\frac{r_0}{\alpha}
&=\frac{2G}{b\alpha}\frac{\Mbh}{c\st{s}^2} \\
%&=\frac{2G}{b\alpha}4\pi\rho_0\alpha^3\phi\st{BH}\frac{n}{(n+1)K\rho_0^\frac{1}{n}}\\
%&=\frac{2n}{b}\alpha^2\phi\st{BH}\frac{4\pi G}{(n+1)K\rho_0^{\frac{1}{n}-1}} \\
&=\frac{2G}{b\alpha}4\pi\rho_0\alpha^3\phi\st{BH}\frac{n}{(n+1)K}\rho_0^{-\frac{1}{n}}\\
&=\frac{2n}{b}\alpha^2\phi\st{BH}\frac{4\pi G}{(n+1)K}\rho_0^{1-{\frac{1}{n}}} \\
&=\frac{2n}{b}\phi\st{BH}\text{.}
\end{align}
Similarly, from equation (\ref{mcavint}), $M\st{cav}$ is expressed by
% \begin{align}
% \phi\st{cav}
% &=\frac{\frac{8\pi}{3}\rho_0(\alpha\xi_0)^3}{4\pi\rho_0\alpha^3}\\
% &=\frac{2}{3}\xi_0^3\text{.}
% \end{align}
\shorteq{\phi\st{cav}
=\frac{\frac{8\pi}{3}\rho_0(\alpha\xi_0)^3}{4\pi\rho_0\alpha^3}
=\frac{2}{3}\xi_0^3\text{.}}
The dimensionless mass and radius boundary conditions are now related
by
\shorteq{\phi_0\equiv\phi(\xi_0)=\frac{b}{2n}\xi_0+\frac{2}{3}\xi_0^3\text{.}}
Thus, for a given polytropic index $n$, we can choose a
value $\xi_0$ and integrate the equations.  

\begin{figure}\begin{center}
\includegraphics{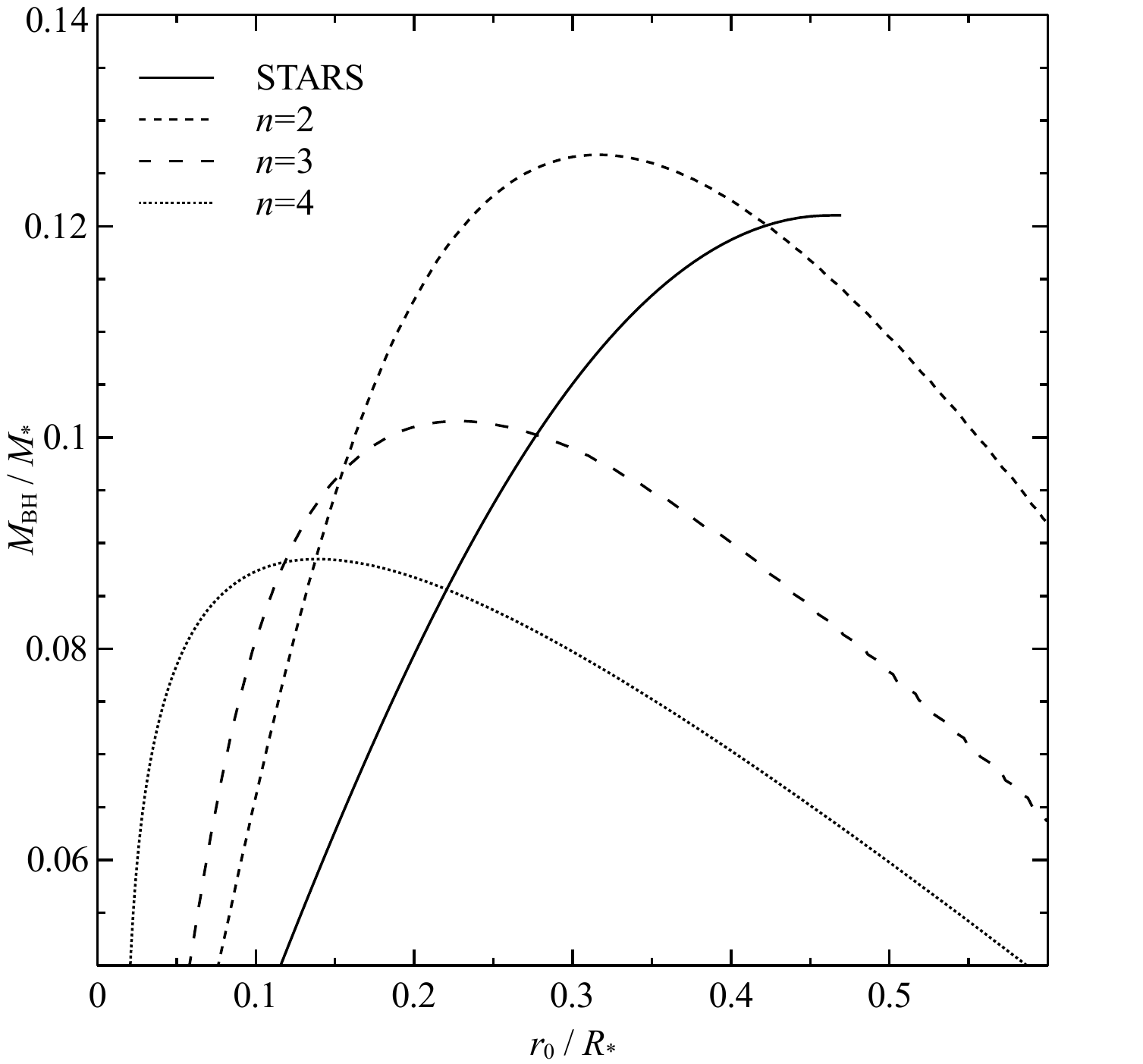}
\caption[Ratios of BH mass to total mass for polytropic quasi-stars
and the fiducial evolution, demonstrating a maximum ratio in all
cases.]{Plot of the ratio of BH mass against total mass against inner
  envelope radius to outer envelope radius.  The short-dashed,
  long-dashed and dotted lines correspond to polytropic solutions with
  $n=2,3$ and $4$.  The solid line represents the fiducial evolution.
  The polytropic models show an upper limit to the BH mass ratio.  The
  fiducial results reach a similar limit but the mass ratio cannot
  decrease so the evolution terminates.}
\label{f234stars}
\end{center}\end{figure}

I integrated a sequence of solutions for $n=3$ and found that there
exists a maximum value of $\phi\st{BH}/\phi_*=0.102$, where $\phi_*$
is the total dimensionless mass of the quasi-star.  The maximum occurs
when $\xi_0=0.995$.  Similar limits appear for polytropic indices
between $2$ and $4$.  For $n=2$, the maximum is
$\phi\st{BH}/\phi_*=0.127$ when $\xi_0=1.012$ and for $n=4$ the
maximum is $\phi\st{BH}/\phi_*=0.089$ at $\xi_0=0.968$.
Fig.~\ref{f234stars} shows plots of the curves of the ratio of inner
to outer envelope radius ($\xi_0/\xi_*$ in dimensionless variables,
where $\xi_*$ is the outer radius of the envelope) against the
fractional BH mass ($\phi\st{BH}/\phi_*$ in dimensionless variables)
for $n=2$, $3$ and $4$ together with our results for the fiducial
model.  The maximum mass ratio is clear in each curve.  In principle,
further hydrostatic solutions exist along the sequence computed by
\stars{} but they require that the BH mass decreases.\footnote{I tried
  to compute hydrostatic solutions on the other side of the maximum
  mass ratio but never succeeded.}

The existence of a maximum inner mass ratio is therefore a robust
feature of the structure equations.  Maximum ratios exist for all
choices of inner radius and cavity mass in this chapter.  In
particular, a maximum also exists when the cavity mass is ignored
completely.  The cause of the maximum mass ratio is not generally
clear but in Chapter \ref{cscl} we show that, without the cavity gas,
the maximum exists for the same reason as the
Sch\"onberg--Chandrasekhar limit.  The cavity gas modifies the inner
boundary condition, and therefore the envelope solution, but the same
mechanism is at work.

\subsection{Post-quasi-star evolution}

At the end of the fiducial run the cavity contains $3311\Msun$.  Under
our assumptions, some of this material is already moving towards the
BH and may become part of it.  If the BH accretes all the mass in the
cavity, its final mass would be $\Mbh\approx4505\Msun$, nearly half of
the total mass of the original quasi-star.  Presuming the BH accretes
at its Eddington-limited rate, this growth would take about $51\Myr$.

What actually happens to the material in the cavity after the end of
the hydrostatic evolution? It appears that the entire envelope might
be swallowed but it is not certain that this should be the case.  The
accretion flow is at least partly convective so there must be a
combination of inward and outward flowing material within the Bondi
radius.  In the theoretical limit for a purely convective flow, the
accretion rate is zero \citep{quataert00} and half of the material is
moving inward and half outward.  If the flow is sustained, we might
expect at least half of the cavity mass to be accreted.  On the other
hand the flow structure might change completely.  The infalling
material could settle into a disc and drive disc winds or jets so that
the overall gain in mass is relatively small.  Hydrostatic equilibrium
is probably failing in the envelope so its dynamics might also change
drastically.

\citet{johnson+10} modelled the accretion on to massive BHs formed
through direct collapse.  They assumed that the BH accretes from a
multi-colour black-body disc after its quasi-star phase and found
that, once the BH mass exceeds about $10^4\Msun$, the accretion rate
decreases owing to radiative feedback.  This result supports the case
for a substantial decrease in the BH growth rate if a thin disc forms
after the quasi-star phase but additional growth can occur during the
transition to a new structure.

\section{Parameter exploration}
\label{smore1}

In Section \ref{sfid1}, I established the basic qualitative structure
and evolution of the quasi-star envelope.  In this section, I explore
their dependence on some of the parameters of the model.  I begin in
Section \ref{ssepseta} with the radiative and convective efficiencies
$\epsilon$ and $\eta$.  I then vary the choice of cavity mass in
Section \ref{sscav}, surface mass loss and gain in Section \ref{ssenv}
and the total initial envelope mass in Section \ref{ssmi}.  Finally,
in Section \ref{ssr0} I vary the inner radius.

\subsection{Radiative and convective efficiencies}
\label{ssepseta}

\begin{figure}\begin{center}
\includegraphics{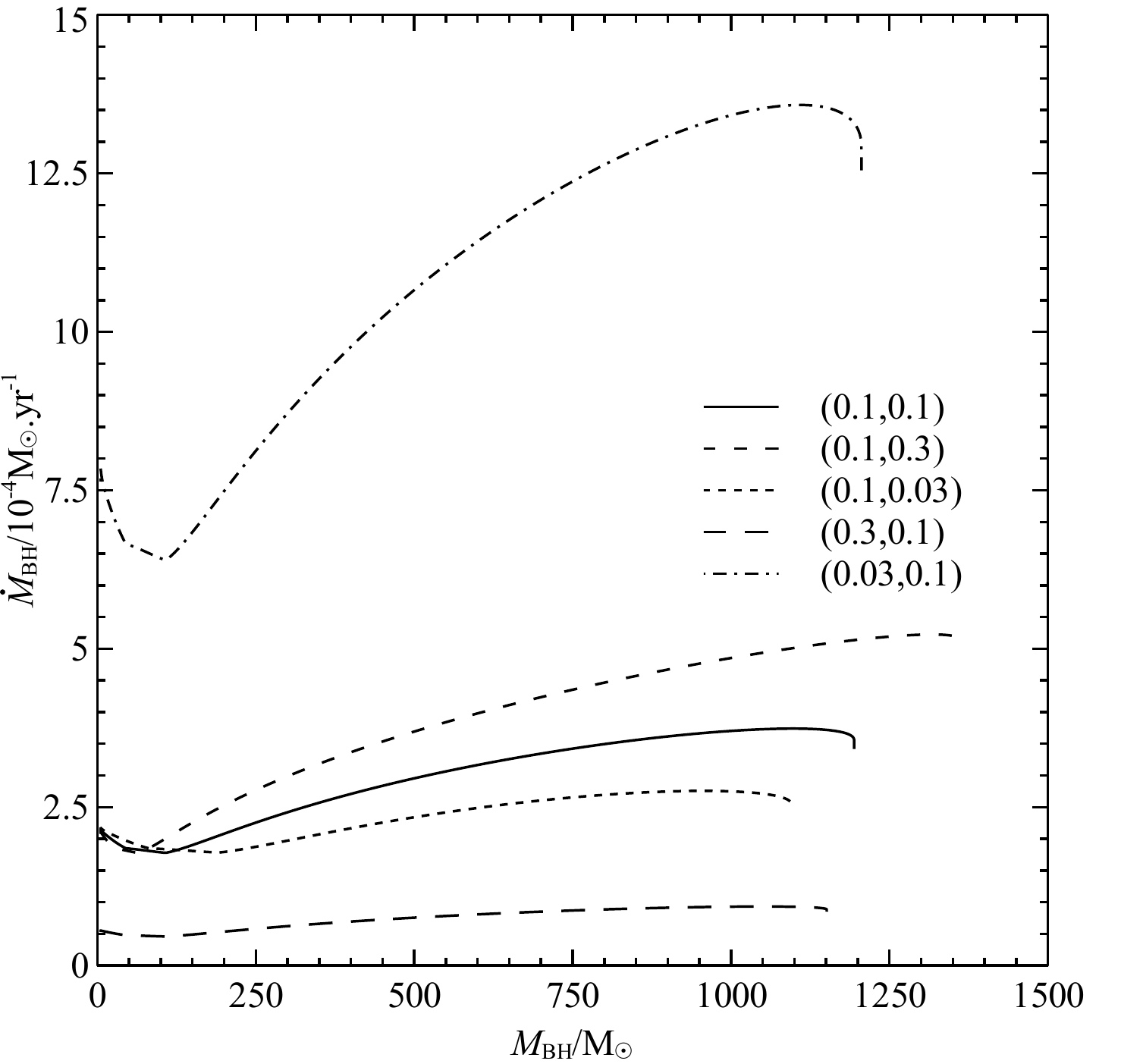}
\caption[BH accretion rate versus BH mass for various pairs of
radiative and convective efficiencies.]{Plot of BH accretion rate
  $\dMbh$ against BH mass $\Mbh$ for various pairs of radiative and
  convective efficiencies ($\epsilon,\eta$).  The fiducial values
  ($0.1,0.1$) correspond to the solid line.  A change to the radiative
  efficiency changes $\dMbh$ but leaves the overall structure
  unaffected.  A decrease of $\eta$ causes the envelope to be hotter
  and denser in order to achieve the same luminosity.  This the
  discontinuities in the gradient of the accretion rate to later times
  and leaves a smaller final BH.}
\label{fepseta}
\end{center}\end{figure}

Fig.~\ref{fepseta} shows the accretion history against BH mass for a
number of choices of radiative efficiency $\epsilon$ and convective
efficiency $\eta$.  Because the luminosity always settles on the same
convection-limited rate, a change in $\epsilon$ only rescales the
accretion rate through equation (\ref{lum1}) and does not affect the
structure.  The final BH mass and intermediate properties are the
same.  The only difference is that the evolution takes longer for
larger values of $\epsilon$.

A larger convective efficiency $\eta$ allows a greater flux to be
radiated by the accretion flow and therefore a larger accretion rate
for given interior conditions.  To establish the same overall
luminosity, the envelope must be less dense so the discontinuities in
the gradient of the accretion rate appear later.  In addition, the
lower density means that, although the final total inner masses are
similar, the cavity mass is smaller and the BH mass larger.
Conversely, for smaller values of $\eta$, the envelope is denser, the
discontinuities appear earlier and the final BH mass is smaller.

\subsection{Cavity properties}
\label{sscav}

\begin{figure}\begin{center}
\includegraphics{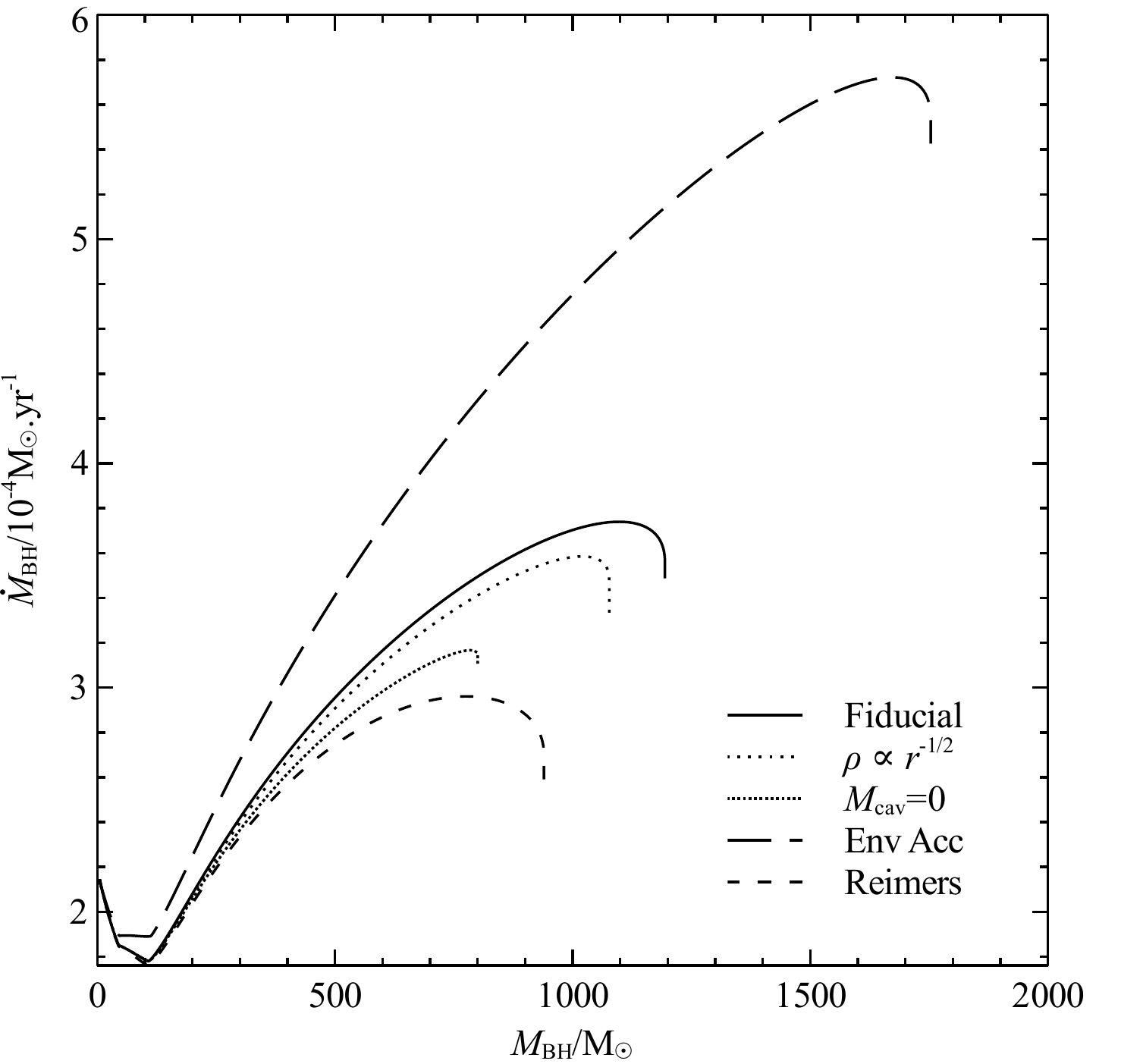}
\caption[Evolution of quasi-stars with accretion, mass loss and
different cavity masses.]  {Plot of BH accretion rate $\dMbh$ against
  BH mass $\Mbh$ for the fiducial run and runs with a shallower radial
  dependence of the interior density, % (``$\rho\propto r^{-1/2}$''),
  no cavity mass, %(``$M\st{cav}=0$''),
  constant accretion on to the surface of the star, %(``Env Acc'')
  and a \citet{reimers75} mass loss rate. % (``Reimers'').
  A shallower radial dependence of the interior density leads to a
  smaller cavity mass and thus a smaller interior density and
  accretion rate.  The accreting envelope gives final quasi-star and
  BH masses of $14\,511\Msun$ and $1753\Msun$, which give a similar
  mass ratio of $0.121$.  With a Reimers mass-loss rate, the final
  quasi-star and BH masses are $7754\Msun$ and $940\Msun$, which also
  give a ratio of $0.121$.}
% {Plot of BH accretion rate $\dMbh$ against BH
%   mass $\Mbh$ for the fiducial run, a run with constant accretion on
%   to the surface of the star (``Env Acc''), a run with a
%   \citet{reimers75} mass loss rate (``Reimers''), a run with a
%   shallower radial dependence of the interior density (``$\rho\propto
%   r^{-1/2}$'') and a run with no cavity mass (``$M\st{cav}=0$'').
%   With a Reimers mass-loss rate, the final quasi-star and BH masses
%   are $7754\Msun$ and $940\Msun$, which give a similar mass ratio of
%   $0.121$.  The accreting envelope gives final quasi-star and BH
%   masses of $14\,511\Msun$ and $1753\Msun$, again with a similar ratio
%   of $0.121$.  A shallower radial dependence of the interior density
%   leads to a smaller cavity mass and therefore a smaller interior
%   density and accretion rate.}
\label{fcavenv}
\end{center}\end{figure}

If the inward transport of angular momentum by advection and
convection is greater than the outward transport by magnetic fields
and other sources of viscosity, the density profile in the cavity
tends towards $\rho(r)\propto r^{-\frac{1}{2}}$
\citep{quataert00,narayan00}.  Recomputing the cavity mass from
equation (\ref{mcav}), we find
\shorteq{M\st{cav}=\frac{8\pi}{5}\rho_0r_0^3\text{.}}  
The corresponding quasi-star evolution is shown in Fig.~\ref{fcavenv}.
The smaller interior mass leads to a less evident density spike at all
times.  The decreased interior mass is subject to a lower mass limit
for the BH and leaves a BH of $1077\Msun$.  Although the change to the
cavity properties affects the numerical results, there is no
qualitative change to the quasi-star evolution, which still terminates
owing to the same physical mechanism as the fiducial run.

The maximum mass ratio is not principally caused by the cavity mass.
Fig.~\ref{fcavenv} also shows a run with $M\st{cav}=0$.  The BH
achieves a final mass of $800\Msun$.  A maximum mass ratio still
exists, though with a different numerical value, and a similar limit
is found for polytropic models with no cavity mass.  In this case, the
maximum mass ratio exists for the same reason as the
Sch\"onberg--Chandrasekhar limit.  The connection between the mass
limits is demonstrated in Chapter \ref{cscl}.

\subsection{Envelope mass loss and gain}
\label{ssenv}

To illustrate the effect of net accretion on to the surface of the
envelope, Fig.~\ref{fcavenv} shows the evolution of the fiducial run
if the envelope accretes at a constant rate of $\sci{2}{-3}\Msunpyr$.
Although this rate initially exceeds the quasi-star's Eddington limit,
such rapid infall is believed to occur as long as the bars-within-bars
mechanism transports material towards the centre of the pregalactic
cloud.  Accreted mass is %simply
added to the %surface value of the 
surface 
mass co-ordinate and no additions
are made to any other equations.  In particular, we do not include a
ram pressure at the surface.

The only qualitative change to the evolution is that it takes longer
than if the total mass had been kept constant at the same final value
of $14\,511\Msun$.  The final BH mass is subject to the same ratio
limit and the larger final envelope permits a larger final BH mass of
$1753\Msun$.

To investigate the effect of mass loss we use a Reimers rate
\citep{reimers75}, an empirical relation that describes mass
loss in red giants.  The mass-loss rate is
\shorteq{\dot{M}\st{loss}=\sci{4}{-13}\frac{L_*R_*}{M_*}
  \frac{\Msun}{\Lsun\Rsun}\Msunpyr\text{.}}  
Fig.~\ref{fcavenv} shows the evolution for a quasi-star envelope with
this prescription.  The mass loss is significant but, again, there is
no qualitative change in the results.  The limit holds and the BH has
a smaller mass of $940\Msun$.  This is in proportion with the decrease
in the total mass of the quasi-star to $7754\Msun$.

\subsection{Initial envelope mass}
\label{ssmi}

\begin{figure}\begin{center}
\includegraphics{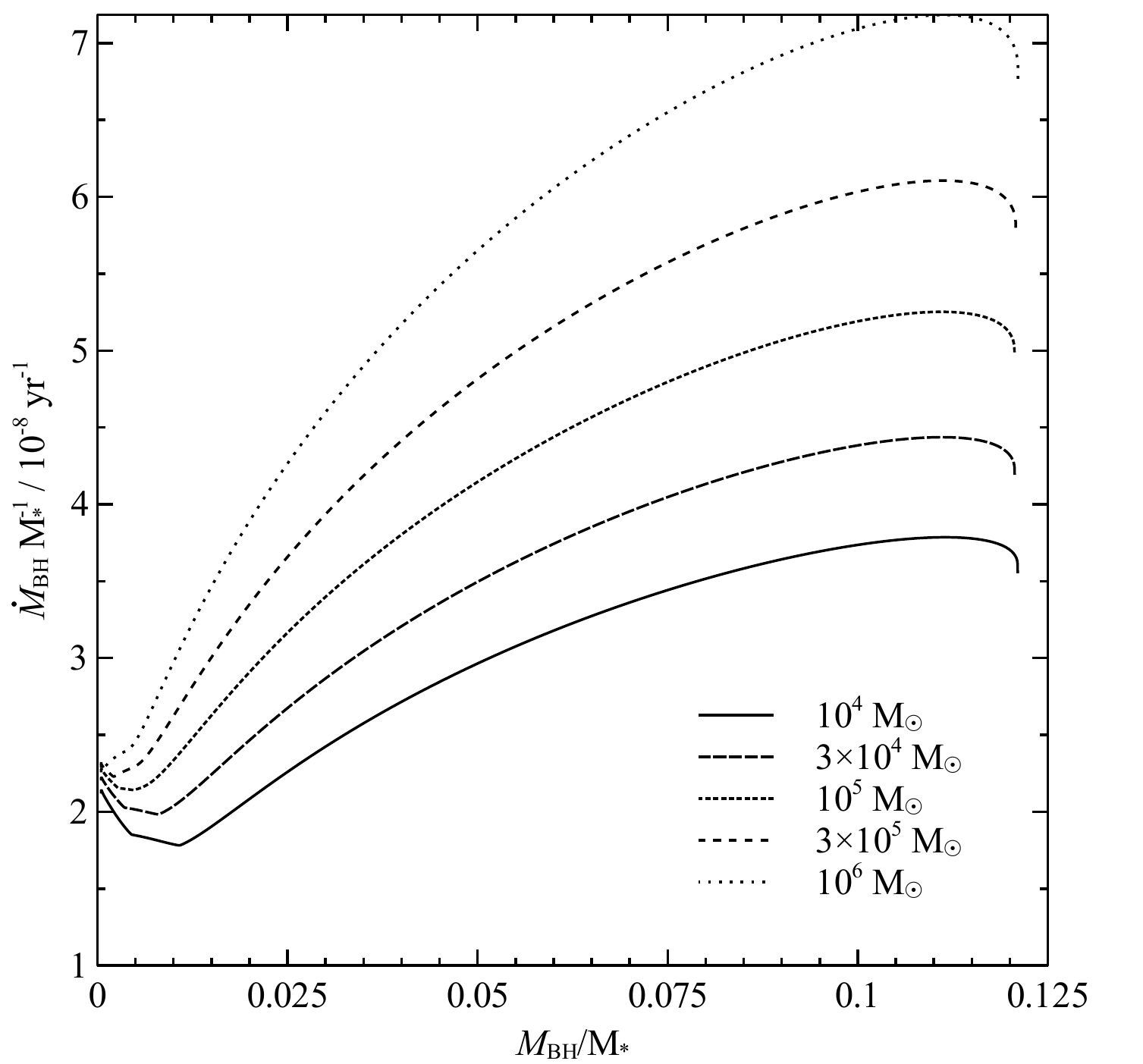}
\caption[Evolution of quasi-stars of initial masses
$M_*/\Msun=10^4,\sci{3}{4},10^5,\sci{3}{5}$ and $10^6$.]{Plot of the
  evolution of quasi-stars of total initial masses
  $M_*/\Msun=10^4,\sci{3}{4},10^5,\sci{3}{5}$ and $10^6$.  The BH
  mass and accretion rates have been divided by the total quasi-star
  mass to illustrate the consistency of the upper mass ratio limit of
  $0.121$ and the slight dependence of the accretion rate with
  quasi-star mass.  Because larger quasi-stars permit greater scaled
  accretion rates, they have shorter hydrostatic lifetimes.}
\label{fmi}
\end{center}\end{figure}

The structure of the envelope appears to be chiefly dependent on the
ratio of envelope mass to BH mass.  Fig.~\ref{fmi} shows the evolution
of the the BH mass and of the accretion rate divided by total mass for
quasi-stars of total initial masses
$M_*/\Msun=10^4,\sci{3}{4},10^5,\sci{3}{5}$ and $10^6$.  The
fractional upper BH mass limit holds for all the quasi-star masses in
this range though the BH-envelope mass ratio after which the entire
envelope is convective depends on the envelope mass.

For a given mass ratio, once the entire envelope has become
convective, the following approximate relations hold for the
properties of two quasi-stars of different masses (denoted by
subscripts 1 and 2).
% \shorteq{\bracfrac{T_{0,1}}{T_{0,2}}=\bracfrac{M_{*,1}}{M_{*,2}}^{-0.04}\text{,}\label{T0scal}}
% \shorteq{\bracfrac{T_{\text{eff},1}}{T_{\text{eff},2}}=
%    \bracfrac{M_{*,1}}{M_{*,2}}^{0.01}\text{,}\label{Teffscal}}
% \shorteq{\bracfrac{R_{*,1}}{R_{*,2}}=\bracfrac{M_{*,1}}{M_{*,2}}^{0.54}\text{,}\label{Rscal}}
% \shorteq{\bracfrac{\rho_{0,1}}{\rho_{0,2}}=
% \bracfrac{M_{*,1}}{M_{*,2}}^{-0.66}\text{,}\label{rhoscal}}
\begin{gather}
\bracfrac{T_{0,1}}{T_{0,2}}=\bracfrac{M_{*,1}}{M_{*,2}}^{\sq-0.04}\text{,}\label{T0scal}
\end{gather}\begin{gather}\bracfrac{T_{\text{eff},1}}{T_{\text{eff},2}}
  =\bracfrac{M_{*,1}}{M_{*,2}}^{\sq0.01}\text{,}\label{Teffscal} \\
\bracfrac{R_{*,1}}{R_{*,2}}=\bracfrac{M_{*,1}}{M_{*,2}}^{\sq0.54}\text{,}\label{Rscal} \\
\bracfrac{\rho_{0,1}}{\rho_{0,2}}=\bracfrac{M_{*,1}}{M_{*,2}}^{\sq-0.66}\text{,}\label{rhoscal}
\end{gather}
and
\shorteq{\bracfrac{\dot{M}_{\text{BH},1}}{\dot{M}_{\text{BH},2}}
=\bracfrac{M_{*,1}}{M_{*,2}}^{\sq1.14}\text{.}\label{dMdtscal}}
For example, compared with a quasi-star of mass $10^4\Msun$ at the 
same BH-envelope mass ratio, a $10^5\Msun$ quasi-star roughly has
an interior temperature that is
$10^{0.04}=1.10$ times smaller, a surface temperature that is
$10^{0.01}=1.02$ times greater, an outer radius that is 
$10^{0.54}=3.47$ times greater, an interior density that is
$10^{0.66}=4.57$ times smaller and a mass accretion rate that is
$10^{1.14}=13.8$ times greater.
Equations (\ref{T0scal}) and (\ref{Teffscal}) imply that %
the temperature profiles of the envelopes depend almost exclusively on
the BH-envelope mass ratio.  The variation in inner or surface
temperature, for a given mass ratio, with respect to the total mass of
the quasi-star is less than 0.1 per factor of ten in the total mass.
The density and radius profiles are more strongly dependent on the
mass of the envelope.

The final relation (equation \ref{dMdtscal}) implies that the lifetime
of a quasi-star scales roughly as $\tau\st{QS}\propto M_*^{-0.14}$ so
larger quasi-stars have slightly shorter hydrostatic lifetimes.  By
fitting a straight line to the $\log\tau\st{QS}$--$\log M_*$ relation
for the five models described here, we find %that the lifetimes scale
%as $\tau\st{QS}\propto M_*^{-0.13}$.  More precisely, we find
\shorteq{\log_{10}(\tau\st{QS}/\Myr)=-0.127\log_{10}(M_*/\Msun)+1.13\text{.}}
Note that $M_*$ here denotes the initial mass of the quasi-star.  In
all other relations $M_*$ decreases slowly during the quasi-star's
evolution owing to the mass-energy lost as radiation in the accretion
flow.

\subsection{Inner boundary radius}
\label{ssr0}

\begin{figure}\begin{center}
\includegraphics{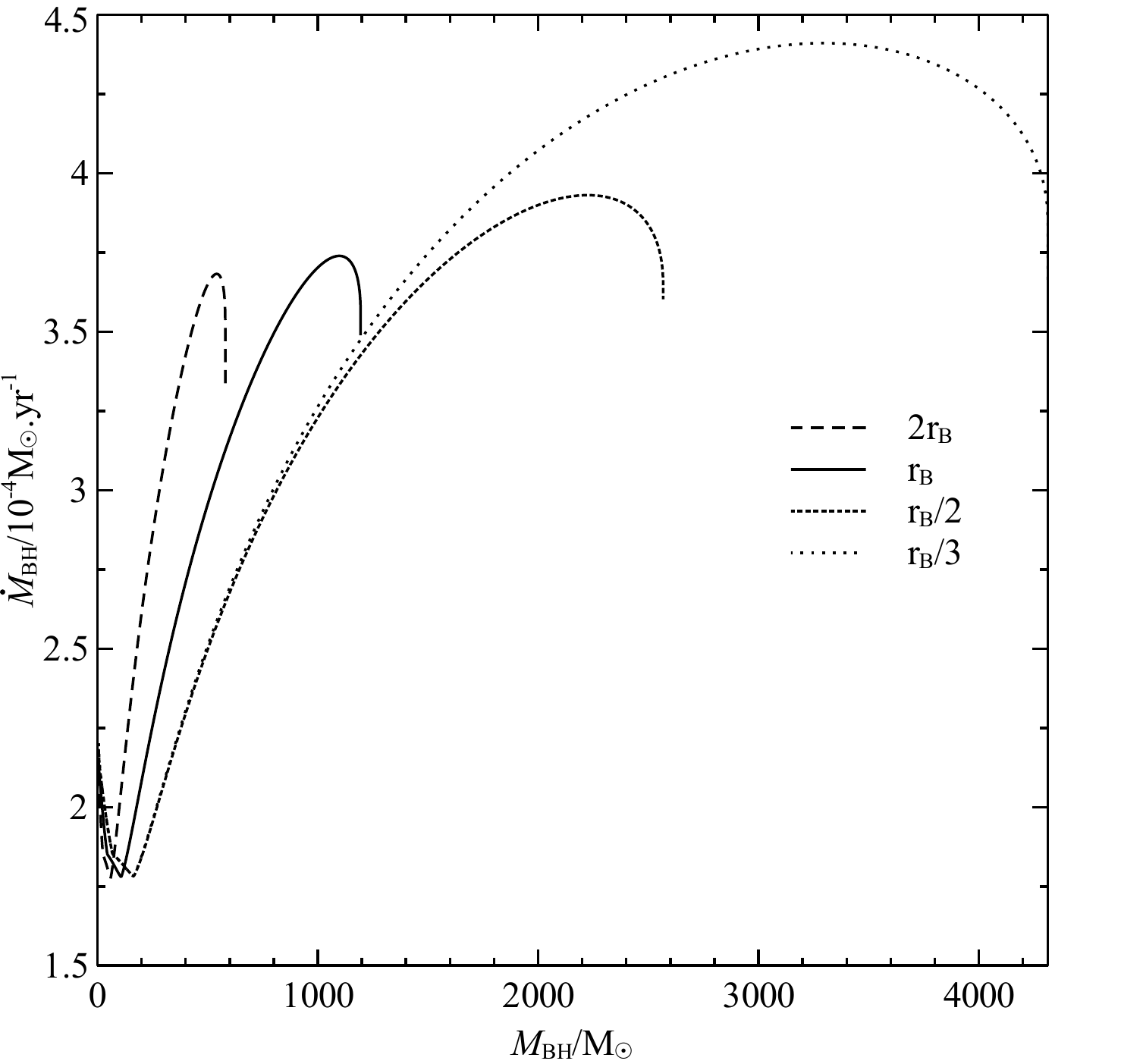}
\caption[Evolution of quasi-stars for different choices of the inner
radius.]{Plot of BH accretion rate $\dMbh$ against BH mass $\Mbh$ for
  different inner radius parameters $b=1/3$, $1/2$, $1$ and $2$.  The
  various values of $r_0$ lead to qualitatively similar results but
  the quantitative evolution is strongly affected.  The final BH mass
  is approximately $b^{1.11}$ times the fiducial value of
  $1194\Msun$.}
\label{fr0}
\end{center}\end{figure}

Fig.~\ref{fr0} shows the evolution of quasi-stars where the inner
radius was changed to a third, half and twice the Bondi radius through
the parameter $b$.  The final BH mass is strongly affected although
the evolution is qualitatively similar.  The results of our polytropic
analysis are affected in a consistent manner.  I further found that I
could not construct model envelopes with $r_0\lesssim0.3\rb$ and that
this is reflected in the polytropic models.  Although reasonable, our
choice of inner radius is somewhat arbitrary and critical in deciding
the quantitative evolution of the BH.

Despite this, the total mass inside the cavity is broadly similar
across the models.  For models with $b=3$, $2$, $1$ and $1/2$, the
total inner masses (BH and cavity) are approximately $M_0/\Msun=5253$,
$4904$, $4505$ and $4328$.\footnote{The last figure is interpolated
  from the model profiles, which were not recorded near the last model
  before convergence began to fail.} %
If the BH ultimately accretes most of the material in the cavity after
its quasi-star phase ends, %then the BH
it reaches a similar mass for any choice of $b$.

Begelman (2010, private communication) pointed out that, in the
presence of a substantial mass of gas in the cavity, the Bondi radius
should be defined for the total mass inside $r_0$, not just the BH
mass.  Presuming the gas has a density distribution $\rho(r)\propto
r^{-\frac{3}{2}}$ inside the cavity, this gives the equation
\shorteq{r_0=\frac{2G}{bc\st{s}^2}
  \left(\Mbh+\frac{8\pi}{3}\rho_0r_0^3\right)\text{.}}  For an $n=3$
polytrope, we can substitute for dimensionless variables (see Section
\ref{sslimit}) to obtain
\shorteq{\phi\st{BH}+\frac{2}{3}\xi_0^3-\frac{1}{6}b\xi_0=0\text{,}}
which only has a real positive root if
$\phi\st{BH}<b^\frac{3}{2}/(18\sqrt{3})$.  For maximum $\phi\st{BH}$,
the ratio of BH mass to total mass is $0.017$ for $b=1$ and for $b=3$
it is $0.126$.

\section{Comparison with Begelman et al. (2008)}
\label{sbra}

BRA08 estimated some envelope properties by presuming that the
envelope is described by an $n=3$ polytrope.  They employed an overall
accretion efficiency parameter\footnote{This should not be confused
  with $\alpha$ defined in \mbox{Section \ref{sslimit}}.}
$\alpha\st{BRA}$, which is determined by numerical factors in the
accretion rate including the radiative efficiency, convective
efficiency and adiabatic index.  I calibrated $\alpha\st{BRA}$
through the BH luminosity (equation 3 of BRA08) and chose
$\alpha\st{BRA}=0.257$ for the fiducial run.  The analytical BH
luminosity is then accurate to within 0.2 per cent over the entire
evolution.  I also compared a run with $b=3$ by using
$\alpha\st{BRA}=0.0287$, which is similarly accurate to within 0.3
per cent.

\begin{figure}\begin{center}
\includegraphics{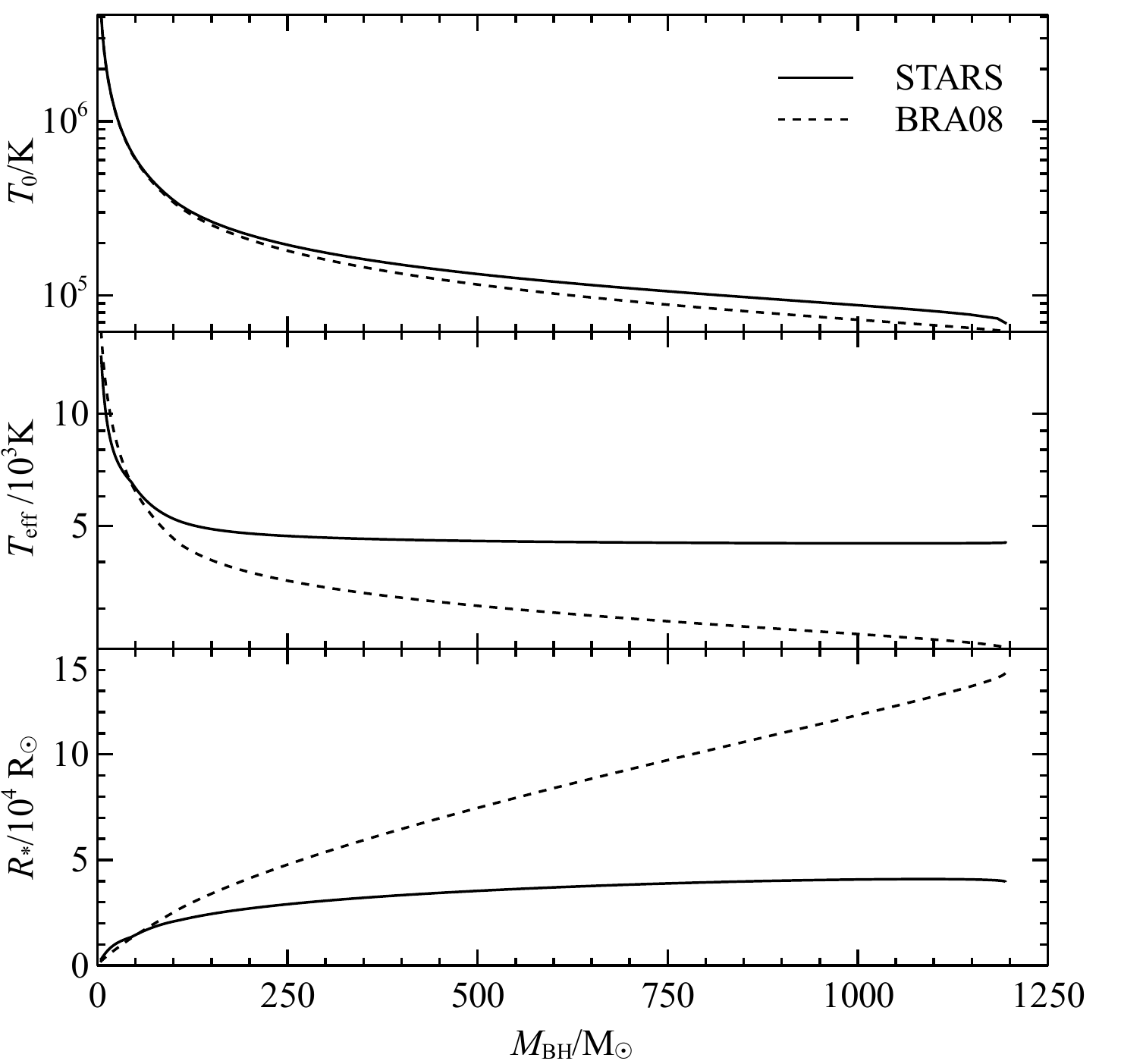}
\caption[Comparison of the fiducial run with analytic estimates for
interior temperature, surface temperature and envelope radius by
BRA08.]{Comparison for the fiducial run (solid lines) of analytic
  estimates for interior temperature (top), surface temperature
  (middle) and envelope radius (bottom) by BRA08 (dashed lines)
  against results.  BRA08's estimate of the interior temperature is
  accurate but those for the photospheric temperature and envelope
  radius become increasingly inaccurate as the BH grows.}
\label{fbra1}
\end{center}\end{figure}

\begin{figure}\begin{center}
\includegraphics{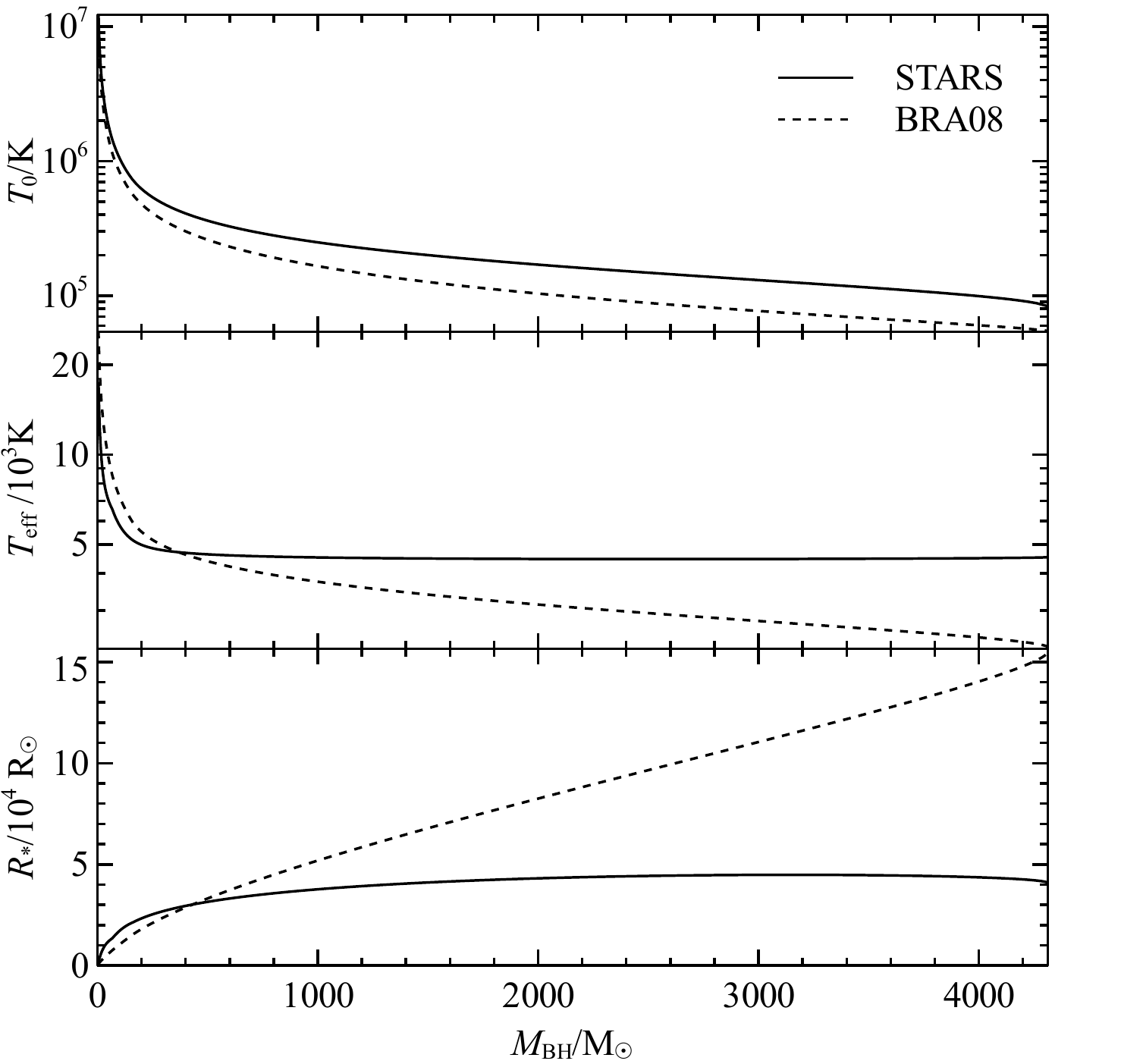}
\caption[Comparison of a run with $b=3$ with analytic estimates for
interior temperature, surface temperature and envelope radius by
BRA08.]{Comparison for a run with $b=3$ (solid lines) of analytic
  estimates for interior temperature (top), surface temperature
  (middle) and envelope radius (bottom) by BRA08 (dashed lines)
  against results.  BRA08's estimate of the interior temperature is
  less accurate than before but the photospheric temperature is
  closer.}
\label{fbra3}
\end{center}\end{figure}

BRA08 provide the following analytic estimates for the inner
temperature, photospheric temperature and envelope radius (equations
7, 11 and 10 in their paper, respectively).
% \shorteq{T_0=\sci{1.4}{4}\left(\frac{L}{L\st{Edd}}\right)^\frac{2}{5}
% \left(\alpha\st{BRA} \frac{\Mbh^2}{\mathrm{M}_\odot^2}\right)^{-\frac{2}{5}}
% \left(\frac{M_*}{\Msun}\right)^\frac{7}{10}\K\text{,}\label{T0bra}}
% \shorteq{T\st{eff}=\sci{1.0}{3}\left(\frac{L}{L\st{Edd}}\right)^\frac{9}{20}
% \left(\alpha\st{BRA} \frac{\Mbh^2}{\mathrm{M}_\odot^2}\right)^{-\frac{1}{5}}
% \left(\frac{M_*}{\Msun}\right)^\frac{7}{20}\K\label{Teffbra}}
\begin{gather}
T_0=\sci{1.4}{4}\left(\frac{L}{L\st{Edd}}\right)^{\sq\frac{2}{5}}%\sq
\left(\alpha\st{BRA} \frac{\Mbh^2}{\mathrm{M}_\odot^2}\right)^{\sq-\frac{2}{5}}%\sq
\left(\frac{M_*}{\Msun}\right)^{\sq\frac{7}{10}}\K\text{,}\label{T0bra} \\
T\st{eff}=\sci{1.0}{3}\left(\frac{L}{L\st{Edd}}\right)^{\sq\frac{9}{20}}%\sq
\left(\alpha\st{BRA} \frac{\Mbh^2}{\mathrm{M}_\odot^2}\right)^{\sq-\frac{1}{5}}%\sq
\left(\frac{M_*}{\Msun}\right)^{\sq\frac{7}{20}}\K\label{Teffbra}
\end{gather}
\noindent and
\shorteq{R_*=\sci{4.3}{14}\left(\frac{L}{L\st{Edd}}\right)^{\sq-\frac{2}{5}}%\sq
\left(\alpha\st{BRA} \frac{\Mbh^2}{\mathrm{M}_\odot^2}\right)^{\sq\frac{2}{5}}%\sq
\left(\frac{M_*}{\Msun}\right)^{\sq-\frac{1}{5}}\cm\text{.}\label{Rbra}}
Here, $L\st{Edd}=4\pi GcM_*/\kappa$ is the Eddington luminosity.
BRA08 computed this using the opacity at the boundary of the
convective zone but such estimates differ by a factor of the order of
$\kappa/\kappa\st{es}$ when compared with our results.  The comparison
is thus made using the Eddington limit with opacity
$\kappa\st{es}=0.34\cmsq\pg$.

The three estimates are plotted against the fiducial run in
Fig.~\ref{fbra1}.  The estimate for the interior temperature is
accurate to within 20 per cent.  The deviation grows as the
approximation of the envelope to an $n=3$ polytrope becomes
increasingly poor.  The estimate for the photospheric
temperature is not accurate.  At the end of the run the estimated
photospheric temperature is about $2400\K$ compared with the model
result of about $4500\K$.  Because the BH luminosity estimate is
accurate, it follows that the envelope radius is not because the
surface luminosity is $L_*=\pi acR_*^2T\st{eff}^4$.  This is
confirmed in the bottom panel of Fig.~\ref{fbra1}.
BRA08 used an inner boundary with $b=3$ so the analytic estimates are
compared with such a run in Fig.~\ref{fbra3}.  The interior
temperature estimate is worse but the photospheric temperature
estimate is better.  The final surface temperature remains %is still
remains
underpredicted and the surface radius %remains 
inaccurate.

The analytic estimates by BRA08 can be compared with the scaling
relations found in Section \ref{ssmi}.  By fixing $\Mbh/M_*$ and
$L/L\st{Edd}$, equations (\ref{T0bra}), (\ref{Teffbra}) and
(\ref{Rbra}) can be re-arranged to give
% \shorteq{T_0\propto M_*^{-\frac{1}{10}}\text{,}}
% \shorteq{T\st{eff}\propto M_*^{-\frac{1}{20}}\text{,}}
\begin{gather}
T_0\propto M_*^{-\frac{1}{10}}\text{,} \\
T\st{eff}\propto M_*^{-\frac{1}{20}}\text{,}
\end{gather}
 and
\shorteq{R_*\propto M_*^{\frac{3}{5}}\text{.}}
These compare reasonably with the scaling relations derived from the
\stars{} models.  Written similarly as power-laws in $M_*$, equations
\ref{T0scal}, \ref{Teffscal} and \ref{Rscal} have indices $-0.04$,
$0.01$ and $0.54$, respectively.

BRA08 argue that quasi-star evolution terminates owing to the
opacity at the edge of the convection zone increasing.  An increased
opacity causes the envelope to expand and the opacity increases further.  
The envelope then expands even more and the process is claimed 
to run away.  BRA08 refer to this process as the \emph{opacity crisis}.  Our
results do not terminate for this reason.  Similar behaviour does occur at the
beginning of the evolution while the photospheric temperature is
greater than $10^4\K$ but it does not disperse the quasi-star.  For
most of a quasi-star's evolution, the opacity at the convective
boundary is already beyond the H-ionization peak and is decreasing as
the BH grows.

Instead, my models terminate because of a basic property of mass
conservation and hydrostatic equilibrium.  I have shown that further
models do not exist in the sequence by showing that the same
fractional mass limit exists in polytropic models.  Direct calculation
of the polytropic mass limits shows that they are similar to the
results found with the \stars{} code.  In Chapter \ref{cscl}, I show
how the fractional mass limit found here for quasi-stars is directly
related to the Sch\"onberg--Chandrasekhar limit as well as several
related limits in the literature.

\section{Conclusion}

The models presented in this chapter provide several important
results.  The first is the existence of a robust upper limit on the
ratio of inner BH mass to the total quasi-star mass, equal to about
$0.121$.  The limit is reflected in solutions of the Lane--Emden
equation, modified for the presence of a point mass interior to some
specific boundaries.  All the evolutionary runs here terminate once
the limit is reached and it is difficult to say what happens after the
hydrostatic evolution ends.  Some of the material within the Bondi
radius has begun accelerating towards the BH so we expect that it can
be captured by the BH.  The remaining material may be accreted or
expelled, depending on the liberation of energy from the material that
does fall inwards.  After the BH has evolved through the quasi-star
phase, it is probably limited to accrete at less than the Eddington
rate for the BH alone and thus much less than the accretion rate
during the quasi-star phase.

These results suggest that quasi-stars produce BHs that are on the
order of at least $0.1$ of the mass of the quasi-star and around $0.5$
if all the material within the inner radius is accreted.  For
conservative parameters, this growth occurs within a few million years
after the BH initially forms.  Realistic variations in the parameters
(e.g. larger initial mass, lower radiative efficiency) lead to shorter
lifetimes.  Such BHs could easily reach masses in excess of
$10^9\Msun$ early enough in the Universe to power high-redshift
quasars.

However, it is also clear that the models are critically sensitive to
the choice of inner boundary radius and the results should be treated
with due caution.  While the Bondi radius used here is reasonable, it
is worthwhile to consider other sets of boundary conditions.  In the
next chapter, I construct models that use a qualitatively different
set of boundary conditions, which lead to very different quasi-stars
than the Bondi-type models.
 % Bondi-type quasi-stars
\clearpage\thispagestyle{tocstyle}
\begin{savequote}[80mm]
  With four parameters I can fit an elephant, and with five I can make
  him wiggle his trunk.  \qauthor{John von Neumann, attrib.}
\end{savequote}

\chapter{CDAF-ADAF quasi-stars}
\label{cqs2}

In the work reported in Chapter \ref{cqs1}, the structure of
quasi-stars is calculated under the assumption that the dynamics in
the cavity around the black hole (BH) are described by Bondi
accretion.  Although the corresponding models are reasonable and
self-consistent, they are strongly sensitive to the choice of the
inner radius.  In this chapter, I construct models with a different
set of boundary conditions and find that the corresponding models
evolve very differently.

The model of the accretion flow, described in Section \ref{sbc2},
depends critically on the presence of rotation in the envelope. The
inner boundary of the envelope becomes sufficiently small that
relativistic effects cannot be ignored as they were in Chapter
\ref{cqs1}.  In Section \ref{ssrot}, I introduce \rose{}, a variant of
the \stars{} code that incorporates rotation in the structure
equations.  In Section \ref{ssrel}, I describe additional relativistic
corrections I added to the code based on the work of \citet{thorne77}.

In Section \ref{sfidevol2}, I present a fiducial set of results that
demonstrate the main qualitative features of the evolution of the
models.  The most important result is that the BH ultimately accretes
nearly the whole envelope.  This is in stark contrast with the results
for the Bondi-type quasi-stars of Chapter \ref{cqs1}, where the BH was
subject to a robust fractional mass limit.  In addition, the accretion
rates in the models presented here are about an order of magnitude
greater.  I vary the free parameters of the model in Section
\ref{smore2} and find that the evolution remains qualitatively similar
to the fiducial run.  There are, however, selections of parameters for
which the evolution can only be modelled if we begin with larger BH
cores.  Finally, there are also parameters for which so much energy is
trapped in the accretion flow that the envelope cannot be supported
and the initial models fail to converge.

\section{The convective-advective boundary}
\label{sbc2}

Accretion on to a black hole admits solutions in which a substantial
fraction of energy is trapped in the infalling gas and never radiated
away.  The advection of energy across the event horizon is possible
when the material is so optically thick that radiation takes too long
to scatter out through the gas or the gas is so sparse that radiation
is inefficient.  The loss of energy to the BH permits
\emph{advection-dominated} accretion flows \citep[ADAFs,][]{narayan94},
which are unique to BHs \citep[see][for a review]{narayan08}.

It is understood that these flows are convectively unstable in the
radial direction.  \citet*{narayan00} and \citet{quataert00}
independently found self-similar accretion flows with convection where
the outward transport of angular momentum by viscosity is precisely
balanced by the inward transport by convection.  Because the structure
is defined by convection, such flows were called
\emph{convection-dominated} accretion flows (CDAFs).

Self-similar CDAFs extending to zero radius have zero net accretion
and are in effect convective envelopes around BHs \citep{narayan00}.
In reality, the BH has a finite mass and radius and the structure of
the flow admits a small but finite accretion rate.
\citet[][hereinafter AIQN02]{abramowicz+02} showed both analytically
and numerically that a real convective flow must surround an inner
advection-dominated flow.  The boundary between the two regimes can be
estimated by equating the rate at which energy is advected into the BH
and the convective luminosity in the CDAF.  \citet*[][hereinafter
LLG04]{lu04} confirmed these results by integrating the
vertically-averaged structure equations of the accretion flow.

\begin{figure}\begin{center}
\includegraphics[angle=90,width=13.8cm]{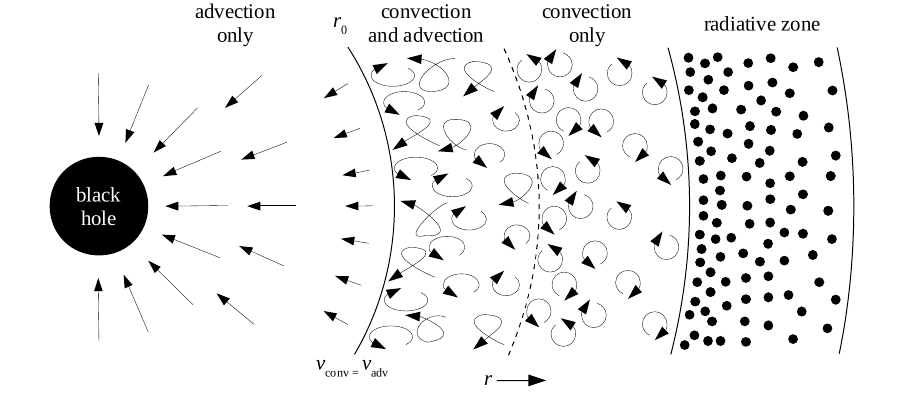}
\caption[Model of the radial structure of CDAF-ADAF quasi-stars.]
{Diagram of the basic model adopted for the radial structure of
  CDAF-ADAF quasi-stars.  The radial co-ordinate increases to the
  right but the structure is not drawn to scale.  At the inner radius
  $r_0$, the inward advective velocity is equal to the convective
  velocity and outward convective energy transport balances inward
  advective transport.  Outside but near to $r_0$, the convective
  envelope loses energy to advection.  As in Bondi-type quasi-stars,
  the convective envelope is necessarily surrounded by a radiative
  atmosphere.}
\label{fqsd4}
\end{center}\end{figure}

With \rose{}, I model the accretion flow on to the BH in a quasi-star
as a massive CDAF that surrounds an ADAF and central BH.  In contrast
to the Bondi-type quasi-stars described in Chapter \ref{cqs1}, I refer
to these models as \emph{CDAF-ADAF} quasi-stars.  Below, I use the
results of AIQN02 and LLG04 to construct a new set of boundary
conditions.  The basic structure of the models is shown in
Fig.~\ref{fqsd4}.  The inner radius $r_0$ is placed where the inward
advective velocity is equal to the convective velocity.  Inside this
radius, all material necessarily flows inward and there is no outward
transport of energy.  Beyond $r_0$, the structure is calculated by
\rose{}.  The amount of energy is advected inwards is substantial at
small radii but declines as $r$ increases.  The advected energy is
correctly computed but the pressure does not incorporate the gradual
increase of the bulk velocity as material approaches the inner radius.
As in the Bondi-type quasi-stars, the surface is necessarily radiative
because of the surface boundary conditions.

\subsection{Black hole mass and accretion rate}

As in Chapter \ref{cqs1}, the mass boundary condition is taken to be
the mass of the BH and any gas inside the inner radius $r_0$.
According to AIQN02 and LLG04, the inner radius should be of
the order of a few tens of Schwarzschild radii
$\rs=2G\Mbh/c^2\approx3(\Mbh/\Msun)\km$ so the cavity mass
should be negligible.  Regardless, the mass boundary condition (see
Section \ref{ssmbc1}) \shorteq{M_0=\Mbh+\frac{8\pi}{3}\rho_0r_0^3\text{,}}
where $\rho_0$ is the density at $r_0$, is implemented in the code.

The boundary between the CDAF and ADAF is characterised by the point
at which the inward velocity of the flow $v\st{adv}=\dMbh/4\pi
r^2\rho$ is equal to the convective velocity $v\st{con}$ (see Fig.~1
of AIQN02), where $\dMbh$ is the BH accretion rate, $r$ is the radial
co-ordinate and $\rho$ is the density at $r$.  At this point, the
total velocity of any parcel of gas must be directed inwards because
the advective velocity is greater.  In other words, the inward
transport of energy by advection overwhelms the outward transport by
convection.

The accretion rate is, by definition, 
\shorteq{\dMbh=4\pi r^2\rho v\st{adv}\text{.}}  
Replacing the velocity we obtain the accretion rate 
\shorteq{\dMbh=4\pi r^2\rho v\st{con}\text{,}} 
which is determined entirely in terms of variables calculated by the
\stars{} code.

\subsection{Luminosity conditions}

The remaining boundary conditions are determined by %considering
the energetics of the accretion flow at the CDAF-ADAF boundary.
The total luminosity throughout the flow has three components:
the advective luminosity $L\st{adv}$,
the viscous luminosity $L\st{vis}$ and
the convective luminosity $L\st{con}$.

Consider first the convective outer flow.  The viscous flux depends on
the total angular momentum flux $\dot{J}$, which is a sum of
contributions from convection and advection.  A defining
characteristic of the CDAF is the balance between these angular
momentum fluxes so the net flux is presumed to be exactly zero
(AIQN02).  Hence, there is no viscous flux in the CDAF.  The
convective luminosity is the luminosity variable $L$ determined by the
code and the advective luminosity is in effect calculated by the
thermal energy generation rate 
\shorteq{\pdif{L}{r}=\dMbh T\pdif{s}{r}\text{,}} 
where $L$ is the local luminosity in the envelope, $T$ the temperature
and $s$ the specific entropy of the gas \citep{markovic95}.

In the advective inner flow, convection no longer functions so
$L\st{con}$ is set to zero.  The gas possesses angular momentum that
must be removed by viscous processes.  As in Section \ref{sslbc1}, the
finite viscous luminosity is taken as
\shorteq{L\st{vis}=\epsilon\dMbh c^2\text{,}} where $c$ is the speed
of light and $\epsilon$ the radiative efficiency.
%, the fraction of accreted rest mass that is released as energy.  
As in Section \ref{sslbc1}, the fraction of mass that is converted
into energy is lost from the quasi-star and its total mass decreases
over time.  Across the CDAF-ADAF boundary, the advective luminosity is
presumed to be continuous and therefore equal on both sides.  For the
total luminosities to be equal, the convective luminosity in the
envelope $L\st{con}$ must be the same as the viscous luminosity in the
ADAF $L\st{vis}$.  The corresponding boundary condition is
\shorteq{L=\epsilon\dMbh c^2\text{,}\label{lbc2}} 
% where the convective luminosity $L\st{con}$ has been replaced with the
% code variable $L$ that represents it.
where I have used the code variable $L$ that represents the convective
luminosity $L\st{con}$.

% The boundary conditions are completed by equating the advective
% luminosity in the ADAF and convective luminosity in the CDAF.
% \citet{abramowicz+02} use this condition to analytically estimate the
% location of the CDAF-ADAF boundary.  In both their integrations and
% those of LLG04, the advective luminosity is taken as
% $L\st{adv}=\dMbh B$, where the Bernoulli number $B$ is essentially
% the total specific energy of the gas and determined by
% \shorteq{B=u+Gm/(r-\rs)+(v\st{adv}^2+r^2\Omega^2)/2}
% evaluated at the innermost meshpoint.  Here, $u$ is the specific
% internal energy of the gas and the post-Newtonian potential
% $\Phi=Gm/(r-\rs)$ \citep{paczynski80} has been used to
% approximate the effects of general relativity.  Equating $L\st{adv}$
% and $L\st{con}$ obtains the boundary condition 
% \shorteq{\epsilon c^2=B\text{.}}

The boundary conditions are completed by balancing the luminosities
inside the ADAF.  As stated, there is no convective luminosity.  The
total flux is the sum of the advective and viscous luminosities
$L\st{adv}$ and $L\st{vis}$.  The total flux is the fraction of
viscous flux that is not advected on to the BH.  In their integrations
AIQN02 and LLG04 take the advective luminosity to be
$L\st{adv}=\dMbh B$, where the Bernoulli function $B$ is in essence the
total specific energy of the gas and is determined by
\shorteq{B=u+Gm/(r-\rs)+(v\st{adv}^2+r^2\Omega^2)/2} evaluated at the
innermost meshpoint.  Here, $u$ is the specific internal energy of the
gas and the post-Newtonian potential $\Phi=Gm/(r-\rs)$
\citep{paczynski80} has been used to approximate the effects of
general relativity.  I presume that some fraction $\eta$, which I call
the \emph{advective efficiency}, of the viscous flux is lost to
advection and use the boundary condition 
\shorteq{\dMbh B=\eta\epsilon\dMbh c^2\text{.}}  
The fiducial value of $\eta=0.8$ is estimated from the work of LLG04
(Fig.~2) and this parameter is varied in Section \ref{sseta2}.

To summarize, the three boundary conditions that replace the
usual stellar conditions $r,\,m,\,L_r=0$ are
% \shorteq{M_0=\Mbh+\frac{8\pi}{3}\rho_0r_0^3\text{,}}
% \shorteq{L_0=\epsilon\dMbh c^2} and
\begin{gather}
M_0=\Mbh+\frac{8\pi}{3}\rho_0r_0^3\text{,} \\
L_0=\epsilon\dMbh c^2
\end{gather}
and
\shorteq{B=\eta\epsilon c^2\text{.}}
The third boundary condition is approximate because of both
the form of the gravitational potential and the use of the
Bernoulli function in the advective luminosity of the ADAF.

\section{Additional physics}
\label{sadph}

Based on the calculations of AIQN02 and LLG04, the inner radius
of CDAF-ADAF quasi-star models is expected to be of the order of tens
of Schwarzschild radii.  In Chapter \ref{cqs1}, rotation was ignored
on the grounds that even Keplerian rotation at a few hundred
Schwarzschild radii is dynamically insignificant at the Bondi radius
when convection enforces constant specific angular momentum.  The
compact surroundings of the BH are now incorporated into the region
modelled by the code so rotation can no longer be ignored.  Its
effects are included by building on the variant of the \stars{} code
developed by \citet*{potter12}, \rose{}.  In Section \ref{ssrot}, I
summarize the modifications to the structure equations and derive the
boundary condition for the rotation variable.

In addition, general relativity introduces effects on the order of
$\rs/r$ and these must also be incorporated.  This is achieved by
including corrections to the structure variables, described by
\citet{thorne77}.  In Section \ref{ssrel}, I restate the original
correction factors and describe their implementation in \rose{}.  The
corrections of \citet{thorne77} also cater for special relativity and
it should be noted that the CDAF-ADAF models do not properly treat
relativistic rotation.  Based on the largest rotational velocities
encountered in the models, the effects of relativistic rotation are
less than a few per cent.  A fully relativistic description of
rotation is not warranted in the presence of other approximations and
the exploratory nature of this work.

\subsection{Rotation}
\label{ssrot}

\citet{potter12} based the rotating stellar evolution code \rose{} on
the Cambridge \stars{} code as described in Chapter \ref{cstars}.  The
effects of rotation are introduced by modifying the structure
variables and equations according to prescriptions of \citet{endal78}
and \citet{meynet97}.  Let $S_p$ be the area of a surface of
constant pressure $p$, $V_p$ the volume it contains and $r_p$
the radius of a sphere with the same volume.  In the following, some
quantities are averaged over the surface and we define
\shorteq{\mean{q}=\frac{1}{S_p}\oint_{S_p}q\mathrm{d}\sigma\text{,}}
where $d\sigma$ is a surface area element.  The main assumption of the
structure model is the \emph{Roche approximation}: the gravitational
potential at radius $r_p$ is the same as if all the mass inside
the surface $S_p$ were spherically distributed within $r_p$.

The mass conservation equation becomes
\shorteq{\pdif{m_p}{r_p}=4\pi r_p^2\rho\text{,}} 
where $\rho$ is the density on the constant-pressure surface $S_p$ and
$m_p$ the mass it encloses.  The equation of hydrostatic equilibrium
changes to
\shorteq{\pdif{p}{m_p}=-\frac{Gm_p}{4\pi r_p^4}f_p\text{,}}
where\vskip-10mm
\shorteq{f_p=\frac{4\pi r_p^4}{Gm_pS_p}
 \mean{g\st{eff}^{-1}}^{-1}\text{.}}
Here, $g\st{eff}$ is the magnitude of the effective gravity, which is
a sum of the gravitational and centrifugal accelerations.

The rotational profile of convective zones is controlled by a
parameter in \rose{}.  Mixing-length theory implies that a buoyant
parcel of gas retains its specific angular momentum until it
disperses.  This %process
drives the rotational profile to a state of
constant specific angular momentum.  Magnetic fields may %are suspected to
modify the angular momentum distribution but, because they are
disregarded here, the convective zones are presumed to tend towards a
state of constant specific angular momentum.  Note that convection is
thought strong enough that modifications owing to rotation can be
ignored.  The criterion for convective instability is implemented
unmodified from the \stars{} code.

Several %A number of 
diffusion coefficients for 
angular momentum mixing %the mixing of angular momentum
are implemented in \rose{}.  For CDAF-ADAF quasi-stars, 
%the rotational diffusion coefficient in the convective zones was
we chose $D\st{MLT}=v\st{con}\mixl/3$, where $\mixl$ is the
mixing length determined by \rose{} in convective zones.  In
radiative zones, the shear component of the diffusion coefficient is
determined by the model of \citet{talon+97} and the horizontal
component by the model of \citet{maeder03}.  The parameter used 
in this chapter corresponds to case 4 of \citet{potter12}.

The boundary condition for the specific angular momentum $\Omega r^2$,
which is now a variable in the code, is determined by conservation of
angular momentum and the stipulation that convective zones are driven
to constant specific angular momentum.  Angular momentum
redistribution is implemented as a diffusion process, which can be
used to replace the rate of change of the specific angular momentum.
Almost all of the envelope is fully convective throughout the
evolution so the rate of change of total angular momentum of the
envelope $J_*$ can be written as

\begin{align}
\pdif{J_*}{t}
&=\pdif{}{t}\int_{M_0}^{M_*}\sq\Omega r^2\mathrm{d}m \\
&=\dot{M}_*\Omega_*R_*^2-\dMbh \Omega_0r_0^2+\int_{M_0}^{M_*}\pdif{}{t}(\Omega r^2)\mathrm{d}m \\
&=\dot{M}_*\Omega_*R_*^2-\dMbh \Omega_0r_0^2+\int_{r_0}^{R_*}\sq\frac{1}{\rho r^2}\pdif{}{r}\left(\rho D\st{con}r^2\pdif{}{r}(\Omega r^2)\right)4\pi r^2\rho \mathrm{d}r \\
&=\dot{M}_*\Omega_*R_*^2-\dMbh \Omega_0r_0^2+\left[4\pi\rho D\st{con}r^2\pdif{}{r}(\Omega r^2)\right]_{r_0}^{R_*}\sq\text{.}
\end{align}
Above, $\dot{M}_*$ is the rate of change of the mass of the envelope
at the surface, either by accretion or mass-loss, and $D\st{con}$ is
the angular momentum diffusion coefficient in the convective zones, as
calculated in \rose{}.  The subscript $0$ indicates variables
evaluated at the innermost meshpoint, $\Omega_*$ is the angular
velocity at the surface and $R_*$ is the outer radius of the
envelope.  To conserve angular momentum in the envelope, the final term
must be equal to zero.  In normal stars, the photosphere is presumed
radiative and the surface boundary condition is
%$\pdif{\Omega}{r}=0$.  
$\qdif{\Omega}{r}=0$.  
% The modified boundary condition for the rotation variable,
% \shorteq{\pdif{}{r}(\Omega r^2)=0\text{,}} is applied at the innermost
% meshpoint.
At the innermost meshpoint, I apply the modified boundary condition
for the rotation variable, 
\shorteq{\pdif{}{r}(\Omega r^2)=0\text{.}}

\subsection{Special and general relativity}
\label{ssrel}

In his formulation of the equations of relativistic stellar structure,
\citet{thorne77} introduced two new variables and two associated
differential equations to compute them.  The mass variable $m$
represents the total rest mass inside radius $r$.  The total mass
variable $m\st{tr}$ is given by the total mass inside a radius $r$,
including the contributions of the material's nuclear binding energy,
internal energy and gravitational potential.  The gravitational
potential $\Phi$ is related to the metric tensor.

In terms of these variables and the standard structure variables (see
Chapter \ref{cstars}), \citet{thorne77} defined five correction
factors for the structure equations.  They are
% \shorteq{\mathcal{R\st{T}}=\exp(\Phi/c^2)\text{,}}
% \shorteq{\mathcal{V\st{T}}=1/\sqrt{1-\frac{2Gm\st{tr}}{c^2r}}\text{,}}
% \shorteq{\mathcal{G\st{T}}=\frac{1}{m}(m\st{tr}+\frac{4\pi r^3p}{c^2})\text{,}}
% \shorteq{\mathcal{E\st{T}}=1+\frac{u}{c^2}\text{}} and
% \shorteq{\mathcal{H\st{T}}=1+\frac{u+p/\rho}{c^2}\text{.}}
% \begin{align}
% &\mathcal{R\st{T}}=\exp(\Phi/c^2)\text{,} \\
% &\mathcal{V\st{T}}=1/\sqrt{1-\frac{2Gm\st{tr}}{c^2r}}\text{,} \\
% &\mathcal{G\st{T}}=\frac{1}{m}(m\st{tr}+\frac{4\pi r^3p}{c^2})\text{,} \\
% &\mathcal{E\st{T}}=1+\frac{u}{c^2}\text{}
% \end{align}
\begin{gather}
\mathcal{R\st{T}}=\exp(\Phi/c^2)\text{,} \\
% \mathcal{V\st{T}}=1\left/\sqrt{1-\frac{2Gm\st{tr}}{c^2r}}\right.\text{,} \\
\mathcal{V\st{T}}=\left(1-\frac{2Gm\st{tr}}{c^2r}\right)^{\sq-1/2}\sq\text{,} \\
\mathcal{G\st{T}}=\frac{1}{m}\left(m\st{tr}+\frac{4\pi r^3p}{c^2}\right)\text{,} \\
\mathcal{E\st{T}}=1+\frac{u}{c^2}\text{}
\end{gather}
and
\shorteq{\mathcal{H\st{T}}=1+\frac{u+p/\rho}{c^2}\text{.}}
The gravitational acceleration $g$ is replaced with
$\mathcal{G\st{T}V\st{T}}g$, the pressure scale height $H_p$ with
$H_p/\mathcal{H}\st{T}$ and the radiative gradient $\nabla\st{rad}$
with $\nabla\st{rad}\mathcal{H\st{T}G\st{T}V\st{T}}
+(1-\mathcal{E}\st{T}/\mathcal{H}\st{T})$.  The additional structure equations
are
\shorteq{\pdif{m\st{tr}}{m}=\frac{\mathcal{E}\st{T}}{\mathcal{V}\st{T}}}
and
\shorteq{\pdif{\Phi}{m}=\frac{Gm}{4\pi r^4\rho}\mathcal{G\st{T}V\st{T}}\text{.}}
The mass conservation equation becomes
\shorteq{\pdif{m}{r}=4\pi r^2\rho\mathcal{V}\st{T}}
and hydrostatic equilibrium is written
\shorteq{\pdif{p}{m}=-\frac{Gm}{4\pi r^4}\mathcal{G\st{T}H\st{T}V\st{T}}\text{.}}
Here, the correction factors are simplified by taking
$m\st{tr}=(\mathcal{E}/\mathcal{V})m$ and
%$\Phi=(1/2)c^2\log\left|1-2Gm/c^2r\right|$, 
\shorteq{\Phi=(1/2)c^2\log\left|1-2Gm/c^2r\right|\text{,}}
which removes the need for
the additional differential equations.  The correction factors are
implemented as
% \shorteq{\mathcal{R}=\sqrt{1-\frac{2Gm}{c^2r}}=\mathcal{V}^{-1}\text{,}}
% \shorteq{\mathcal{G}=1+\frac{4\pi r^3p}{mc^2}\text{,}}
% \shorteq{\mathcal{E}=1+\frac{u}{c^2}} and
\begin{gather}
\mathcal{R}=\sqrt{1-\frac{2Gm}{c^2r}}=\mathcal{V}^{-1}\text{,} \\
\mathcal{G}=1+\frac{4\pi r^3p}{mc^2}\text{,} \\
\mathcal{E}=1+\frac{u}{c^2}
\end{gather}
and
\shorteq{\mathcal{H}=1+\frac{u+p/\rho}{c^2}=1+\mathcal{E}+\frac{p}{\rho c^2}\text{.}}
They are applied to the structure variables as described above.  The
relativistic corrections are thus not exact but still represent an
improvement over the Newtonian treatment (see Section \ref{ssnorogr}).  

\section{Fiducial model}
\label{sfidevol2}

In this section, I present the evolution of a quasi-star with total
mass $M_*=10^4\Msun$, initial BH mass $84\Msun=0.0084M_*$ and a
uniform composition of $0.7$ hydrogen and $0.3$ helium by mass.  The
initial BH mass is larger than that of the models described in Chapter
\ref{cqs1} because none evolved smoothly from states with interior
radiative zones into a state of total convection.  The advective and
radiative efficiencies $\eta$ and $\epsilon$ are $0.8$ and $0.04$
respectively.  The envelope initially has a constant specific angular
momentum profile.  The specific angular momentum at the inner envelope
boundary $j_0=\sci{1.83}{17}\cmsq\ps$ is chosen so that a BH with the
same specific angular momentum would have a dimensionless spin
parameter $a_*=J\st{BH}/(G\Mbh^2/c)=0.5$ (see Section \ref{ssspin}).

The strong advective luminosity in the quasi-star envelope constitutes
a departure from thermal equilibrium.  Because the evolving models are
not thermally relaxed, it is impossible to initialize a model in
thermal equilibrium. 
%, as for the Bondi-type quasi-stars.
Instead, models are initialized with a constant energy generation term
of $\epsilon\st{c}=\sci{5}{4}\ergps\pg$, which adds a total luminosity
of about two thirds of the Eddington luminosity for a given object.
The extra energy declines exponentially like $\exp(-0.008t/\yr)$ so
that $\epsilon\st{c}<10^{-2}\ergps\pg$ after less than $2000\yr$.

\begin{figure}\begin{center}
\includegraphics{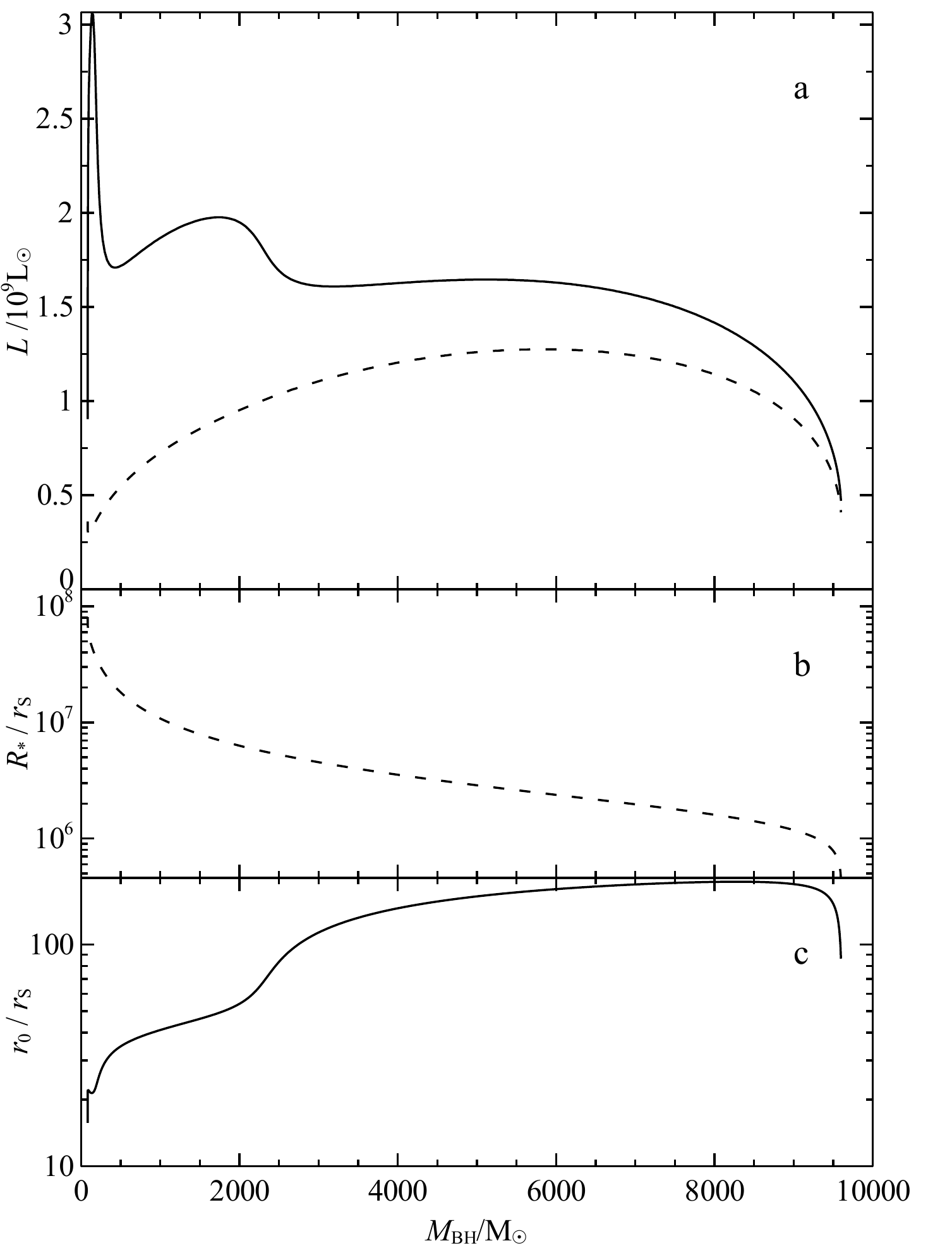}
\caption[Evolution of the BH and surface luminosities (a) and inner
and surface radii in Schwarzschild units (b, c) for the fiducial run.]
{The top plot (a) shows the evolution of the BH luminosity $\Lbh$
  (solid line) and surface luminosity $L_*$ (dashed line) in the
  fiducial run.  The inner luminosity is divided into three phases but
  these are not reflected in the surface luminosity.  A varying and
  initially substantial fraction of the BH luminosity is lost to
  advection and not radiated from the surface.  The lower plots (b, c)
  show the evolution of the inner radius $r_0$ and surface radius
  $R_*$ in units of the BH's Schwarzschild radius $\rs$.  The surface
  radius reflects the smooth behaviour of the surface luminosity but
  the inner radius shows changes that are connected to the changing BH
  luminosity and accretion rate.  Although the changes are correlated,
  it is unclear which are causal.}
\label{ffidev2}
\end{center}\end{figure}

\subsection{Evolution}

Fig.~\ref{ffidev2} shows the evolution of the BH and surface
luminosities as a function of BH mass and contains all the qualitative
behaviour present in any of the model sequences.  The BH luminosity
exhibits three humps of decreasing maximum magnitude and increasing
width.  The peaks correspond to luminosities $L_*/10^9\Lsun=3.07$,
$1.98$ and $1.65$ when $\Mbh/\Msun=143$, $1738$ and $5115$.  The BH
luminosity is defined by equation (\ref{lbc2}) and is related to the
accretion rate by $\dMbh=\sci{1.696}{-3}(L_*/10^9\Lsun)\Msunpyr$ so
the accretion rate has peaks of $\dMbh/10^{-3}\Msunpyr=5.21$, $3.35$
and $2.79$.  I refer to these local maxima, in both the BH luminosity
and accretion rate, as the \emph{luminosity humps}.

The evolution terminates after $3.64\Myr$ at which time $11.8\Msun$
remains in the envelope and the BH has grown to $9592\Msun$.  Thus,
the BH accretes all but 0.12 per cent of the available mass and this
fraction varies very little between the models presented in this
chapter.  The nearly complete accretion of the envelope on to the BH
is the most important feature of the models in this chapter.  Neither
is there an opacity crisis nor is the BH limited to the fractional
mass limit found in Chapter \ref{cqs1}.

It remains unclear why the accretion rate exhibits the three-humped
behaviour.  Just as the surface luminosity in Fig.~\ref{ffidev2}, none
of the surface properties shows any humps so they must be caused
either at the inner boundary or deep inside the envelope.  It is,
however, difficult to disentangle the possibilities.  The humps are
certainly not caused by changes in the location of convective zones.
Throughout the evolution, the only convectively stable region is
adjacent to the surface and extends over less than $100\Msun$ by mass.
I suspect that the humps are related to changes in the opacity or
ionization state of the gas but I have not been able to demonstrate
this.  They are possibly also affected by further departures from
thermal equilibrium because the thermal timescale $t\st{KH}=Gm^2/rL$
increases both with time and towards the inner boundary.  During the
final hump, the thermal timescale at $r_0$ is a few $10^5\yr$.

Figs \ref{ffidev2}b and \ref{ffidev2}c show the locations of the inner
and surface radii in units of $\rs$ as the BH mass grows.  The inner
radius has a typical magnitude of a few tens of times the Schwarzschild
radius.  This is consistent with the estimates and integrations by
AIQN02 and LLG04.  Some differences can be attributed to their
results being calculated for thick discs with finite opening angles
whereas the quasi-star envelopes are spherical.  Variations in
the rate of change of the inner radius co-incide with similar shifts
in the BH luminosity and accretion rate but it is unclear whether
changes in the luminosity or accretion rate affect the location of the
inner radius or vice versa.

\begin{table}\begin{center}
\caption[Properties of the fiducial model as $\Mbh$ increases.]
{Properties of the fiducial model as $\Mbh$ increases.  The first and
  last entries correspond to the initial and final models in the run,
  respectively.  Luminosity profiles are plotted in Fig.~\ref{fadv}
  and density profiles in Fig.~\ref{fprof2}.}
\begin{tabular}{@{}crccr@{}lr@{}lr@{}lr@{}l@{}}
\toprule[1pt]
$t$ &$\Mbh$ &$\dMbh$ &$L_0$&\mchead{$T_0$}&\mchead{$r_0$} \\
/$10^6\yr$&/\Msun&/$10^{-3}\Msunpyr$&/$10^9\Lsun$
&\mchead{/$10^5\K$}&\mchead{/$\Rsun$}& \\
\midrule
0.00  &84     &1.53&0.90    &168&   &0&.00769\\
0.29  &1000   &3.17&1.87    & 39&.9 &0&.175 \\
0.59  &2000   &3.30&1.95    & 24&.4 &0&.463 \\
1.65  &5000   &2.79&1.65    &  8&.26&3&.49  \\
3.64  &9592   &0.86&0.51    &  6&.96&3&.96  \\
\midrule
 & &$\rho_0$& $L_*$  &\mchead{$T\st{eff}$}& \mchead{$R_*$} \\
 & &/\gpcm & /$10^8\Lsun$ &\mchead{/$10^3\K$}& \mchead{/$10^4\Rsun$} \\
\midrule
0.00  &84     &$\sci{1.27}{-5}$ &3.51&4&.76&2&.76 \\
0.29  &1000   &$\sci{5.09}{-8}$ &7.27&4&.43&4&.59 \\
0.59  &2000   &$\sci{7.69}{-9}$ &9.54&4&.39&5&.35 \\
1.65  &5000   &$\sci{1.19}{-10}$&12.6&4&.41&6&.09 \\
3.64  &9592   &$\sci{5.39}{-11}$&4.36&5&.79&2&.08 \\
\bottomrule[1pt]
\end{tabular}
\label{tprof2}
\end{center}\end{table}

\begin{figure}\begin{center}
\includegraphics{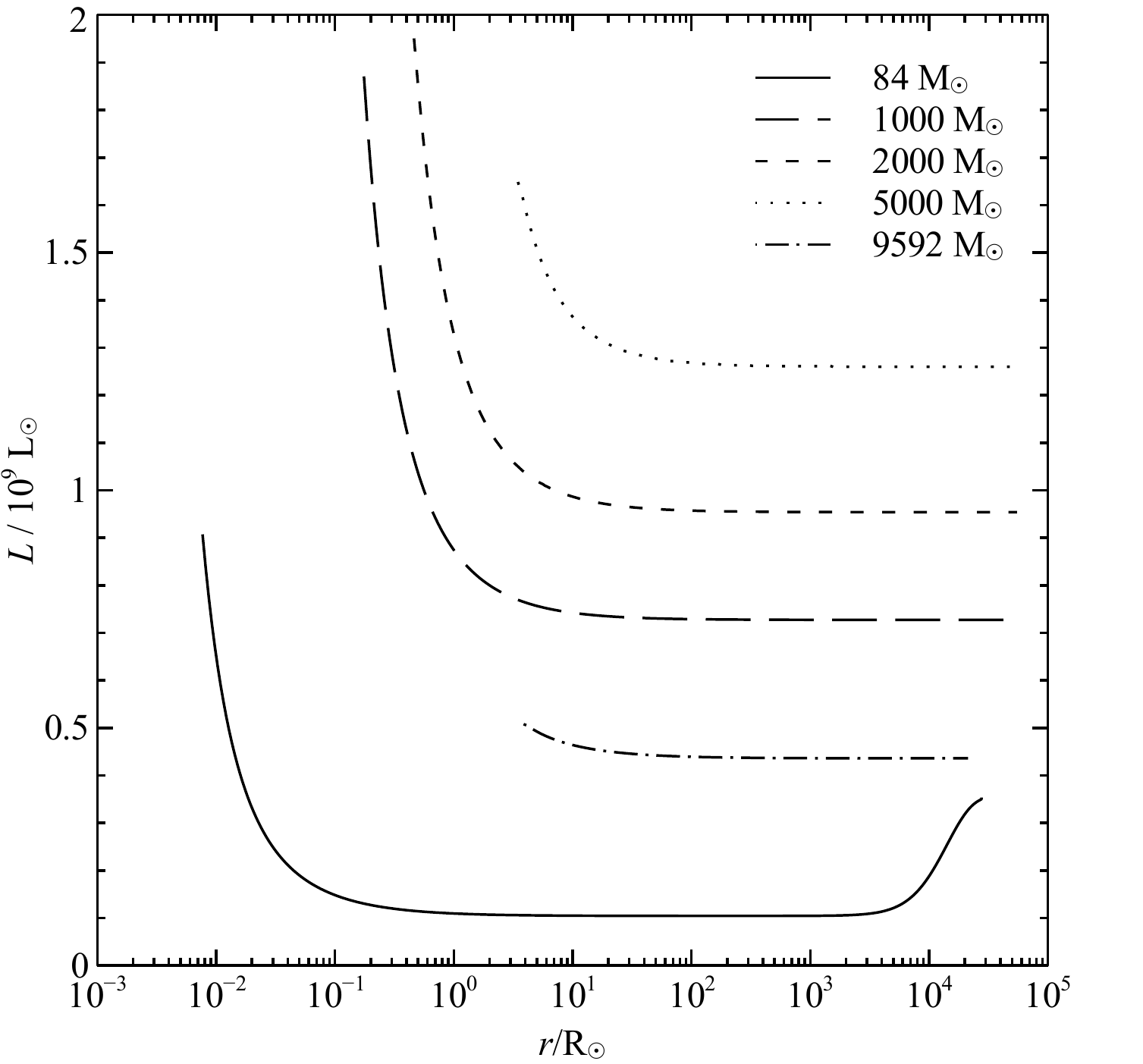}
\caption[Luminosity profiles of the fiducial quasi-star.]  {Plot of
  luminosity against radius for models in the fiducial sequence with
  $\Mbh/\Msun=84$, $1000$, $2000$, $5000$ and $9592$ (see Table
  \ref{tprof2}).  In the outer envelope, the luminosity is roughly
  constant. Near the inner boundary, advection affects the luminosity.
  Advection carries energy inwards and acts like a negative thermal
  energy generation so the luminosity decreases outwards. The initial
  model has a constant energy generation rate of $\sci{5}{4}\ergps\pg$
  and the luminosity consequently increases outwards at the outer
  edge.}
\label{fadv}
\end{center}\end{figure}

\subsection{Structure}

The variation of luminosity with radius for the models listed in Table
\ref{tprof2} is shown in Fig.~\ref{fadv}.  The portion of the envelope
that experiences strong advection is visible in each profile.  In
addition, the decreasing relative strength of the advection for
increasing BH mass can be seen.  This phenomenon is also apparent in
Fig.~\ref{ffidev2} where the difference between the BH and surface
luminosities broadly decreases as the BH grows.  Note that the first
luminosity profile shows a large trough because the initial constant
energy generation $\epsilon\st{C}$ is still present.
Owing to the low density throughout the envelope, the region of the
quasi-star with a significant advective luminosity is small in mass.
In all the models in the fiducial run, the thermal energy generation
$\epsilon\st{th}$ is less than $1000\ergps\pg$ beyond $60\Msun$ and
less than $200\ergps\pg$ outside of the first $220\Msun$ of the
envelope.  

\begin{figure}\begin{center}
\includegraphics{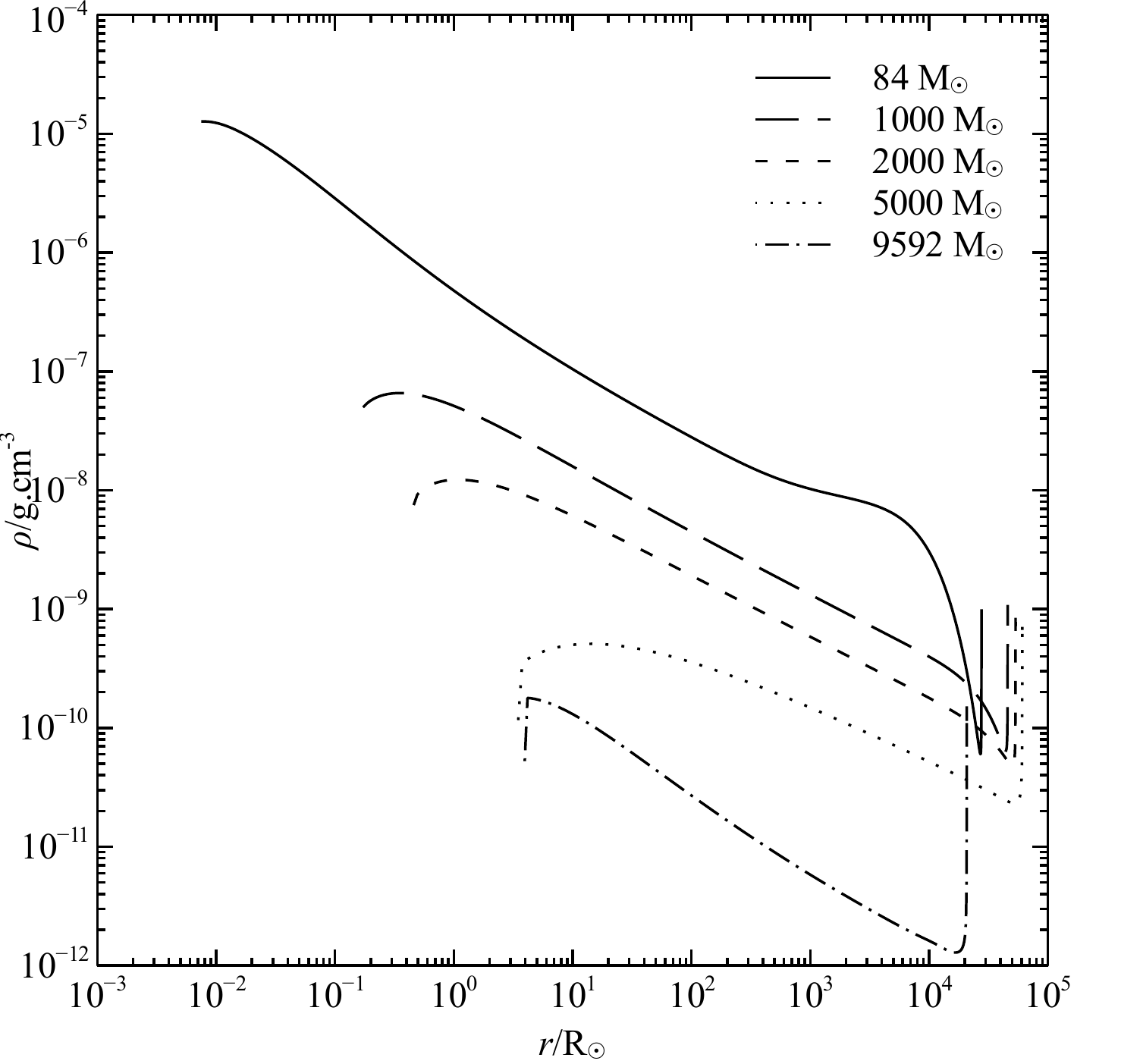}
\caption[Density profiles of the fiducial quasi-star.]  {Plot of
  density against radius for models in the fiducial run with
  $\Mbh/\Msun=84$, $1000$, $2000$, $5000$ and $9592$ (see Table
  \ref{tprof2}).  The models show an increasingly strong density
  inversion at the centre.  This is formally
  unstable %in a steady model
  but the large advection velocity might depress the density to
  conserve mass.}
\label{fprof2}
\end{center}\end{figure}

The density profiles of the models in Fig.~\ref{fprof2} exhibit an
inversion near the inner boundary.  That is, the density increases
outwards.  This is theoretically unstable to the Rayleigh--Taylor
instability.  Density inversions are understood to appear near the
surfaces of one-dimensional models of red supergiants.  However, the
implications of an inversion near the centre of the CDAF-ADAF
quasi-stars are unclear.  The models are not in a steady state, so a
a sufficient increase in the advection velocity may be represented by
a corresponding decrease in the density in order to keep the mass
accretion rate constant throughout the model.

\begin{figure}\begin{center}
\includegraphics{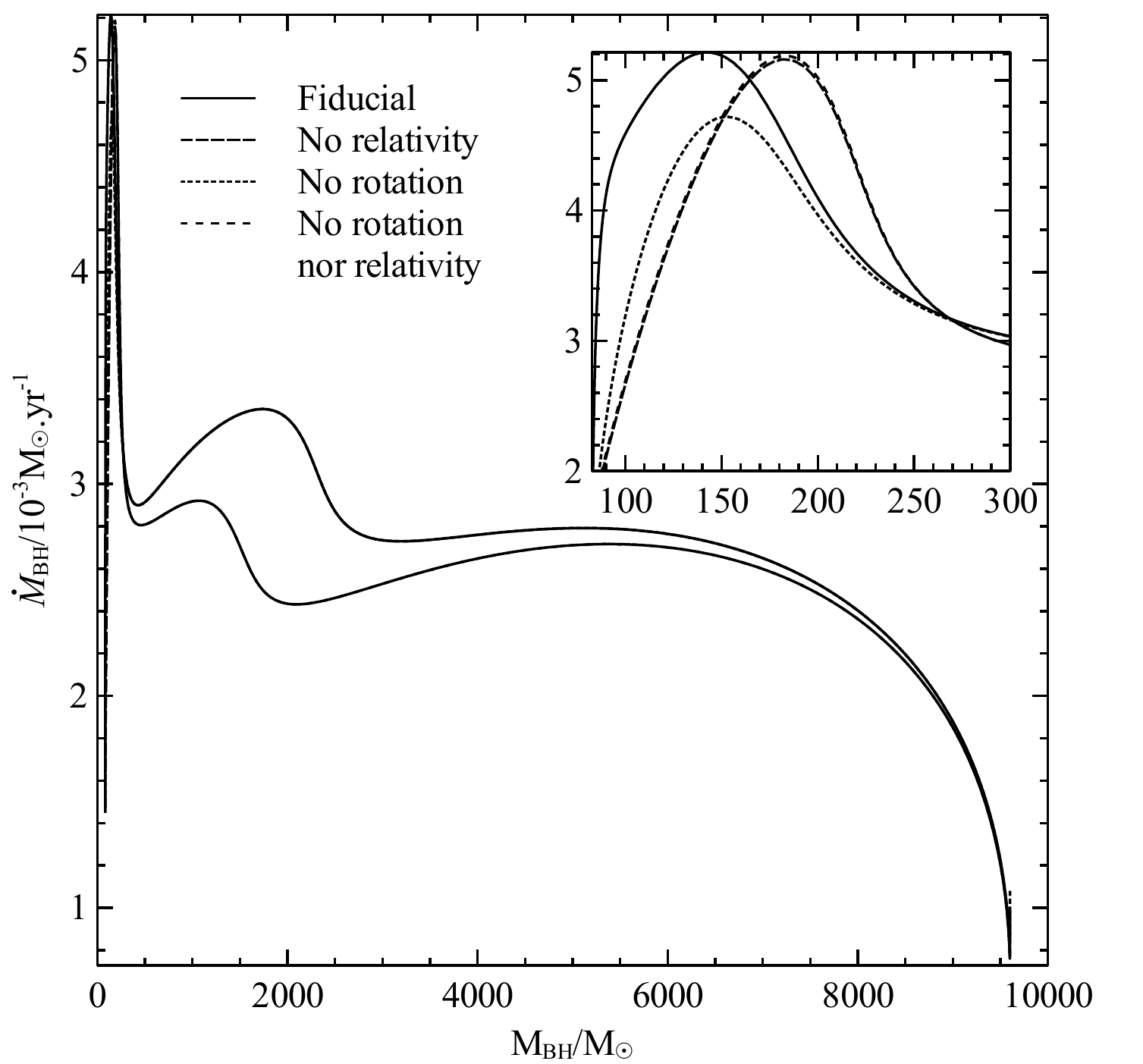}
\caption[Evolution of quasi-stars with or without rotation and
relativistic effects.]  {Plot of the evolution of quasi-stars with or
  without rotation or relativistic effects.  The inset has the same
  scales on the axes.  For $\Mbh\gtrsim500\Msun$, the rotating and
  non-rotating models are indistinguishable. The relativistic models
  have smaller inner radii, larger advective luminosities and so
  larger accretion rates.}
\label{fnorogr}
\end{center}\end{figure}

\subsection{Non-rotating and non-relativistic models}
\label{ssnorogr}

To evaluate the importance of rotation and relativity in the models,
three additional runs were conducted with no rotation, no corrections
due to relativity or both.  The %corresponding 
evolution for each case
is plotted in Fig.~\ref{fnorogr}.
Rotation only makes a significant difference during the first %inner
luminosity hump with relativity included.  It otherwise plays a small
role in the structure of the envelope. % and the evolution of the BH.
This is not surprising.  The inner radius increases almost
monotonically with $\rs$, which is itself %a function of 
proportional to the BH mass.
The specific angular momentum $\Omega r^2$ is %remains 
roughly constant
so the angular velocity falls %off 
at least as $1/\Mbh^2$ and %even
faster if the inner radius increases as a multiple of the
Schwarzschild radius.  %By comparison, the 
The Keplerian velocity
$\Omega\st{K}=\sqrt{G\Mbh/r^3}$ falls off more slowly, as $1/\Mbh$.
Thus the ratio $\Omega_0/\Omega\st{K}$ decreases over time, as %and so 
do the structural effects of rotation.

The relativistic corrections introduce the contribution of the energy
density to the gravitational potential, which is therefore steeper
than its Newtonian counterpart.  The inner radius is smaller, the
advected luminosity greater and the accretion rate larger.  At later
times, when the inner radius exceeds $100\rs$, relativity has little
effect on the structure and the accretion rates %effectively 
converge.

All the runs achieve roughly the same final BH mass of about
$9592\Msun$.  Both the non-relativistic rotating and the
non-relativistic non-rotating models have lifetimes of $3.87\Myr$.
The non-rotating relativistic model achieves its final BH mass after
$3.65\Myr$, which is very similar to the lifetime of the fiducial
model.  The difference in the lifetimes exists mainly because of the
higher accretion rates in the relativistic models.

\section{Parameter exploration}
\label{smore2}

Having established the basic properties of CDAF-ADAF quasi-stars, I
now present models with different choices of the free parameters, as
in Section \ref{smore1}.  I begin by varying the speed of rotation in
Section \ref{ssspin} and discussing the evolution of the BH spin.  In
Sections \ref{sseps2} and \ref{sseta2}, I vary the radiative and
advective efficiencies and, in Section \ref{ssmi2}, I consider changes
to the total mass of the quasi-star, both in the initial model and
during the evolution.

\subsection{Rotation}
\label{ssspin}

\begin{figure}\begin{center}
\includegraphics{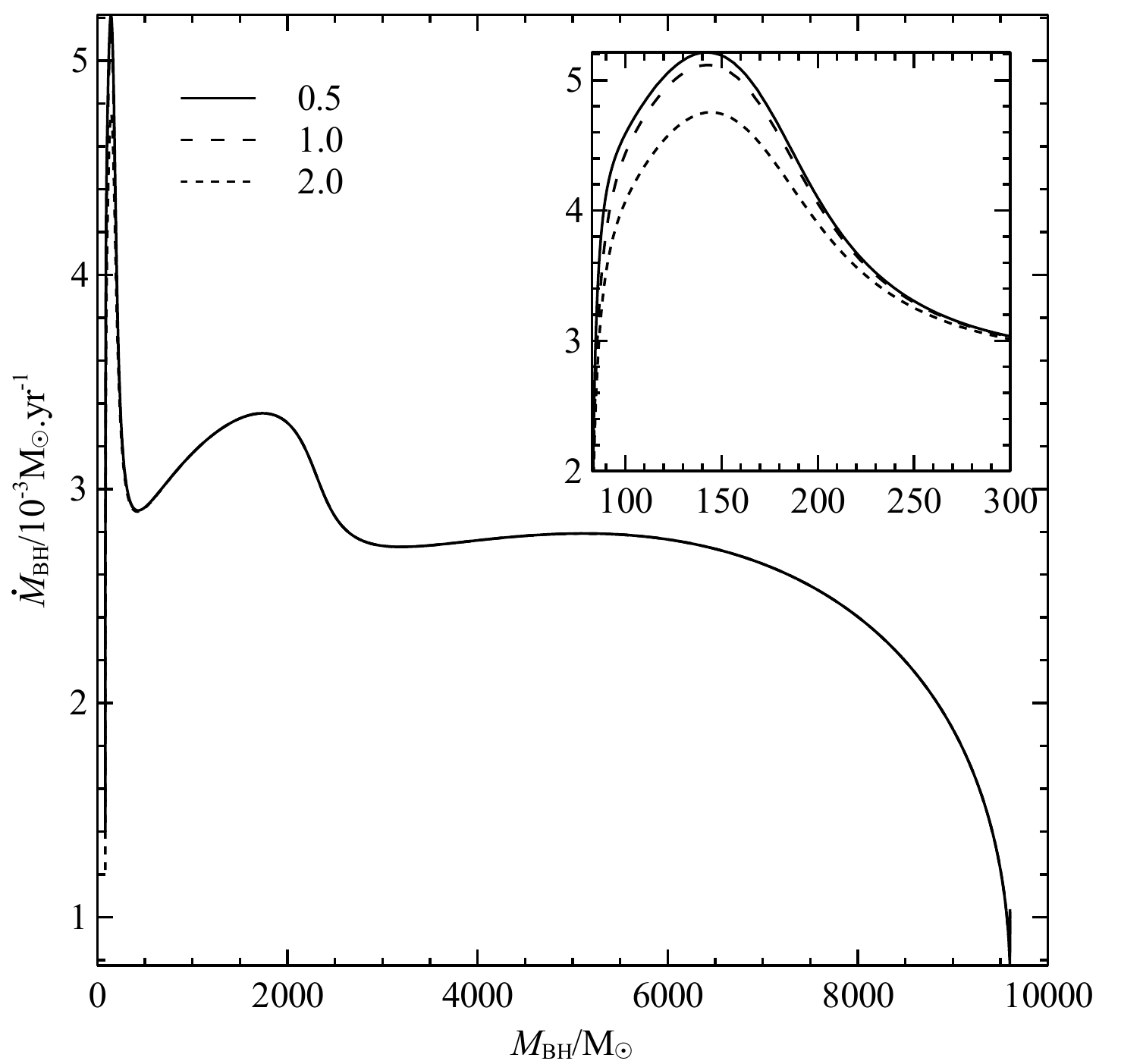}
\caption[Evolution of quasi-stars with different rotation rates.]
{Plot of the evolution of quasi-stars with different rotation
  rates. The inset has the same scales on the axes. The models have
  constant specific angular momenta and are parametrized by the spin
  of a BH with the same specific angular momentum and of the same
  initial mass as in the relevant evolution. The initial BH masses in
  these runs are $84\Msun$ and the spins $a_*=0.5$, $1$ and $2$
  correspond to specific angular momenta $\Omega_0
  r_0^2/10^{17}\cmsq\ps=1.83$, $3.66$ and $7.31$. The rotation rate
  falls off rapidly and does not affect the evolution beyond
  $\Mbh\approx300\Msun$.}
\label{fomega}
\end{center}\end{figure}

Reasonable values for the rotation of the envelope are chosen such
that a BH with the same specific angular momentum has a realistic
spin.  If the BH has spin parameter $a_*$, its specific angular momentum
$j\st{BH}$ is given by
\begin{align}
j\st{BH}
&=\frac{J\st{BH}}{\Mbh}
=\frac{a_*G\Mbh^2}{c\Mbh}
=\frac{a_*G\Mbh}{c} \\
&=\sci{4.43}{15}\bracfrac{\Mbh}{\Msun}a_*\cmsq\ps\text{,}
\end{align}
where $J\st{BH}$ is the total angular momentum of the BH.  Thus, the
fiducial choice $j_0=\sci{1.83}{17}\cmsq\ps$ corresponds to a spin
parameter $a_*=0.5$ for an $84\Msun$ BH.  Fig.~\ref{fomega} shows the
evolution of central luminosity against BH mass for $j_0$
corresponding to initial spin parameters $a_*=0.5$, $1$ and $2$.
Choices of $a_*$ larger than $1$ can be regarded as cases where the BH
was born with a smaller specific angular momentum than the envelope.
For a given angular velocity, a rigidly-rotating sphere with constant
density would have $0.6$ times and a rigidly-rotating $n=1$ polytrope
about $0.37$ times the angular momentum of a sphere with constant
specific angular momentum.  More centrally-condensed objects would
have even smaller values so if the BH was born from a gas with such a
profile it would have a smaller spin parameter to that inferred from
the envelope under the presumption of constant specific angular
momentum.

\begin{figure}\begin{center}
\includegraphics{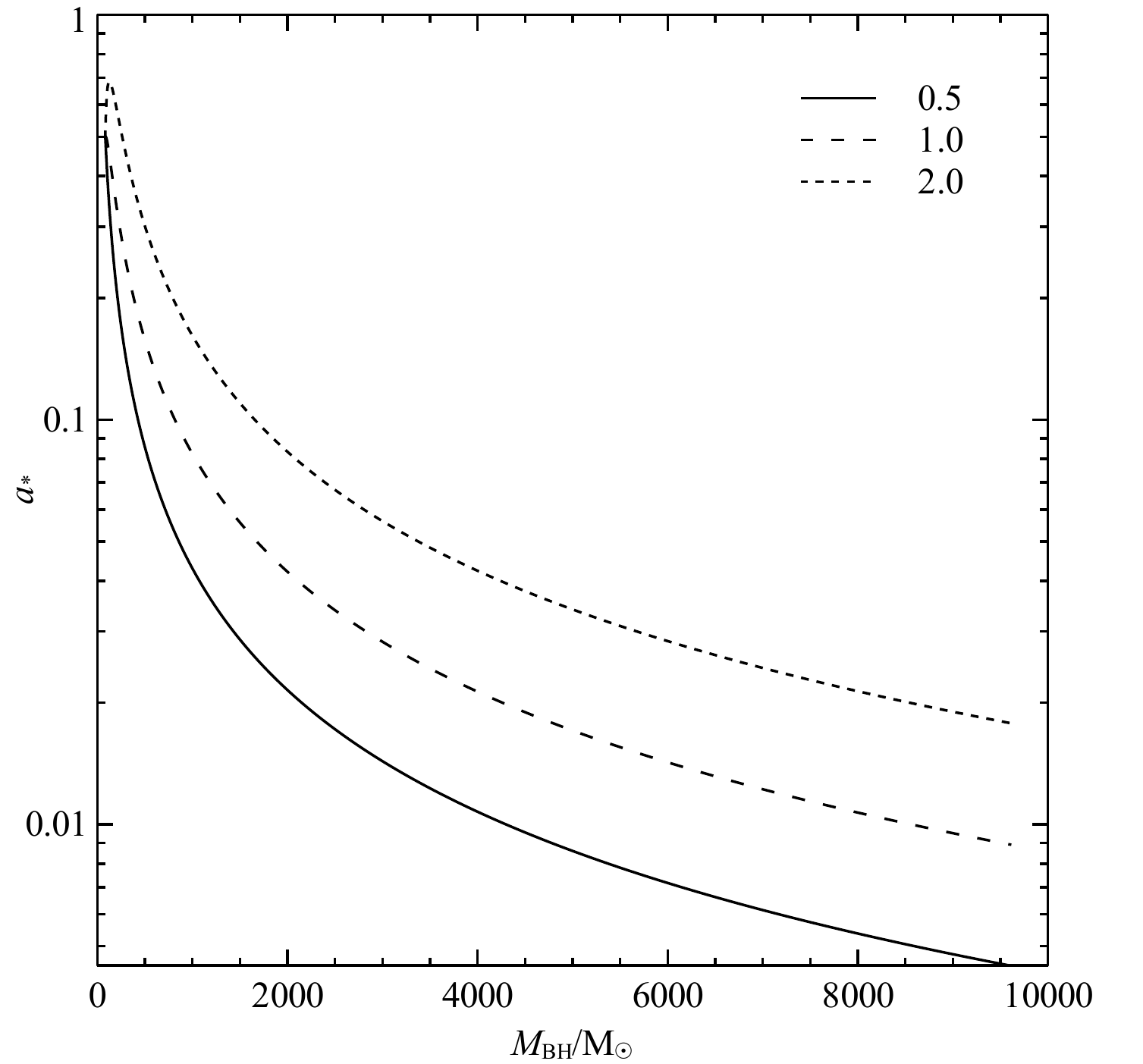}
\caption[Evolution of BH spin for quasi-stars with different rotation
rates.]{Plot of BH spin as a function of BH mass for the quasi-star
  models in Fig.~\ref{fomega}. In each model, the spin parameter $a_*$
  is inversely proportional to the BH mass to high accuracy. The spin
  thus decreases rapidly as the BH grows.}
\label{fspin}
\end{center}\end{figure}

% \comment{Check the calculation for an $n=1$ polytrope.}

Because the angular velocity in the envelope declines rapidly, as
noted in Section \ref{ssnorogr}, different rotation speeds only matter
near the beginning of the evolution.  After the first luminosity hump,
the inner radius increases significantly and, for
$\Mbh\gtrsim300\Msun$, the evolutionary sequences are nearly
indistinguishable.  All the runs end with BH masses of about
$9592\Msun$.

Fig.~\ref{fspin} shows the evolution of the spin parameters $a_*$ for
the quasi-stars in Fig.~\ref{fomega}.  The spin of the BH is
calculated by recording the total angular momentum lost at the inner
boundary and presuming that it becomes part of the BH.  In all three
cases, the spin parameters are proportional to $1/\Mbh$ and this is
easily explained.  If a BH is born from and accretes matter with some
constant specific angular momentum $j$, its angular momentum
increases as $\dot{J}\st{BH}=j\dMbh$.  Convective transport of angular
momentum maintains a constant specific angular momentum profile.  The
BH's total angular momentum is the sum of its initial angular momentum
and whatever it accretes and is therefore
\shorteq{J\st{BH}=jM_{\text{BH},i}+j(\Mbh-M_{\text{BH},i})=j\Mbh\text{,}}
where a subscript $i$ represents an initial value.  The BH's spin
parameter is
% \shorteq{a_*=\frac{J\st{BH}c}{G\Mbh^2}=\frac{c}{G\Mbh}\frac{a_{*,i}G\Mbh{_{,i}}}{c}
%   =a_{*,i}\frac{\Mbh{_{,i}}}{\Mbh}\text{.}}  Because the initial
\begin{align}a_*&=\frac{J\st{BH}c}{G\Mbh^2}=\frac{c}{G\Mbh}\frac{a_{*,i}G\Mbh{_{,i}}}{c}\\
  &=a_{*,i}\frac{\Mbh{_{,i}}}{\Mbh}\text{.}\end{align} 
Because the initial values are constant, it follows that the spin
parameter is inversely proportional to the BH mass.

\begin{figure}\begin{center}
\includegraphics{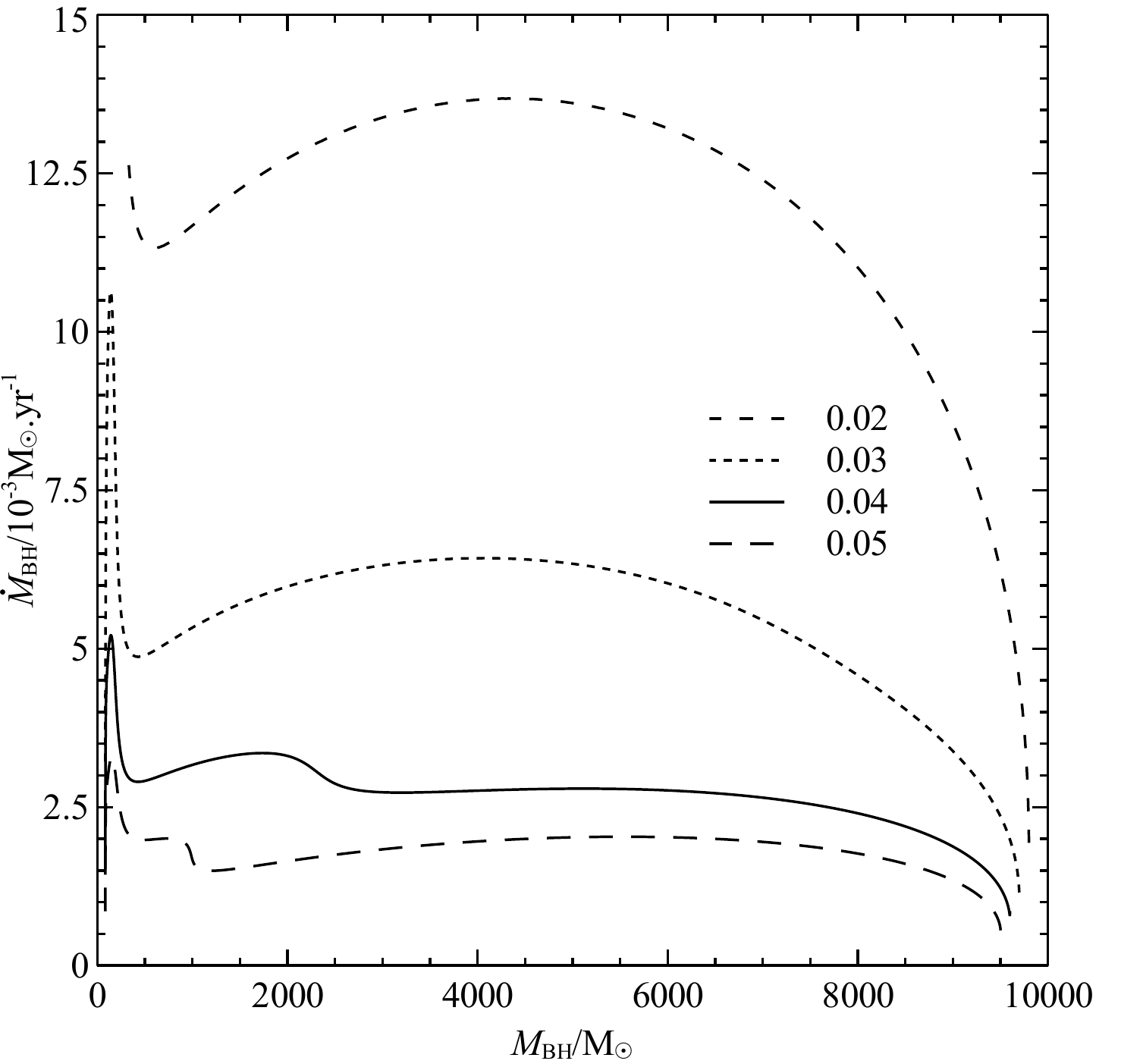}
\caption[Evolution of quasi-stars with different radiative
efficiencies.]  {Plot of the evolution of quasi-stars for radiative
  efficiencies $\epsilon=0.02$, $0.03$, $0.04$ and $0.05$. For
  $\epsilon=0.02$, models could only be constructed with
  $\Mbh\gtrsim300\Msun$ and the initial BH mass was $331\Msun$.  In all
  cases, larger $\epsilon$ leads to smaller accretion rates and
  narrower luminosity humps. For $\epsilon=0.02$ and $0.03$, the humps
  are broader than the total quasi-star mass so there is one fewer
  hump.  Models with larger values of $\epsilon$ also achieve slightly
  larger final BH masses because less mass is destroyed to produce
  energy.}
\label{feps2}
\end{center}\end{figure}

\subsection{Radiative efficiency}
\label{sseps2}

The evolution of quasi-stars with radiative efficiencies $0.02$,
$0.03$ and $0.05$ is plotted in Fig.~\ref{feps2} along with the
fiducial evolution in which $\epsilon=0.04$.  The accretion rate
increases for smaller $\epsilon$ because a larger accretion rate is
required to achieve the same surface luminosity.  The dependence is
non-linear and also related to the properties of the advective
luminosity.  Decreasing the radiative efficiency by one quarter to
$0.03$ roughly doubles the accretion rate over the evolution.  The
surface luminosities are nearly the same so the advection luminosity
must be much greater.

The non-linear effect of the radiative efficiency is understood as
follows.  A smaller radiative efficiency requires a larger accretion
rate and the envelope contracts and becomes denser and therefore
hotter.  The advected luminosity is also greater and therefore even
more accretion is required.  If the advective luminosity were a fixed
fraction of the total in the envelope, the relationship would be
linear.  However, the advective luminosity increases for smaller
$\epsilon$ so a larger accretion rate is required than would be
just to achieve the same surface luminosity.

This also explains the non-existence of any converged models for
radiative efficiencies below some minimum.  For the fiducial run, I
could not evolve quasi-stars with radiative efficiencies below
$0.021$.  In short, such models advect all of the luminosity inwards.
Without any energy to support the envelope, it collapses on to the BH.
The advection luminosity is more modest after the first BH luminosity
hump so models with smaller radiative efficiencies can be constructed
if the initial BH mass is larger.  For this reason, the evolution with
$\epsilon=0.02$ begins with initial BH mass $331\Msun$ but its
subsequent evolution follows the same trends as the models with the
same initial BH mass as the fiducial run.

The qualitative evolution of the runs is similar but the major
quantitative difference between them is that the luminosity humps are
narrower in $\Mbh$ for larger values of $\epsilon$.  For
$\epsilon=0.02$ and $0.03$, the first hump is wider than the total
mass so there is one fewer hump in those sequences.  This is probably
because the envelopes are hotter and denser for smaller values of
$\epsilon$ and the microphysical change that causes the three-humped
behaviour is delayed.

The radiative efficiency $\epsilon$ is defined by the inner luminosity
boundary condition (equation \ref{lbc2}).  We can also define an
overall radiative efficiency $\epsilon_*=L_*/\dMbh c^2=\epsilon
L_*/\Lbh$.  It has already been shown that $L_*/\Lbh$ varies during
the evolution of a CDAF-ADAF quasi-star and the overall efficiency
therefore varies too.  Notably, when advection is strongest, the
surface efficiency is smallest.  In the runs here, the overall
efficiency reaches a low of $0.15\epsilon$ during the first luminosity
hump.  During the second hump, the ratio is between about $0.3$ and
$0.5$.  Thereafter, where applicable, the ratio is between $0.7$ and
$0.85$.  The non-existence of models with $\epsilon<0.21$ for initial
BH masses less than about $100\Msun$ can be thought of as the overall
efficiency $\epsilon_*$ being negative.

\begin{figure}\begin{center}
\includegraphics{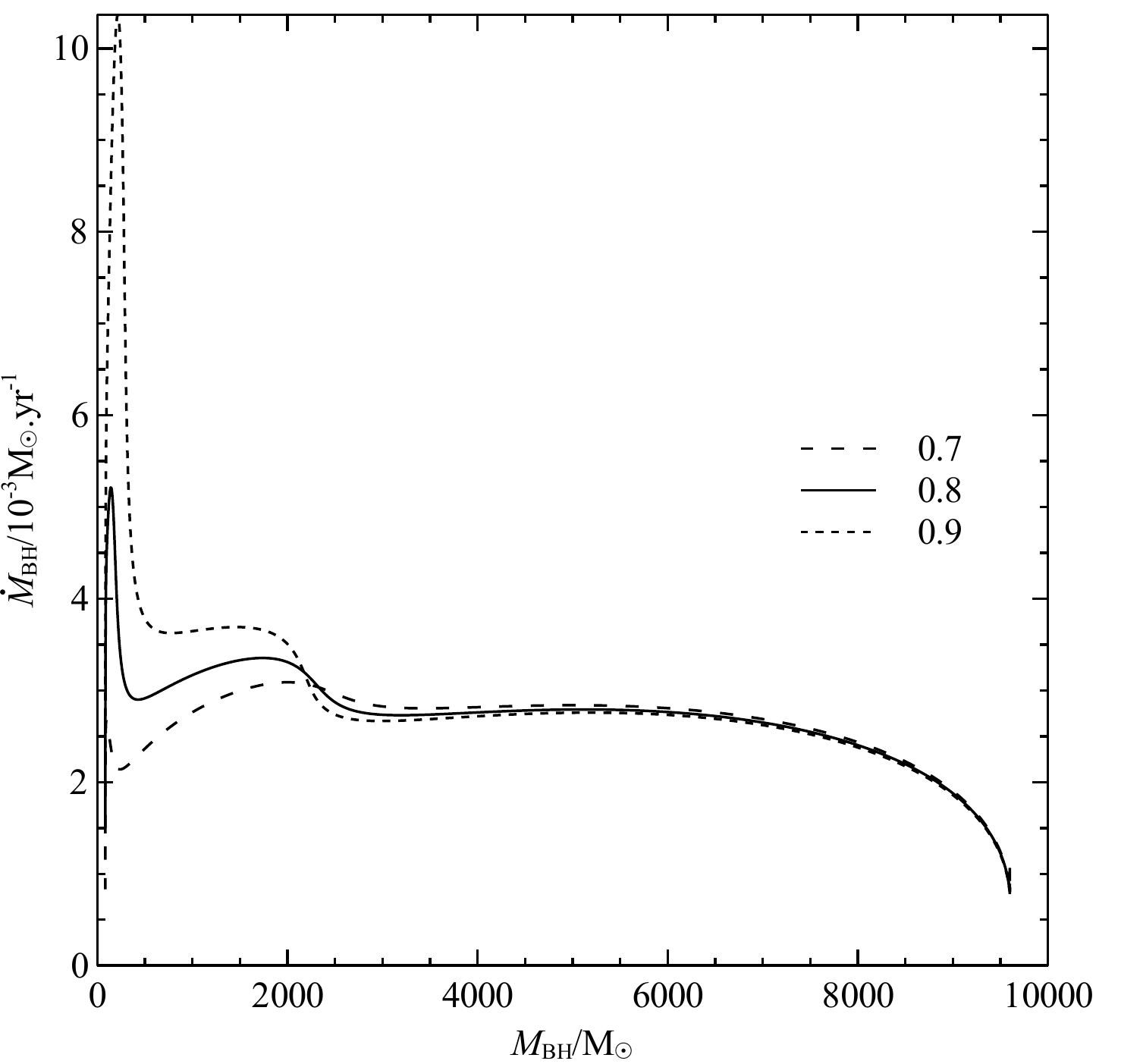}
\caption[Evolution of quasi-stars with different advective
efficiencies.]{Plot of the evolution of quasi-stars for advective
  efficiencies $\eta=0.7$, $0.8$ and $0.9$. Stronger advection
  requires larger accretion rates to support the envelope. Once the
  effect of advection has declined, the models converge to similar
  accretion rates.}
\label{feta2}
\end{center}\end{figure}

\subsection{Advective efficiency}
\label{sseta2}

The advective efficiency $\eta$ represents the fraction of the
luminosity generated in the advected zone that falls on to the BH
without being radiating away.  Larger values of $\eta$ correspond to
less energy being released and should require a higher accretion rate
to provide the same overall luminosity.  Fig.~\ref{feta2} confirms
this.  The fiducial run is shown with runs where $\eta=0.7$, which has
a lower accretion rate, and $\eta=0.9$, which has a higher rate.  The
accretion rates converge at larger BH masses once the role of the
advective luminosity has declined.  The luminosity humps occur at
roughly the same points in each run but with different amplitudes.

\begin{figure}\begin{center}
\includegraphics{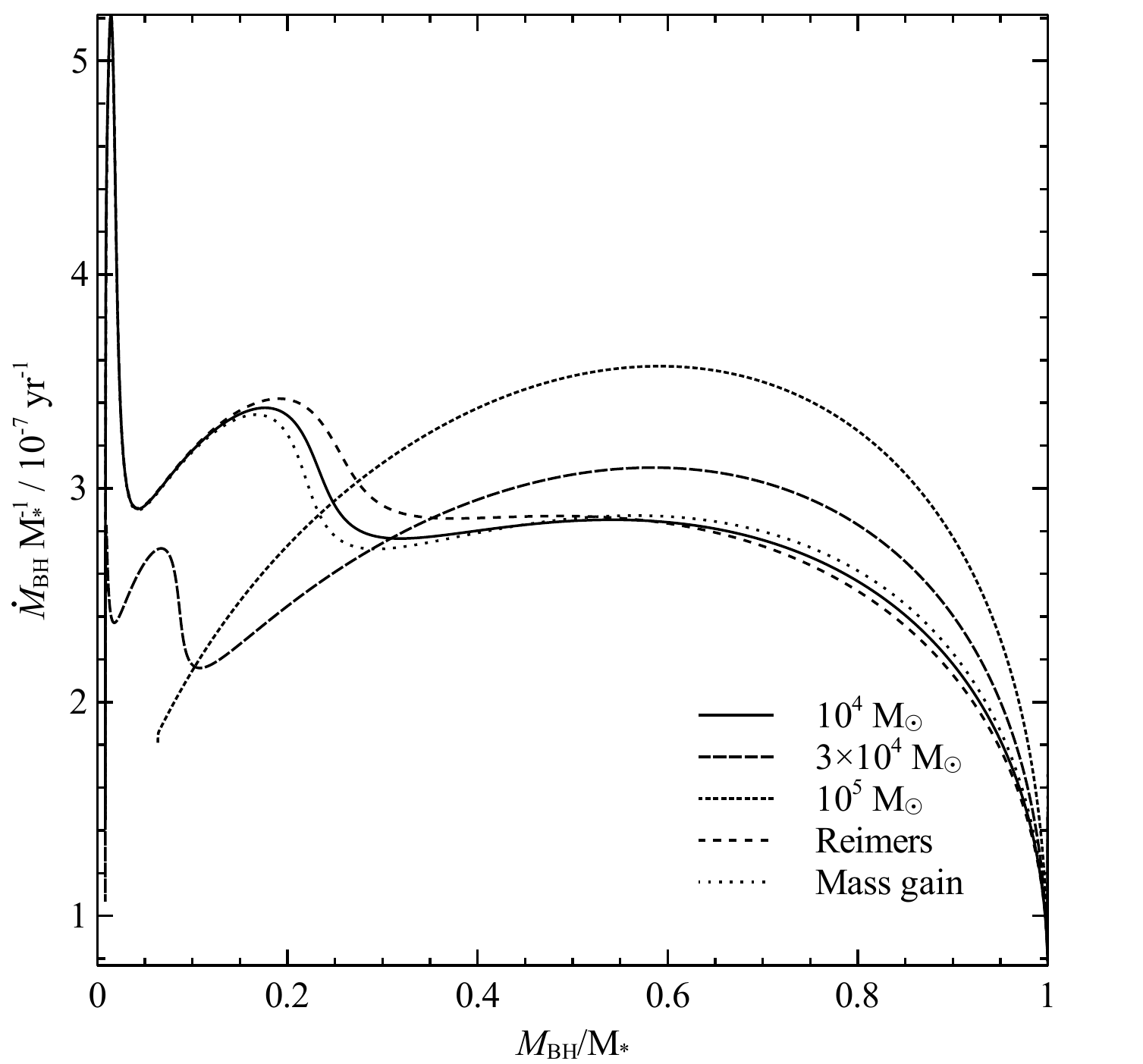}
\caption[Evolution of quasi-stars with various initial masses and
surface mass loss or gain.]  {Plot of evolution of quasi-stars with
  total initial masses $M_*/\Msun=10^4$, $\sci{3}{4}$ and $10^5$, as
  well as models with surface mass loss and gain. For larger total
  masses, the first two luminosity humps are narrower in $\Mbh$ and
  smaller in magnitude when scaled by the total mass. The final phase
  of accretion is more rapid. In addition, more massive models achieve
  higher fractional BH masses.}
\label{fmi2}
\end{center}\end{figure}

\subsection{Initial mass and total mass loss or gain}
\label{ssmi2}

The final free parameter that is varied is the total mass of the
quasi-star.  Evolutionary sequences for quasi-stars with total initial
masses $M_*/\Msun=10^4$, $\sci{3}{4}$ and $10^5$ are plotted in
Fig.~\ref{fmi2}.  The BH masses and accretion rater are divided by the
total masses so that the different runs can be compared.  In general,
I found it difficult to find suitable parameters for successful models
with larger total masses.  The $10^5\Msun$ quasi-star is initialized
with a BH mass of $6356\Msun\approx0.064\,M_*$.  This is nearly an
order of magnitude larger than the initial fractional BH mass in the
fiducial run.  I was unable to construct models with larger total
masses and could not determine why. \\

More massive quasi-stars have smaller scaled accretion rates before
and larger scaled accretion rates during the final luminosity hump.
The humps are narrower in fractional mass but the first two are of
roughly the same width (about $2500\Msun$) in absolute BH mass.  The
complicated dependence of the accretion rate on total mass means there
is not a simple scaling relation for quasi-stars' lifetimes as there
was for the Bondi-type quasi-stars.  Measuring the lifetime from the
initial fractional mass of the $10^5\Msun$ model, the fiducial model
evolved for $3.48\Myr$, the $3\times10^4\Msun$ model for $3.49\Myr$
and the $10^5\Msun$ model for $3.13\Myr$.  The trend towards larger
scaled accretion rates during the final luminosity hump appears to
continue to higher masses so I expect that larger quasi-stars have
shorter lifetimes than the most massive model presented here.

Fig.~\ref{fmi2} also shows evolution where the quasi-star loses mass
at a \citet{reimers75} rate,
\shorteq{\dot{M}\st{loss}=\sci{4}{-13}\frac{L_*R_*}{M_*}
  \frac{\Msun}{\Lsun\Rsun}\text{,}}%
or gains mass at a constant rate of
$\dot{M}\st{gain}=\sci{2}{-3}\Msunpyr$.  The trends identified above
hold for the total masses.  At the end of the run, the model with mass
loss contains a BH of mass $5321\Msun$ inside an envelope of
$1.53\Msun$.  The model that gains mass finishes with a BH of mass
$14063\Msun$ and an envelope of $0.677\Msun$.  As in the models that
do not have additional mass loss or gain from the surface, the BH is
able to consume almost all of the envelope.

\section{Discussion}
\label{sdisc2}

What, if anything, do CDAF-ADAF quasi-stars have in common with the
Bondi-type models of Chapter \ref{cqs1}? Regarding the structure, very
little.  The accretion rates of the CDAF-ADAF models are about an
order of magnitude larger and the BHs ultimately accrete almost all of
the available gas in the envelope. The mass limit found in Chapter
\ref{cqs1} is robust but CDAF-ADAF quasi-stars are not in thermal
equilibrium and therefore the polytropic analysis does not apply to
them.  While the CDAF-ADAF quasi-stars in this chapter are not
necessarily more reliable, they demonstrate that the choice of the
inner boundary is critical in determining the qualitative and
quantitative behaviour of the envelope's structure.

The surface behaviour of the CDAF-ADAF quasi-stars, however, is not so
different from the Bondi-type models.  In both cases, the surface
temperatures are effectively set by the Hayashi limit and are
therefore quite accurately around $4500\K$.  More roughly, the surface
luminosity is on the order of the Eddington luminosity of the
quasi-star.  The surface boundary condition $L_*=4\pi R_*^2\sigma
T\st{eff}^4$\, requires that the envelope's surface radius is on the
order of $322(M_*/\Msun)^\frac{1}{2}\Rsun\approx1.5(M_*/\Msun)^\frac{1}{2}\AU$.
Although the details of the structure depends critically on selecting
appropriate boundary conditions, the surface properties of the objects
are reasonably similar.  The most important quantitative difference in
the evolution of the observable properties is the total lifetime,
which determines the duty cycle of the quasi-star phase.

For comparison, the location of the Bondi radius can be found in the
CDAF-ADAF models.  In the fiducial run, the Bondi radii (where
$c\st{s}^2=2G\Mbh/\rb$) when $\Mbh/\Msun=200$ and $1000$ are $\rb/\Rsun=436$
and $\sci{1.46}{4}$, respectively.  For the corresponding BH masses in
the fiducial Bondi-type quasi-star, the inner radii are
$r_0/\Rsun=373$ and $\sci{1.15}{4}$, which compare reasonably
well. However, for $\Mbh\gtrsim1200\Msun$, a Bondi radius cannot be
found in the CDAF-ADAF models. This is consistent with the existence
of a BH mass limit when the inner radius is defined through the Bondi
radius but also suggests that Bondi-type models fail to capture the
true behaviour of the models at higher BH masses.

The three-phase behaviour of the CDAF-ADAF models was also not seen in
the Bondi-type quasi-stars and probably occurs because of changes in
the advective luminosity owing to some microphysical property of the
envelope material.  Some models require that the first hump is avoided
entirely.  The high initial accretion rate might indicate that the BH
generally grows at first through a very rapid accretion phase before
the envelope settles into a state described by approximate hydrostatic
equilibrium.

The changing advective luminosity implies that the overall radiative
efficiency of accretion, defined by $\epsilon_*=L_*/\dMbh c^2$, is not
constant.  It may be possible to construct CDAF-ADAF quasi-stars by
regarding $\epsilon_*$, rather than $\epsilon$, as a constant.  LLG04
adopted this approach and found $\epsilon_*=0.0045$.  In the late
phases of the quasi-star evolution, such a low value of $\epsilon_*$
would almost certainly admit a much larger accretion rate and
therefore shorter lifetimes.

% \comment{Discuss reliability of models at late times.  Basically,
%   probably not great.  Deviations from sphericalness probably aren't
%   unstable but might also have larger amplitudes than is usually
%   allowed.}
 
\section{Conclusion}
\label{sconc2}

I have presented quasi-star models that employ boundary conditions
based on the global accretion solutions of AIQN02 and LLG04.  An
advection-dominated flow is presumed to be surrounded by a convective
envelope, the structure of which is computed with approximate
treatments of rotation and relativity.  The behaviour of the models is
qualitatively and quantitatively distinct from the Bondi-type models
described in Chapter \ref{cqs1}.  In my opinion, the most important
conclusion to draw here is that the results computed using
near-spherical hydrostatic quasi-stars are only as good as the inner
boundary conditions that they use.

In CDAF-ADAF quasi-stars, the properties of the advective luminosity
dictate how the central BH evolves. The models indicate a three-phase
behaviour in which the importance of advection gradually decreases.
The initial hump in the accretion rate can be sufficiently large to
disrupt numerical convergence.  For runs with small radiative
efficiencies or large total initial masses, I bypassed this phase
entirely.  It may correspond to a rapid initial growth of the BH
before an approximately hydrostatic configuration can be established.

For all the parameters explored here, the BH ultimately accretes
nearly all of the available material.  The precise fraction varies
very little with the parameters of the model and is more than $0.9988$
in all the models tested here.  The lifetimes of the objects are only
a few millions of years so CDAF-ADAF models create larger BHs more
rapidly than Bondi-type quasi-stars.  Thus, CDAF-ADAF quasi-stars can
certainly leave BHs that are massive enough to reach masses of
$10^9\Msun$ by redshift $z\approx6$ but whether these structures are
accurate models of the quasi-star phase is not yet established.
 % CDAF-ADAF quasi-stars
\clearpage\thispagestyle{tocstyle}
\begin{savequote}[80mm]
%   I am convinced that purely mathematical construction enables us to
%   find those concepts and those lawlike connections between them that
%   provide the key to the understanding of natural
%   phenomena.
%   \qauthor{Albert Einstein, 1933}
%   One should not reproach the theorist who undertakes such a task by
%   calling him a fantast; instead, one must allow him his fantasizing,
%   since for him there is no other way to his goal whatsoever. Indeed,
%   it is no planless fantasizing, but rather a search for the logically
%   simplest possibilities and their consequences.
%   \qauthor{Albert Einstein, 1954}
  As far as the laws of mathematics refer to reality, they are not
  certain; and as far as they are certain, they do not refer to
  reality.
  \qauthor{from \emph{Sidelights on Relativity}, \\
    Albert Einstein, 1922}
\end{savequote}

\chapter[The nature of the Sch\"onberg--Chandrasekhar limit]
{The nature of the \\Sch\"onberg--Chandrasekhar limit}
\label{cscl}

\citet{schoenberg42} showed that, if embedded in a polytropic stellar
envelope with index $n=3$, there is a maximum fractional mass that an
isothermal core can achieve in hydrostatic equilibrium.  This upper
limit is known as the Sch\"onberg--Chandrasekhar (SC) limit.  If the
core is less massive, it remains isothermal while nuclear reactions
continue in a surrounding shell.  If the core is more massive, it
contracts until it is supported by electron degeneracy pressure or
helium begins to burn at the centre.  The idealised result provides an
estimate of the point at which an evolved isothermal core embedded in
an extended envelope begins to contract and is sufficiently accurate
that it has become a well-established element of the theory of the
post-main sequence evolution of stars.

In Chapter \ref{cqs1}, I showed that the black hole (BH) masses of
Bondi-type quasi-stars are subject to a robust fractional limit.  I
have determined why this limit exists in terms of contours of
fractional core mass of solutions when plotted in the space of
homology invariant variables $U$ and $V$.  I further found that the SC
limit can be explained in the same way.

Fractional mass limits have been computed for a number of other
polytropic solutions.  \citet{beech88} calculated the corresponding
limit for an isothermal core surrounded by an envelope with $n=1$.
\citet*[][hereinafter EFC98]{eggleton98} found that, when $n=1$ in the
envelope and $n=5$ in the core, a fractional mass limit exists if the
density decreases discontinuously at the core-envelope boundary by a
factor exceeding 3.  They further proposed conditions on the
polytropic indices of the core and envelope that lead to fractional
mass limits.  I refer to all these limits, including the original
result of \citet{schoenberg42} as \emph{SC-like} limits.

In this chapter, I present my analysis, which unifies SC-like limits
and indicates that they exist in a wider range of circumstances than
the handful of cases discussed in the literature.  In Section
\ref{scont}, I present the new interpretation of SC-like limits.  In
Section \ref{scons}, I provide a description that captures all the
SC-like limits of Section \ref{scont} and I consider broad classes of
models that must also exhibit SC-like limits.  I close by discussing
some implications for real stellar models when they reach SC-like
limits.

\section{Fractional mass contours in the \uvp{}}
\label{scont}

The analysis presented here employs the plane of homology invariant
variables $U$ and $V$.  A thorough exposition of its features are
provided in Appendix \ref{auvp} but I briefly review here the most
important.  An equation of state is polytropic if it obeys the
relation \shorteq{p=K\rho^{1+\frac{1}{n}}\text{,}} where $p$ is the
pressure, $\rho$ the density, $n$ the polytropic index and $K$ a
constant of proportionality.  By defining the dimensionless
temperature $\theta$ by $\rho=\rho_0\theta^n$, the equations of mass
conservation, 
\shorteq{\tdif{m}{r}=4\pi r^2\rho\text{,}} and
hydrostatic equilibrium,
\shorteq{\tdif{p}{r}=-\frac{Gm\rho}{r^2}\text{,}} can be written in
the dimensionless forms
\shorteq{\tdif{\theta}{\xi}=-\frac{1}{\xi^2}\phi\label{elee1}} and
\shorteq{\tdif{\phi}{\xi}=\xi^2\theta^n\text{,}\label{elee2}} where
$\rho_0$ is the central density, $r$ is the radial co-ordinate, $m$
the mass inside $r$, $\xi=r/\eta$ is the dimensionless
radius\footnote{The scale factor is usually $\alpha$.  We have used
  $\eta$ to avoid confusion with the density jump at the core-envelope
  boundary, which EFC98 called $\alpha$.}  and
$\phi=m/4\pi\eta^3\rho_c$ is the dimensionless mass.  The radial scale
factor $\eta$ is defined by 
\shorteq{\eta^2=\frac{(n+1)K}{4\pi G}\rho_c^{\frac{1}{n}-1}\text{.}}
Equations \ref{elee1} and \ref{elee2} are equivalent to the Lane--Emden
equation (LEE, \ref{LEE}).

Homology invariant variables $U$ and $V$ are defined by 
\shorteq{U= \tdif{\log m}{\log r}=\frac{\xi^3\theta^n}{\phi}}
and
\shorteq{V=-\tdif{\log p}{\log r}=(n+1)\frac{\phi}{\theta\xi}}
and obey the differential equation
\shorteq{\tdif{V}{U}=-\frac{V}{U}
  \left(\frac{U+(n+1)^{-1}V-1}{U+n(n+1)^{-1}V-3}\right)\text{.}}

Consider the problem of fitting a polytropic envelope to a core of
arbitrary mass and radius.  For a given $n<5$, we can regard a given
point $(U_0,V_0)$ in the \uvp{} as the interior boundary of a
polytropic envelope by integrating the LEE from that point to the
surface.  More precisely, we can take interior conditions
\shorteq{\theta_0=1\text{,}}
so that $\rho_0$ is the density at the base of the envelope,
\shorteq{\xi_0=\sqrt{(n+1)^{-1}U_0V_0}\text{}}
% \begin{gather}
% \theta_0=1\text{,} \\ \xi_0=\sqrt{(n+1)^{-1}U_0V_0}\text{}
% \end{gather}
and \shorteq{\phi_0=\sqrt{(n+1)^{-3}U_0V_0^3}}%
and integrate the LEE up to the first zero of $\theta$, where we set
$\xi=\xi_*$.  This point marks the surface of a polytropic envelope,
at which $\phi_*$ is the total dimensionless mass of the solution,
including the initial value $\phi_0$.  The ratio $q=\phi_0/\phi_*$ is
the fractional mass of a core that occupies a dimensionless radius
$\xi_0$.  By associating each point in the \uvp{} with the value of
$q$ for a polytropic envelope that starts there, we define a surface
$q(U,V)$.  We use the contours of this surface to characterise the SC
limit.

\begin{figure}\begin{center}
\includegraphics{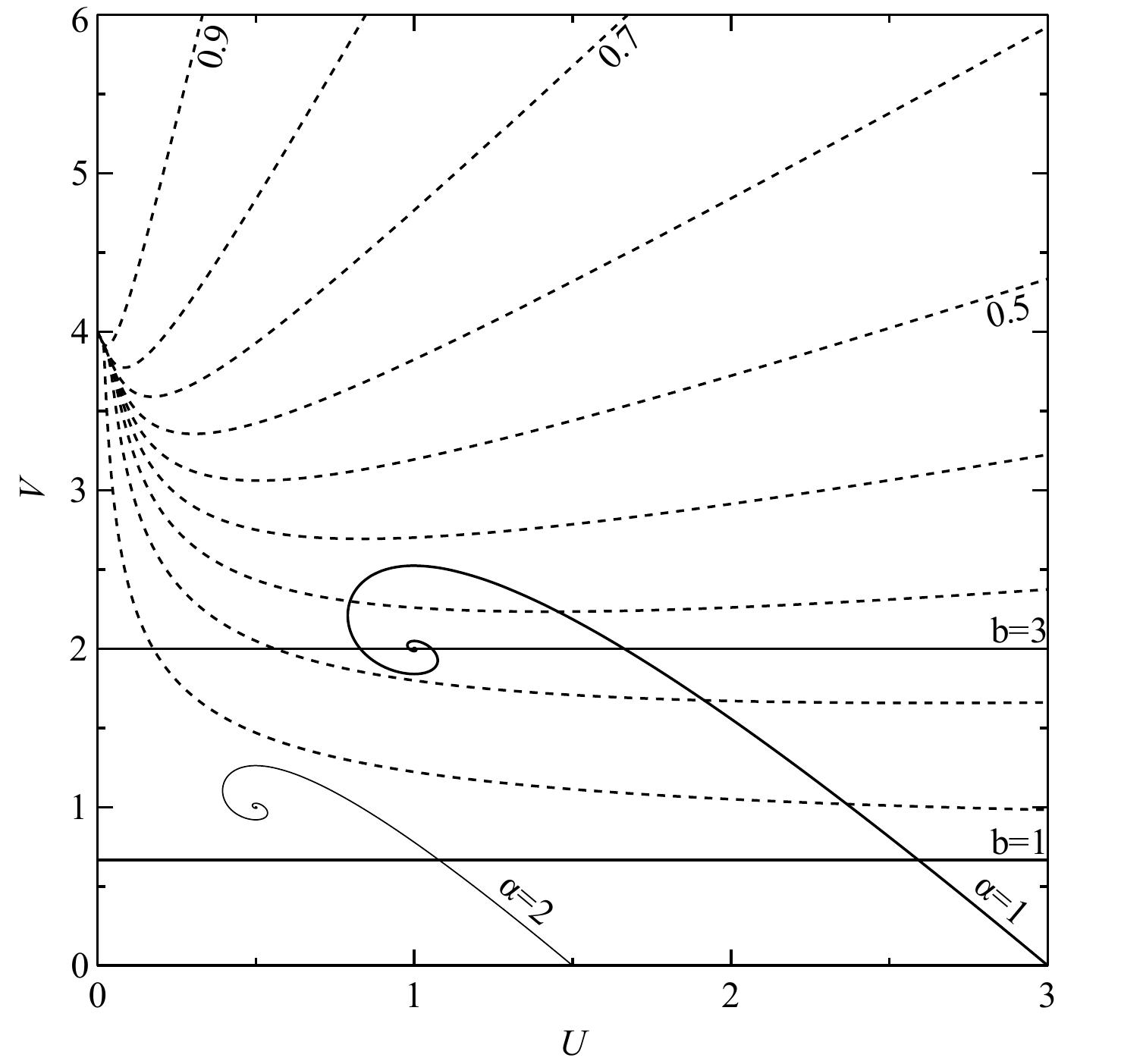}
\caption[Fractional mass contours and SC-like limits for $n=3$.]  {The
  dashed lines are contours of a core's fractional mass
  $q=\phi_0/\phi_*$ beneath an envelope with $n=3$.  They increase in
  steps of $0.1$ from $0.1$, at the bottom, to $0.9$, at the top.  The
  larger solid spiral is the isothermal core with $\alpha=1$.  The
  smaller spiral represents an isothermal core when $\alpha=2$.  The
  upper and lower straight lines represent the inner boundaries for
  quasi-stars with %$n=3$ and
  $b=3$ and $b=1$ respectively (see Section \ref{ssqs}).}
\label{fcont3}
\end{center}\end{figure}

\begin{figure}\begin{center}
\includegraphics{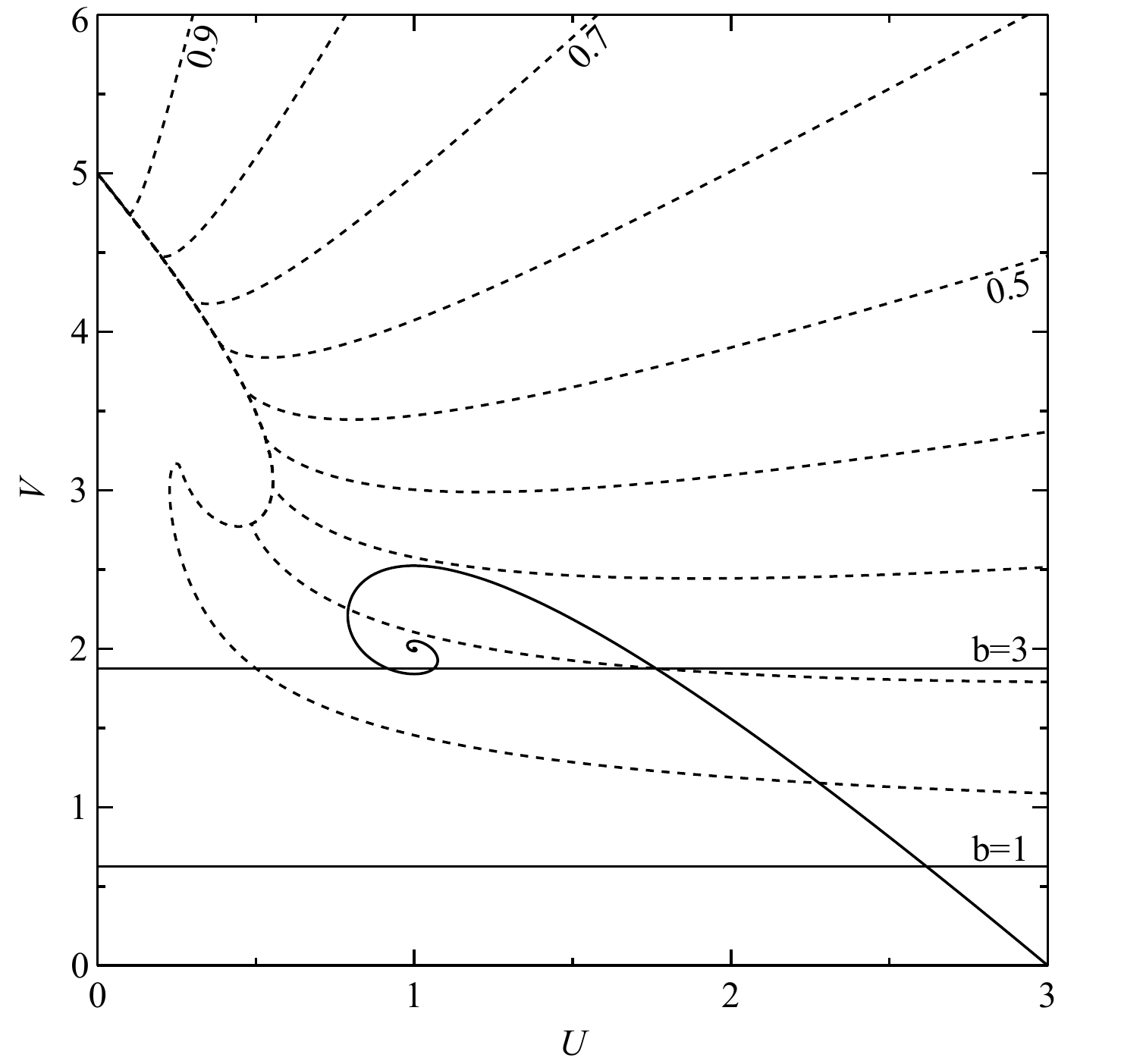}
\caption[Fractional mass contours and SC-like limits for $n=4$.]  {The
  dashed lines are contours of fractional mass $q=\phi_0/\phi_*$ for
  $n=4$.  The solid spiral is again the isothermal core solution.  The
  upper and lower straight lines represent the inner boundaries for
  quasi-stars with %$n=3$ and
  $b=3$ and $b=1$ respectively.  The behaviour of the contours is
  dominated by the critical point $G_s=(1/3,10/3)$.}
\label{fcont4}
\end{center}\end{figure}

Figs \ref{fcont3}, \ref{fcont4} and \ref{fcont1} show contours of $q(U,V)$ for
polytropic envelopes with $n=3$, $4$ and $1$ respectively, along with a
selection of interior solutions that lead to SC-like limits.  For $n<3$
the contours are dominated by the critical point $V_s=(0,n+1)$ and for
$n>3$ by $G_s=(\frac{n-3}{n-1},2\frac{n+1}{n-1})$.  Away from $V_s$ or
$G_s$ all the contours at first curve away from the $U$-axis and then
tend towards straight lines.

\subsection{The Sch\"onberg--Chandrasekhar limit}

\citet[][p.  203]{kw90} discuss the SC limit in terms of fractional
mass contours.  \citet{cannon92} also explicitly described the SC limit
in terms of fractional mass contours, although he employed a different
set of homology-invariant variables.  Fig.~\ref{fcont3} shows the
isothermal solution along with the fractional mass contours for $n=3$
envelopes.  The SC limit exists because the isothermal solution only
intersects fractional mass contours up to a maximum
$q_\text{max}=0.359$ when $\alpha=1$.  In other words, along the
isothermal solution, the function $q(U,V)$ achieves a maximum of
$0.359$.

%A maximum $V$ for the isothermal curve exists because, for a given
%mass, there is a finite maximum pressure that a core can exert.  
The SC limit is usually derived by defining the core pressure using
virial arguments and maximizing it with respect to the core radius
\citep[e.g.][p.  285]{kw90}.  Such an explanation partly describes the
SC limit but our interpretation makes clear that the existence of the
SC limit has as much to do with the behaviour of the envelope
solutions as the isothermal core.  For example, changing the polytropic
index of the envelope changes the mass limit.

If the density jumps by a factor $\alpha$ at the core-envelope
boundary, $U$ and $V$ must be transformed at the edge of the core
to find the base of the envelope in the \uvp{}.  That is, if
$\rho\to\alpha^{-1}\rho$, then $(U,V)\to\alpha^{-1}(U,V)$.  The
contraction of the isothermal core for $\alpha=2$ is included in
Fig.~\ref{fcont3}.  The inner boundary of the envelope shifts to a
smaller fractional mass of about $0.09$ so the SC limit falls too.

The argument presented here implies that SC-like limits exist whenever
an envelope is matched to a core that only intersects fractional mass
contours of that envelope up to some maximum.  We now use this to
explain the existence of other SC-like limits.

\subsection{Quasi-stars}
\label{ssqs}

In Chapter \ref{cqs1}, I showed that the fractional BH mass limit for
Bondi-type quasi-stars also exists in polytropic models.  Using
fractional mass contours in the \uvp{}, I now show how the limit is
essentially the same as the SC limit, at least in the case of zero
mass in the cavity.  When the cavity mass is included, the connection
between the SC limit and the fractional BH mass limit is not clear but
it is qualitatively the same and a similar mechanism is %likely to be
at work.

The interior boundary condition for the quasi-star models can be
written as
\shorteq{r_0=\frac{1}{b}\frac{2Gm_0}{c\st{s}^2}\label{r0def}\text{,}}
where $b$ is a scale factor, $m_0$ the mass interior to $r_0$ and
$c\st{s}=\sqrt{\gamma p/\rho}$ the adiabatic sound speed.  The
boundary condition is a fraction $1/b$ of the Bondi radius, where
$mc\st{s}^2/2=Gm/r$.  \citet*{begelman08} used $b=3$; the models in
Chapter \ref{cqs1} used $b=1$.  Accretion on to the central BH
supports the envelope by radiating near the Eddington limit of the
entire object so the envelope is strongly convective and the pressure
is dominated by radiation.  The envelope is approximately polytropic
with index $n=3$.  Now, at the interior boundary,
$c\st{s}^2=K\rho_0^{1/n}(n+1)/n$, $m_0=4\pi\eta^3\rho_0\phi_0$, and
$r_0=\eta\xi_0$ so
\shorteq{\label{qbc}\phi_0=\frac{b}{2n}\xi_0\text{.}}%
Transforming to $U$ and $V$ gives $U_0=2n\xi_0^2/b$ and
$V_0=b(n+1)/2n$.  Varying $\xi_0$ traces a straight line, parallel to
the $U$-axis.  Fig.~\ref{fcont3} includes the straight lines $V_0=2/3$
and $2$, which correspond to $b=1$ and $3$, respectively, for $n=3$.
The line of $V_0$ does not intersect all the contours of fractional
core mass because many of them are positively curved.  Thus, a mass
limit exists, as in previous cases.  For larger values of $b$, $V_0$
is also larger and intersects more of the contours.  The mass limit is
therefore larger.

\begin{figure}\begin{center}
\includegraphics{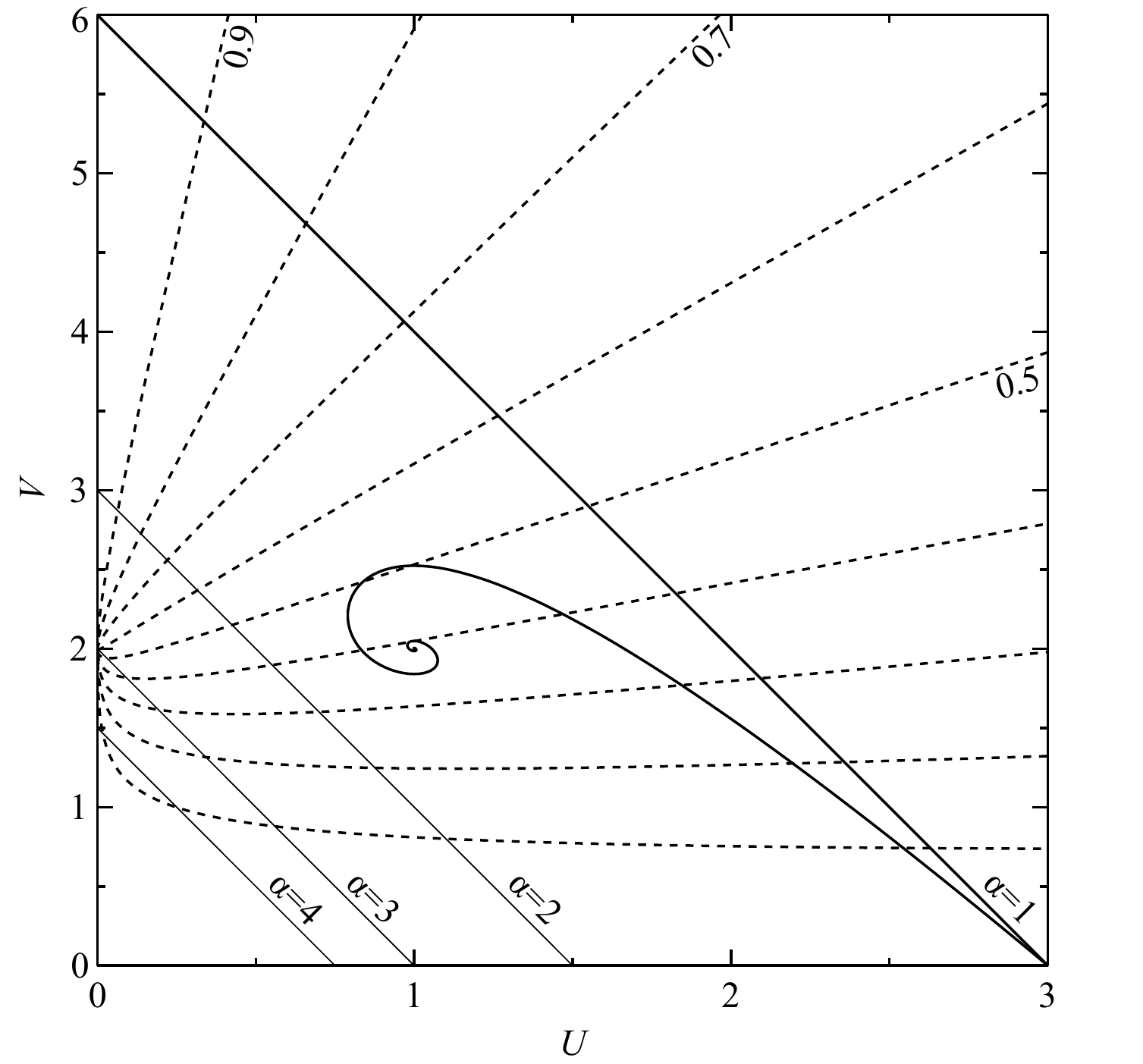}
\caption[Fractional mass contours and SC-like limits for $n=1$.]  {The
  dashed lines are contours of fractional mass $q=\phi_0/\phi_*$ for
  $n=1$.  The solid spiral is again the isothermal core solution.  The
  top-most diagonal line is the polytrope of index $5$ with
  $\alpha=1$.  The other diagonal lines are, from top to bottom,
  core-envelope boundary conditions for the envelope when
%  $\alpha=2,3,4$ for $n=5$ as shown by \citet[][Fig.~3]{eggleton98}.}
  $\alpha=2,3,4$ for $n=5$ as shown by EFC98.}
\label{fcont1}
\end{center}\end{figure}

We have limited ourselves to the case where $n=3$.  The fractional
mass contours in Figs \ref{fcont3} and \ref{fcont1} show similar
behaviour.  Convective envelopes are approximately adiabatic and have
effective polytropic indices between $3/2$ and $3$, depending on the
relative importance of gas and radiation pressures.  All such
envelopes possess fractional mass contours that are similar to the two
cases here and we conclude that a fractional mass limit for the BH
exists in all realistic cases.  Envelopes with $3<n<5$ have more
complicated fractional mass contours so we cannot immediately draw
similar conclusions.  For $n=4$, Fig.~\ref{fcont4} shows that the
fractional mass limits for quasi-star cores behave as described above
only for $b\lesssim4.4$.

In trying to move $r_0$ inwards, I found I could not construct models
with the \stars{} code with $b\geq3.8$ in equation (\ref{r0def}).  The
reason for this can be determined from the behaviour of the polytropic
limit.  As $b$ increases, $V_0$ increases and eventually passes the
critical point $V_s$ when the contours change from increasing along
$V_0$ as $U$ increases to decreasing along $V_0$ as $U$ increases.  In
other words, if a polytropic envelope is integrated from small $U_0$
and $V_0<n+1$ it has $q\approx0$.  If the envelope is instead
integrated from $V_0>n+1$, it has $q\approx1$ and $q$ decreases if
$U_0$ increases. When $V_0>n+1$, $b>2n$ so small inner masses
correspond to envelopes with negligible envelope mass.  It becomes
impossible to embed a small BH inside a massive envelope.  The mass
limit becomes a \emph{minimum} inner mass limit.  For the models
presented in Chapter \ref{cqs1}, the finite mass of the BH corresponds
to a finite value of $U_0$ that displaces the envelope slightly from
the $V$-axis.  The fractional mass contours are closely packed near
$V_s$ so a small value of $U_0$ introduces a minimum inner mass limit
for some $b<2n$.
% Thus, for a fractional inner mass, the maximum $b$ is smaller than $2n$.

\subsection{Other polytropic limits}

Fig.~\ref{fcont1} shows the fractional mass contours for $n=1$, the
isothermal solution and $n=5$ polytropes with $\alpha=1,2,3$ and $4$
as used by EFC98.
% The result of \citet{beech88} is analogous to the SC limit for an
% $n=1$ envelope.
\citet{beech88}\footnote{There appears to be an error in equation (17)
  of \citet{beech88}.  The right-hand side, which should be
  dimensionless, has dimensions $(\gpcm)^\frac{1}{2}$.  I suspect that
  the appearance of $\rho_c$ in the denominator should be should
  actually be $\rho_c^\frac{1}{2}$.  The mistake does not appear to
  propagate into subsection equations.} calculated a SC-like limit of
about $0.27$ for an isothermal core embedded in a polytropic envelope
with $n=1$.  Because the behaviour of fractional mass contours is
similar for $n=1$ and $n=3$, the existence of the limit is now no
surprise.  Inspection of Fig.~\ref{fcont1} suggests
$q\st{max}\approx0.5$, which differs from the value found by
\citet{beech88}.  This is at least partly because he included the
radiation pressure of the isothermal core.  The value determined here
can be partly reconciled by replacing $p$ with
$p+p\st{r}=p(1+p\st{r}/p)$, where $p\st{r}=aT^4/3$ is the radiation
pressure with the radiation constant $a$.  The pressure appears in the
definition of $V$, which is therefore replaced with $V/(1+p\st{r}/p)$.
The isothermal cores of \citet{beech88} have central temperature
$\sci{2}{7}\K$ and $\rho_0=10^3\gpcm$.  The density at the
core-envelope boundary is about $20\gpcm$ so $p\st{r}/p$ is about
$0.016$ and $V$ is reduced by a similar fraction.  The inclusion of
radiation pressure thus explains a small fraction of the difference
between the result presented here and that of \citet{beech88}.  I am
still unsure where the remainder of the difference is found.

The conclusions of EFC98 are also accommodated.  The
critical point $V_s=(0,n+1)$ separates solutions, and thus contours,
with $q\approx0$ from those with $q\approx1$.  EFC98
concluded that, for $n=1$ envelopes, cores with $n<5$ are never
subject to a SC limit; those with $n>5$ always are; and those with
$n=5$ constitute the marginal case for which the limit exists when
$\alpha>3$.  The three cases are demonstrated in Fig.~\ref{fcont1}.
When $\alpha=1$ or $2$ the core solution intersects all the contours.
When $\alpha=4$ the core solution is everywhere below the point
$(0,2)$ and only intersects contours up to $q\st{max}\approx0.15$.
The marginal case is $\alpha=3$, for which the core solution
intersects $(0,2)$ exactly and separates the values of $\alpha$ for
which the core is or is not subject to a SC-like limit.  Because
these conclusions are based on the critical behaviour of the
solutions, which is reflected in the behaviour of the contours, the
same results follow here.  We have shown how they are characterised by
the contours in the same way as other limits and are a particular
example of our broader result.  That is, we have shown that SC-like
limits exist whenever the core solution fails to intersect all
fractional mass contours.  The cases identified by EFC98
fall within this description.

\section{Consequences for stellar evolution}
\label{scons}

I complete this chapter by discussing four points that connect the
polytropic results above to real stellar evolution.  First, I identify
broad classes of polytropic models that are similar to real models and
subject to SC-like limits.  Secondly, I discuss how stellar structure
responds to a SC-like limit and, thirdly, how SC-like limits are
related to the instability described by \citet{ebert55},
\citet{bonnor56} and \citet{mccrea57}.  Finally, I introduce a
theoretical test of whether a composite polytrope is at a SC-like
limit.

\subsection{General limits}
\label{ssgenlim}

\begin{figure}\begin{center}
\includegraphics{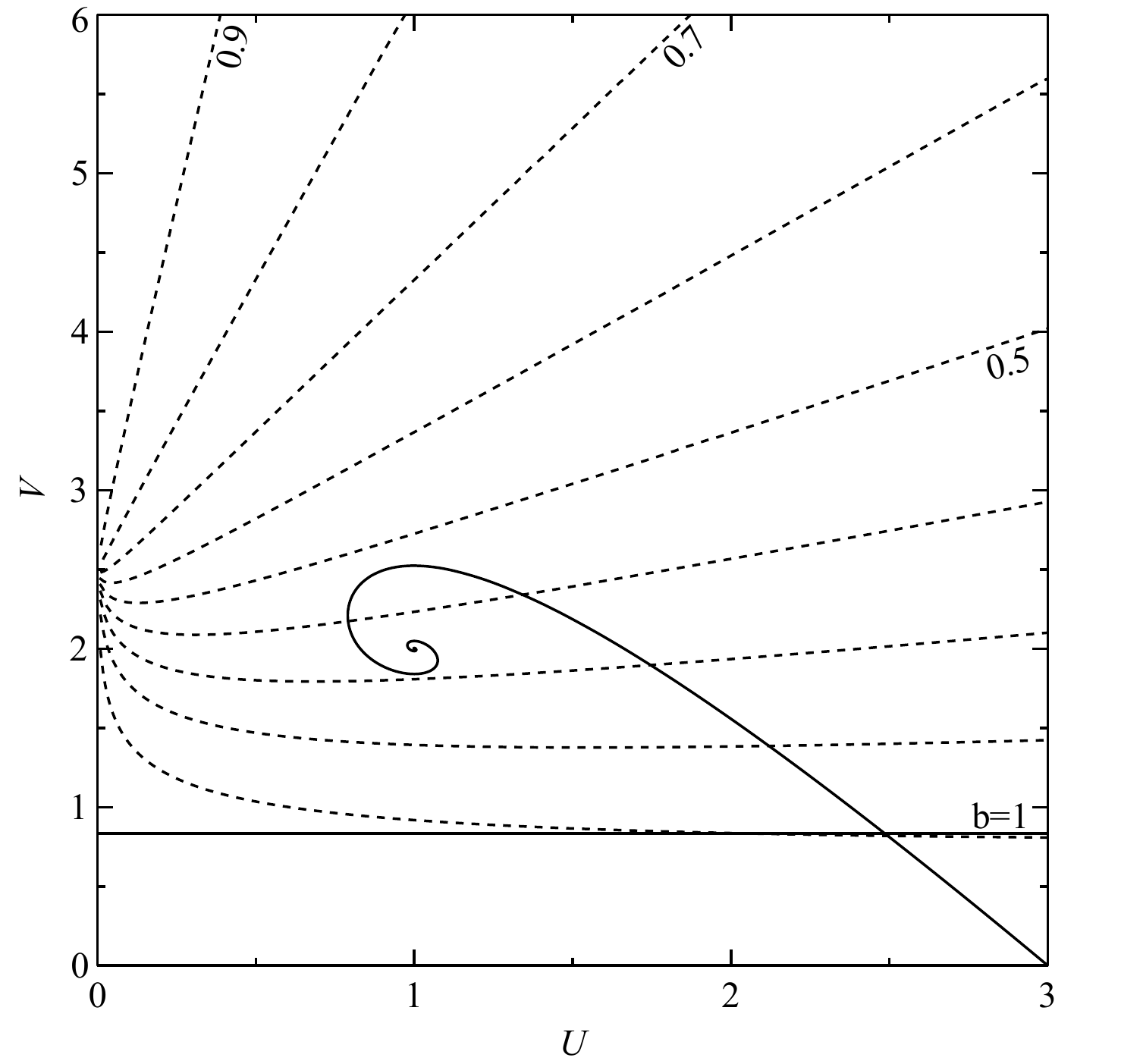}
\caption[Fractional mass contours and SC-like limits for $n=3/2$.]
{The dashed lines are contours of fractional mass $q=\phi_0/\phi_*$
  for $n=3/2$.  The solid spiral is the isothermal core solution and
  the horizontal straight line is the Bondi radius of a quasi-star
  with $b=1$.  The values of SC-like limits for an isothermal core
  inside an $n=3/2$ envelope and a polytropic quasi-star with $n=3/2$
  can be estimated by inspection to be roughly $0.46$ and $0.10$.}
\label{fcont15}
\end{center}\end{figure}

The SC-like limits discussed above all exist because each locus of
core-envelope boundaries only intersects fractional mass contours with
$q$ smaller than some $q\st{max}$.  This condition is generally
satisfied whenever the curve defining the inner edge of the envelope
touches but does not cross some fractional mass contour.  That is, the
condition is satisfied when the core solution is tangential to the
contour with $q=q\st{max}$.  This condition allows one to identify
immediately the existence of additional SC-like limits and estimate
their value.  For example, Fig.~\ref{fcont15} shows fractional mass
contours for polytropic envelopes with $n=3/2$ as well as the
isothermal core and the Bondi radius of a quasi-star with $b=1$.  By
inspection, a SC-like limit exists for an isothermal core embedded in
an $n=3/2$ envelope and the corresponding limit is roughly $0.46$.
\citet{yabushita75} used such composite polytropes to model neutron
stars and computed their maximum total and core masses.  Although he
did not explicitly calculate the maximum fractional core mass, a rough
value of $0.46$ can be determined by fitting the core and total mass
curves in his Fig.~1.  The precise value of the maximum fractional
core mass is $0.464$.  Similarly, a SC-like limit must exist for
quasi-stars with $n=3/2$ envelopes.  By inspection, the limit is about
$0.10$ for $b=1$, which compares well with the value $0.105$
determined by integrating polytropic envelopes.

The condition for the existence of SC-like limits also identifies
large classes of core solutions to which the limits apply.  For
example, a SC-like limit must exist whenever the core solution has
everywhere $V<n+1$, where $n<5$ is the polytropic index of the
envelope, because there are always contours that have everywhere
$V>n+1$.  Any composite polytrope with $n>5$ in the core and $n<5$ in
the envelope satisfies this condition and is subject to a SC-like
limit.  Cores described by $n=5$ polytropes with $\alpha>6/(n+1)$
envelopes, as discussed by EFC98, fail to intersect all contours.  An
example is plotted in Fig.~\ref{fcont1}, where the core solution with
$n=5$ and $\alpha=4$ clearly does not intersect contours with
$q\gtrsim0.15$ for envelopes with $n=1$.  SC-like limits exist when
there is a layer with $n\gg5$ at the base of the envelope, on top of a
polytropic core with $n<5$.  In fact, the analysis is not restricted
to polytropic models so even contrived core solutions, such as
$V=1-(U-2)^2$, exhibit SC-like limits.

The envelope solution can also vary as long as its fractional mass
contours allow the above condition to be satisfied.  Although the
contours are qualitatively similar for $n<5$, it is not obvious that
there are useful non-polytropic envelopes that have contours conducive
to the existence of SC-like limits.  One useful case, at least, is the
convective model described by \citet{henrich41}.  This is adiabatic
with varying contributions from radiation and gas pressures.
Envelopes calculated with his model behave like polytropic envelopes
with $n=3$ near the inner boundary and $n=3/2$ near the surface.

In real stars, nuclear burning shells can meet the criteria for the
existence of a SC-like limit.  Nuclear reactions usually depend
strongly on temperature and flatten the temperature gradient.  Regions
of a star where nuclear reactions are taking place therefore tend to
have larger values of $n$ so SC-like limits can be present.  The
criteria become even stronger if the density gradient at the
core-envelope boundary becomes steeper or the mean molecular weight
jump becomes more pronounced.  These effects shift the limiting region
to smaller $U$ and $V$ where the fractional mass contours correspond
to smaller fractional masses.

\subsection{Evolution beyond the limit}

What happens when an isothermal core exceeds a SC-like limit? To
remain in hydrostatic equilibrium, its effective polytropic index must
change.  This can happen in two ways.  Under suitable conditions, the
inner part of the core becomes degenerate.  Degenerate matter is
described by a polytropic equation of state with $n=3$ or $n=3/2$ in
relativistic or non-relativistic cases, respectively.  The inner core
can tend to such an equation with an isothermal layer further from the
centre.  A SC-like limit still exists but the isothermal layer is
displaced upwards in $V$ so a larger core mass is possible.

Alternatively, the core can develop a steeper temperature gradient by
departing further from thermal equilibrium.  For an ideal gas the
stellar structure can be described by a varying polytropic index such
that $\udif{\log p}{\log\rho}=1+1/n=1+\udif{\log T}{\log\rho}$ so an
increase in the temperature gradient decreases the effective
polytropic $n$.  As long as $n\gg5$, a SC-like limit persists but, as
in the previous case, it corresponds to a larger fractional core mass.
In the original SC limit, the core is isothermal so any outward
temperature gradient is enough to relax the limit.  For SC-like limits
in general, the core can be mass-limited even if it is not in thermal
equilibrium.

The structure of the envelope offers some respite from the constraints
imposed by the core.  For radiative envelopes, where $n$ is not much
greater than $3$, the fractional mass contours are similar to those
shown in Fig.~\ref{fcont3}.  If the envelope becomes convective, the
effective polytropic index varies between $3/2$ and $3$.  An SC-like
limit still exists but the behaviour of the fractional mass contours
is less extreme near $V_s$ for smaller values of $n$ (compare Figs
\ref{fcont3} and \ref{fcont1}).  Away from $G_s$, contours run along
smaller values of $V$ for smaller $n$.  Equivalently, $q(U,V)$ is
larger at a given point $(U,V)$ for smaller values of $n$.  For
example, for $n=3$, $q(2,4)=0.528$, whereas for $n=3/2$,
$q(2,4)=0.562$ and for $n=1$, $q(2,4)=0.576$.  Thus, a smaller
polytropic index in the envelope permits a larger fractional core
mass.

\subsection{Stability of polytropic cores}

It is known that an isothermal sphere, of mass $M$ and temperature
$T$, is unstable if its radius is smaller than the critical value
\shorteq{r\st{c}=0.41\frac{GM}{kT}\mu m\st{p}\text{,}} %
where $m\st{p}$ is the mass of a proton
\citep{ebert55,bonnor56,mccrea57}.  The sphere is unstable in the
sense that, if its radius decreases, its surface pressure decreases
too.  In the presence of an external pressure, the sphere is
compressed further and the contraction runs away.  Equivalently, if
the surface pressure is increased by an external agent, the sphere is
doomed to collapse.  The marginally stable solution is known as a
\emph{Bonnor--Ebert} sphere.  \citet{bonnor58} extended his analysis
to polytropes with index $n>0$ and determined that a polytropic core
with $n>3$ is unstable whenever the expression
\shorteq{\frac{1-\frac{n-3}{2}\left(\tdif{\theta}{\xi}\right)^{\sq2}\theta^{-(n+1)}}
  {1-\frac{n-3}{n-1}\xi^{-1}\theta^{-n}\tdif{\theta}{\xi}}\label{bonnor1}}
is negative.  In terms of homology-invariant variables, this is
equivalent to
\shorteq{\frac{U+\frac{U_G}{V_G}V}{U-U_G}\text{,}\label{bonnor2}}
where $U_G=(n-3)/(n-1)$ and $V_G=2(n+1)/(n-1)$ are the $U$ and $V$
co-ordinates of $G_s$.  The homology-invariant form of the %stability
criterion was independently derived by \citet{hiroshi79}.  The
numerator and denominator of expression (\ref{bonnor2}) define
straight lines in the \uvp{}.  The line that corresponds to the
numerator passes through the origin and $G_s$.  The line that
corresponds to the denominator is a vertical line through $G_s$.  For
given $n$, the critically stable polytropic core can be determined by
finding the first intersection of the core with one of the two %straight
lines because infinitesimally small cores are stable.  For the
isothermal case, the point of intersection defines the Bonnor--Ebert
sphere and similar solutions for all $n>3$ can be found in the same
way.

The perturbations that lead to instability require only that, during
the perturbation, the core has constant mass, remains in hydrostatic
equilibrium and satisfies the same polytropic relation.  These
conditions are satisfied by the cores of composite polytropes so it is
natural to ask whether the existence of Bonnor--Ebert spheres for
$n>3$, demonstrated by \citet{bonnor58}, is related to the existence
for SC-like limits for $n>5$, demonstrated here.  I contend that they
are \emph{not} for several reasons.  First, the limits are invoked at
different points along the core solution.  In the case of an
isothermal core embedded in an $n=3$ envelope, the fractional mass of
the Bonnor--Ebert sphere would be $q=0.341$ whereas the SC limit is
found at $q=0.359$.  Secondly, the statements have different
characters.  The work of \citet{bonnor58} regards \emph{stability}
whereas SC-like limits regard the \emph{existence} of solutions.
Thirdly, the stability of the isothermal core is independent of the
envelope unlike SC-like limits.  Although SC-like limits and
Bonnor--Ebert stability are different concepts, they occur at similar
points in a sequence of composite polytropes with growing cores.  All
else being equal, a core that reaches the critical Bonnor--Ebert mass
achieves a SC-like limit shortly thereafter.  Thus, if the core
becomes unstable and begins to contract, it soon finds itself unable
to accommodate additional mass.

% \citep{lynden-bell68} identify a similar instability that occurs in an
% ideal gas that is immersed in a heat bath with constant temperature.
% If the ratio of central to surface density is greater than $32.2$, the
% outer parts of the gas are unable to heat up as rapidly as the central
% regions do via contraction.  The temperature gradient grows ever
% steeper, the core contracts further and the process runs away.  This
% is the \emph{gravo-thermal catastrophe}.  \citet{lynden-bell68}
% identify similarities between the SC limit and the gravo-thermal
% catastrophe and \citet{iben93} claim that they are indeed the same.

\begin{figure}\begin{center}
\includegraphics{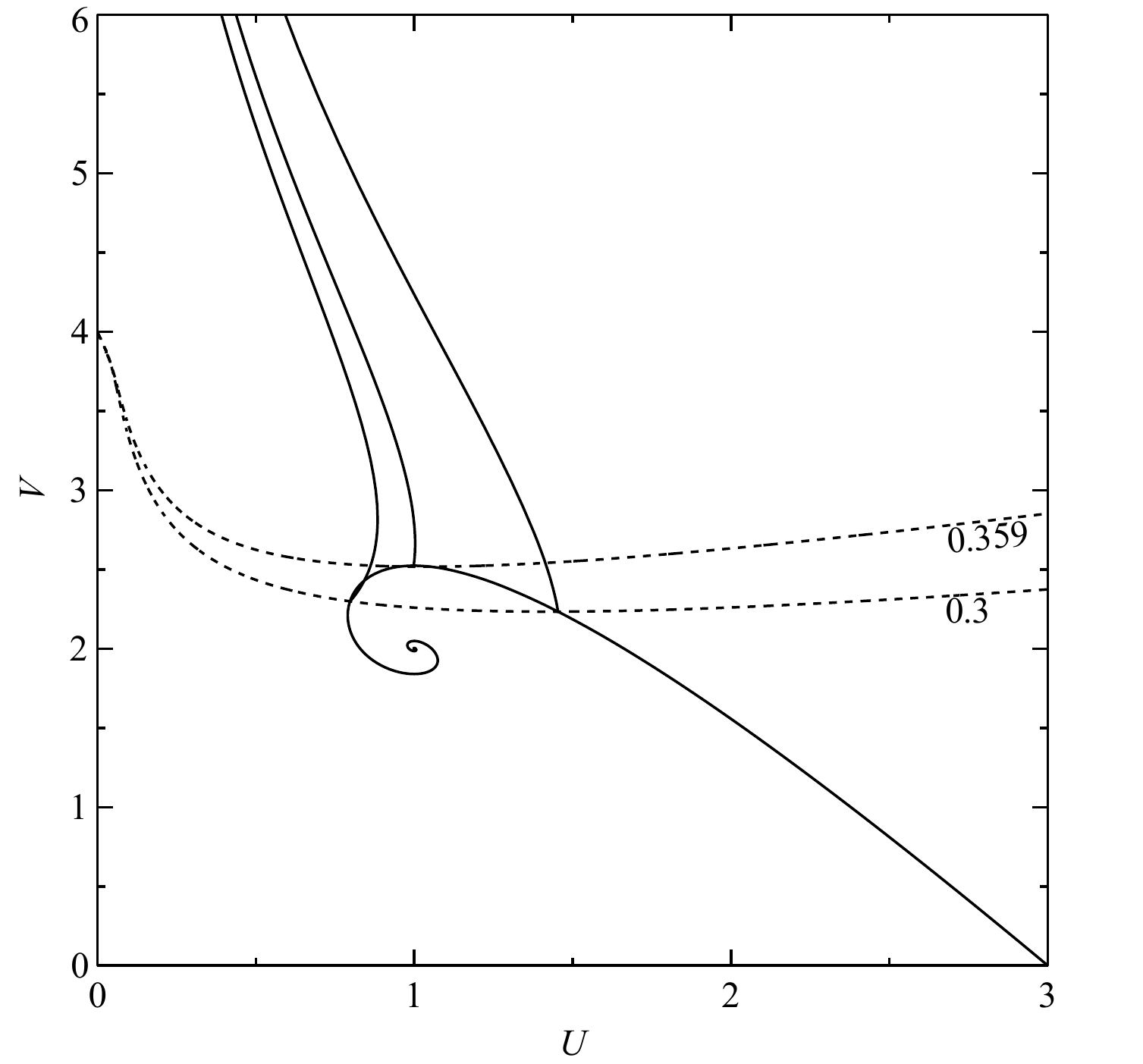}
\caption[Demonstration of the SC-like limit test on the original SC
limit.]  {An application of the SC-like limit test for an isothermal
  core and polytropic envelope with $n=3$, as in the original SC
  limit.  The contours corresponding to the fractional masses of each
  core solution are plotted in dashed lines.  The envelopes to the
  right and left are not tangential to the relevant contours and
  therefore fail the test.  The middle envelope is exactly tangential
  and at a SC-like limit.  As expected, this model has the maximum
  fractional mass originally found by \citet{schoenberg42}.}
\label{ftest}
\end{center}\end{figure}

\subsection{Identifying mass-limited models}
\label{sstest}

It is possible to test whether a composite polytrope is at a SC-like
limit.  If the core solution touches but does not cross the contour
corresponding to the given fractional core mass, the model is at a
SC-like limit.  That is, given the polytropic model and the index in
the envelope, the fractional mass at the core-envelope boundary can be
computed and the whole solution plotted with the relevant fractional
contour for the envelope.  The condition can be applied to determine
whether the model is limited.  If the condition is satisfied,
extending or contracting the core can only admit a smaller fractional
core mass.  Fig.~\ref{ftest} demonstrates the test applied to a model
at its SC limit.  The core-envelope solution touches and is curved
away from the contour with fractional mass $q=0.359$ so our test
confirms that the model is at a limit.  The other models, with
$q=0.3$, are not tangential to the contours and are therefore not at a
SC-like limit.

The test can be applied to realistic stellar models and allows us to
connect the purely theoretical results of this chapter with real
stellar models.  It is, however, difficult to identify stars that have
reached a SC-like limit because the inner edge of the envelope is not
clearly defined and the effective polytropic index varies throughout
the envelope.  In the next chapter, the test is applied to realistic
stellar models from the Cambridge \stars{} code and the results
suggest that there is a connection between stars reaching SC-like
limits and subsequently evolving into giants.

 % The Schönberg-Chandrasekhar limit
\begin{savequote}[80mm]
%   At terrestrial temperatures matter has complex properties which are
%   likely to prove most difficult to unravel; but it is reasonable to
%   hope that in the not too distant future we shall be competent to
%   understand so simple a thing as a star.
%   \qauthor{Arthur Eddington, 1926} 
  I would not recommend anyone, particularly if they are in a
  temporary research position, to waste any time looking for a
  sufficient condition.
  \qauthor{Peter Eggleton, 2000}
\end{savequote}

\chapter[Fractional mass limits and the structure of
giants]{Fractional mass limits \\and the structure of giants}
\label{crg}

When the core of a main-sequence star exhausts its supply of hydrogen,
it undergoes a number of qualitative changes.  Nuclear burning moves
from the centre to a shell around the inert core and the mean
molecular weight gradient at the core-envelope boundary steepens.  At
the same time, the envelope begins to expand increasingly quickly.
With the total luminosity roughly constant, the surface temperature
drops until the star reaches the Hayashi track and a deep convective
zone penetrates inwards from the surface, marking the beginning of the
red giant branch.  The rapid decrease in surface temperature means
that few stars are found in the so-called Hertzsprung gap.  During
this phase, a star's envelope expands greatly, potentially by orders
of magnitude, and continues to do so as it ascends the red giant
branch.

Though the evolution of main-sequence stars into giants has in essence
been reproduced since the pioneering calculations by \citet{hoyle55},
the cause of a star's substantial expansion after leaving the main
sequence remains unknown.  This \emph{red giant problem} continues to
receive occasional attention.  Some authors have claimed both to solve
the problem and explain why previous solutions were incomplete only to
have the same claim made against them by others.  %One might say
It appears that whether or not the red giant problem is an open
question is itself an open question.

Some clarity is achieved by more precisely specifying the problem.  In
my opinion, there are in fact \emph{two} red giant problems that are
discussed and accusations of incomplete solutions have been directed
at authors who were actually answering a different question.  The
first question is ``why do stellar cores necessarily contract after
leaving the main sequence?''  The second question is ``why does a
modest contraction of the core lead to extensive expansion of the
envelope?''  
% Put differently, the questions are ``why does a star
% begin to become a giant?'' and ``why do giants' envelopes expand so
% much more than their cores contract?''

Many potential solutions have been posed (and an approximately equal
number contested) to both of these problems.  \citet{eggleton91}
showed that if the effective polytropic index of a star is everywhere
less than some $n_\text{max}<5$, it is less centrally condensed (i.e.
%$\rho_c/\bar\rho$ 
the ratio of central to mean density is smaller) than the polytrope of
index $n_\text{max}$.  \citet{eggleton00} further conjectured that the
evolution of dwarfs into giants during shell burning therefore
requires that a significant part deep in the envelope has an effective
polytropic index $n\gg5$.  Because these conditions are similar to
those under which SC-like limits exist, I explore in this chapter the
possible connection between SC-like limits and giant formation.  In
Section \ref{shist}, I summarize some ideas on the red giant problem
from the last thirty years.  Then, in Section \ref{sscreal}, I apply
the test for SC-like limits devised in Section \ref{sstest} to
realistic stellar models to determine if there is a deeper connection
between SC-like limits and the evolution of dwarfs into giants.  I
close the chapter by discussing, in Section \ref{srealdisc}, the
results with respect to the two questions above.

\section{A brief history of the red giant problem}
\label{shist}

\citet{eggleton81} summarized and refuted a number of potential causes
for the formation of giants.  They first assert that the red giant
problem is \emph{not} resolved by the development of convective
envelopes, the development of degenerate cores, the Virial Theorem or
the SC limit.  The first two points are discussed below.  The Virial
Theorem is discounted because it regards the whole star and says
nothing of the density or temperature distribution within the star.
The role of the SC limit is the subject of this chapter.
\citet{eggleton81} assert that the evolution of stars into giants is
caused by the formation of a molecular weight gradient and the shift
from core burning to shell burning.  They also concede that the
evidence is complicated.  First, the molecular weight gradient
develops before hydrogen is exhausted in the core and only reaches a
factor of two even though the envelope expands by several orders of
magnitude.  Secondly, despite developing a ``strong'' shell source,
low-mass helium stars do not become giants.  Finally, the burning
shells in massive stars are not strong at first.  Thus, to explain the
red giant problem with the molecular weight gradient or shell burning,
one must also explain why some stars do not become giants even when
one or both of the conditions is present.

In this section, I review a handful of contributions to the literature
since the summary by \citet{eggleton81}.  Some are corroborations of
the points they raised.  Others are novel ideas.  Ultimately, any
discussion of the red giant problem is a contemporary snapshot of an
unresolved problem.  The discussion demonstrates the complicated
nature of the problems of isolating cause and effect in full stellar
calculations and in connecting simplified models with real physical
behaviour.

% The effective polytropic index
% of the envelope is unimportant because giant models can be
% constructed for a range of $n$ beyond what is observed.  Finally, the
% transition of nuclear reactions to burning in a shell around the
% core is also not the cause.  Pure helium stars with masses in the
% approximate range $0.3\Msun$ to $0.55\Msun$ possess burning shells
% but, instead of evolving into giants, proceed directly onto the
% white-dwarf cooling sequence.

% In particular, we are considering the hypothesis that stars that
% become SC-limited evolve into giants.  Fig.\,\ref{fm9popI} shows the
% evolution of a $9\Msun$ star at solar metallicity in the
% \uvp{}.  Even before exhausting hydrogen at the centre, the star
% appears to reach a SC-like limit.  At the same time, the star
% expands and the effective temperature decreases, marking its
% transition towards giant-like structure.  As it crosses the
% Hertzsprung gap, the star develops a growing loop in the \uvp{}.  It
% no longer appears to be at a SC-like limit.  , although its core
% solution is continued beyond where the SC-like limit would apply.

\subsection{Convection, degeneracy and thermal stability of the envelope}

If the ongoing discussion shows a lack of consensus on what causes
stars to become giants, at least there is some consensus on physical
changes that do \emph{not} lead to giant formation, neither in the
sense of explaining why the core contracts nor why the expansion of
the envelope is disproportionately large.  First, the development of a
deep convective zone adjacent to the stellar surface is not relevant.
Secondly, the increasing contribution of electron degeneracy to the
pressure in the core is also unrelated.  Massive stars begin to expand
towards giant proportions before either condition is true and low-mass
stars evolve into giants even if convection is artificially suppressed
in numerical models \citep{stancliffe+09}.

\citet{renzini84} suggested that the expansion of the envelope is
caused by a runaway thermal instability of the envelope.  He used the
diagnostic $W=\udif{\log L\st{rad}}{\log r}$ to decide whether the
envelope is unstable to an increased incident luminosity at the edge
of the shell.  His parameter is related to the thermal conductivity
and therefore opacity, so he argued that the runaway expansion of the
envelope is a result of radiation being locked in the envelope by
increasing opacity.  Only the onset of convection on the Hayashi track
assuages the instability.

Already, \citet{weiss89} questioned this approach by demonstrating
that the approximate formula for $W$ used by \citet{renzini84} is
generally inaccurate, often including its sign, and there exist
regions of the envelope that are supposedly unstable by this criterion
but are actually stable.  \citet{renzini+92} later restated Renzini's
(1984) solution to the red giant problem with little detailed
discussion of the work of \citet{weiss89}.  \citet{iben93} also
pointed out that thermal instability is not a necessary condition
because stars still become giants when overall thermal equilibrium is
artificially enforced.  For low-mass stars, there is little difference
between models with and without thermal equilibrium.  Thus, the
thermal instability in the envelope cannot be the cause of a star's
evolution to gianthood.

The final word in the conversation of papers appears to belong to
\citet{renzini94}, who argue that, while stars expand into giants when
the opacity is constant or thermal instability is artificially
suppressed, in these cases the core luminosity must exceed a critical
value.  Low-metallicity massive stars only reach sufficient luminosity
after core-helium burning begins. \citet{renzini94} claim that this
supports their argument.  First, this appears to be a concession that
their envelope instability does not account for at least some giants.
Secondly, it is not clear what the critical luminosity is or why it
should exist.  In fact, massive low-metallicity stellar models
successfully evolve into giants at lower luminosities if nuclear
transformations are stopped (which fixes the mass location of
shell-burning) and only energy from expansion, contraction and
hydrogen-burning is included (so that core-helium burning cannot
restore thermal equilibrium).  The observation of thermal instability
is probably a confusion of cause and effect.  Stellar envelopes expand
increasingly quickly in the Hertzsprung gap and, whatever the cause,
the expansion is necessarily reflected in an increasing thermal
imbalance between core and envelope.

% \subsection{Molecular weight gradients, shell burning and the polytropic index}
\subsection{Molecular weight gradients, shell burning and polytropic indices}

\citet{eggleton81} already claimed that stars expand to giant
dimensions because the molecular weight gradient between the core and
the envelope steepens and the nuclear reactions proceed in a shell
rather than the core.  \citet{eggleton81} also acknowledge that these
two qualities are insufficient when isolated.  A molecular weight
gradient develops during the main sequence but stars only evolve into
giants after hydrogen is exhausted in the core.  Low-mass (less than
about $0.7\Msun$) helium stars have strong burning shells but do not
evolve into giants, so the burning shell alone does not cause giant
behaviour.  These observations are corroborated by
\citet{stancliffe+09}, who show that, if the molecular weight gradient
is suppressed, a $1\Msun$ Pop I stellar model does not become a giant
but a $5\Msun$ model does.

\citet{eggleton00} expanded on the connection of these two effects in
terms of the effective polytropic index $n$ of a stellar model, where
$n$ is defined by 
\shorteq{1+\frac{1}{n}=\tdif{\log p}{\log\rho}\text{.}}  
In a perfect gas, an increase in the mean molecular weight causes an
increase in $n$.  Nuclear burning has a thermostatic effect and
flattens the temperature gradient where it occurs, which also
increases $n$.  \citet{eggleton00} demonstrated that, using the
variation of $n$ with $\log p$ in a stellar model, one can distinguish
dwarf-like models from giant-like models by the presence of
substantial regions where $n$ is much greater than $5$.  The
shortcoming of this argument is that it only says that a model without
any such regions is dwarf-like.  It is not clear how large $n$ must be
or in how large a region for a star to be giant-like.

\citet{sugimoto00} quote similar conditions under which stellar models
in the \uvp{} cross the line where $\udif{V}{U}=V/U$.  Their claim
that curves that cross this line have divergent radii is refuted by
\citet{faulkner05} as a mathematical misinterpretation.  It is true,
however, that solutions that have loops in the \uvp{} cross this line
twice.  Such solutions can accommodate a greater fractional mass in
the envelope and this is a characteristic of giants.  However, the
presence of loops has not been demonstrated to be the \emph{cause} of
a star's evolution to a giant.  \citet{faulkner05} contests that giant
structure occurs because the stellar core cannot accommodate more mass
without first inflating the envelope.  This is similar to the
situations identified by SC-like limits.

Based on the calculation of fractional mass limits in Chapter
\ref{cscl}, the conditions described by \citet{eggleton00} correspond
well with the conditions under which fractional mass limits are likely
to exist.  Certainly, when there is a layer between the core and
envelope where $n\to\infty$, a SC-like limit exists.  Whether or not a
stellar model reaches the limit depends on the extent of that region.
The purpose of the remainder of this chapter is to evaluate when
stellar models appear to reach SC-like limits and whether it
corresponds to the evolution of a star into a giant.

\section{Fractional mass limits in realistic stellar models}
\label{sscreal}

Realistic stellar models are now tested for the presence of SC-like
limits.  In particular, I analyse $U$--$V$ profiles of stellar models
with masses between $3$ and $15\Msun$ at solar metallicity (Section
\ref{sspopIhi}), between $0.5$ and $1\Msun$ at solar metallicity
(Section \ref{sspopIlo}), and between $0.5$ and $1\Msun$ with helium
instead of hydrogen (Section \ref{sshe}).  No convective overshooting
or mass loss was included and the models are initially of homogeneous
composition.

The fractional mass limit test is applied by searching for the first
model in the evolution for which its $U$--$V$ profile %in the \uvp{} 
contains a
segment that is tangential to %touches and is parallel to : LBfix
a fractional mass contour.  
%In all cases, the fractional mass contour presumes 
Models shortly before and after the limit are also tested to show that
the behaviour is similar to the SC limit.  The fractional mass
contours presume that the
envelope's %has an effective polytropic index of $3$.
effective polytropic index is $3$.  In a given model, a larger
effective polytropic index would admit a smaller fractional mass
earlier in a star's evolution.  The approximation of $n=3$ in the
envelope is usually reasonable but some models have polytropic indices
as low as $2$ or as high as $4$.  In these cases, the value $3$ is
justified because the models ultimately exceed the limits for other
choices of $n$ and the fractional mass is approximate at a similar
level to errors introduced by changing the envelope's polytropic
index.

\begin{figure}\begin{center}
\includegraphics{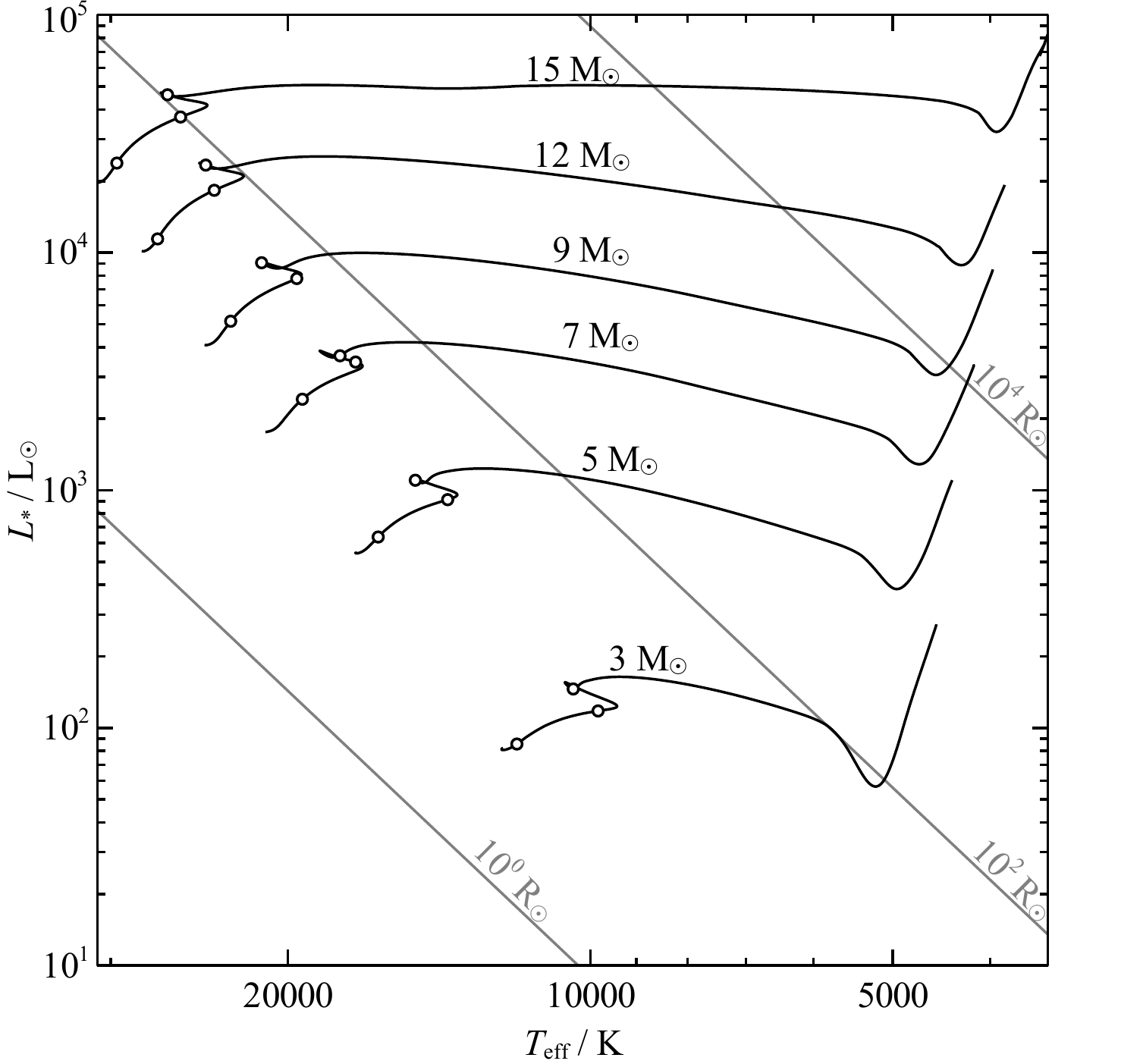}
\caption[Evolutionary tracks of intermediate-mass Pop I stars in the
HRD.]{Evolutionary tracks of intermediate-mass Pop I stars in the
  theoretical Hertzsprung--Russell Diagram.  The models are labelled by
  their total masses and are run from the main sequence up to the tip
  of the red giant branch.  The open circles mark the models whose
  profiles are plotted in Fig.~\ref{fpopIhi-grid}.}
\label{fpopIhi-hr}
\end{center}\end{figure}

\begin{figure}\begin{center}
\includegraphics{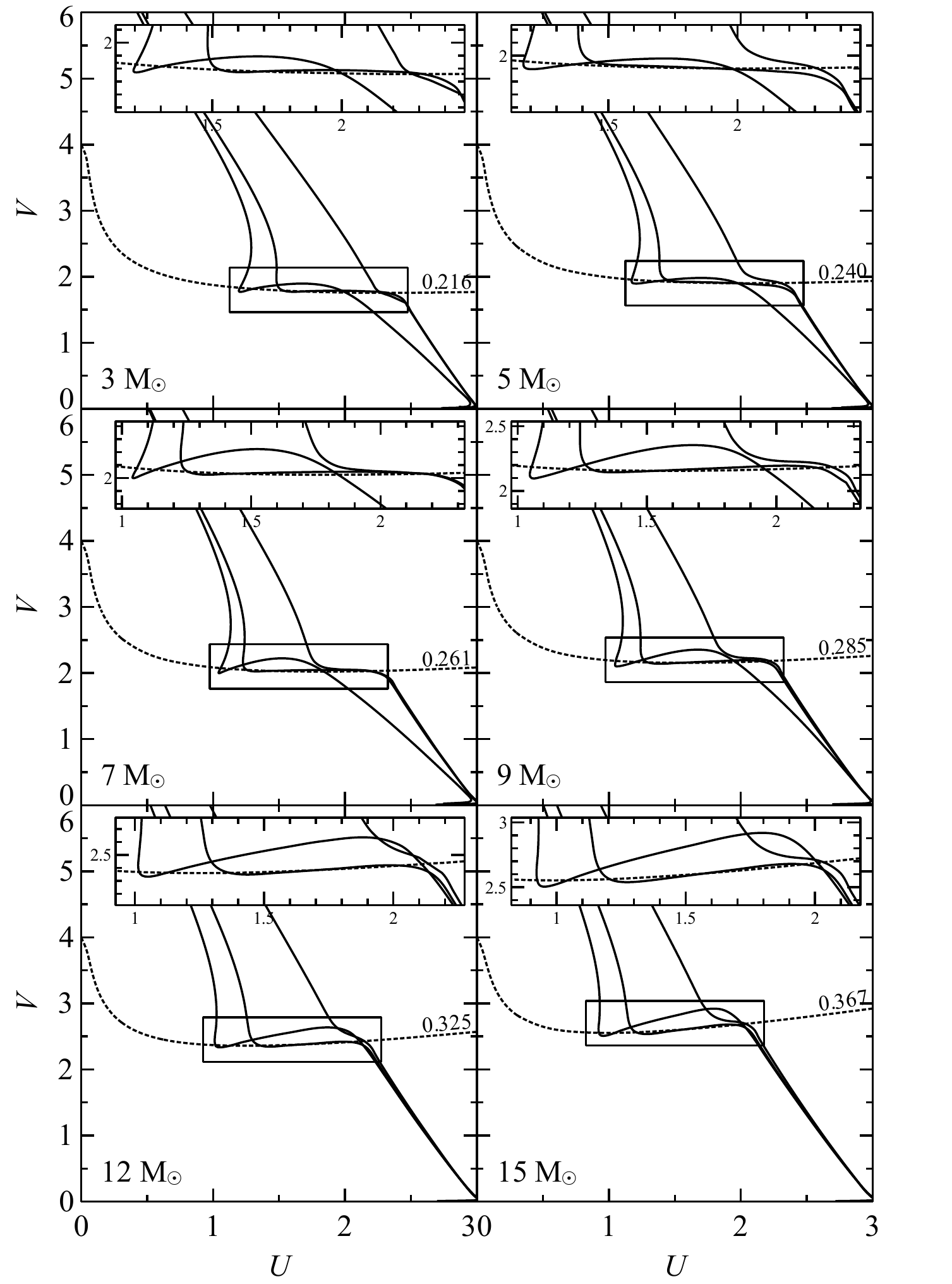}
\caption[Grid of intermediate-mass Pop I stars in the \uvp{}.]  {The
  solid lines are profiles of the intermediate-mass Pop I stars in the
  \uvp{} when the stars are at points marked by open circles in
  Fig.~\ref{fpopIhi-hr}.  The dashed lines are fractional mass
  contours of the indicated fractional masses for polytropic envelopes
  of index $3$.  The models evolve towards the left for $V>7/2$.  The
  middle model in each panel was selected by finding by eye the first
  model where a segment of the profile appears to reach a fractional
  mass limit and the earlier and later models were selected to show
  the behaviour of the models on either side of any apparent SC-like
  limit.  The boxes indicate the regions of interest, which are
  magnified in the subplots.}
\label{fpopIhi-grid}
\end{center}\end{figure}

\subsection{Intermediate-mass Population I giants}
\label{sspopIhi}

Fig.~\ref{fpopIhi-hr} shows evolutionary tracks for stars of solar
metallicity at masses $3$, $5$, $7$, $9$, $12$ and $15\Msun$.
% At the beginning of each run, the composition first varies to
% establish an equilibrium abundance of carbon, nitrogen and oxygen
% through the CNO cycle.
As each star consumes its core hydrogen, it becomes slightly brighter
and redder, moving up and to the right in the Hertzsprung--Russell
diagram (HRD).  During the main sequence evolution, the core is
convective but the convective boundary slowly retreats, leaving a
composition gradient outside the convective core.  When it exhausts
its hydrogen, the whole core does so simultaneously and there is a
phase of contraction before hydrogen shell-burning commences.  The
shrinking envelope briefly becomes bluer before moving redward across
the Hertzsprung gap.  The models were followed to the top of their
ascent on the red giant branch.  Note that all of these stars become
red giants in the sense that they ultimately cross the Hertzsprung gap
completely but the $12\Msun$ and $15\Msun$ stars ignite helium in the
core before they reach the base of the red giant branch.

Fig.~\ref{fpopIhi-grid} shows the $U$-$V$ profiles of the models
marked with open circles in Fig.~\ref{fpopIhi-hr}.  The middle model
in each case is at a SC-like limit.  The molecular weight gradient
extends from $q\approx0.1$ to $q\approx0.4$ so the composition profile
does not define a clear core-envelope boundary.  The fractional core
mass is determined by finding, by eye, the first envelope contour that
is tangential to a segment of the model.  This process clearly
identifies a particular contour and the fractional core masses are
roughly consistent with those found by other measures such as the
centre of the composition gradient. % or the nuclear burning region.
The models are compared to the same fractional mass contour because
the core mass does not appear to change noticeably between them.  In
all cases, the result of the test is similar: the models approach,
reach and exceed the %relevant
mass derived from the polytropic contour.  The masses are larger than
traditional values of the core mass because the limiting region is now
the base of the envelope.  The limiting mass includes the shell where
hydrogen shell-burning begins, which is initially quite
thick.% in mass.

The point %location 
in the HRD at which models reach a SC-like limit is always %around the point of 
near core hydrogen depletion.  There are two
important observations to be made.  First, this is earlier than the
point at which most authors say a star reaches the SC limit.  A
$5\Msun$ star, for example, is usually thought to reach the SC limit
while it crosses the Hertzsprung gap.  Secondly, stars are usually
thought to reach the SC limit at different points.  Broadly,
higher-mass stars reach the SC limit earlier in their evolution.  
The results presented here 
%These results 
suggest that all intermediate-mass stars reach mass
limits %as soon as 
once hydrogen is depleted in the core.  The core structure
is subject to a SC-like limit during the whole 
%throughout its 
evolution into a giant, irrespective of whether the
core (or any other part of the star) is in thermal equilibrium.

\begin{figure}\begin{center}
\includegraphics{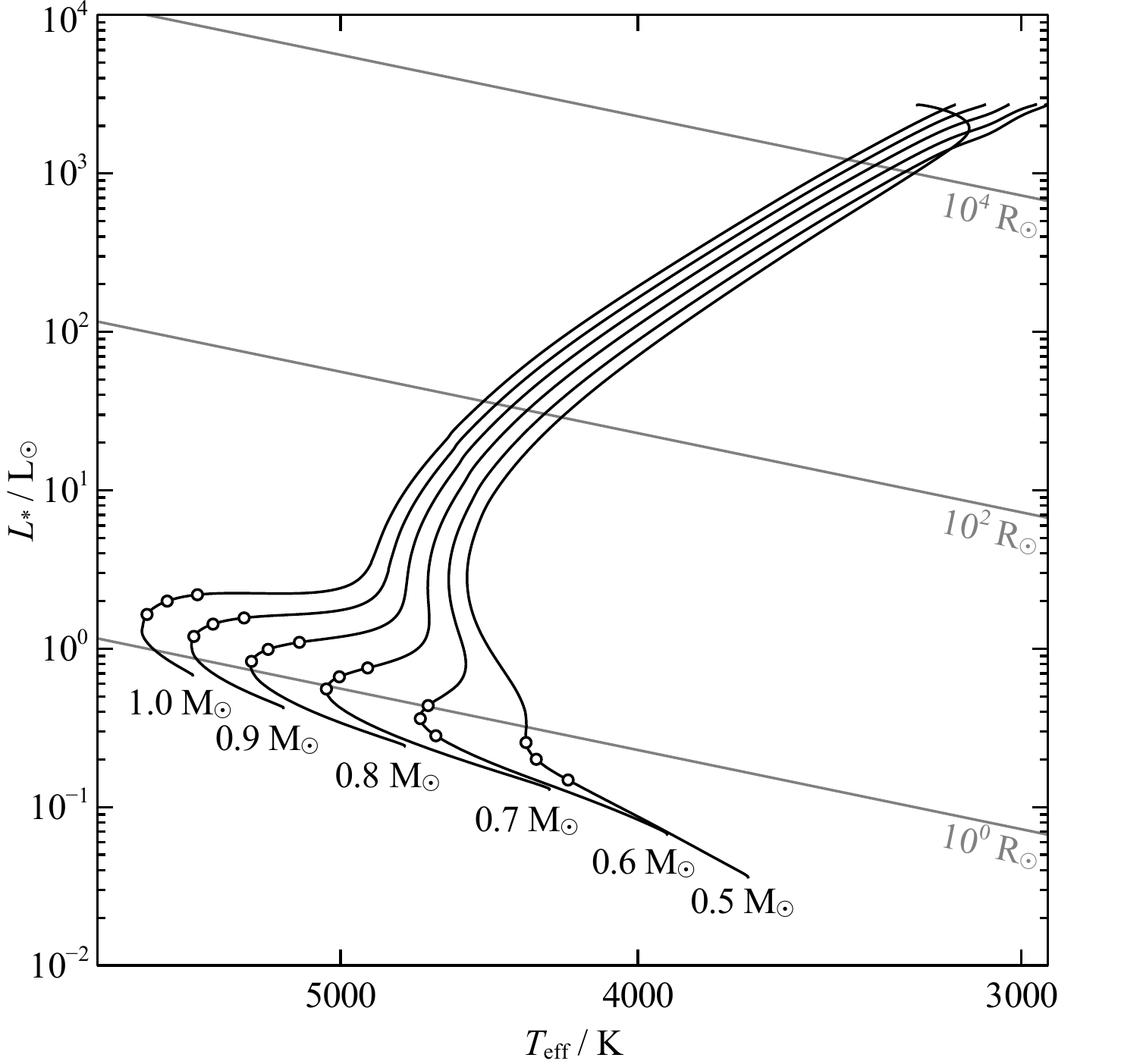}
\caption[Evolutionary tracks of low-mass Pop I stars in the HRD.]
{Evolutionary tracks of low-mass Pop I stars in the theoretical
  Hertzsprung--Russell Diagram.  The models are labelled by their
  total masses and are run from homogeneous composition near the main
  sequence up to the tip of the red giant branch.  The open circles
  mark the models whose profiles are plotted in
  Fig.~\ref{fpopIlo-grid}.}
\label{fpopIlo-hr}
\end{center}\end{figure}

\begin{figure}\begin{center}
\includegraphics{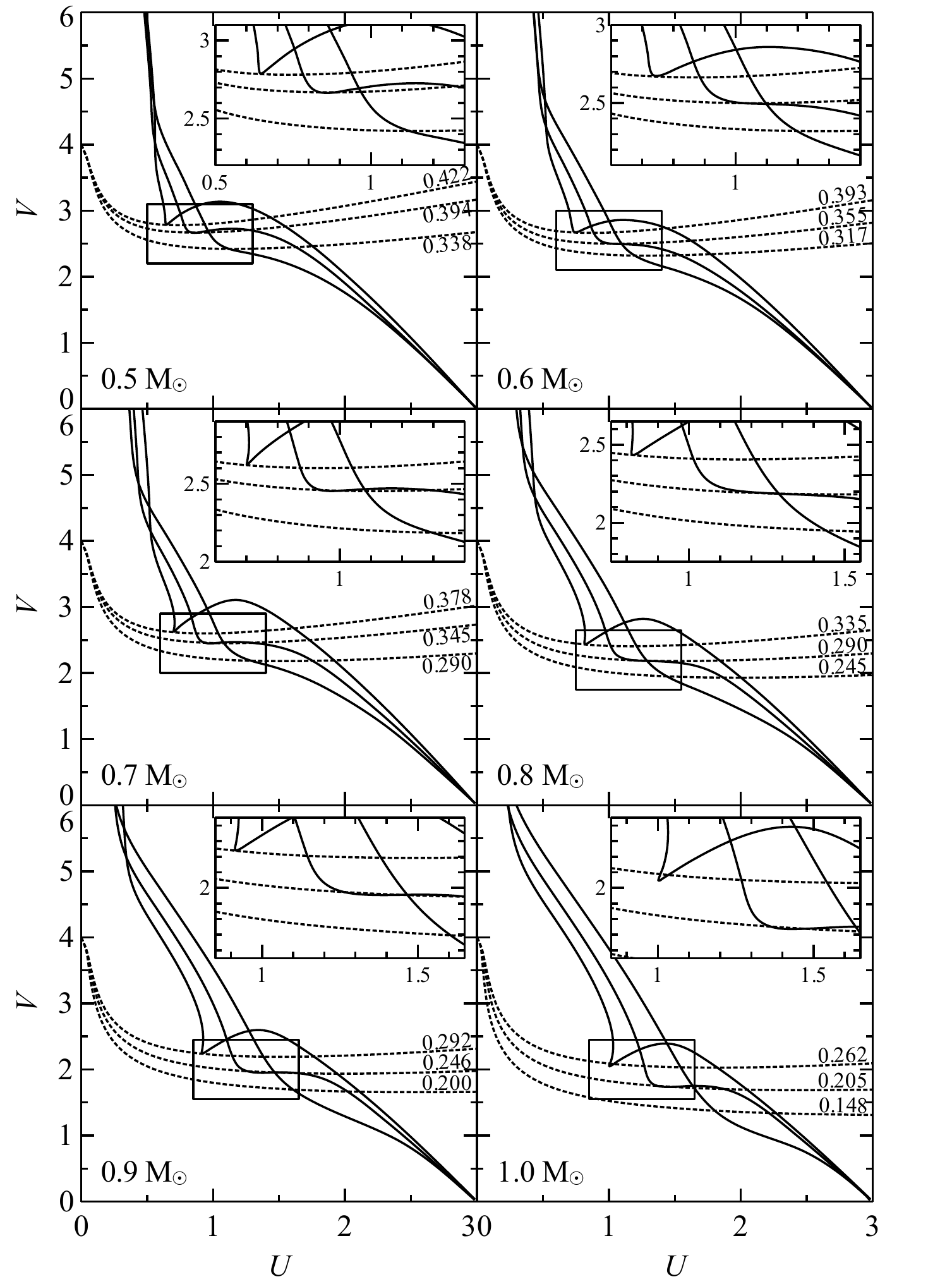}
\caption[Grid of low-mass Pop I stars in the \uvp{}.]  {The solid
  lines are profiles of the low-mass Pop I stars in the \uvp{} when
  the stars are at points marked by open circles in
  Fig.~\ref{fpopIlo-hr}.  The dashed lines are fractional mass
  contours of the indicated fractional masses for polytropic envelopes
  of index $3$.  The models evolve towards the left for $V=7/2$.  The
  middle model in each panel was selected by finding by eye the first
  model where a segment of the profile appears to reach a fractional
  mass limit and the earlier and later models were selected to show
  the behaviour of the models on either side of any apparent SC-like
  limit.  The boxes indicate the regions of interest, which are
  magnified in the subplots.}
%The models were selected by finding by eye the first model where a
%  segment of the profile appears to reach a fractional mass limit.}
\label{fpopIlo-grid}
\end{center}\end{figure}

\subsection{Low-mass Population I giants}
\label{sspopIlo}

Fig.~\ref{fpopIlo-hr} shows evolutionary tracks for stars of solar
metallicity at masses $0.5$, $0.6$, $0.7$, $0.8$, $0.9$ and $1\Msun$.
The cores of these stars are convectively stable during the main
sequence so there is a gradual progression from core-burning to
shell-burning.  Specifically, there are no blueward hooks, as in the
previous section.  The evolution was followed up to core helium
ignition.  In low-mass stars, the core becomes electron degenerate and
roughly isothermal before core helium-burning begins.  As a result,
helium is ignited almost simultaneously across the core.  The sudden,
rapid and unstable increase in luminosity is known as a \emph{flash}.
Such flashes are a known stumbling block of the \stars{} code.

\begin{figure}\begin{center}
\includegraphics{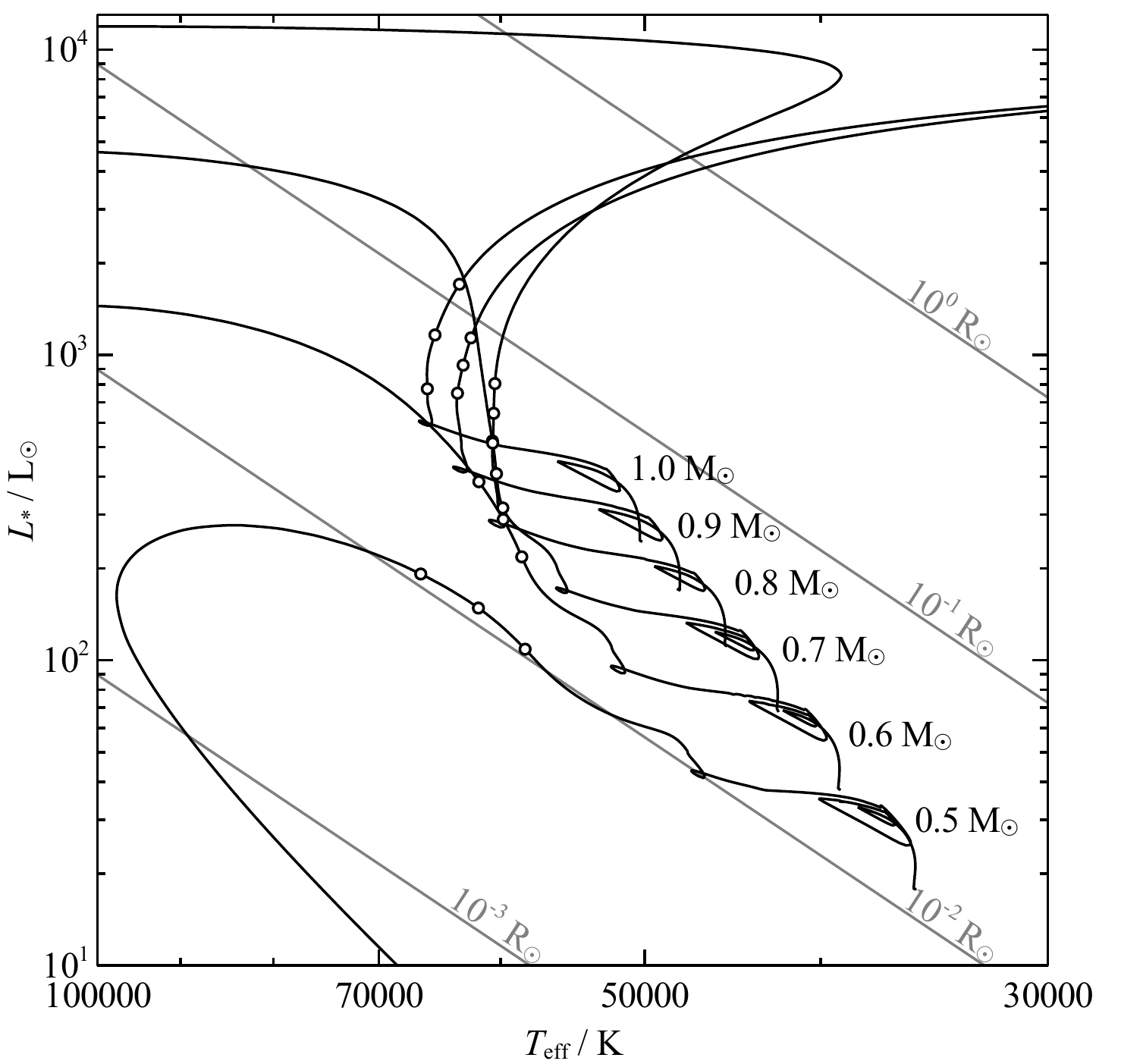}
\caption[Evolutionary tracks of helium stars in the HRD.]
{Evolutionary tracks of helium stars in the theoretical
  Hertzsprung--Russell Diagram.  The models are labelled by their total
  masses and are run from homogeneous composition near the main
  sequence up to the tip of the red giant branch.  The open circles
  mark the models whose profiles are plotted in Fig.~\ref{fhe-grid}.}
\label{fhe-hr}
\end{center}\end{figure}

\begin{figure}\begin{center}
\includegraphics{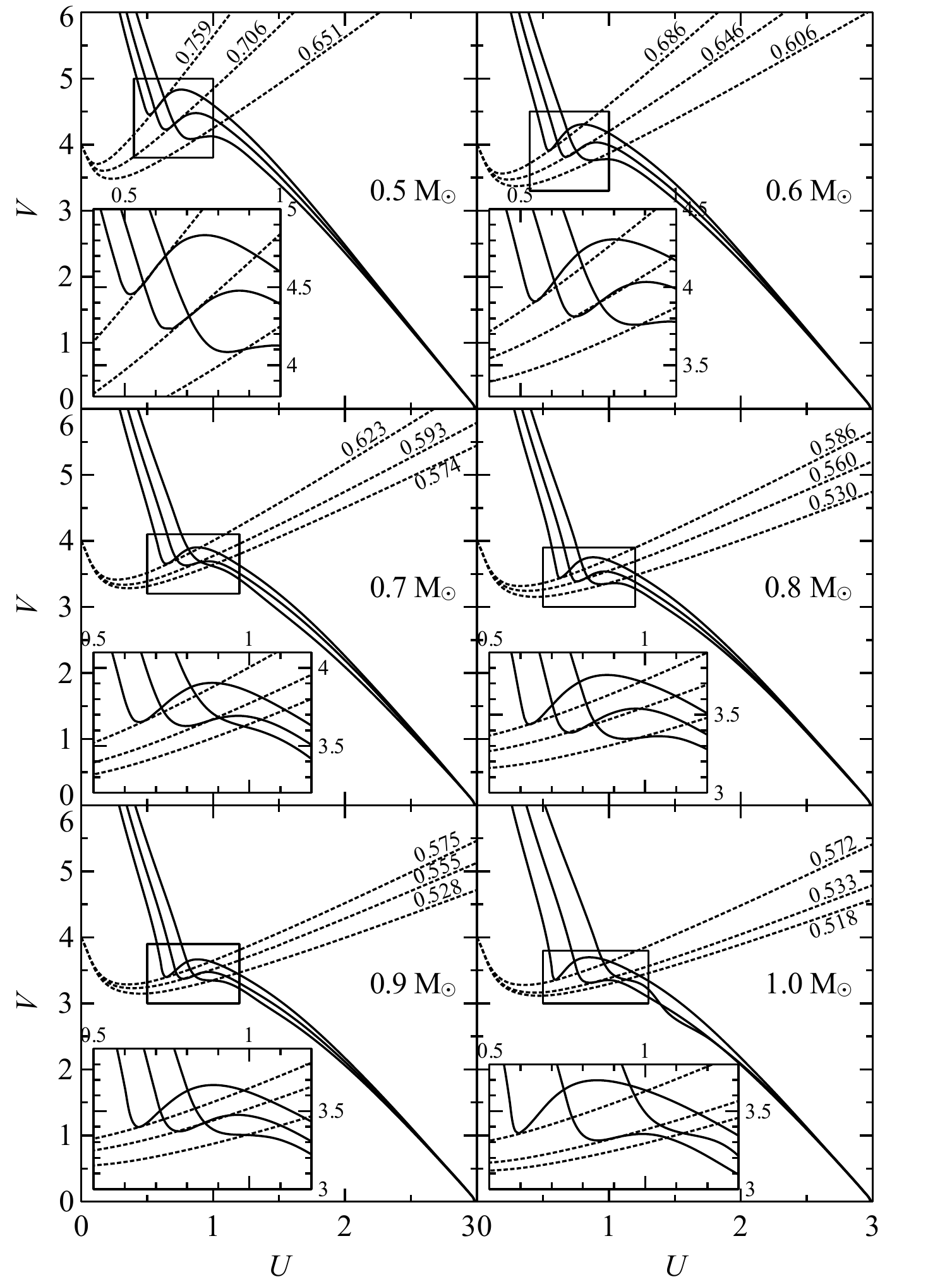}
\caption[Grid of helium stars in the \uvp{}.]  {The solid lines are
  profiles of the helium stars in the \uvp{} when the stars are at
  points marked by open circles in Fig.~\ref{fhe-hr}.  The dashed
  lines are fractional mass contours of the indicated fractional
  masses for polytropic envelopes of index $3$.  The models evolve
  towards the left for $V>5$.  The middle model in each panel was
  selected by finding by eye the first model where a segment of the
  profile appears to reach a fractional mass limit and the earlier and
  later models were selected to show the behaviour of the models on
  either side of any apparent SC-like limit.  The boxes indicate the
  regions of interest, which are magnified in the subplots.}
\label{fhe-grid}
\end{center}\end{figure}

As in Section \ref{sspopIhi}, Fig.~\ref{fpopIlo-grid} shows the
$U$-$V$ profiles of the models marked with open circles in
Fig.~\ref{fpopIlo-hr} and the middle profile is thought to have
reached a SC-like limit.  The profiles bend at large $V$ because the
polytropic index of the envelope changes to $3/2$ in the surface
convection zone, which is always present.  Outside the surface
convection zones, the envelopes have larger polytropic indices than
the intermediate-mass stars, so the fractional masses are probably a
few hundredths smaller and occur when the luminosities are smaller by
a few per cent.  Aside from this approximation, the models all appear
to reach SC-like limits shortly after hydrogen is exhausted in the
core.

The $0.5\Msun$ track turns towards higher surface temperatures near
the end of the run because the unburned hydrogen envelope contains
less that 5 per cent of the total mass.  The run halts because of
degenerate helium ignition.  Slightly less massive models do not
ignite helium and evolve to the helium white dwarf cooling sequence.
The fact that the $0.5\Msun$ model does not truly finish its life as a
giant does not undermine any effort to identify why it became a giant
in the first place.

\subsection{Helium giants}
\label{sshe}

Fig.~\ref{fhe-hr} shows evolutionary tracks for helium stars with Pop
I metal content and masses $0.5$, $0.6$, $0.7$, $0.8$, $0.9$ and
$1\Msun$.  The stars have convective cores during the core
helium-burning phase.  As the helium nears depletion, the core
convective zone extends further out, mixing in unburned helium.  This
extends the core burning phase.  The episodes of ingestion cause loops
in the theoretical HRD that are known as \emph{breathing pulses} after
the repeated contraction and expansion of the core.  Once the core
evolves past the last breathing pulse, it depletes helium completely
and convection stops.  Helium-shell burning begins at the small loop
to the left of the breathing pulses.  Models were followed either
until their effective temperatures began to decrease on a white dwarf
cooling track or until a surface convection zone formed, indicating
that the star arrived on the helium red giant branch.

Fig.~\ref{fhe-grid} again shows the $U$-$V$ profiles of the models
marked with open circles in Fig.~\ref{fhe-hr}.  These models also
appear to reach SC-like limits but only after a substantial amount of
mass has been added to the core during shell burning.  The connection
between SC-like limits and gianthood is weaker here.  Although the
models appear to reach limits, they do not clearly exceed them as the
hydrogen-burning models did.  That is, for masses less than
$0.8\Msun$, the models never have $U$-$V$ profiles where the gradient
$\partial V/\partial U$ is \emph{clearly} greater than a contour that
passes through the nuclear burning shell.  Indeed these models do not
become giants although the $0.6$ and $0.7\Msun$ models do undergo some
expansion.  The $0.8\Msun$ model expands like the $0.9$ and $1\Msun$
models but it turns back towards the white dwarf cooling tracks when
the surface temperature reaches about $40\,000\K$.  As with the
$0.5\Msun$ Pop I star, the envelope mass becomes negligible and the
star is almost entirely described by the core.

The helium stars presented here indicate that stars appear to expand
when they are at or near a SC-like limit.  Weaker nuclear burning
shells and shallow molecular weight gradients reduce the effective
polytropic index and diminish the prospects of a star being subject to
a SC-like limit.  The response of the envelope may restore the
stellar profile to a state where it is not subject to a limit, in
which case it continues to become bluer until it finally moves on to a
white dwarf cooling track.

%\section{Stability at fractional mass limits}
\section{Discussion}
\label{srealdisc}

The analysis presented above suggests that stars begin to evolve into
giants when they reach a SC-like limit.  In this section, I discuss
the combined evidence with respect to the two red giant problems.  In
Section \ref{sscore}, I address whether SC-like limits explain the
contraction of the core.  In Section \ref{ssexp}, I consider how
SC-like limits are related to the drastic expansion of the envelope.

\subsection{Core contraction}
\label{sscore}

Traditional discussions of the relevance of the SC limit compare the
mass of the exhausted core to a fractional mass limit of
$q\approx0.1$.  Stars of different masses reach this limit at
different stages of evolution.  Here, all the stellar models that
begin to evolve into giants appear to reach a SC-like limit just
before depleting the nuclear fuel in their cores.  This implies three
important distinctions from the usual interpretation of the SC limit.
First, fractional mass limits like the SC limit are relevant earlier
than usually thought.  Secondly, the point in a star's evolution at
which the mass limit is relevant is roughly the same independent of
mass.  Thirdly, the limiting mass co-ordinate is located in the
burning shell and does not only include the exhausted core.  That all
stars that evolve into giants achieve a limiting fractional mass at
the same point suggests a connection between SC-like limits and
evolution into giants but the issue is far from clear cut.

In the low-mass Pop I models and helium stars, SC-like limits do
correlate with expansion of the envelope.  Models that reach and
exceed a SC-like limit start to expand even though the expansion
sometimes ceases.  There are many possible reasons for this.  In the
simplest cases, shell-burning continues to increase the core mass and
the envelope becomes vanishingly small by mass.  As the shell-burning
declines, the limiting region (with $n\gg5$) vanishes and
the %relevant
SC-like limit is relaxed.  The $0.7\Msun$ helium star and $0.5\Msun$
Pop I model demonstrate such behaviour.  It can be argued that these
stars started to become giants but the envelope was transformed into
core material before the transition was accomplished.  Another option
is that the microphysics of the limiting region responds to the
expansion of the envelope such that a SC-like limit is avoided and
never encountered again.  For example, the $0.6\Msun$ helium star
expands slightly and briefly after core helium exhaustion but shell
burning is weakened by the expansion and the envelope is processed by
nuclear reactions before another limit is encountered.  The $0.5\Msun$
star never appears to exceed a SC-like limit although it does roughly
reach one.  It expands only slightly during the shell-burning phase
but soon turns back towards the white dwarf cooling sequence.

The intermediate-mass Pop I models also reach a limit as they exhaust
their core hydrogen supplies but they briefly contract rather than expand. %instead of expanding.  
The envelopes do expand once shell-burning begins but not
before.  This suggests that SC-like limits are connected to the
contraction of the core but not necessarily the expansion of the
envelope.  %It is worth mentioning that 
The model %profiles 
adhere to the same SC-like limit during the contraction so the core
may be able to sustain the limit until shell-burning begins.

These conclusions are all subject to the approximations of the test.
There is no well-defined core-envelope boundary at the end of the main
sequence so it is not surprising that there are differences between
the core masses determined by the test and masses determined by other
methods.  The polytropic index is certainly not constant and in some
models deviates quite substantially from $3$, which was used to test
the models for fractional mass limits.  A variable polytropic index
might change the shape of the fractional mass contours depending on
how it varies through the envelope.  Despite this, the overall shape
of the envelope profiles is broadly consistent with some SC-like limit
even if it occurs at a slightly different point in evolution.

% To conclude the discussion, I consider briefly the stability of models
% at SC-like limits.  For a given core solution and polytropic index of
% the envelope, there are potentially two solutions either side of an
% SC-like limit that correspond to the same fractional core mass.  The
% evolutionary sequence of static models determines which solutions
% occur in reality.  In the case of the SC limit, it is known
% \citep[e.g.][]{kw90} that stars with cores that exceed the limit are
% unstable.  Interestingly, though I calculated models of polytropic
% quasi-stars beyond the relevant SC-like limit, I could not compute
% \stars{} models along the same sequence.  If stellar models beyond the
% SC limit are unstable, is the same true of SC-like limits in general?
% If so, once a model exceeds a SC-like limit, it begins a race between
% a post-limit instability and any response of the envelope that might
% quench it.

\begin{figure}\begin{center}
\includegraphics{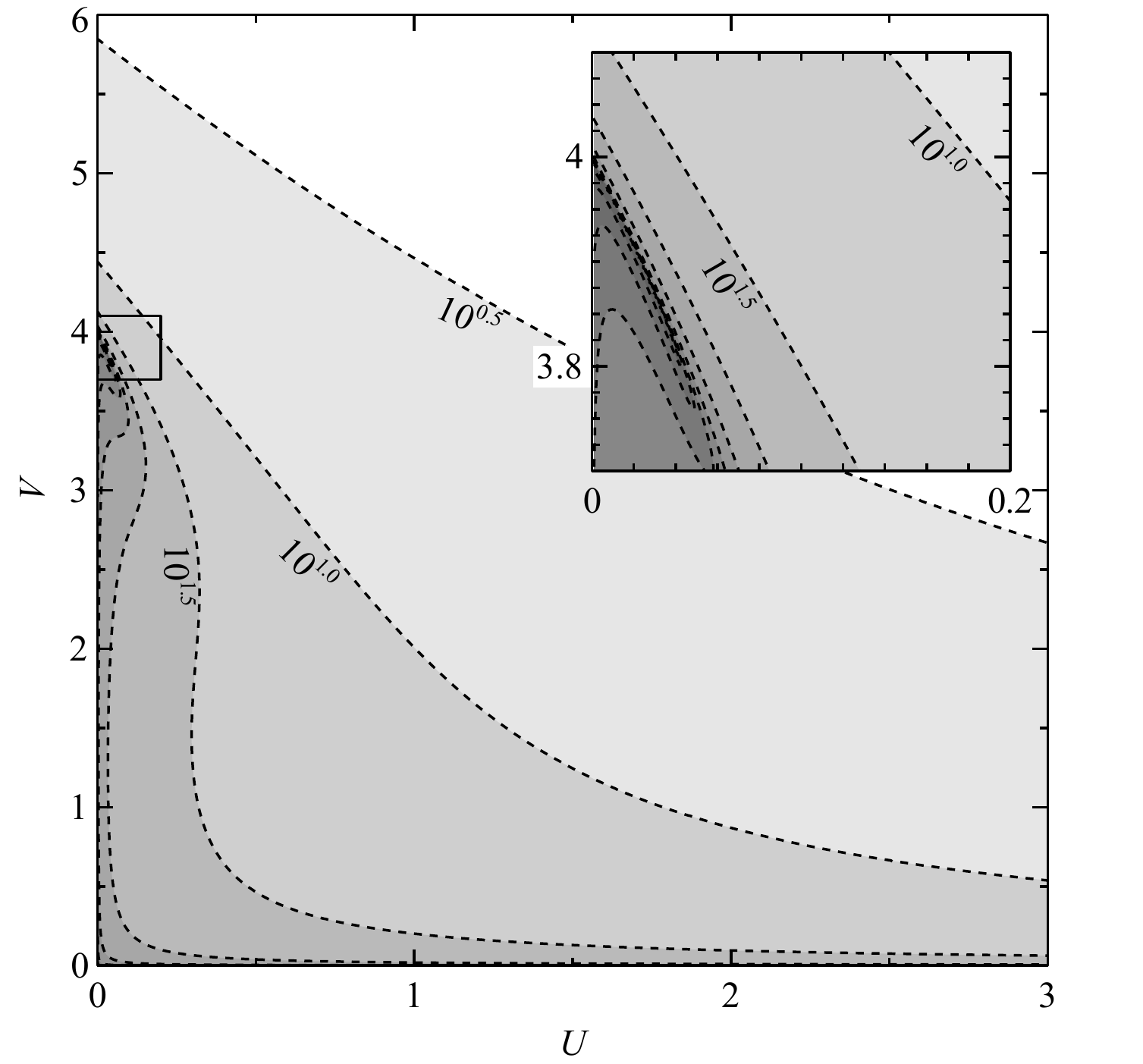}
\caption[Contours of constant surface-core radius ratio for $n=3$.]
{The dashed lines are contours of the ratio of total radius to core
  radius $R_*/r_0$ for an envelope with $n=3$.  They are spaced
  logarithmically in steps of $0.5$ dex from $0.5$, at the top, to
  $5.0$, near the critical point $G_s=(0,4)$ (see Section
  \ref{ssexp}).  The inset is a magnification of the boxed region and
  shows the extreme behaviour of the contours near the critical
  point.}
\label{frcont}
\end{center}\end{figure}

\subsection{Envelope expansion}
\label{ssexp}

Let us now consider the other problem: why are the envelopes of red
giants so extended?  Fig.~\ref{frcont} shows contours of constant
ratio of total radius to core radius $R_*/r_0$ for polytropic
envelopes with $n=3$.  The figure is created in the same way as the
figures showing contours of constant fractional mass.  Here, the
inverse of the fractional ratio is shown and the contours are spaced
logarithmically.  As an example, if the core radius is fixed, an
envelope with an inner boundary along the $10^{1.0}$ curve is
$\sqrt{10}\approx3.16$ times smaller than an envelope with an inner
boundary along the $10^{1.5}$ curve.  The plot shows that the
fractional surface radius increases as $U$ or $V$ decreases.  The
behaviour is particularly extreme near the critical point $G_s=(0,4)$.

Given a composite polytrope for which a SC-like limit exists, there
are usually two models with the same fractional core mass.
Fig.~\ref{ftest} shows this for isothermal cores with fractional mass
$q=0.3$ embedded in envelopes with $n=3$.  The two solutions have
different fractional radii so, for fixed core radius, the envelope is
more extended in the model for which $U$ is smaller at the
core-envelope boundary.  The more extended model is giant-like and the
other is dwarf-like.  

Realistic models are initially in the dwarf-like configuration.  The
feature that forces the core-envelope boundary to smaller $U$ is a
large effective polytropic index between the core and the envelope.
The large polytropic index means the solution tries to orbit the
critical point $G_s$ in the \uvp{}.  Hence, if a star has a region
with $n\gg5$ between the core and envelope, its radius is larger than
if this region did not exist.  As discussed in Section \ref{ssgenlim},
shell burning and a molecular weight gradient increase the effective
polytropic index and they are both connected to the large surface
radii of giants.  Convection drives the effective polytropic index
towards $3/2$, for which the critical point $V_s=(0,5/2)$ is closer to
the point around which isothermal solutions spiral, $G_s=(1,2)$.  On
the other hand, polytropic envelopes with $n=3/2$ behave less
extremely near $V_s$ because $V_s$ and $G_s$ do not co-incide as they
do for $n=3$.  It is not clear whether or not the envelope's expansion
is more extreme once convection develops.

The argument above offers an explanation of why the envelopes of
giants are so large.  It does not explain why
% evolution requires that
the core-envelope boundary evolves 
%in a direction that forces the envelope to become 
such that the envelope becomes 
even larger.  Do SC-like limits play a
role?  It has already been noted that stars expand at the same time as
they reach SC-like limits.  All else being equal, the core cannot
accommodate more mass by continuing to larger radii because the
effective polytropic index at the core-envelope boundary is too large.
A change in the core's structure is required, be it the central
density, the effective polytropic index or something else, and this
change might be unstable.  For example, if the core contracts, its
mean density increases.  If the presence of a nuclear burning shell
prevents the density at the core-envelope boundary from increasing by
the same amount, $U$ decreases, potentially enough that the 
envelope's expansion %expansion of the envelope 
is greater than the %contraction of the core.  
core's contraction. 
It is thus possible that, once a SC-limit has been exceeded and the
core adjusts %its structure
to incorporate more mass, the additional mass forces it to adjust
further but I have not been able to demonstrate that this is the case.

\clearpage

\section{Conclusion}
\label{sconc}

Three sets of stellar models have been tested for the presence of
SC-like limits.  All stars that become giants appear to reach a limit
shortly before the end of their core-burning phases.  When the
transition to shell-burning is continuous, the limit is associated
with expansion of the envelope.  That is, the star begins to become a
giant.  Whether the star continues to expand depends on whether the
rest of the envelope is transformed into core material and how the
burning-shell responds to the envelope's expansion.  Stars that are
convective during core-burning contract until shell-burning begins,
after which the envelope expands.  Some stars never appear to exceed a
SC-like limit and proceed directly to a white dwarf cooling sequence
without any period of significant expansion.

Though the test is approximate, it suggests that SC-like limits are
relevant immediately at the end of the main sequence in all stars and
are determined within the burning shell or molecular weight gradient.
Stars that become giants must have previously reached a SC-like limit.
As with all possible explanations of the red giant problem, it is
difficult to isolate cause and effect.  The presence of loops, as
identified by \citet{sugimoto00}, already requires that giants reach
SC-like limits at some point in their evolution but it is unclear
which phenomenon, if either, is causal.  Nevertheless, we have
demonstrated that the original SC limit is a particular case of a
broader phenomenon, that SC-like limits apply earlier in a star's
evolution than previously thought and that there is evidence for a
connection between exceeding these limits and evolving into a giant.
 % The red giant problem
\begin{savequote}[80mm]
  Now, this is not the end. It is not even the beginning of the
  end. But it is, perhaps, the end of the beginning.
  \qauthor{Sir Winston Churchill, 1942}
\end{savequote}

\chapter{Conclusions}
\label{ccon}

In this final chapter, I summarize the work presented in this
dissertation, give it contemporary context and propose directions for
future work.  The initial aim of this dissertation was to examine
models of quasi-stars with accurate microphysics.  These models were
presented in Chapters \ref{cqs1} and \ref{cqs2} and my findings are
discussed in Section \ref{send1}.  In Chapter \ref{cscl}, I presented a
detailed analysis of the fractional mass limit found in Chapter
\ref{cqs1} and I demonstrated that it is of the same nature as the
Sch\"onberg--Chandrasekhar limit.  In Chapter \ref{crg}, I assessed
whether realistic stellar models are subject to the limit and found
evidence that the limits are connected to the evolution of starts into
giants.  I discuss these results in Section \ref{send2}.

\section{The formation of supermassive black holes}
\label{send1}

In Chapters \ref{cqs1} and \ref{cqs2}, I presented models of
quasi-stars constructed with the Cambridge \stars{} code.  The two
sets of models made use of distinct inner boundary conditions to
describe conditions at the base of the giant-like envelope.  In
Chapter \ref{cqs1}, I followed the work of \citet*{begelman08} and
placed the base of the envelope at the Bondi radius.  In Chapter
\ref{cqs2}, I instead drew on the work of \citet{abramowicz+02} and
\citet*{lu04} and placed the inner boundary at a theoretical
transition from a convection-dominated accretion flow to an
advection-dominated flow.  I refer to the models of Chapters
\ref{cqs1} and \ref{cqs2} as \emph{Bondi-type} and \emph{CDAF-ADAF}
quasi-stars, respectively.

The evolution of the Bondi-type and CDAF-ADAF quasi-stars is
qualitatively and quantitatively very different.  Black holes (BHs)
embedded in Bondi-type quasi-stars accrete 
% at a rate on the order of $10^{-8}\,M_*\yr^{-1}$ 
about $10^{-8}$ of the total mass of the quasi-star each year and
respect a robust upper fractional mass limit of about $0.1$, depending
on the choice of parameters.  In all cases, the quasi-star phase ends
after a few million years.  The fate of the remaining material is
unclear.  A further $40$ per cent or so of the quasi-star's mass is
within the Bondi radius.  If the BH accretes this material, it
probably does so on the order of its own Eddington-limited rate.  If
the material is in some way dispersed, it could return to the BH and
be accreted later.  Whatever the details, quasi-stars are certainly a
possible mechanism by which early BHs could grow quickly enough to
become the supermassive objects that power high-redshift quasars.

The BHs in CDAF-ADAF quasi-stars accrete roughly an order of magnitude
faster than their Bondi-type cousins.  This is because a substantial
fraction of the BH luminosity is lost to the bulk flow of material
towards the BH.  More importantly, the BHs are not subject to any
upper mass limit and ultimately accrete all of the available gas under
all circumstances.  They do so through a series of peaks in the
accretion rate that appear to be related to the interaction of
microphysical processes, such as ionization, with the inward advection
of energy.  The nature of these peaks depends on the choices of
parameters in the boundary conditions but they do not depend closely
on the rotation of the quasi-star envelope.

Despite the differences between Bondi-type and CDAF-ADAF models, both
predict that quasi-stars leave BHs with masses on the order of a tenth
of the total quasi-star mass and that the quasi-star phase lasts just
a few million years.  In addition, all models have luminosities
roughly equal to the Eddington luminosity of the whole quasi-star,
surface radii on the order of $100\AU$ and effective temperatures that
are consistently around $4500\K$.  It thus appears that quasi-stars
have similar outward properties regardless of their inner structure or
even the details of their evolution.  However, although their
observable properties are not sensitive to the choice of inner
boundary conditions, quasi-stars like those modelled in this
dissertation are short-lived and would be difficult to find.

The most important conclusion that I draw from the work on quasi-stars
summarized above is that, when regarded as spherical, hydrostatic
objects, their detailed structure hinges on the conditions imposed at
the inner boundary.  The boundary conditions presented in Chapters
\ref{cqs1} and \ref{cqs2} are reasonable but there are several ways in
which modelling the inner region could be improved.  First,
incorporating the bulk velocity of the envelope material as a
structural variable, as is the pressure or luminosity, should allow
models to be followed close to the inner sonic point of the accretion
flow.  \citet{markovic95} modelled a small black hole at the centre of
the Sun with a similar technique but he solved the structure equations
with a shooting method and did not integrate all the way to the Bondi
radius.

Another possibility is modelling a quasi-star with three-dimensional
fluid simulations.  In general, solving the multi-dimensional problem
is much more difficult and more computationally demanding but I
believe it is a realistic possibility in the near future.
\citet*{barai11} used the smoothed particle hydrodynamics code
\textsc{gadget-3} to model Bondi accretion on to a supermassive BH
with both the sonic radius and Bondi radius resolved in the
simulation.  The main difference between their model and a quasi-star
envelope is that the mass of gas in the accretion flow was much
smaller that the mass of the BH.  If this situation can be reversed,
it should be possible to simulate the dynamical structure of a
quasi-star for a brief moment in its evolution.  Details derived from
such a model could be used to corroborate or improve boundary
conditions for spherically symmetric models.

The recent discovery of a bright quasar at redshift $z\approx7$
\citep{mortlock+11} and new simulations that suggest the first stars
were less massive than previously thought
%\citep{clark+11,hosokawa+11,stacy12} 
\citetext{\citealp{clark+11}; \citealp{hosokawa+11}; \citealp*{stacy12}} 
both aggravate the problem of luminous quasars at high redshift.  The
solution must now explain how BHs grew even more rapidly from even
smaller seeds.  But the solution needs only a small number of such
objects.  The comoving space density of bright high-redshift quasars
is only a few per $\Gpc^3$ and simulations of sufficiently large
volumes indicate that, given sufficiently large BH seeds, there is
enough material to feed a growing BH \citep{dimatteo+12}.

The direct collapse of massive protogalactic clouds remains an
appealing option for a solution to the problem provided that
fragmentation can be prevented.  \citet{dijkstra+08} estimate that a
fraction between $10^{-8}$ and $10^{-6}$ of sufficiently large dark
matter halos are illuminated by stars and stellar mass BHs in nearby
protogalaxies.  Molecular hydrogen formation can be suppressed and
fragmentation thus prevented.  The fate of the collapsing gas then
depends on the accretion rate of the central object.  If it is slow
enough for accreted material to relax into thermal equilibrium, a
supermassive star might form and collapse directly into a massive BH.
If the accretion is so rapid that only the central stellar core is
relaxed, a quasi-star can form.  In either case, a BH on the order of
$0.1$ of the total mass is left.

This is not to say that other mechanisms for massive BH formation do
\emph{not} occur.  Many protogalactic clouds can fragment into stars
and, even if the initial mass function of metal-free stars is less
top-heavy than previously thought, the most massive of these stars can
leave BH remnants.  As the hierarchical growth of large-scale
structure continues, these BHs grow and migrate towards the centres of
their host galaxies via dynamical friction.  There might exist a
currently undetected population of smaller BHs that power less
luminous quasars.  For the problem of the bright high-redshift objects
that are already known, however, I believe the direct collapse
mechanism is currently the strongest candidate solution.

\section{The red giant problem}
\label{send2}

The quasi-star models presented in Chapter \ref{cqs1} exhibit a robust
upper limit to the fractional mass of the inner BH.  In Chapter
\ref{cscl}, I showed how this limit is, in essence, the same as the
Sch\"onberg--Chandrasekhar limit.  I further demonstrated that both
limits are particular examples of the same general phenomenon.  I
refer to all such limits as \emph{SC-like} limits.  By considering the
contours of constant fractional mass of polytropic envelopes, it is
straightforward to determine whether a given core solution corresponds
to a SC-like limit, even when the core solution itself is not
polytropic.  An additional product of my analysis is that, given a
composite polytrope, one can test whether the model is at a SC-like
limit.  In Chapter \ref{crg}, I applied this test, albeit
approximately, to realistic stellar models and found evidence that
there is a connection between a star reaching a SC-like limit and
evolving into a giant.  In a sense, Chapter \ref{cscl} contains a
strong theoretical result and Chapter \ref{crg} a weaker practical
one.

The difficulty in resolving the red giant problem lies in
distinguishing cause and effect.  Any star that becomes a red giant
must have previously reached a SC-like limit because loops are always
present in the $U$--$V$ profiles of giants.  However, this sheds no
light on whether a star expands into a giant \emph{because} of the
SC-like limit.  Furthermore, the test that is applied in Chapter
\ref{crg} suffers from two drawbacks. First, the effective polytropic
index in the envelope of a realistic stellar model varies.  Secondly,
the models appear to reach the limit before a clear core-envelope
boundary has developed.  Despite these flaws, my analysis does
indicate a connection between SC-like limits and gianthood.  Stars
that expand appear to do so when they reach SC-like limits.  Whether a
star continues to expand and evolve into a giant depends on the
detailed response of envelope to its own expansion.  For example, if
the density in the burning shell decreases too much, nuclear reactions
cease and a dwarf-like structure is restored.  The SC-like limit might
therefore address the question of why a star begins to become a giant
but it offers no clear answers to the question of why giants become so
extended while the contraction of the core remains moderate.

Some of these shortcomings could be addressed with further work.  The
test could be applied to realistic models for a range of indices
simultaneously.  I doubt, however, that this would be enlightening
given that the behaviour of the envelope solutions is qualitatively
similar across all the polytropic indices encountered in realistic
giant envelopes.  It may be more useful to investigate the behaviour
of the contours when the effective polytropic index is allowed to vary
continuously.  These envelope solutions would be a better
approximation to realistic envelopes and would more accurately
identify the point at which a star reaches a SC-like limit.

Confronted with a long-standing problem that has already drawn and
still occasionally draws detailed study, it is difficult to imagine a
straightforward solution.  I believe that progress is only achieved by
a meticulous study of stars that do and do not become giants.  Such a
survey should include both realistic models and contrived models where
specific physical effects or evolutionary processes are numerically
isolated or removed.  If a new hypothesis is formed, it should be
thoroughly tested in the cases where it makes predictions, even if the
predictions are only relevant in contrived or unphysical models.  The
red giant problem will not be solved easily but the work presented in
this dissertation at least provides new clues to the answer.

\clearpage

% I presently believe that there is growing evidence that the evolution
% of a star into a red giant is caused by a region where the effective
% polytropic index greatly exceeds $5$.  The work presented in Chapter
% \ref{crg}, by \citet{eggleton00} and by \citet{sugimoto00} all support
% this hypothesis even if they all differ on the cause of a star's
% expansion.  But the evidence drawn from the \uvp{} does not 
% It's all about the macroscopic fluid equations responding to the
% microscopic material properties.
 % Conclusions

\singlespacing

\pagestyle{tocstyle}
%\backmatter
%\bibliographystyle{plainnat}
{
\footnotesize
\bibliographystyle{thesis}
\bibliography{thesis}}
\clearpage

\onehalfspacing
\appendix
\pagestyle{mainstyle}
\begin{savequote}[60mm]
  ... by means of that gruesome tool: the \uvp{}!
  \qauthor{Martin Schwarzschild, 1965}
\end{savequote}

\chapter{Polytropes and the \uvp{}}
\label{auvp}

The analysis in Chapters \ref{cscl} and \ref{crg} makes extensive use
of solutions of the Lane--Emden equation in the plane of
homology-invariant variables $U$ and $V$.  In this appendix, 
I provide the background material for that analysis.
% This work rests on the behaviour of solutions in the \uvp{} so we
% begin with a review of its features.  
I derive the Lane--Emden equation (LEE) from hydrostatic equilibrium
and mass conservation, introduce homology invariant variables $U$ and
$V$ and explain their physical meaning, present the homology invariant
transformation of the equation and study %the behaviour of 
its solutions in the \uvp{}.
%through critical points
I hope that, by presenting concisely the details of the \uvp{} in a
context where it is usefully applied, I might partly remove its stigma
as `that gruesome tool'.\footnote{\citet{faulkner05} explains that
  Martin Schwarzschild described the \uvp{} as such in a referee's
  report in 1965.  The same quote is presumably the citation by
  \citet{eggleton98} of `(Schwarzschild 1965, private
  communication)'.}
%\comment{Rewrite introduction.}

\section{The Lane--Emden equation}
\label{slee}

Consider a star that obeys everywhere a polytropic equation of state,
\shorteq{p=K\rho^{1+\frac{1}{n}}\text{,}}
where $p$ is the pressure, $\rho$ the density, $n$ the polytropic
index and $K$ a constant of proportionality.  A polytropic equation of
state approximates a fluid that is between the adiabatic and
isothermal limits.  Shallower temperature gradients correspond to
larger effective polytropic indices and the isothermal case (zero
temperature gradient) corresponds to $n\to\infty$.  In this case, the
equation of state must be transformed differently but the limit is
well-defined when working in the with homology invariant variables in
the \uvp{}.

Real stars are locally polytropic in the sense that an effective
polytropic index $n$ can be defined at each point by
\shorteq{\tdif{\log p}{\log\rho}=1+\frac{1}{n}\text{.}}  
When the star is globally polytropic, this definition reproduces the
polytropic index.  Certain conditions correspond to certain values of
$n$.  In convective zones, the temperature gradient is approximately
adiabatic, so an ideal gas without radiation has $n=3/2$ and pure
radiation has $n=3$.  Real stars are more radiation-dominated towards
the centre and $n$ varies between these limiting values in convection
zones.  In radiative zones, $n$ can also depend on the opacity or
energy generation rate.  It can be shown, for example, that for a
polytropic model with uniform energy generation and a Kramer's opacity
law, $n$ ranges from $13/4$ for a pure ideal gas to $7$ for pure
radiation \citep[][p. 556]{horedt04}.  Nuclear burning shells and
ionisation regions have shallow temperature gradients and therefore
large values of $n$.  Thus, the effective polytropic index can vary
widely within a star.

Consider the equations of mass conservation,
\shorteq{\label{dmdr}\frac{dm}{dr}=4\pi r^2\rho\text{,}} and
hydrostatic equilibrium,
\shorteq{\label{dpdr}\frac{dp}{dr}=-\frac{Gm\rho}{r^2}\text{,}} 
where $r$ is the distance from the centre of the star and $m$ is the mass
within a concentric sphere of radius $r$.  One can define the dimensionless
temperature\footnote{This is by the analogy to an ideal gas, for which
  $T\propto p/\rho$.}  $\theta$ by $\rho=\rho_0\theta^n$, where
$\rho_0$ is the density at the centre of the star.  By re-arranging and 
differentiating the equation of hydrostatic equilibrium, we obtain
\shorteq{\tdif{}{r}\left(\frac{r^2}{\rho}\tdif{p}{r}\right)=-G\tdif{m}{r}\text{.}}
By using mass conservation to replace $\mathrm{d}m/\mathrm{d}r$ and
dividing both sides by $r^2$ we obtain
\shorteq{\frac{1}{r^2}\tdif{}{r}\left(\frac{r^2}{\rho}\tdif{p}{r}\right)
=-4\pi G\rho\text{.}}
Substituting the pressure via the polytropic equation of state,
replacing the density with the dimensionless temperature and defining
the dimensionless radius\footnote{The scale factor is usually denoted
  by $\alpha$.  Here, the variable $\eta$ is used to avoid confusion
  with the density jump at the core-envelope boundary, which
  \citet{eggleton98} called $\alpha$.} $\xi=r/\eta$, we write
\shorteq{\frac{\eta^2}{\xi^2}\tdif{}{\xi}
  \left(\xi^2K\rho_0^\frac{1}{n}(n+1)\tdif{\theta}{\xi}\right)
  =-4\pi G\rho_0\theta^n\text{,}}
which we render dimensionless by choosing
\shorteq{\eta^2=\frac{(n+1)K}{4\pi G}\rho_0^{\frac{1}{n}-1}\text{.}}
The dimensionless differential equation is the LEE,
\shorteq{\label{LEE}\frac{1}{\xi^2}\frac{d}{d\xi}
  \left(\xi^2\frac{d\theta}{d\xi}\right)=-\theta^n\text{.}} 

\begin{figure}\begin{center}
\includegraphics{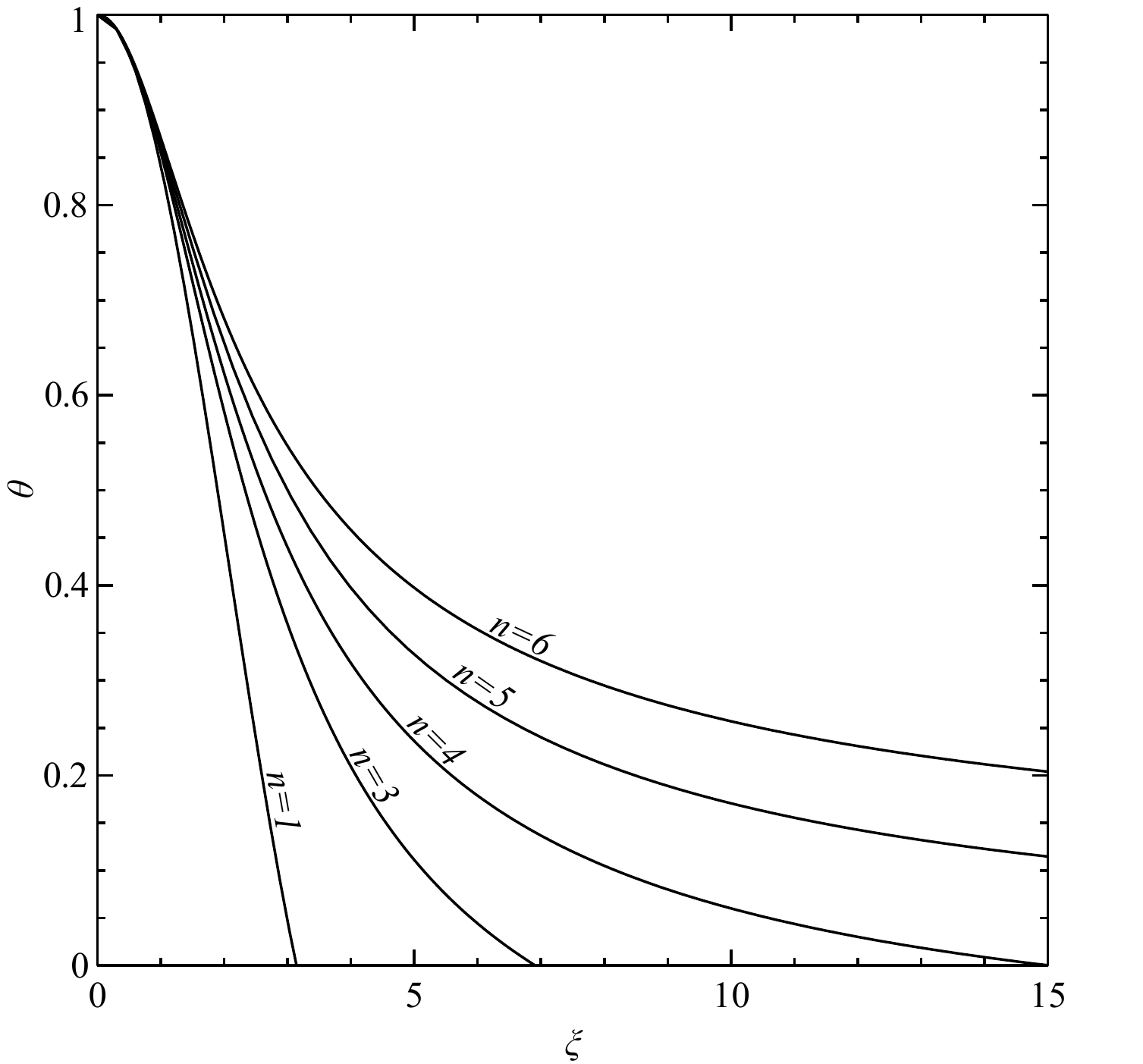}
\caption[Dimensionless temperature $\theta$ versus dimensionless
radius $\xi$ for polytropes with $n=1$, $3$, $4$, $5$ and $6$.] {Plot
  of dimensionless temperature $\theta$ versus dimensionless radius
  $\xi$ for polytropes with $n=1$, $3$, $4$, $5$ and $6$.  Note that
  the surface radius increases with $n$.  For $n\ge5$, there is no
  surface and the polytrope extends to infinite $\xi$.}
\label{fleet}
\end{center}\end{figure}

\begin{figure}\begin{center}
\includegraphics{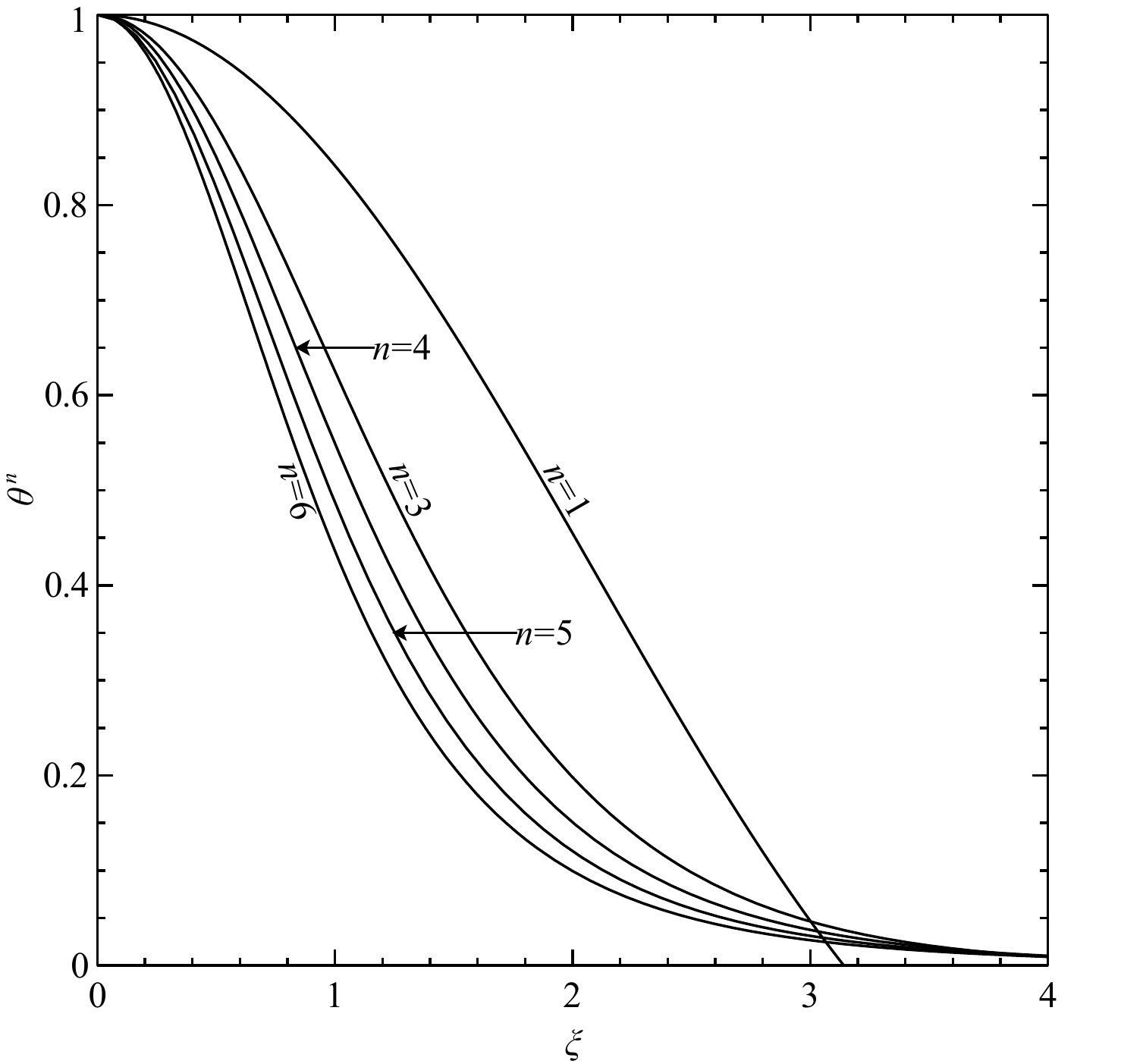}
\caption[Normalized density $\theta^n$ versus dimensionless radius
$\xi$ for polytropes with $n=1$, $3$, $4$, $5$ and $6$.] {Plot of
  normalized density $\theta^n$ versus dimensionless radius $\xi$ for
  polytropes with $n=1$, $3$, $4$, $5$ and $6$, from right to left.
  Note that as $n$ increases, so the polytrope is more centrally
  condensed.}
\label{fleerho}
\end{center}\end{figure}

Let us define the dimensionless mass $\phi=m/4\pi\eta^3\rho_0$.
Introducing the dimensionless mass, temperature and radius directly
into equations (\ref{dmdr}) and (\ref{dpdr}) allows us to write
\shorteq{\label{dphdxi}\frac{d\phi}{d\xi}=\xi^2\theta^n} and
\shorteq{\label{dthdxi}\frac{d\theta}{d\xi}=-\frac{1}{\xi^2}\phi\text{.}}
The expression of the LEE as two first-order equations (\ref{dphdxi}
and \ref{dthdxi}) preserves the physical meaning of the equations and
easily permits arbitrary boundary conditions for the inner mass and
radius.  

Solutions of the LEE that are regular at the centre have
$\xi_0=\phi_0=0$.  The subset of solutions that extend from the centre
to infinite radius or the first zero of $\theta$ are \emph{polytropes}
of index $n$.  I refer to solutions that are regular at the centre but
truncated at some finite radius as \emph{polytropic cores}.
Conversely, solutions that extend from a finite radius to infinity or
the first zero of $\theta$ are \emph{polytropic envelopes}.  Models in
which polytropic cores are matched to polytropic envelopes are
referred to as \emph{composite} polytropes.
% Realistic polytropes take $n$ in the range $3/2$ to infinity.  Some of
% the results we reproduce are calculated for the analytic solutions
% when $n=1$, so we restrict our discussion to cases with $1\leq n<\infty$.  
For $n<5$ polytropes are finite in both mass and radius
while for $n>5$ they are infinite in mass and radius.  The case $n=5$
represents the threshold between the two: it has a finite mass but
infinite radius. \\
% An immediate conclusion is that realistic stellar
% models composed of polytropic solutions are either polytropes with
% $n<5$ or have polytropic envelopes with $n<5$.

\section{Homology and homology-invariant variables}
\label{shom}

If, given a solution to a differential equation, further solutions can
be found by rescaling the given solution, the differential
equation is said to admit an \emph{homology transformation}
\citep[][p. 102]{chandra39}.  Two solutions related in this way are
\emph{homologous}; the similarity between them is %called
\emph{homology}.  The LEE admits an homology and the appropriate
scaling relation is now derived.

Consider the usual second-order form of the LEE (equation \ref{LEE})
with a solution $\theta(\xi)$.  If there is a further solution of the
form $\theta'(\xi')=C^k\theta(C\xi)$, it must satisfy the equation
\shorteq{\frac{1}{\xi^2}\frac{d}{d\xi}\left(\xi^2\frac{d\theta'(\xi')}{d\xi}\right)
  =-\theta'^n(\xi')\text{.}}
The left-hand side can be manipulated as follows.

\begin{align}
 &\frac{1}{\xi^2}\tdif{}{\xi}\left(\xi^2\tdif{\theta'(\xi')}{\xi}\right) \\
=&\frac{1}{\xi^2}\tdif{}{\xi}\left(\xi^2\tdif{}{\xi}C^k\theta(C\xi)\right) \\
=&C^{2+k}\frac{1}{(C\xi)^2}\tdif{}{(C\xi)}\left((C\xi)^2\tdif{}{(C\xi)}C^k\theta(C\xi)\right) \\
=&-C^{2+k}\theta^n(C\xi)
\end{align}
If the last line is to be the same as
$-\theta'^n(\xi')=-(C^k\theta(C\xi))^n$, we must have $2+k=kn$
and hence $k=2/(n-1)$.  The homology transformation of the LEE for
$\theta$ is therefore $\theta(\xi)\to C^{2/(n-1)}\theta(C\xi)$.

To find the transformation for $\phi$, we seek $\phi'(\xi')=C^h\phi(C\xi)$
that satisfies
\shorteq{\frac{d\phi'(\xi')}{d\xi}=\xi^2\theta'^n(\xi')}
and
\shorteq{\frac{d\theta'(\xi')}{d\xi}=-\frac{1}{\xi^2}\phi'(\xi')\text{.}}
The left-hand sides of the two equations can be considered separately
and simultaneously.
% \begin{align}
%  &\tdif{}{\xi}\theta'(\xi') \\
% =&\tdif{}{\xi}C^\frac{2}{n-1}\theta(C\xi) \\
% =&C^{\frac{2}{n-1}+1}\tdif{}{(C\xi)}\theta(C\xi) \\
% =&C^{\frac{2}{n-1}+1}\left(-\frac{\phi(C\xi)}{(C\xi)^2}\right) \\
% =&C^{\frac{2}{n-1}-1-h}\left(\frac{-C^h\phi(C\xi)}{\xi^2}\right) \\
% =&-C^{\frac{3-n}{n-1}-h}\frac{\phi'(\xi')}{\xi^2}
% \end{align}
% %If last line equals $-C^h\phi(C\xi)/\xi^2$, $2/(n-1)-1-h=0\Rightarrow h=(3-n)/(n-1)$.
% \begin{align}
%  &\tdif{}{\xi}\phi'(\xi') \\
% =&\tdif{}{\xi}C^h\phi(C\xi) \\
% =&C^{h+1}\tdif{}{(C\xi)}\phi(C\xi) \\
% =&C^{h+1}(C\xi)^2\theta^n(C\xi) \\
% =&C^{h+3-2n/(n-1)}\xi^2C^{2n/(n-1)}\theta^n(C\xi) \\
% =&C^{h-\frac{3-n}{n-1}}\xi^2\theta'(\xi')
% \end{align}
% %If last line equals $C^{2n/(n-1)}\theta^n(C\xi)$, then $h=(3-n)/(n-1)$.
\begin{align}
 &\tdif{}{\xi}\phi'(\xi')                    & &\tdif{}{\xi}\theta'(\xi') \\
=&\tdif{}{\xi}C^h\phi(C\xi)                  &=&\tdif{}{\xi}C^\frac{2}{n-1}\theta(C\xi) \\
=&C^{h+1}\tdif{}{(C\xi)}\phi(C\xi)            &=&C^{\frac{2}{n-1}+1}\tdif{}{(C\xi)}\theta(C\xi) \\
=&C^{h+1}(C\xi)^2\theta^n(C\xi)               &=&C^{\frac{2}{n-1}+1}\left(-\frac{\phi(C\xi)}{(C\xi)^2}\right) \\
=&C^{h+3-2n/(n-1)}\xi^2C^{2n/(n-1)}\theta^n(C\xi)&=&C^{\frac{2}{n-1}-1-h}\left(\frac{-C^h\phi(C\xi)}{\xi^2}\right) \\
%=&C^{h-\frac{3-n}{n-1}}\xi^2\theta'(\xi')    &=&-C^{\frac{3-n}{n-1}-h}\frac{\phi'(\xi')}{\xi^2}
=&C^{h-\frac{3-n}{n-1}}\xi^2\theta'(\xi')       &=&C^{\frac{3-n}{n-1}-h}\left(\frac{-\phi'(\xi')}{\xi^2}\right)
\end{align}
In both cases, the remaining exponent of $C$ must be zero and thus
$h=(3-n)/(n-1)$.  That is, the homology transformation for $\phi$ is
$\phi(\xi)\to C^{(3-n)/(n-1)}\phi(C\xi)$.

By choosing variables that are invariant under the homology
transformation, the LEE can be formulated as a single first-order
equation that captures all essential behaviour.  The variables used
here are
\shorteq{\label{Ugen}U=\frac{d\log m}{d\log r}=\frac{3\rho}{\bar\rho}}
and 
\shorteq{\label{Vgen}V=-\frac{d\log p}{d\log r}
  =\frac{Gm}{r}\frac{\rho}{p}\text{,}}
where $\bar\rho=3m/4\pi r^3$ is the mean density of the material
inside $r$.  Although $U$ and $V$ have been defined to reduce the
order of the LEE, the corresponding physical definitions make them
meaningful for discussions of any stellar model.  A brief excursion is
made to explain these definitions.

\begin{figure}\begin{center}
\includegraphics{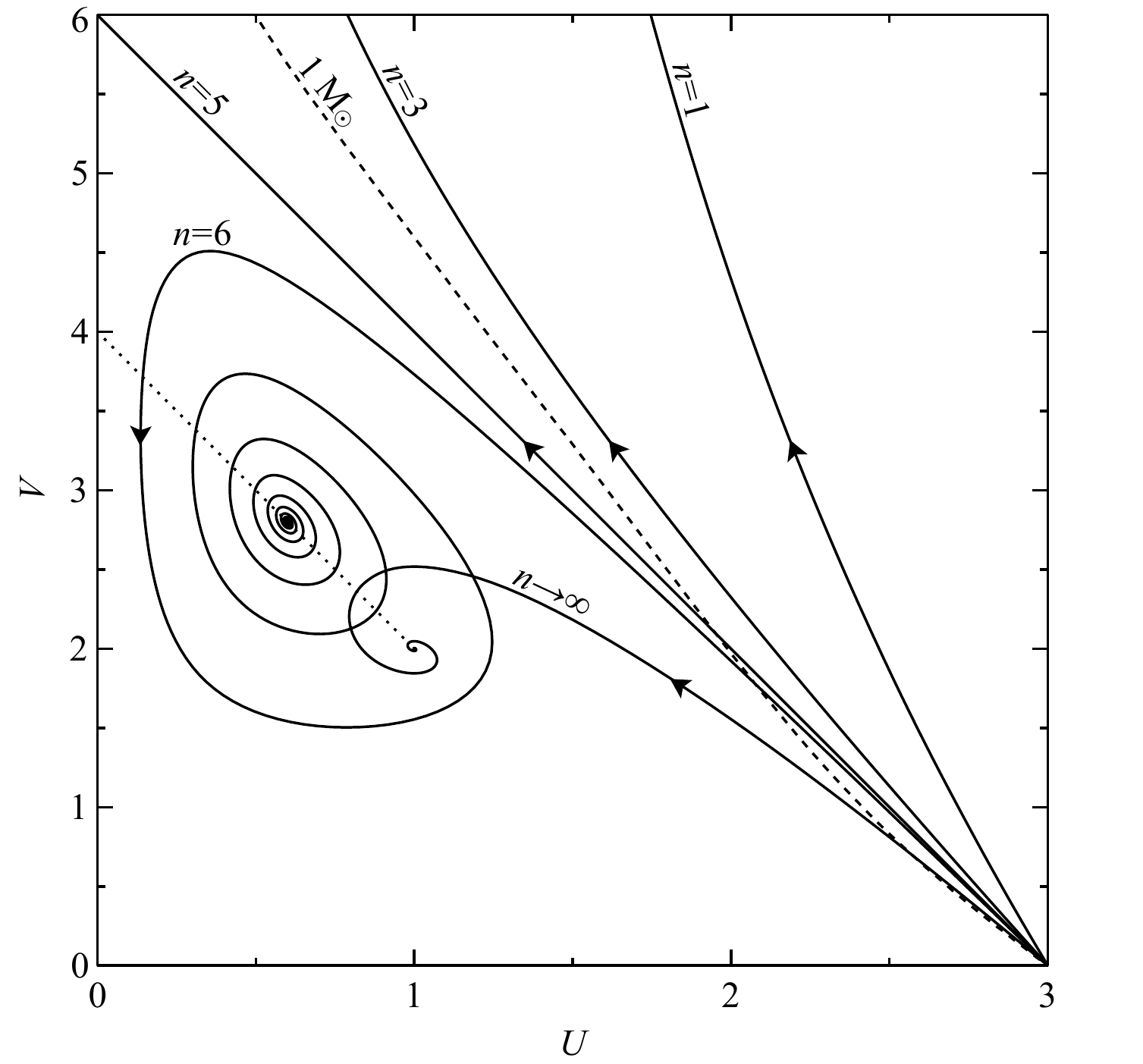}
\caption[General features of solutions in the \uvp{}.]  {Some general
  features of the \uvp{}.  The solid lines are, from top right,
  polytropes of index $1$, $3$, $5$, $6$ and $\infty$.  The arrows
  point in the direction of increasing $\xi$.  The dashed line is a
  \stars{} model of a $1\Msun$ star of solar metallicity when its
  radius is $1.012\Rsun$ and its luminosity $0.974\Lsun$.  The dotted
  line shows the locus of the critical points $G_s$.  The locus begins
  in the plane at $(0,4)$ when $n=3$ and tends to (1, 2) as n
  increases to $\infty$.  A polytrope with $n=4$ is not plotted to
  avoid cluttering the Sun-like model.}
\label{fhlee}
\end{center}\end{figure}

The physical variables are all positive so only the first quadrant of
the \uvp{} is of interest.  The variable $U$ is three times the ratio
of the local density to the mean density inside that radius.  As
$r\to0$, so also $\rho\to\bar\rho$ and thus $U\to3$.  We expect the
density of a stellar model to decrease with radius, so
$\rho/\bar\rho<1$ and hence $U<3$.  The variable $V$ is related to the
ratio of specific gravitational binding energy to specific internal
energy.  At the centre, $p$ and $\rho$ are finite and $m\sim
\frac{4\pi}{3}\rho r^3$, so $V\to0$.  Thus, in all physical solutions
that extend to $r=0$, the centre corresponds to $(U,V)=(3,0)$.  If an
interior solution has $V$ everywhere smaller than the appropriate
polytrope, it behaves as if it has a finite point mass at its centre.
\citet{huntley75} referred to similar models, integrated outwards from
a finite radius, as \emph{loaded polytropes}.  If a solution has $V$
everywhere greater than the polytrope, it reaches zero mass before
zero radius.  Such solutions have a massless core with finite radius,
which is unphysical.  At the surface, $\rho\to0$ so $U\to0$ too.  On
the other hand, $Gm/r$ takes a finite value but $p/\rho\propto T\to0$
so $V\to\infty$.  All realistic models, be they polytropes, 
% composites of a polytropic core and envelope, 
composite polytropes
or output from a detailed
calculation, must adhere to these central and surface conditions in
the \uvp{}.  They therefore extend from $(3,0)$ towards $(0,\infty)$.
% Finite stellar
% models, be they polytropes, composites of a polytropic core and
% envelope or output from a detailed calculation, extend from the point
% $(3,0)$ along the $U$-axis towards the $V$-axis and infinite
% $V$.  
Fig.~\ref{fhlee} shows this behaviour for polytropes of indices $1$,
$3$, $5$, $6$ and $\infty$.  The $n=1$ and $n=3$ models extend
properly to the surface.  The $n=5$, $6$ and $\infty$ polytropes do
not and therefore cannot represent real stars.  In addition, we have
plotted a $1\Msun$ model produced by the Cambridge \stars{} code to
show that it also satisfies the boundary conditions described above.
Note that this model has not been calibrated to fit the Sun precisely.

For a composite polytrope,
% where a polytropic core is joined to an envelope, 
the pressure, mass and radius are continuous at the join.  If the
density decreases by a factor $\alpha$ \citep[c.f.][]{eggleton98},
$U$ and $V$ decrease by the same factor.  In other words, if
$\rho\to\alpha^{-1}\rho$ then $(U,V)\to\alpha^{-1}(U,V)$.  The
corresponding point on the \uvp{} is contracted towards the origin by
the factor $\alpha$.  Such a jump occurs %, for example, 
if there is a
discontinuity in the mean molecular weight $\mu$ between the core and
the envelope.  In this case, $\alpha=\mu_c/\mu_e$.

Let us now return to the polytropic solutions for which we defined $U$
and $V$ in the first place.  From the definitions above,
\shorteq{\label{Upol}U=\tdif{\log\phi}{\log\xi}=\frac{\xi^3\theta^n}{\phi}}
and
\shorteq{\label{Vpol}V=-(n+1)\tdif{\log\theta}{\log\xi}
  =(n+1)\frac{\phi}{\xi\theta}\text{.}}  
%The homology invariance of the variables is straightforwardly verified.
To demonstrate that $U$ and $V$ are homology invariant, note that the
homology transformations $\theta'(\xi')=C^{2/(n-1)}\theta(C\xi)$ and
$\phi'(\xi')=C^{(3-n)/(n-1)}\theta(C\xi)$ are
equivalent to $\theta'(\xi/C)=C^{2/(n-1)}\theta(\xi)$ and 
$\phi'(\xi/C)=C^{(3-n)/(n-1)}\theta(\xi)$.  Making the homology transformation
with $\xi\to\xi/C$,
\begin{align}
% U'&=\frac{\xi^3\theta'^n(\xi')}{\phi'(\xi')}                  
%                          &V'&=(n+1)\frac{\phi'(\xi')}{\xi\theta'(\xi')} \\
%   &=C^{\frac{2n}{n-1}-\frac{3-n}{n-1}}\frac{\xi^3\theta^n(\xi')}{\phi(\xi')} 
%                          &  &=(n+1)C^{\frac{3-n}{n-1}-\frac{2}{n-1}}\frac{\phi(\xi')}{\xi\theta(\xi')} \\
%   &=\frac{(C\xi)^3\theta^n(\xi')}{\phi(\xi')}
%                          &  &=(n+1)\frac{\phi(\xi')}{(C\xi)\theta(\xi')}   \\
%   &=\frac{\xi'\theta^n(\xi')}{\phi(\xi')}=U
%                          &  &=(n+1)\frac{\phi(\xi')}{\xi'\theta(\xi')}=V\text{.}
U'&=\frac{\xi^3}{C^3}\frac{\theta'^n(\xi/C)}{\phi'(\xi/C)}                  
                         &V'&=(n+1)\frac{C}{\xi}\frac{\phi'(\xi/C)}{\theta'(\xi/C)} \\
  &=C^{\frac{2n}{n-1}-\frac{3-n}{n-1}-3}\frac{\xi^3\theta^n(\xi)}{\phi(\xi)} 
                         &  &=(n+1)C^{\frac{3-n}{n-1}-\frac{2}{n-1}+1}\frac{\phi(\xi)}{\xi\theta(\xi)} \\
  &=\frac{\xi^3\theta^n(\xi)}{\phi(\xi)}=U
                         &  &=(n+1)\frac{\phi(\xi)}{\xi\theta(\xi)}=V\text{,}
\end{align}
which proves that $U$ and $V$ are homology invariant.

Let us differentiate $\log U$ and $\log V$ as they are defined for
polytropes.  This gives
\shorteq{\label{dU}\frac{1}{U}\tdif{U}{\xi}=\frac{1}{\xi}(3-n(n+1)^{-1}V-U)}
and
\shorteq{\label{dV}\frac{1}{V}\tdif{V}{\xi}=\frac{1}{\xi}(-1+U+(n+1)^{-1}V)\text{.}}
The ratio of these two equations yields the first-order equation
\shorteq{\label{HLEE}\tdif{V}{U}=-\frac{V}{U}
  \left(\frac{U+(n+1)^{-1}V-1}{U+n(n+1)^{-1}V-3}\right)} 
in which the dependence on $\xi$ has been eliminated.  We refer to
equation (\ref{HLEE}) as the homologous Lane--Emden equation (HLEE).
The SC-like limits we wish to reproduce are shared by polytropic
models so we now explore the behaviour of these solutions in the plane
defined by $U$ and $V$.

\section{Topology of the homologous Lane--Emden equation}
\label{stop}

The behaviour of solutions of the HLEE may be described in terms of
its critical points, where $\udif{U}{\log\xi}$ and $\udif{V}{\log\xi}$
both tend to zero.  \citet{horedt87} conducted a thorough survey of
the behaviour of the HLEE, including the full range of $n$ from
$-\infty$ to $\infty$ in linear, cylindrical and spherical
geometries.\footnote{Readers should note that the definition of $V$
  used by \citet{horedt87} differs by a factor $n+1$.}  Below, we use
his convention for naming the critical points but consider only
spherical cases with $n\geq1$.  Though realistic polytropes take $n$
in the range $3/2$ to infinity, we extend it to accommodate SC-like
limits discussed in the literature for polytropic envelopes with
$n=1$.

From the numerator of equation (\ref{HLEE}), we see that
$\mathrm{d}V/\mathrm{d}U=0$ when $V=0$ or $U+V/(n+1)=1$.  The former
indicates that solutions that approach the $U$-axis proceed along it
until they reach infinity or a critical point.  The latter defines a
straight line in the \uvp{} along which solutions are locally
horizontal.  Following \citet{faulkner05} we refer to this line as the
\emph{line of horizontals}.  Similarly, from the denominator, we find
$\mathrm{d}U/\mathrm{d}V=0$ when $U=0$ or $U+nV/(n+1)=3$.  Again, the
first locus implies that solutions near the $V$-axis have trajectories
that are nearly parallel to it, while the second gives another
straight line, this time along which solutions are vertical,
hereinafter referred to as the \emph{line of verticals}.  The critical
points of the HLEE are located at the intersections of these curves.
Below, we consider the stability of the critical points as $n$ varies.

From equations (\ref{dU}) and (\ref{dV}), we find
\shorteq{\tdif{U}{\log\xi}=-U(U+n(n+1)^{-1}V-3)} and
\shorteq{\tdif{V}{\log\xi}=V(U+(n+1)^{-1}V-1)\text{.}}  
This is an \emph{autonomous} system of equations: the derivatives
depend only on the dependent variables $U$ and $V$.  The linear
behaviour of such systems around the critical points can be
characterised by the eigenvectors and eigenvalues of the Jacobian
matrix
\begin{align}J&=\left(\begin{array}{cc}
\pdif{}{U}\tdif{U}{\log\xi} & \pdif{}{V}\tdif{U}{\log\xi} \\
\pdif{}{U}\tdif{V}{\log\xi} & \pdif{}{V}\tdif{V}{\log\xi}
\end{array}\right)\\
&=\left(\begin{array}{c@{}c}
3-2U-\frac{n}{n+1}V & -\frac{n}{n+1}U \\ \\
V                   & -1+U+\frac{2}{n+1}V
\end{array}\right)\text{,}\end{align}
at the critical point in question \cite[e.g.][p. 150]{strogatz94}.  In
particular, if the real component of an eigenvalue is positive or
negative, solutions tend away from or towards that point along the
corresponding eigenvector.  Such points are \emph{sources} or
\emph{sinks}.  When the point has one positive and one negative
eigenvalue it is a \emph{saddle}.  If the eigenvalues have imaginary
components then solutions orbit the point as they approach or recede.
We describe these as \emph{spiral} sources or sinks.  If the
eigenvalues are purely imaginary, solutions form closed loops
around that point, which we call a \emph{centre}.  The choice of
independent variable, in this case $\log\xi$, is not relevant in such
analysis.

% \begin{figure}\begin{center}
% \includegraphics{apA/ftopgen}
% \caption[Qualitative topology of the homologous Lane--Emden equation.]
% {A diagram depicting the qualitative behaviour of the critical points
%   of the homologous LEE for various $n$ between $1$ and $\infty$.  The
%   dotted line is the locus of $G_s$.  The arrows indicate the
%   directions, but not magnitudes, of the eigenvectors.  The behaviour
%   at $O_s$ is the same for all $n$.  At $U_s$, the behaviour is fairly
%   similar.  The location of $V_s$ varies with $n$ but one eigenvector
%   always points down along the $V$-axis.  The other eigenvector is
%   away from $V_s$ for $n<3$ and towards it for $n>3$.  For $n=3$, one
%   eigenvector is zero and $V_s$ co-incides with $G_s$, which makes its
%   first appearance in the \uvp{}.  As $n$ increases, $G_s$ is a
%   non-spiral source until $n=(11+8\sqrt{2})/14$.  The real component
%   decreases such that $G_s$ is a centre at $n=5$ and thereafter a
%   spiral sink.}
% \label{ftopgen}
% \end{center}\end{figure}

\begin{figure}\begin{center}
    \includegraphics{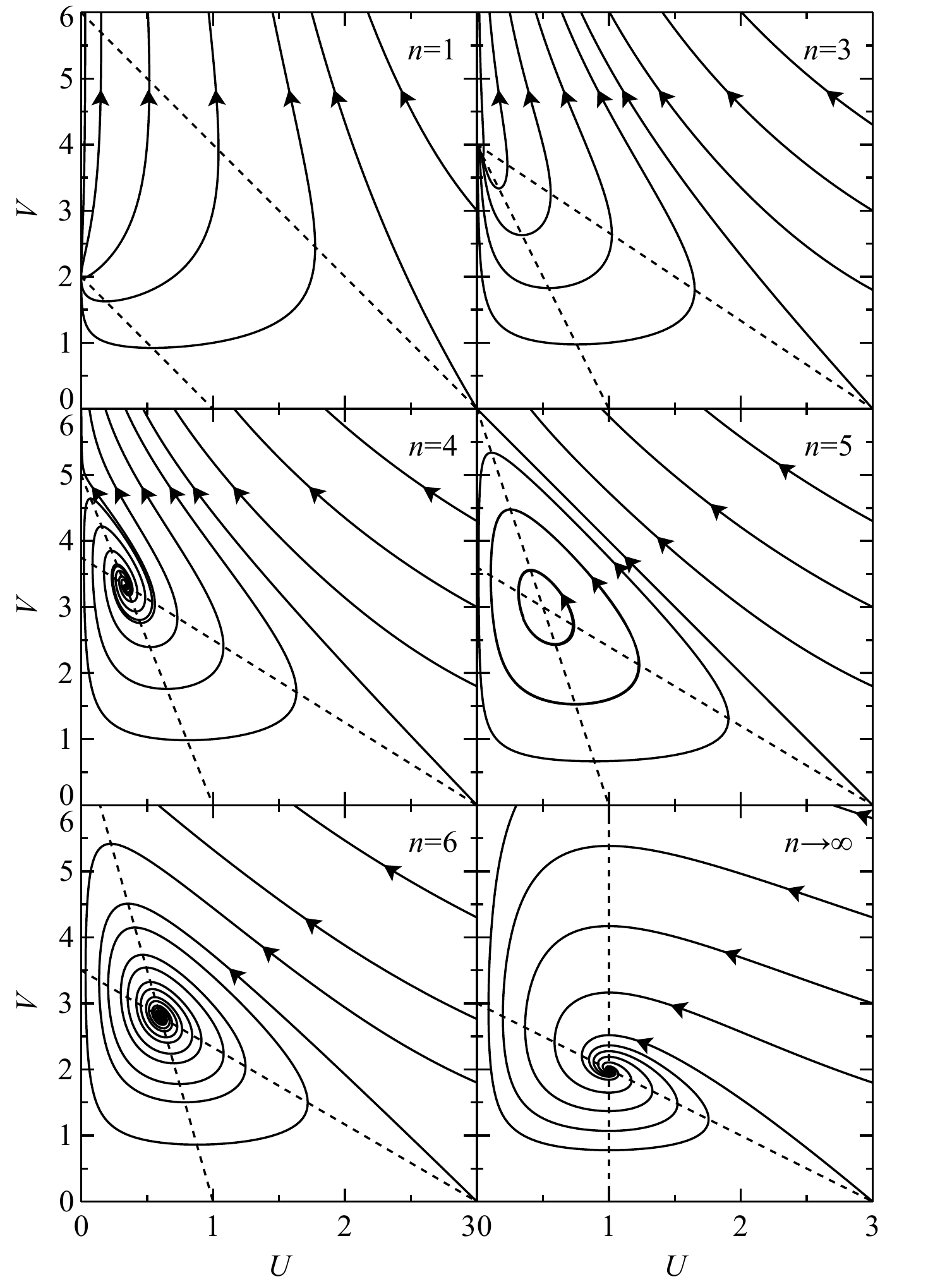}
    \caption[Topology of the homologous Lane--Emden equation for $n=1$,
    $3$, $4$, $5$, $6$ and $\infty$.]{Topology of the homologous
      Lane--Emden equation for $n=1$, $3$, $4$, $5$, $6$ and $\infty$.
      The value of $n$ is indicated in the top-right corner of each
      plot.  The solid lines are solutions of the equation with the
      arrows indicating the direction of increasing $\xi$.  Each
      solution has only one arrow.  The dashed lines are the lines of
      horizontals and verticals.  The critical point $G_s$ is at the
      intersection of these two lines.  For $n<3$, $G_s$ is at $U<0$
      so does not appear in the first quadrant.  When $n=3$ it
      co-incides with the critical point $V_s$ on the $V$-axis.  Once
      in view, $G_s$ is at first a spiral source (e.g.  $n=4$).  For
      $n=5$, it is a centre and the \uvp{} adopts a particular
      topology.  As $n$ continues to increase, $G_s$ becomes an
      increasingly strong spiral sink.  In the limit $n\to\infty$, all
      solutions spiral onto $G_s$.}
\label{ftopgrid}
\end{center}\end{figure}

\begin{table}\begin{center}
    \caption[Critical points of the homologous Lane--Emden
    equation.]{Critical points of the HLEE.
      $\Delta_n=\sqrt{1+n(22-7n)}$.}
\begin{tabular}{cccccc}
\toprule[1pt]
\mchead{Critical point} & \mchead{Eigenvalues} & \mchead{Eigenvectors}  \\
\midrule
$O_s$ & $(0,0)$         &   $3$     &  $-1$   & $(1,0)$ &   $(0,1)$    \\
$U_s$ & $(3,0)$         &   $-3$    &  $2$    & $(1,0)$ & $(-3n,5+5n)$ \\
$V_s$ & $(0,n+1)$       &   $1$     &  $3-n$  & $(0,1)$ & $(2-n,1+n)$  \\
$G_s$ & $\left(\frac{n-3}{n-1},2\frac{n+1}{n-1}\right)$ & 
 \mchead{$\frac{n-5\pm\Delta_n}{2-2n}$} & \mchead{$(1-n\mp\Delta_n,4+4n)$}  \\
\bottomrule[1pt]
\end{tabular}\end{center}
\label{thlee}
\end{table}

Table \ref{thlee} shows the eigenvalues and eigenvectors for the
critical points %in Section \ref{stop} 
as functions of $n$.  The
%qualitative
topologies of the HLEE in the \uvp{} for the cases $n=1$, $3$, $4$,
$5$, $6$ and $n\to\infty$ are plotted in Fig.~\ref{ftopgrid}.
% O_s=(0,0)
The origin is the first critical point.  It is a saddle with solutions
on the $V$-axis approaching and those on the $U$-axis escaping.
Solutions near the origin move down and to the right in the \uvp{}.
% U_s,V_s
There is a further critical point on each of the axes.  On the
$U$-axis, $U_s=(3,0)$ is a saddle for all values of $n$.  It is stable
along the $U$-axis and unstable across it.  This point coincides with
the regular centre of realistic stellar models that was discussed
previously.  Along the $V$-axis, $V_s=(0,n+1)$ is also a critical
point.  For $n<3$ it is a source and for $n>3$ a saddle.
% G_s
%The intersection of the lines of horizontals and verticals is the
%final critical point.  It is located at
The intersection of the lines of horizontals and verticals is the 
final critical point $G_s$.  For each $n$,
$G_s=(\frac{n-3}{n-1},2\frac{n+1}{n-1})$.  
% The character of this point varies with $n$.  Its locus is shown in
% Fig.~\ref{fgen}.
The character of these points varies with $n$.  
% Their locus is shown in Fig.~\ref{ftopgen}.

The behaviour of $G_s$ and $V_s$ distinguishes the topology of
solutions into three regimes.  For $n<3$, $V_s$ is a pure source: it
is unstable across and along the $V$-axis.  The point $G_s$ has $U<0$
and therefore does not feature in the first quadrant of the \uvp{} but
approaches the $V$-axis from the left as $n\to3$.  When $n=3$, $V_s$
and $G_s$ co-incide.  The point is marginally stable across the axis.
For $n>3$, $V_s$ and $G_s$ separate.  $V_s$ is a saddle and $G_s$ a
source, gradually moving towards its position at $(1/2,3)$ when $n=5$.
Fig.~\ref{ftopgrid} illustrates some features of the HLEE when $n=3$.
The lines of verticals and horizontals meet at $G_s$, which has just
appeared on the \uvp{} at $(0,4)$.
%Envelope solutions that are
%more concentrated than the polytrope originate from this point at the
%centre.
%Lines that are less concentrated than the polytrope lie above it.

When $n=5$, which separates the cases of finite and infinite
polytropes, the \uvp{} takes on a particular topology, illustrated in
Fig.~\ref{ftopgrid}.  The $n=5$ polytrope is a straight line from
$U_s=(3,0)$ to $V_s=(0,6)$.  The point $G_s$ is a centre, with
solutions forming closed loops around it.  The polytrope separates
solutions that circulate around $G_s$ from those that go from
$(\infty,0)$ to $(0,\infty)$ entirely above the polytrope.  These
solutions have zero mass at non-zero inner radius but, unlike the
polytrope, have a finite outer radius.

As $n$ increases further $G_s$ becomes a spiral sink.  Polytropes
start at $U_s$ and spiral into $G_s$ (see Fig.~\ref{ftopgrid}).  There
is an unstable solution that proceeds from $(\infty,0)$ to $V_s$ above
which solutions extend to $(0,\infty)$.  As $n\to\infty$ we also find
$V_s\to(0,\infty)$ and, in the limiting case of the isothermal sphere,
\emph{all} solutions ultimately spiral into $G_s$ because they cannot
lie above the unstable solution.

% The origin $O_s$ is always
% a saddle, with paths approaching along the $V$-axis and escaping along
% the $U$-axis.  Because $n\ge1$ for realistic or interesting models,
% $U_s$ also keeps the same saddle behaviour in our discussion, with
% points approaching along the axis and escaping along the other
% eigenvector, which always points towards the top left of the
% \uvp{}.  For $n<3$, $V_s$ is a source.  One eigenvector is always along
% the $V$-axis and the other across it but the latter varies from
% pointing up in the \uvp{} to pointing down.  When $n=3$ one eigenvalue
% is zero so that it is a point of marginal stability.  Points along the
% corresponding eigenvector are also stationary in the linear regime.
% As $n$ increases past $3$, $V_s$ becomes a saddle, with points now
% approaching from positive $U$.

% Lastly, $G_s$ displays the most complicated behaviour.  It first
% appears in the \uvp{} when $n=3$.  In this case, it co-incides with
% $V_s$.  As $n$ increases, $G_s$ is at first a source.  When
% $\Delta_n=\sqrt{1+n(22-7n)}=0$, the eigenvalues take on an imaginary
% component, so $G_s$ becomes a spiral source.  Increasing in $n$, the
% special case $n=5$ is reached.  The eigenvalues become purely imaginary
% at $G_s$, so the point is a pure centre (see Fig.~\ref{ftop5}).  For
% $n>5$, the real part of $G_s$ is negative and it becomes a spiral
% sink.  As $n\to\infty$, $V_s$ effectively vanishes and all solutions
% ultimately reach $G_s$.

 % The U-V plane

\cleardoublepage
\thispagestyle{empty}
\vspace*{\fill}
\begin{quotation}
  \emph{We have been going around the workshop in the basement of the
    building of science.  The light is dim, and we stumble sometimes.
    About us is confusion and mess which there has not been time to
    sweep away.  The workers and their machines are enveloped in
    murkiness.  But I think that something is being shaped
    here---perhaps something rather big.  I do not quite know what it
    will be when it is completed and polished for the showroom.  But
    we can look at the present designs and the novel tools that are
    being used in its manufacture; we can contemplate too the little
    successes which make us hopeful.}
\begin{flushright}
from \emph{The Expanding Universe},\\
Arthur S.~Eddington, 1936~~
\end{flushright}
\end{quotation}
\vspace*{\fill}
\vspace*{\fill}

% \cleardoublepage
% \pagestyle{empty}
% \vspace*{\fill}
% \begin{center}\includegraphics[width=\textwidth]{temp/wordle1}\end{center}
% \vspace{\fill}
% \mbox{}
% \vspace*{\fill}

\end{document}